\documentclass[a4paper,11pt]{article}
\usepackage[english]{babel}
\usepackage{graphicx,amssymb,amsmath,amsfonts,epsfig}
\usepackage[usenames,dvipsnames]{color}
\usepackage{subfigure}
\title{Distinguishing  Black Hole and Naked Singularity  in MOG via 
Inertial Frame Dragging Effect}
\author{Parthapratim Pradhan\footnote{pppradhan77@gmail.com}\\ 
{\it Department of Physics}\\
{\it Hiralal Mazumdar Memorial College For Women}\\
{Dakshineswar, Kolkata-700035, India}}
\date{}

\begin{document}

\maketitle

\begin{abstract}
We analyze the generalized spin precession of a test gyroscope 
around a stationary spacetime i.e. for Kerr-MOG black hole~(BH) in scalar-tensor-vector 
gravity or modified gravity~(MOG). A detailed study of generalized spin frequency has been 
done for \emph{non} extremal Kerr-MOG BH, \emph{extremal} Kerr-MOG BH and \emph{naked singularity~(NS)} 
in comparison to non-extremal BH, extremal BH and NS  of Kerr spacetime. The generalized spin frequency 
that {we have} computed could be expressed in terms of {the} BH mass parameter, the angular momentum 
parameter and the MOG parameter. Moreover,  we differentiate the non extremal BH, 
extremal BH and NS via computation of the said precession frequency. The Lense-Thirring~(LT) frequency 
{can} obtain from generalized spin frequency by taking the limit as $\Omega=0$ i. e. {when the} angular 
frequency is set to zero limit. Furthermore, we compute the LT frequency for various {values of} 
angular coordinates i.e.  starting from polar to {the} equatorial plane. We show that the LT frequency 
diverges at the horizon for extremal BH. Finally, we study the accretion disk physics by computing 
three epicyclic frequencies namely the Keplerian frequency, {the} radial epicyclic frequency and 
{the} vertical epicyclic frequency. We also compute the periastron frequency and nodal frequency. 
With the aid of these frequency profiles, {one} can distinguish three compact objects i. e. 
\emph{non-extremal BH, extremal BH} {and} \emph{NS}.
\end{abstract}

%\pacs{04.70.-s, 04.70.Dy}
\maketitle

\section{Introduction:}
The de-Sitter precession~\cite{de} and the Lense-Thirring~(LT) precession~\cite{lt} are two 
extraordinary effects  predicted by Einstein's general theory of 
relativity~\footnote{The possible new experimental test of 
Einstein's general theory of relativity was first proposed by Schiff in 1960~\cite{schiff}.}. 
They are directly measured by Gravity Probe-B experiment:  a space experiment to test the 
general theory of relativity~{(GTR)}~\cite{gpb}. It was initiated and launched by NASA in 2004. 
The satellite consists of four gyroscopes and a telescope orbiting 642 km or around 400 mile 
above the Earth~\cite{prl11}. Another important satellite i. e. Rossi X-ray 
Timing Explorer~(RXTE)~\cite{rossi} was launched by NASA in 1995 which help 
us to confirm the existence of the frame-dragging effect {which} was 
predicted by Einstein's gravity. It was also detected the X-rays from different 
compact objects like BHs, neutron stars and X-ray pulsars etc. The effects of 
frame dragging effect on the {galactic-center} stars and kinematic properties 
was investigated in~\cite{saha}. {The measurement of the gyroscopic precession 
and its implications was discussed in \cite{defelice93}.

There are several important tests~\cite{wheeler,will,ciufo} of {GTR} 
which was carried out by different space mission experiment. 
For example, {(a)} the weak equivalence principle: the basis of geometrical theories of 
gravity  which was tested by Lunar laser ranging of accuracy in the order of $10^{-13}$. 
{(b)} The strong equivalence principle: the cornerstone of the {GTR} which was  tested 
by the lunar laser ranging of accuracy  $<10^{3}$. The gravitational time dilation or 
gravitational red-shift which was experimentally verified by Gravity Probe-A. 
The Shapiro time-delay which was also experimentally tested by very long baseline 
interferometry~(VLBI)~\cite{wheeler}.

Moreover, the perihelion advance of mercury which was  tested by the mercury 
radar ranging method and the accuracy of the order of $10^{-3}$. The periastron advance, 
time dilation, time delay and the rate of change of orbital period which was verified 
by binary pulsar PSR 1913+16. The geodetic precession or de-Sitter effect which was 
tested by the lunar laser ranging of accuracy in the order of $6\times 10^{-3}$ and 
using Gravity Probe-B. Finally, the LT effect or the frame-dragging effect was 
experimentally confirmed by using laser geodynamics satellite~(LAGEOS)~\cite{rio}
~\footnote{It was launched by NASA in 1976.}  and 
LAGEOS2~\footnote{This satellite was launched in 1992 by Italian space agency and NASA.}  
of accuracy in the order of $10^{-1}$.

The de-Sitter precession or geodetic effect is the dragging of a gyroscope due to its motion in a static 
gravitational field. Whereas the {LT} precession or frame-dragging effect of a gyroscope is due 
to the rotation of a massive body. In the weak field limit, such LT precession frequency is defined 
as~\cite{jh}
\begin{eqnarray}
\vec {\Omega}_{LT} &=& \frac{G}{c^2r^3} \left[3 \left(\vec{J}.\hat{r}\right) \hat{r}-\vec{J}\right]
\end{eqnarray}
where, $\hat{r}$ is the unit vector along  {agular momentum $\vec{J}$ direction, 
$G$ is Newton's gravitational constant and $c$ is the speed of light.}  Earlier derivations 
of the LT frequency~\cite{jh} in the literature {was} considered on the weak field approximation 
by assuming $r>M$~($M$ is {Arnowitt-Deser-Misner}~(ADM) mass of any compact object) i. e. 
large distances for the test gyroscope. Now the question should be arise naturally what {would} 
be the LT precession frequency in the strong gravity {regime}? This is  one of the prime 
motivation behind this work.

In the strong gravity limit, the LT frequency for Kerr BH and Kerr-Taub-NUT~(Newman-Unti-Tamburino) 
BH {was} explicitly derived in ~\cite{chp}. In this work, the authors showed the role of NUT parameter in 
inertial frame dragging effect. It was shown there that  the LT frequency does not vanish for Taub-NUT 
spacetime when the value of angular momentum vanishes. Similarly for axisymmetric NUT spacetime, the 
result becomes more {prominent} when the value of ADM mass parameter does vanish. The LT frequency 
for more generalized class of spacetimes like Pleba\'{n}ski and Demia\'{n}ski in the strong field 
limit {was} explicitly discussed in~\cite{chppp}.  {The authors in~\cite{chppp1}} studied 
the inertial frame dragging effect of a rotating traversable wormhole.  {In his work, he} 
also described the behavior of a test gyroscope when it moves towards a spinning traversable wormhole. 
Moreover, they  derived the LT frequency for this wormhole and showed that the LT frequency diverges 
on the ergosphere. Along the pole, the LT frequency is inversely proportional to the spin parameter 
of the wormhole.

In~\cite{ckd}, the authors derived the LT frequency  inside a rotating neutron star. Where the authors 
showed that the LT frequency rate  along the pole decreases from the center to the surface of the 
neutron star. Along the equatorial plane the LT frequency rate decreases initially away from the center 
and approaches a small value in the surface.  Morsink and Stella~\cite{ms99} first showed that the precession 
frequency of a rotating neutron star could be expressed in terms of orbital frequency which was 
observed at infinity. They also computed the precession frequencies of circular orbits around 
rapidly spinning neutron stars for a variety of masses and the equation of state. 
One should consult a good review on inertial frame-dragging effect which would 
be found in Ref.~\cite{ciufo}. Also one {could} found the history of the 
Lense-Thirring effect in Ref.~\cite{herbert}. Gyroscope precession along 
equatorial plane of stable orbits for Kerr BH could be found in~\cite{bini16}.

However, the LT precession frequency of various axisymmetric BHs including the analogue 
spacetime and neutron star~\cite{chp,chppp,chppp1,ckd,ch18} {are} computed so far 
but {till date} the LT frequency and the generalized spin frequency {are} not  
considered for Kerr-MOG(KMOG) BH. {In the present manuscript, we wish to derive the 
genralized spin frequency of KMOG BH.} It is a kind of BH solution in the 
scalar-tensor-vector gravity~(STVG) or simply it is known as {MOG}. 
This theory was first proposed by Moffat~\cite{mf,mf1,mf3,mf4,mf5,mf6}. 
The fundamental features of this theory as follows. It correctly explains 
various kind of astronomical observations: dynamics of galaxies and cluster 
of galaxies, bullet clusters, galaxy rotation curves, the amount of luminous 
matter, the exotic dark matter and the acceleration of the universe etc. 

The MOG gravity is a {type}  of  alternative theory of gravity. It contains scalar 
field and massive vector field. The action in STVG theory consists of scalar action 
and  vector action hence it modifies the Einstein-Hilbert action. The MOG theory is 
formulated via MOND phenomenology in the weak field approximation. Most importantly, 
this theory correctly explains the observations of the solar system~\cite{mfjcap}. 
It could also be used to describe the growth of the structure of the universe, 
the power spectrum of the matter and the acoustical power spectrum of the cosmic 
microwave background~(CMB) data. It should be noted that various features of MOG 
like superradiance, quasinormal modes, thermodynamics, geodesics properties, orbital 
and vertical epicyclic frequencies, Penrose process, gravitational bending of light, 
BH shadow  and BH merger estimates etc. {are} studied 
in~\cite{mf,mfjcap, mf1,mf3,mf4,mf5,mf6,pp18,liu18,ps,epjc19} {respectively}. 

{The fundamental postulate in MOG theory is that the BH charge 
parameter is proportional to the Komar mass i. e. 
${Q}=\sqrt{\alpha G_{N}}M$~\cite{mf5}. Where  $\alpha=\frac{G-G_{N}}{G_{N}}$ 
should be measured deviation of MOG from GR. Using this criterion 
the MOG action is given by 
\begin{eqnarray}
{\cal I}_{MOG} &=& \frac{1}{16\pi G}\int \sqrt{-g}\, R\, d^4x-
\frac{1}{16\pi}\int \sqrt{-g}\, B_{ab}B^{ab}\, d^4x+{\cal I}_{M}
\end{eqnarray}
where $R$ is the Ricci scalar, $B_{ab}$ is the generalized field strength tensor of the vector 
field $\phi^{a}$, $B_{ab}=\partial_{a}\phi_{b}-\partial_{b}\phi_{a}$ and ${\cal I}_{M}$ is the 
matter action. 

The MOG field equations for vanishing matter electro-magnetic tensor ${\cal T}_{Mab}=0$ are 
\begin{eqnarray}
R_{ab} &=& -8\pi G\, {\cal T}_{\phi ab}  ~. \label{mgl}
\end{eqnarray}
where 
\begin{eqnarray}
{\cal T}_{\phi ab} &=& -\frac{1}{4\pi} \left(B_{a}^{c} B_{bc}-
\frac{1}{4} g_{ab} B^{ef}B_{ef} \right)  ~. \label{mgl1}
\end{eqnarray}
Thus vacuum field equations are
\begin{eqnarray}
\frac{1}{\sqrt{-g}} \partial_{a}\left(\sqrt{-g}B^{ab}\right) &=& 0~, \label{mgl2}
\end{eqnarray}
and
\begin{eqnarray}
\nabla_{a}B_{bc}+\nabla_{b}B_{ca}+\nabla_{c}B_{ab} &=& 0~. \label{mgl3}
\end{eqnarray}
where $\nabla_{a}$ denotes covariant derivative with respect to the metric tensor $g_{ab}$. Using these 
criterion one obtains static spherical symmetric metric in MOG 
\begin{eqnarray}
ds^2=\left[1-\frac{2G_{N}(1+\alpha)M}{r}+\frac{G_{N}^2M^2\alpha(1+\alpha)}{r^2}\right]dt^2-
\frac{dr^2}{\left[1-\frac{2G_{N}(1+\alpha)M}{r}+\frac{G_{N}^2M^2\alpha(1+\alpha)}{r^2}\right]}
- r^{2}d\Omega^2 ~. \label{mgl4}
\end{eqnarray}
where $G_{N}$ is modified Newtonian constant which is related to the Newton's constan by
$G=G_{N} (1+\alpha)$ and modified charge parameter is ${\cal Q}=\sqrt{\alpha G_{N}}M$, 
where $\alpha$ is a MOG parameter.  The above metric can be obtained by substituting 
these values in usual Reissner-Nordstr\"{o}m BH solution.  The metric of
the rotating BH solution in MOG depends on the spin parameter  $a=J/M$, where $J$ is the Komar angular 
momentum of the asymptotically flat, axisymmetric, stationary spacetime. By applying Boyer-Lindquist 
coordinates one obtains the metric form which is written in Eq.~(\ref{mg2.1}) and it is quite similar 
structure of Kerr-Newman BH.}

{One of the main goals of this work is to explore  the difference between 
BH and NS via computation of generalized spin frequency in MOG.} 
For NS, the motivation comes from the works
~\cite{ch17,ch17a,pugliese13,defelice78,calvini78,zhang16,pug19,pug20}. How to differentiate 
a BH from a NS, this is {a}  prime aim of the present work. What is a NS? A NS 
is a {type} of gravitational singularity  without an event horizon. Whereas a BH is a {type} of 
gravitational compact object having an event horizon. Although the thermodynamic properties of 
BH is an established subject but the thermodynamics of NS is completely unknown~\cite{zhang16}.
Recently, the image of a BH has been observed in EHT telescope~\cite{eht} {while} for NS  
there is no evidence {could be seen till date}. 

However in the present work,  we  would like to provide a detailed analysis of inertial 
frame dragging effect for \emph{non-extremal situation, extremal situation and NS} in 
KMOG BH. {Earlier mentioned that the {KMOG} BH is a new class of spinning BH 
proposed by Moffat~\cite{mf} and it is constructed by the ADM mass parameter~(${\cal M}$), 
spin parameter~($a$) and a deformation parameter or MOG parameter~($\alpha$) in comparison 
to Kerr BH which is defined only by the ADM mass parameter and the spin parameter.

{First, we derive the generalized spin precession of a test gyroscope around the KMOG 
spacetime. Using this frequency, we differentiate the behavior of three compact objects:
non-extremal BH, extremal BH amd NS. We point out a clear distinction between these three 
compact objects graphically.  Moreover, this result is compared with Kerr BH. Using 
generalized spin frequency versus radial diagram, one can distinguish three compact 
objects for various spin limits. Also, we investigate the generalized spin frequency 
for various angular coordinte values i. e. $\theta=0$, $\theta=\frac{\pi}{6}$, 
$\theta=\frac{\pi}{4}$, $\theta=\frac{\pi}{3}$ and  $\theta=\frac{\pi}{2}$. 
Furthermore, we examine  the generalized spin frequency for ring singularity. 
Lastly, we compare these results with the result of vanishing angular angular 
velocity $\Omega=0$. This is exactly the LT frequency when the other frequencies are excluded.  
From the LT frequency vs. radial diagram, we show  that  the LT frequency is  
influenced by the MOG parameter . When  $\Omega=0$, we also analyze the LT 
frequency for angular values i.e.  $\theta=0$, $\theta=\frac{\pi}{6}$, $\theta=\frac{\pi}{4}$, 
$\theta=\frac{\pi}{3}$ and $\theta=\frac{\pi}{2}$. Each diagram clearly exhibits 
the key difference between three compact objects.}

{Moreover, we compute three fundamental frequencies i.e.  the Keplerian frequency~($\Omega_{\phi}$), 
the radial epicyclic frequency~($\Omega_{r}$) and the vertical epicyclic frequency~($\Omega_{\theta}$).
Using these frequencies we can find the difference between three compact objects. In ~\cite{maselli17}, 
it was mentioned that the epicyclic frequencies are the main  ingredients for the geodesic models of 
quasi-periodic-oscillations~(QPO). Again these QPOs help us to testify the strong gravity in a novel 
way. The geodesic models were described by relativistic precession model~(RPM)~\cite{stella99} and 
epicyclic resonance model~(ERM)~\cite{torok05}. These models indicates that there exist 
low frequency~(LF) QPO and twin high frequency~(HF) QPO. From RPM, we can find that the 
upper and lower HF QPOs meets with the azimuthal frequency, $\Omega_{per}=\Omega_{\phi}-\Omega_{r}$. 
While the LF QPOs are computed by the nodal precession  frequency, $\Omega_{nod}=\Omega_{\phi}-\Omega_{\theta}$. 
These three QPO frequencies $(\Omega_{\phi}, \Omega_{per}, \Omega_{nod})$ could generate 
at the same orbital radius.  Moreover these frequencies could serve as a tool in our investigation 
to study the crucial differences between three compact objects. }

{Furthermore, our technique suggests that a comparative study of stationary, axisymmetric 
KMOG spacetime for various values of spin parameter. The main findings of the present work 
is to highlight crucial differences between \emph{non-extremal} KMOG BH, \emph{extremal} KMOG BH 
and \emph{NS} by using generalized spin precession frquency~($\Omega_{p}$) and LT precession 
frequency~($\Omega_{LT}$) of a test gyro. Also other frequencies like  radial epicyclic 
frequency~($\Omega_{r}$), vertical epicyclic frequency~($\Omega_{\theta}$),  
Keplerian frequency~($\Omega_{\phi}$), periastron precession frequency~($\Omega_{per}$), 
nodal precession frequency~($\Omega_{nod}$), the ratio  $\frac{\Omega_{r}}{\Omega_{\phi}}$, 
the ratio~$\frac{\Omega_{r}}{\Omega_{\theta}}$ and the ratio~$\frac{\Omega_{\theta}}{\Omega_{\phi}}$ 
are also help us to support this result in strong gravity regime.}

{The paper is organized as follows: in  sec.~2, 
we derive the generalized spin frequency in the background of KMOG spacetime for non-extremal, 
extremal and NS cases. A detailed study has been done and visualize the results for all cases. 
Using the spin frequency expression, we differentiate the non extremal BH, 
extremal BH and NS. Moreover in Sec.~3,  we compute the LT frequency  by using the result of 
generalized spin frequency where we have  taken the {value} of angular frequency $\Omega=0$. Also
in the subsequent sub Sec., we have specialized 
the result for extremal spacetime. Furthermore, we study the accretion 
disk physics in Sec.~4,  by deriving the epicyclic frequencies. 
We also derive the periastron frequency and nodal 
frequency. By using the frequency profile, we can distinguish three spacetimes namely the non-extremal 
spacetime, extremal spacetime and NS. In Sec.~5, we have discussed the result.

\section{Generalized Spin precession of a test gyroscope around a stationary spacetime}
In this section we shall provide the basic formalism~(following the Ref.~\cite{ns,chppp}) of spin 
precession of a test gyroscope which is attached to a stationary observer and moves along a Killing 
path in a stationary spacetime with a timelike Killing vector field $K$. Then the spin of such a 
test gyroscope undergoes a Fermi-Walker transport {along }
\begin{equation}
u=\frac{K}{\sqrt{-K^2}},
\end{equation}
{where $K$ is the timelike Killing vector field. For a stationary observers, the four-velocity
can be defined as $u^{\alpha}=(u^{t}, 0, 0, \Omega\,u^{t})$. Where $t$ is time coordinate and 
$\Omega$ is the angular velocity of the observer.} 
It is well-known that in the special case the frequency of the gyroscope may coincide with the 
vorticity field associated with the Killing congruence. This indicates that the said gyroscope is
rotating with respect to a corotating frame along with an angular velocity. This effect is said to 
be a gravitomagnetic precession in gravitational physics because the vorticity vector behaves as the  
magnetic field in the $3+1$ dimension spacetime~\cite{jantzen}. Since our moto is to derive the spin 
precession frequency of a test gyroscope in a strong gravity regime hence the general spin precession 
frequency of a test gyro $\Omega_{s}$ which is actually the rescaled vorticity field of the stationary 
observer can be derived as 
\begin{eqnarray}
\tilde \Omega_{s} &=& \frac{1}{2K^2}*(\tilde K \wedge d\tilde K)\label{wf1}
\\
\mbox{or} &&  \nonumber
\\
\left(\Omega_{s}\right)_{a} &=&\frac{1}{2K^2} \eta_{a}^{~bcd} K_{b}\partial_{c} K_{d}~, ~\label{wf2}
\end{eqnarray}
where $\eta^{abcd}$ {denotes} the component of the volume-form in the spacetime, 
{$*$ denotes the Hodge dual},
and $\tilde K$ \& $\tilde \Omega_{s}$ represent the one-form of $K$ \& $\Omega_{s}$, respectively. The 
parameter $\tilde \Omega_{s}$ will be vanish if and only if $(\tilde K \wedge d\tilde K)$ does vanish.  
This will be happen only in case of a static spacetime. 

It has already been derived in~\cite{ckd,ccb} the LT  precession frequency of a test 
gyro due to the rotation of any stationary and axisymmetric spacetime  
\begin{equation}
\vec{\Omega}_p|_{\Omega=0}=\frac{1}{2\sqrt {-g}} \times
\left[-\sqrt{g_{rr}}\left(g_{0\phi,\theta}
-\frac{g_{0\phi}}{g_{00}} g_{00,\theta}\right)\hat{r}
+\sqrt{g_{\theta\theta}}\left(g_{0\phi,r}-\frac{g_{0\phi}}{g_{00}}
g_{00,r}\right)\hat{\theta}\right]~\label{wf3}
\end{equation}
This result is valid for outside the ergoregion and $\Omega=0$. For arbitrary value 
of $\Omega$ the formalism is derived in Ref.~\cite{ch17}. 
The timelike Killing vector field in generalized spacetime can be written as 
\begin{eqnarray}
K &=& \partial_{0}+\Omega\, \partial_{s}~\label{K1}
\end{eqnarray}
where $\partial_{s}$ is spacelike Killing vector of the said stationary spacetime 
and $\Omega$ is  angular velocity of an observer moving along integral curves of $K$.
The metric of above stationary  spacetime is independent of $x^{0}$ and $x^{s}$ coordinates. 
Thus the corresponding co-vector of $K$ is
\begin{eqnarray}
\tilde{K} &=& g_{0\alpha}~dx^{\alpha}+\Omega~g_{\beta s}\,dx^{\beta},
\end{eqnarray}
where $\beta, \alpha=0,s,2,3$ in {four dimension} spacetime. Now one can 
write $\tilde{K}$ by separating into space and time components as 
\begin{eqnarray}
\tilde{K}=(g_{00}\,dx^0+g_{0s}\,dx^s+g_{0i}\,dx^i)+\Omega\,\left(g_{0s}\,dx^0+g_{ss}\,dx^s+g_{is}\,dx^i\right)~\label{kt}
\end{eqnarray}
where $i=2, 3$. Since we are interested in ergoregion of the said stationary, axisymmetric spacetime 
thus we are neglecting the terms $g_{0i}$ and $g_{is}$ then one obtains
\begin{eqnarray}
\tilde{K}=(g_{00}\,dx^0+g_{0s}\,dx^s)+\Omega\,\left(g_{0s}\,dx^0+g_{ss}\,dx^s\right)~\label{kt1}
\end{eqnarray}
and 
\begin{eqnarray}
d\tilde{K}=\left(g_{00,k}\,dx^k \wedge dx^{0}+g_{0s,k}\,dx^k \wedge dx^s \right)+
\Omega (g_{0s,k}\,dx^k \wedge dx^0+g_{ss,k}\,dx^k \wedge dx^s).~\label{kto}
\end{eqnarray}
Now, Eq.~(\ref{wf2}) can {be} re-writeen as 
\begin{eqnarray}
\tilde{\Omega}_{p}=\frac{1}{2K^2}*(\tilde{K} \wedge d\tilde{K})~\label{be}
\end{eqnarray}
Substituting the values of $\tilde{K}$ and $d\tilde{K}$ in Eq.~(\ref{be}), 
one finds the one-form of the precession frequency 
$$
\tilde{\Omega}_{p} = \frac{\varepsilon_{skl}\,g_{l\mu}\,dx^{\mu}}{2\sqrt {-g}\left(1+2\Omega\,\frac{g_{0s}}{g_{00}}
+\Omega^2\,\frac{g_{ss}}{g_{00}}\right)} \times
$$
\begin{eqnarray}
\left[\left(g_{0s,k}-\frac{g_{0s}}{g_{00}}\, g_{00,k}\right)+\Omega\,\left(g_{ss,k}-\frac{g_{ss}}{g_{00}}\,g_{00,k}\right)
+\Omega^2\,\left(\frac{g_{0s}}{g_{00}}\,g_{ss,k}-\frac{g_{ss}}{g_{00}}\, g_{0s,k}\right)\right] \nonumber
\end{eqnarray}
where we have used $*\left(dx^0 \wedge dx^k \wedge dx^s \right)=\eta^{0ksl}\,g_{l\mu}\,dx^{\mu}
=-\frac{1}{\sqrt{-g}}\varepsilon_{ksl}\,g_{l\mu}dx^{\mu}$ and $K^2=g_{00}+2\Omega\, g_{0s}+\Omega^2\, g_{ss}$. 
The corresponding vector ($\Omega_p$) of the co-vector $\tilde{\Omega}_p$ is
$$
\Omega_{p} = \frac{\varepsilon_{skl}}{2\sqrt {-g}\left(1+2\Omega\,\frac{g_{0s}}{g_{00}}
+\Omega^2\,\frac{g_{ss}}{g_{00}}\right)}
\times
$$
\begin{eqnarray}
\left[\left(g_{0s,k}
-\frac{g_{0s}}{g_{00}} g_{00,k}\right)+\Omega\left(g_{ss,k}
-\frac{g_{ss}}{g_{00}} g_{00,k}\right)+ \Omega^2 \left(\frac{g_{0s}}{g_{00}}g_{ss,k}
-\frac{g_{ss}}{g_{00}} g_{0s,k}\right) \right] \nonumber\\
~\label{elt} 
\end{eqnarray}
For a stationary and axisymmetric spacetime with coordinates $t,r,\theta,\phi$, the above 
equation becomes
\begin{eqnarray}
\vec{\Omega}_{p} &=&  \frac{-\sqrt{g_{rr}}~{\cal X}(r) \hat{r}+\sqrt{g_{\theta\theta}}~{\cal Y}(r) \hat{\theta}}
{2 \sqrt{-g}~{\cal Z}(r)}~\label{n1}
\end{eqnarray}
where 
\begin{eqnarray}
{\cal X}(r)  &=& {\cal A} +{\cal B} \Omega+{\cal C} \Omega^2\\
{\cal Y}(r)  &=& {\cal F}+{\cal G} \Omega+{\cal H} \Omega^2 \\
{\cal Z}(r)  &=& g_{tt}+2 g_{t\phi} \Omega+g_{\phi\phi}\Omega^2
\end{eqnarray}
and
\begin{eqnarray}
{\cal A} &=& g_{tt}~g_{t\phi,\theta}-g_{t\phi}~g_{tt,\theta}\\
{\cal B}  &=& g_{tt}~g_{\phi\phi,\theta}-g_{\phi\phi}~ g_{tt,\theta}\\
{\cal C} &=& g_{t\phi}~ g_{\phi\phi,\theta}-g_{\phi\phi}~g_{t\phi,\theta}\\
{\cal F} &=& g_{tt}~g_{t\phi,r}-g_{t\phi}~g_{tt,r}\\
{\cal G}   &=& g_{tt}~g_{\phi\phi,r}-g_{\phi\phi}~g_{tt,r}\\
{\cal H} &=& g_{t\phi}~g_{\phi\phi,r}-g_{\phi\phi}~g_{t\phi,r}
\end{eqnarray}
The most striking feature of  Eq.~(\ref{n1}) is that it could be applicable to derive the generalized  
spin precession for any stationary axisymmetric BH spacetime which is valid for both outside and inside the 
ergosphere. In the limit $\Omega=0$, one obtains 
\begin{eqnarray}
\vec{\Omega}_{LT} &=&  \frac{-\sqrt{g_{rr}}~{\cal A} \hat{r}+\sqrt{g_{\theta\theta}}~{\cal F} \hat{\theta}}
{2 \sqrt{-g}~g_{tt}}~~\label{n2}
\end{eqnarray}
This formula is applicable only outside the ergoregion. This is the exact LT  precession frequency of a test 
gyro due to rotation of any stationary and axisymmetric spacetime as mentioned in Eq.~(\ref{wf3}). It must be 
noted that $\vec{\Omega_{p}}$ is not the $\vec{\Omega}_{LT}$ of test gyro. Actually $\vec{\Omega_{p}}$ 
describes the overall frequency of test gyro since gyro has non-zero angular velocity while  
$\vec{\Omega}_{LT}$ describes LT frequency of test gyro when the test gyro has zero angular velocity i.e. 
$\Omega=0$.

\subsection{Application to KMOG spacetime}
In this subsection we will present a detailed analysis of the frame-dragging effect for  
KMOG spacetime according to the above formalism. To do this we have to write the  metric 
explicitly for KMOG BH as described in Ref.~\cite{mf}
\begin{eqnarray}
ds^2 = -\frac{\Delta}{\rho^2} \, \left[dt-a\sin^2\theta d\phi \right]^2+\frac{\sin^2\theta}{\rho^2} \,
\left[(r^2+a^2) \,d\phi-a dt\right]^2
+\rho^2 \, \left[\frac{dr^2}{\Delta}+d\theta^2\right] \nonumber\\
~.\label{mg2.1}
\end{eqnarray}
where
\begin{eqnarray}
\rho^2 & \equiv & r^2+a^2\cos^2~\theta \nonumber\\
\Delta & \equiv & r^2-2G_{N}(1+\alpha)Mr+a^2 + G_{N}^2 \alpha(1+\alpha) M^2 ~.\label{m2.1}
\end{eqnarray}
where $G_{N}$ is Newton's gravitational constant and $M$ is  Komar mass. 
We should mentioned that in the metric $c=1$. The above metric is an 
axially-symmetric and stationary spacetime.  The ADM mass and angular momentum are computed 
in~\cite{ps}  as ${\cal M}=(1+\alpha)M$ and $J=a{\cal M}$ ~\footnote{One could determine 
the relation between the Komar mass and ADM mass as $M=\frac{\cal M}{1+\alpha}$. If one could 
consider either the Komar mass or the ADM mass in the LT frequency computation 
then the physics will not be change. We consider here the ADM mass througout 
the paper for our convenience.}. 
Substituting these values in Eq.~(\ref{m2.1}) then $\Delta$ becomes
\begin{eqnarray}
\Delta & = & r^2-2G_{N}{\cal M}r+a^2 + \frac{\alpha}{1+\alpha} G_{N}^2 {\cal M}^2 
\end{eqnarray}
The above metric describes a BH with horizon radii
\begin{eqnarray}
r_{\pm} &=& G_{N}{\cal M} \pm \sqrt{\frac{G_{N}^2{\cal M}^2}{1+\alpha}-a^2} ~. \label{mg2.2}
\end{eqnarray}
where $r_{+}$ is called as event horizon and $r_{-}$ is called as Cauchy horizon. Note that $r_{+}>r_{-}$. 
{The horizon structure could be seen visually in ~Fig.~\ref{hr}. 
\begin{figure}
\begin{center}
{\includegraphics[width=0.45\textwidth]{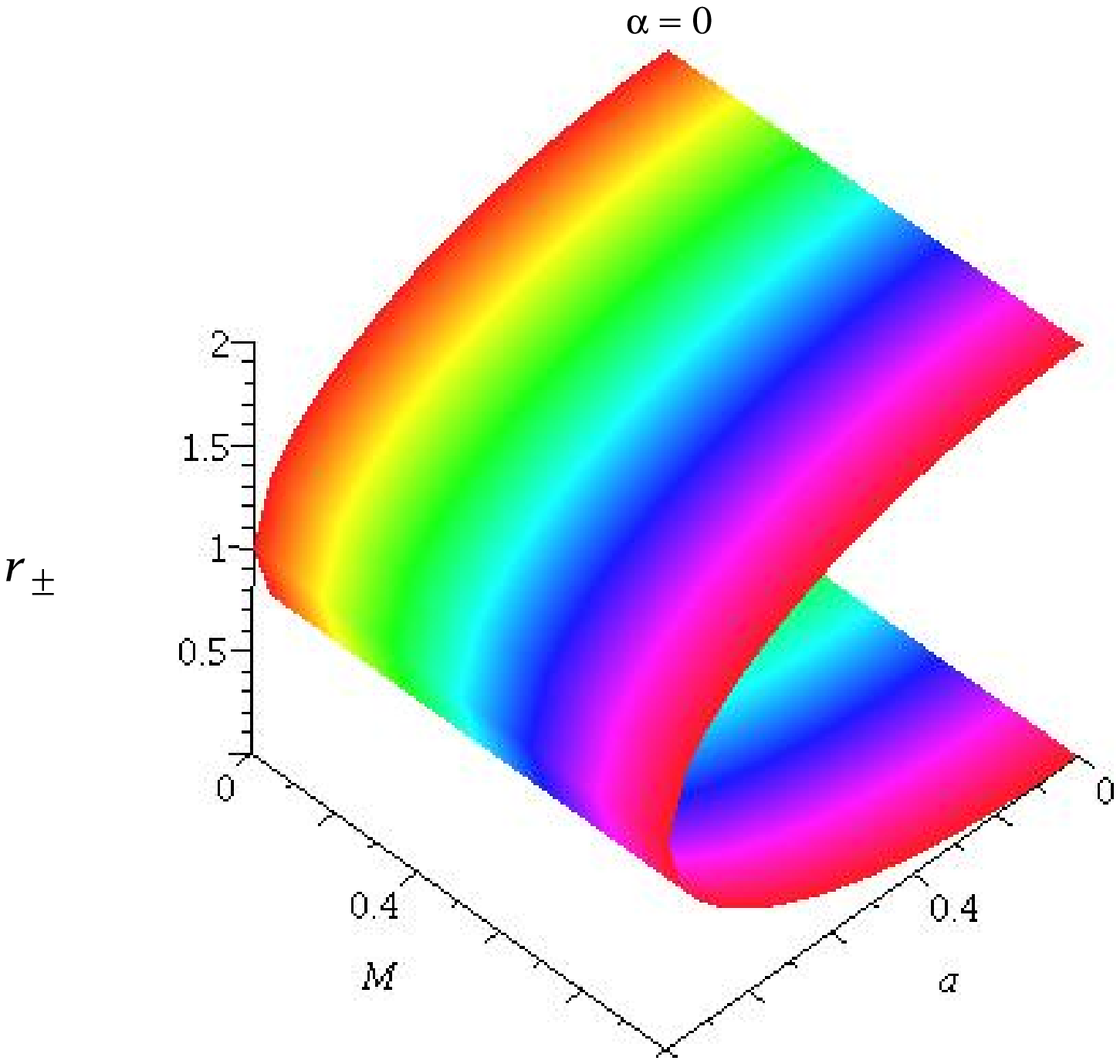}}
{\includegraphics[width=0.45\textwidth]{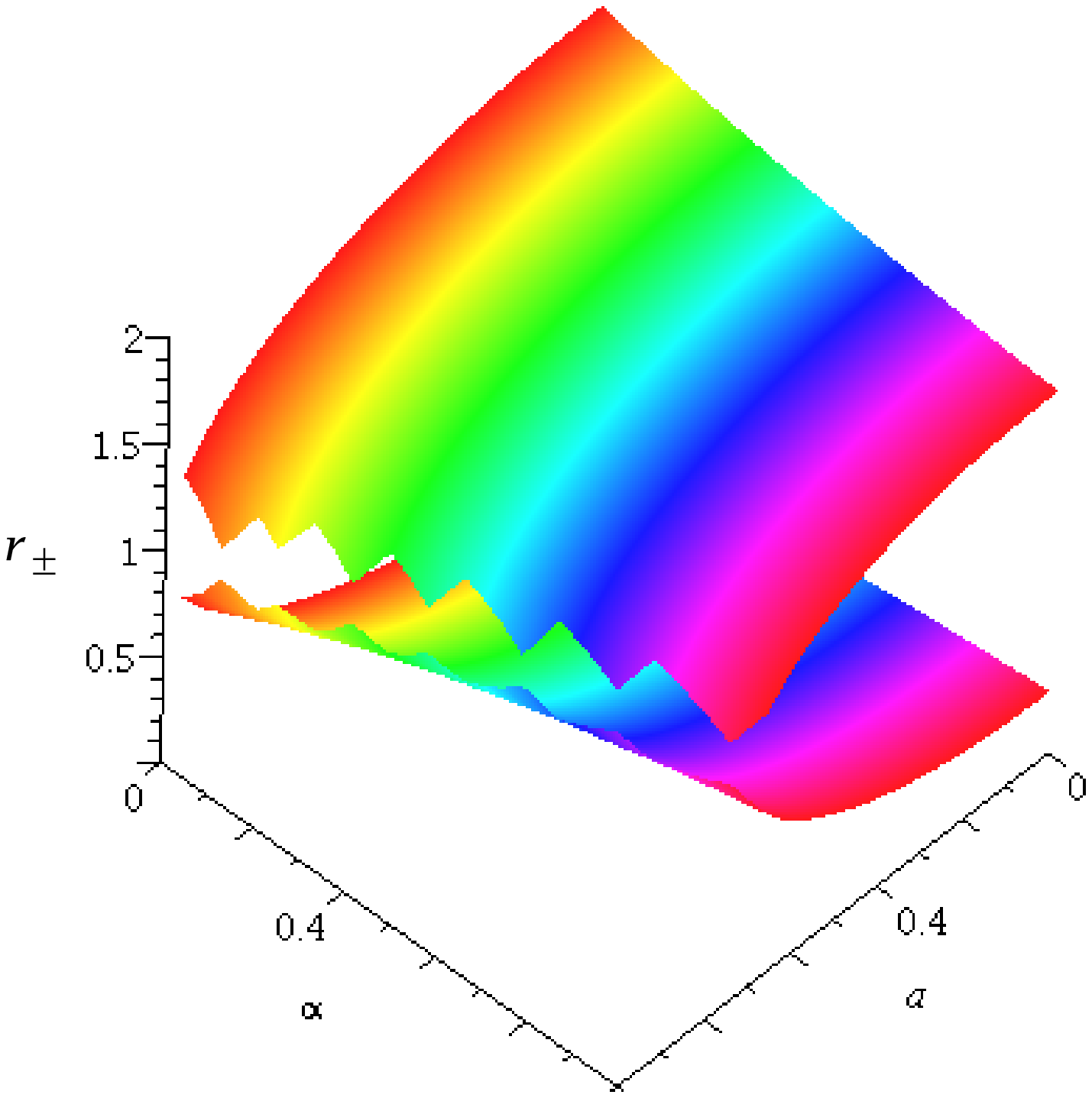}}
\end{center}
\caption{The figure implies the variation  of $r_{\pm}$  with $a$ and $\alpha$ for Kerr BH
and Kerr-MOG BH. Left figure for Kerr BH. Right figure for Kerr-MOG BH. 
The presence of the MOG parameter is deformed the shape of the horizon radii. 
\label{hr}}
\end{figure}
It should be noted that when $\alpha=0$, one gets the horizon radii of Kerr BH. The BH solution exists 
when $\frac{G_{N}^2{\cal M}^2}{1+\alpha} > a^2$. When $\frac{G_{N}^2{\cal M}^2}{1+\alpha}=a^2$, one obtains 
extremal BH. When $\frac{G_{N}^2{\cal M}^2}{1+\alpha} < a^2$, one gets the naked singularity case. 

{For the sake of convenience} and {to show the variation with deformation parameter} 
we can define spin parameter for MOG BH as
\begin{eqnarray}
a_{\alpha}=\frac{a}{G_{N}{\cal M}}=\frac{1}{\sqrt{1+\alpha}} 
\end{eqnarray}
The {new} spin parameter $a_{\alpha}$ varies with MOG parameter {that} could be seen from  
Fig.~(\ref{spn}). From the plot we can infer that there is a restriction on spin 
parameter in MOG as
\begin{eqnarray}
|a_{\alpha}| &<& \frac{1}{\sqrt{1+\alpha}}~~~~~~{\mbox{Non-extremal BH}}\\
|a_{\alpha}| &=& \frac{1}{\sqrt{1+\alpha}}~~~~~~{\mbox{Extremal BH}}\\
|a_{\alpha}| &>& \frac{1}{\sqrt{1+\alpha}}~~~~~~{\mbox{Naked Singularity}}
\end{eqnarray}
Note that the MOG parameter or deformation parameter~($\alpha$) is always positive definite. If we 
invert the above inequality then one gets the restriction on $\alpha$. 

For a Kerr BH, when $a_{\alpha}=\frac{a}{G_{N}{\cal M}}<1$, there exists a BH solution. While 
$a_{\alpha}=\frac{a}{G_{N}{\cal M}}>1$, it is said to be a NS 
and $a_{\alpha}=\frac{a}{G_{N}{\cal M}}=1$ then it is said to be {an} extremal BH solution.
While for a KMOG BH, these limits {can} be reduced in the following way. For example, if we 
take $\alpha=1$ then these limits are defined as $a_{\alpha}=\frac{1}{\sqrt{2}}=0.7$ {which} 
is the extremal limit, $a_{\alpha}>\frac{1}{\sqrt{2}}$ is the 
NS situation and $a_{\alpha}<\frac{1}{\sqrt{2}}$ is the BH solution. Similarly, if we take the MOG 
parameter $\alpha=2$, then these limits are: $a_{\alpha}=\frac{1}{\sqrt{3}}$ {which } is the extremal limit, 
$a_{\alpha}>\frac{1}{\sqrt{3}}$ is the NS situation and $a_{\alpha}<\frac{1}{\sqrt{3}}$ is the BH 
solution.
\begin{figure}
\begin{center}
{\includegraphics[width=0.45\textwidth]{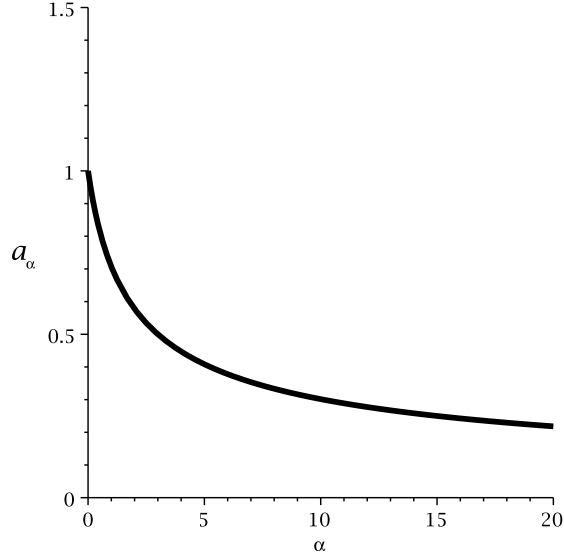}}
\end{center}
\caption{The figure indicates  the variation  of spin parameter $a_{\alpha}$ with $\alpha$. Where 
$a_{\alpha}=\frac{a}{G_{N}{\cal M}}=\frac{1}{\sqrt{1+\alpha}}$. This means that the value of spin 
parameter decreases when MOG parameter increases. Maximum value of MOG parameter is unity. Its value 
gradually decreases when $\alpha$ increases.
\label{spn}}
\end{figure}
The {outer and inner} ergosphere are occur at 
\begin{eqnarray}
r &=& r_{e}^{\pm}(\theta) = G_{N}{\cal M} \pm \sqrt{\frac{G_{N}^2{\cal M}^2}{1+\alpha}-a^2 \cos^2~\theta} 
~. \label{mg2.3}
\end{eqnarray}
{and they satisfied the following inequality $r_{e}^{-}(\theta)\leq r_{-}\leq r_{+}\leq r_{e}^{+}(\theta)$. 
The structure of the outer and inner ergosphere could be seen visually  from Fig.~\ref{erg}.
\begin{figure}
\begin{center}
{\includegraphics[width=0.45\textwidth]{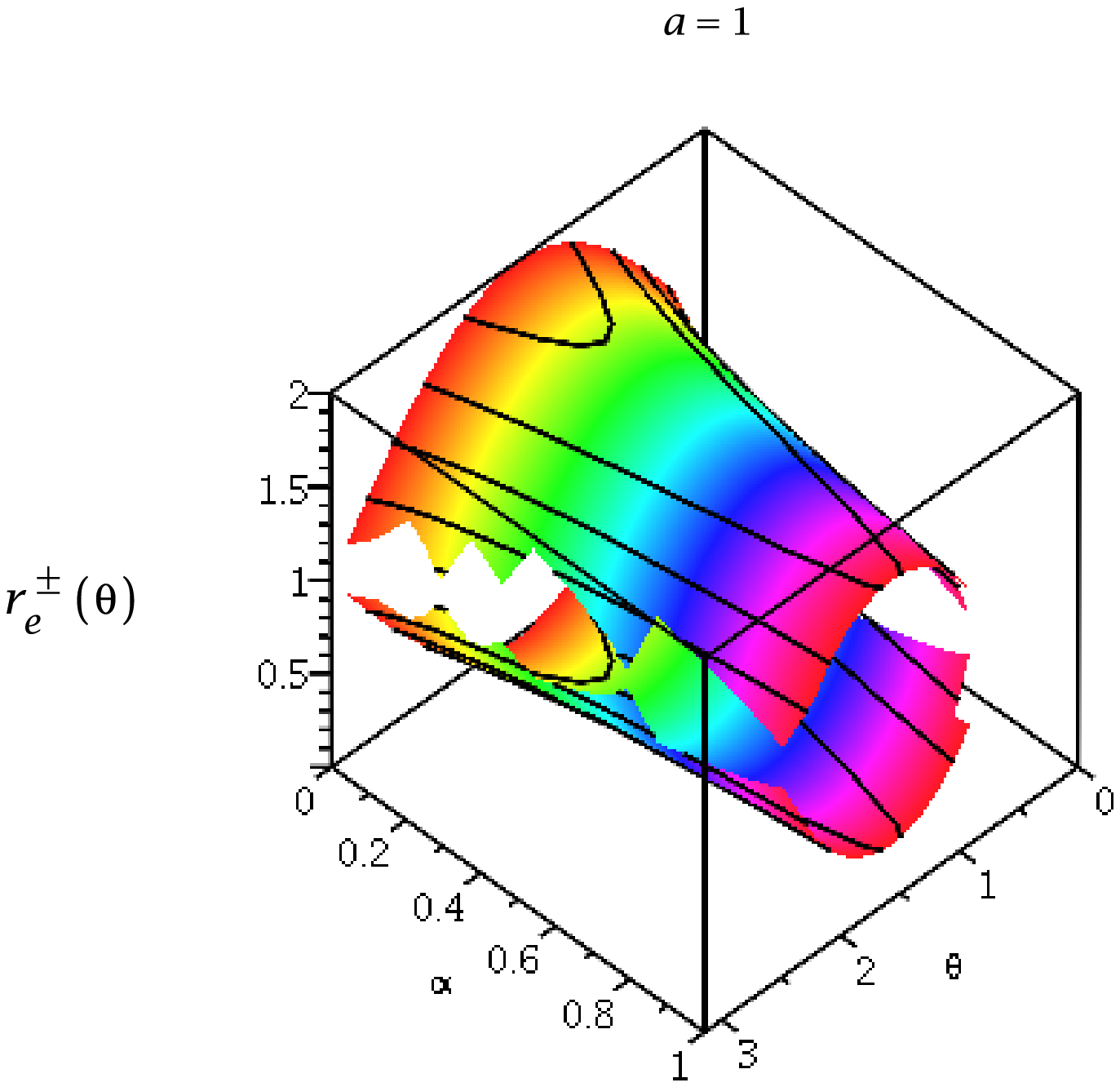}}
{\includegraphics[width=0.45\textwidth]{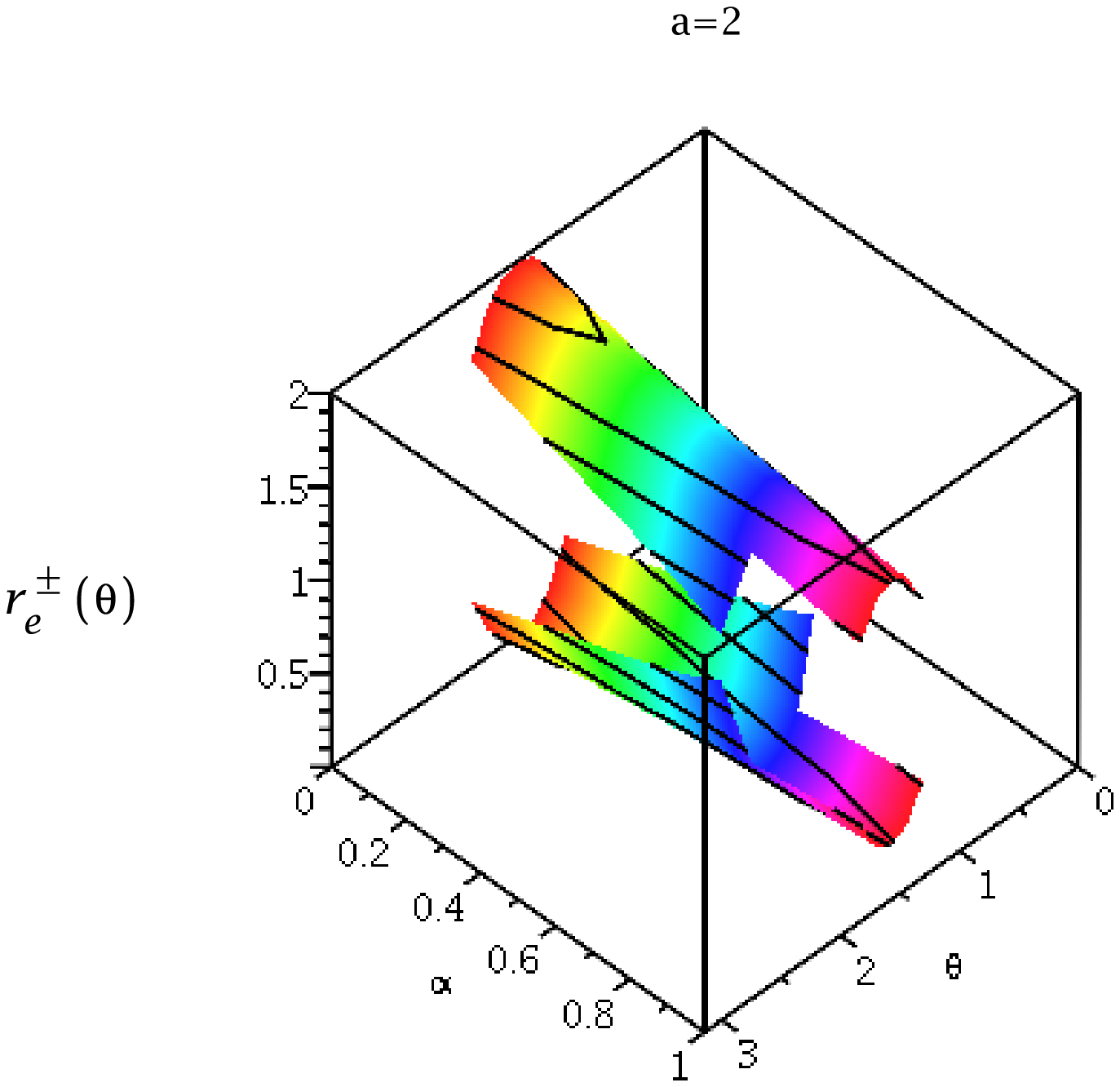}}
{\includegraphics[width=0.45\textwidth]{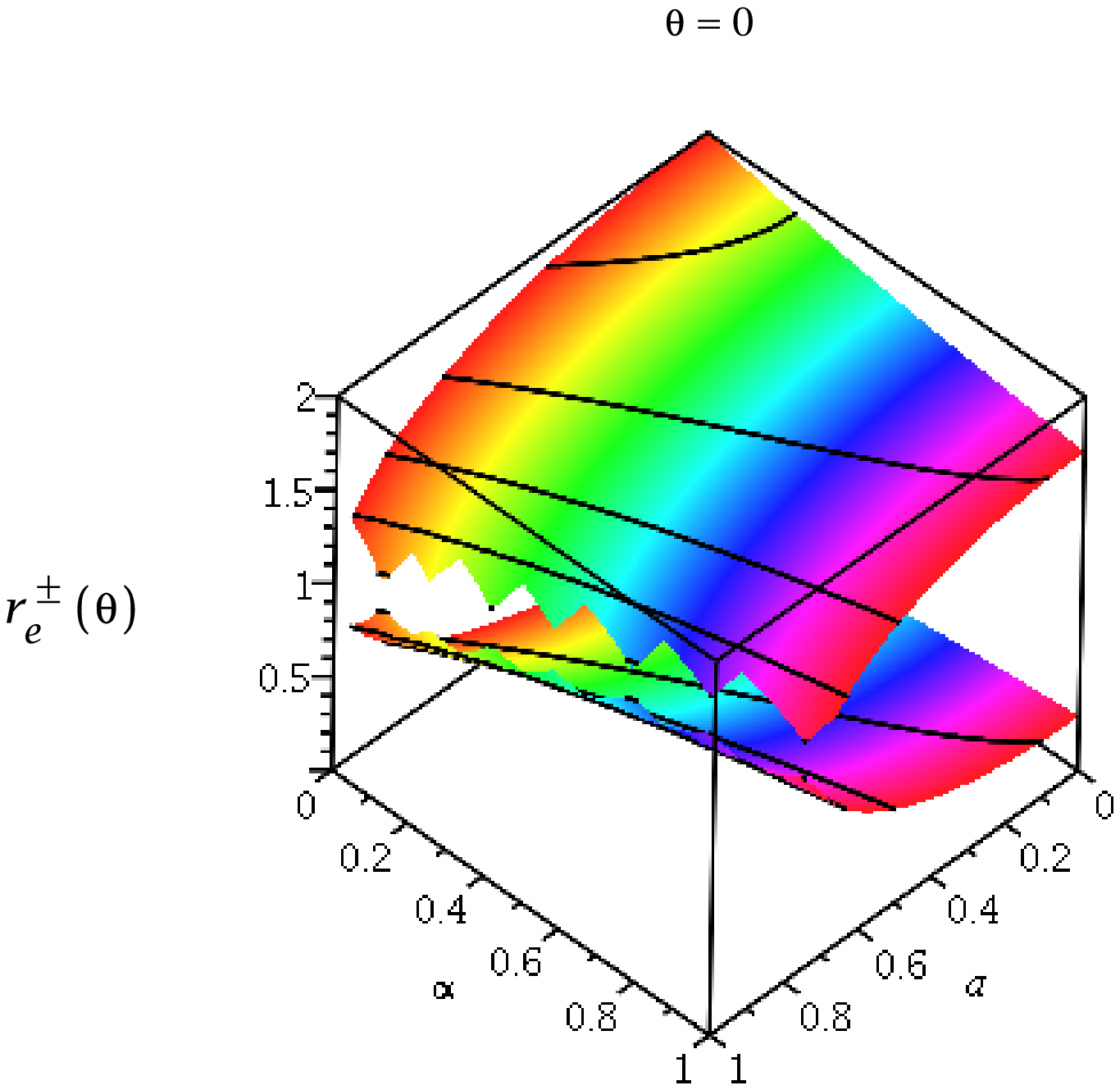}}
{\includegraphics[width=0.45\textwidth]{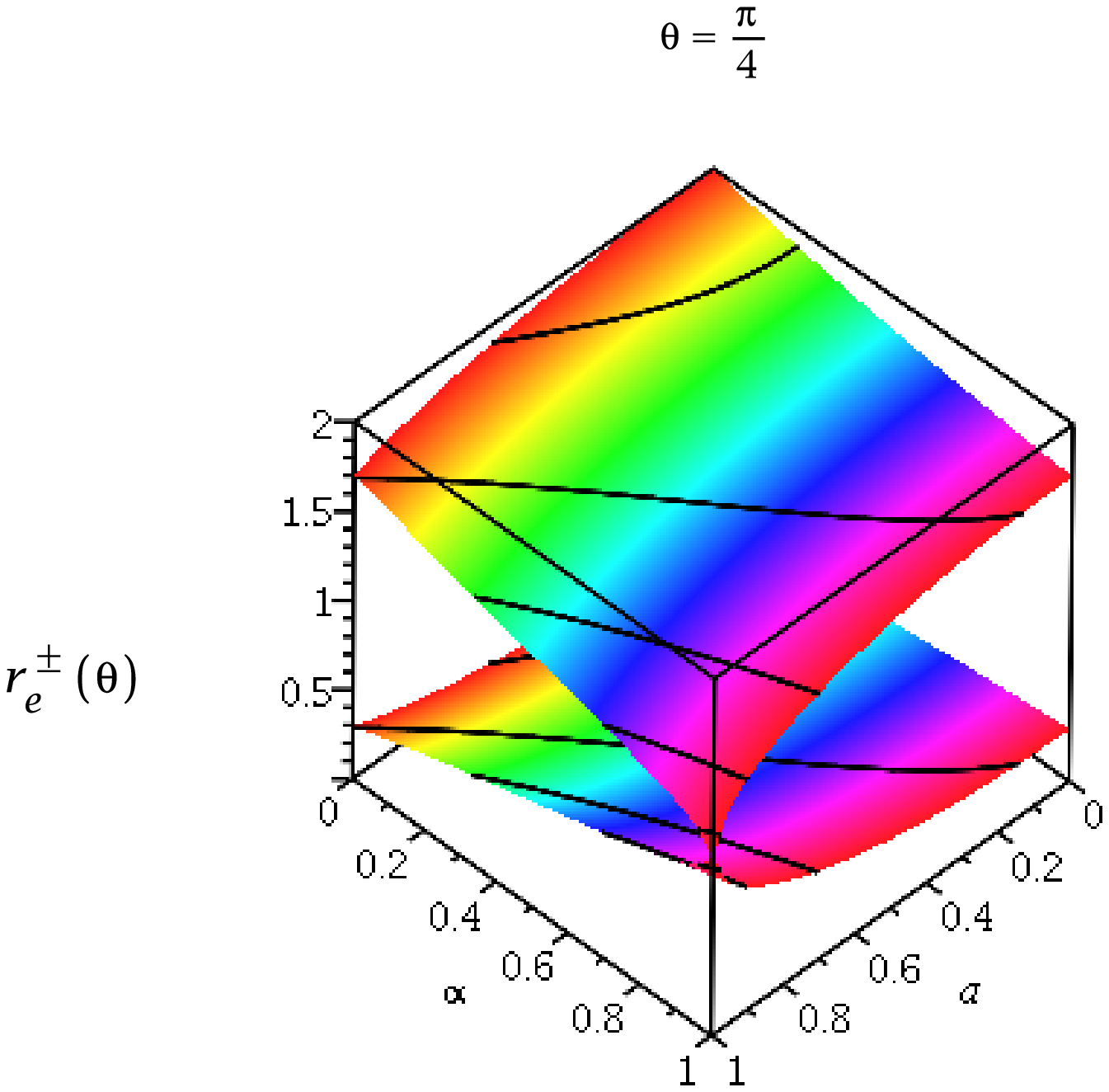}}
{\includegraphics[width=0.45\textwidth]{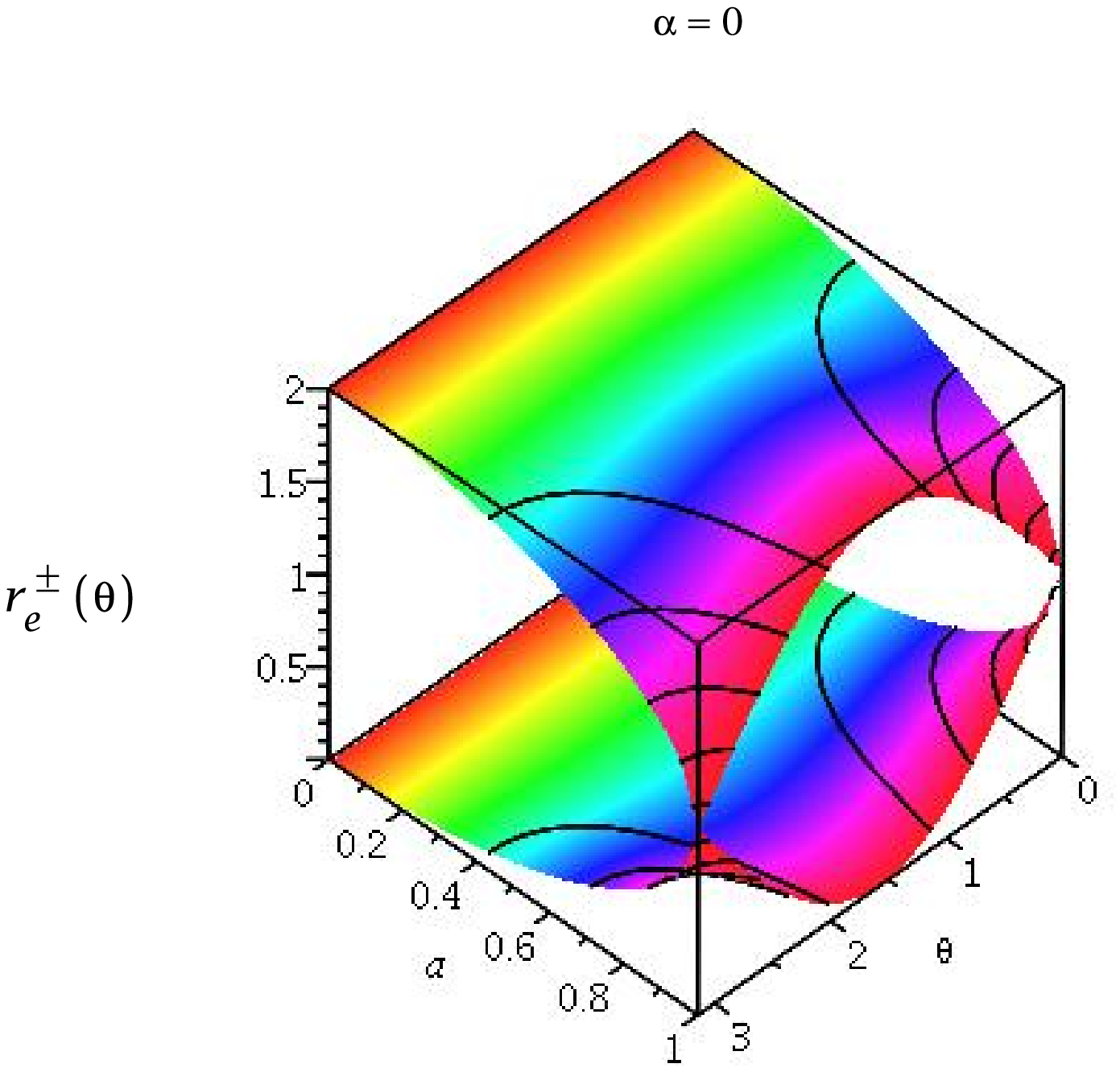}}
{\includegraphics[width=0.45\textwidth]{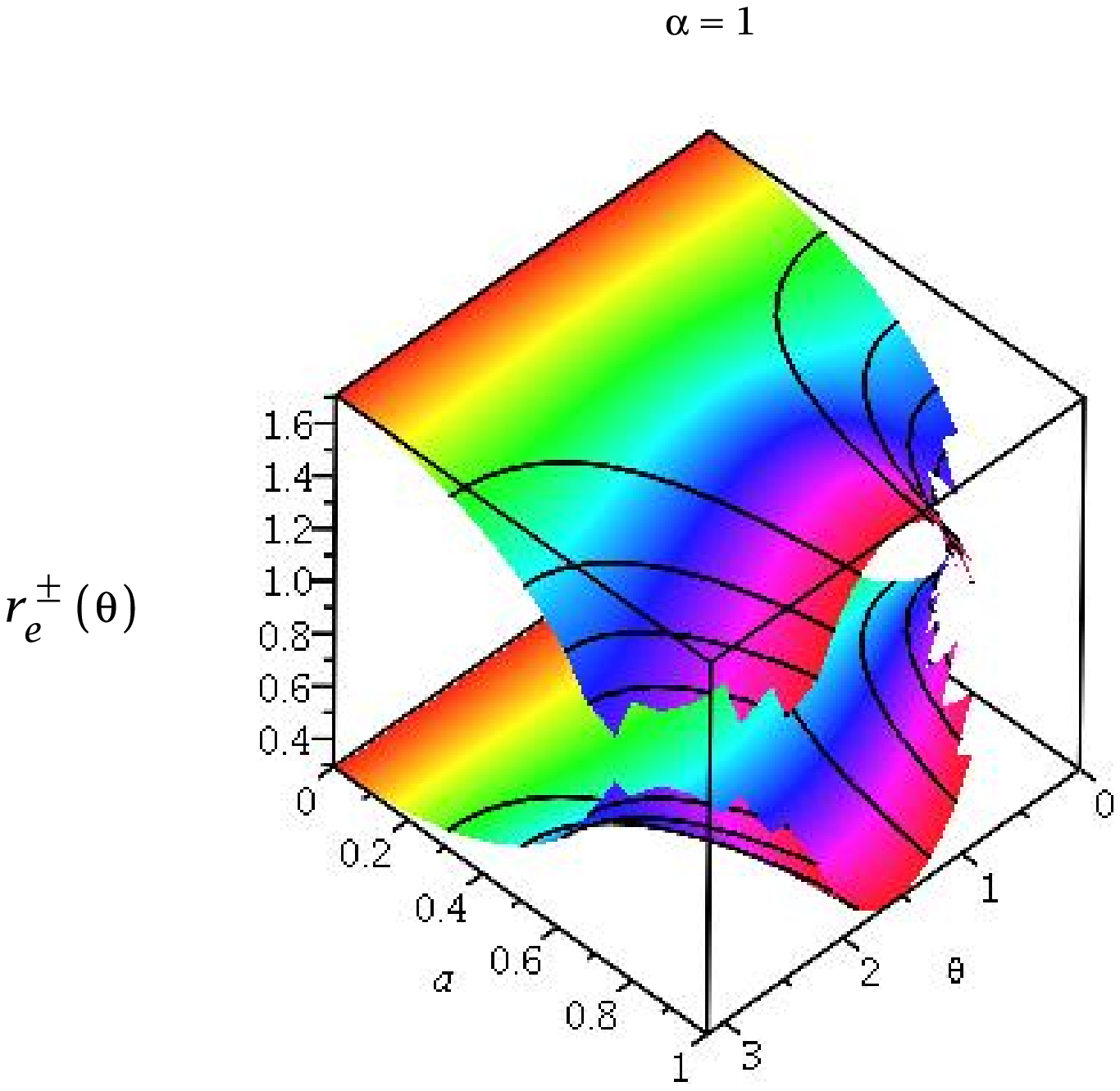}}
\end{center}
\caption{The figure indicates the variation  of $r_{e}^{\pm}(\theta)$ with $a$ and $\alpha$ 
for Kerr BH and Kerr-MOG BH. In each plot upper half
corresponds to $r_{e}^{+}(\theta)$ and lower half corresponds to $r_{e}^{-}(\theta)$.
\label{erg}}
\end{figure}
In the extremal limit, the outer horizon and inner horizon are coincident at $r_{+}=r_{-}= G_{N}{\cal M}$. 
The  outer and inner ergosphere radius reduces to
\begin{eqnarray}
r_{e}^{\pm}(\theta) &=& G_{N}{\cal M} \left(1 \pm \frac{\sin\theta}{\sqrt{1+\alpha}}\right)
~. \label{mg2.4}
\end{eqnarray}
\begin{eqnarray}
r_{e}^{\pm}(\theta)|_{\theta=0} &=& G_{N}{\cal M}=r_{\pm}, \,\,\, \mbox{(on axis)}\\
r_{e}^{\pm}(\theta)|_{\theta=\frac{\pi}{2}} &=& G_{N}{\cal M} \left(1 \pm \frac{1}{\sqrt{1+\alpha}}\right)
=r_{\pm}|_{a=0}\,\,\,\mbox{(equatorial plane)}~. \label{mg2.5}
\end{eqnarray}
The structure of the equatorial ergosphere could be seen visually from Fig.~\ref{erx}.

In the limit $\alpha=0$, 
\begin{eqnarray}
r_{e}^{+}(\theta)|_{\theta=\frac{\pi}{2}} &=& 2G_{N}{\cal M}, \,\,\,\, r_{e}^{-}(\theta)|_{\theta=\frac{\pi}{2}}=0 
~. \label{mg2.6}
\end{eqnarray}

\begin{figure}
\begin{center}
{\includegraphics[width=0.45\textwidth]{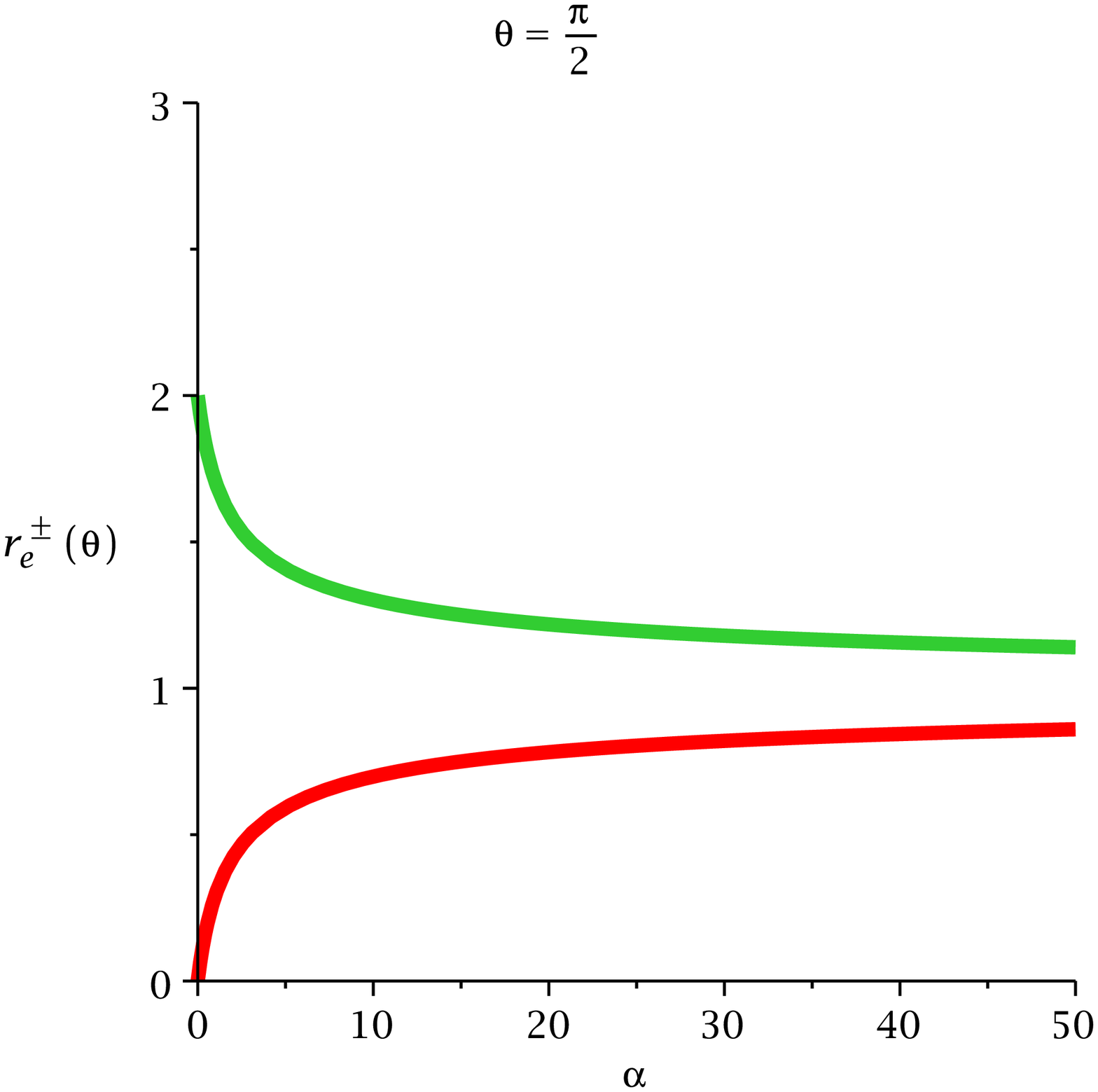}}
{\includegraphics[width=0.45\textwidth]{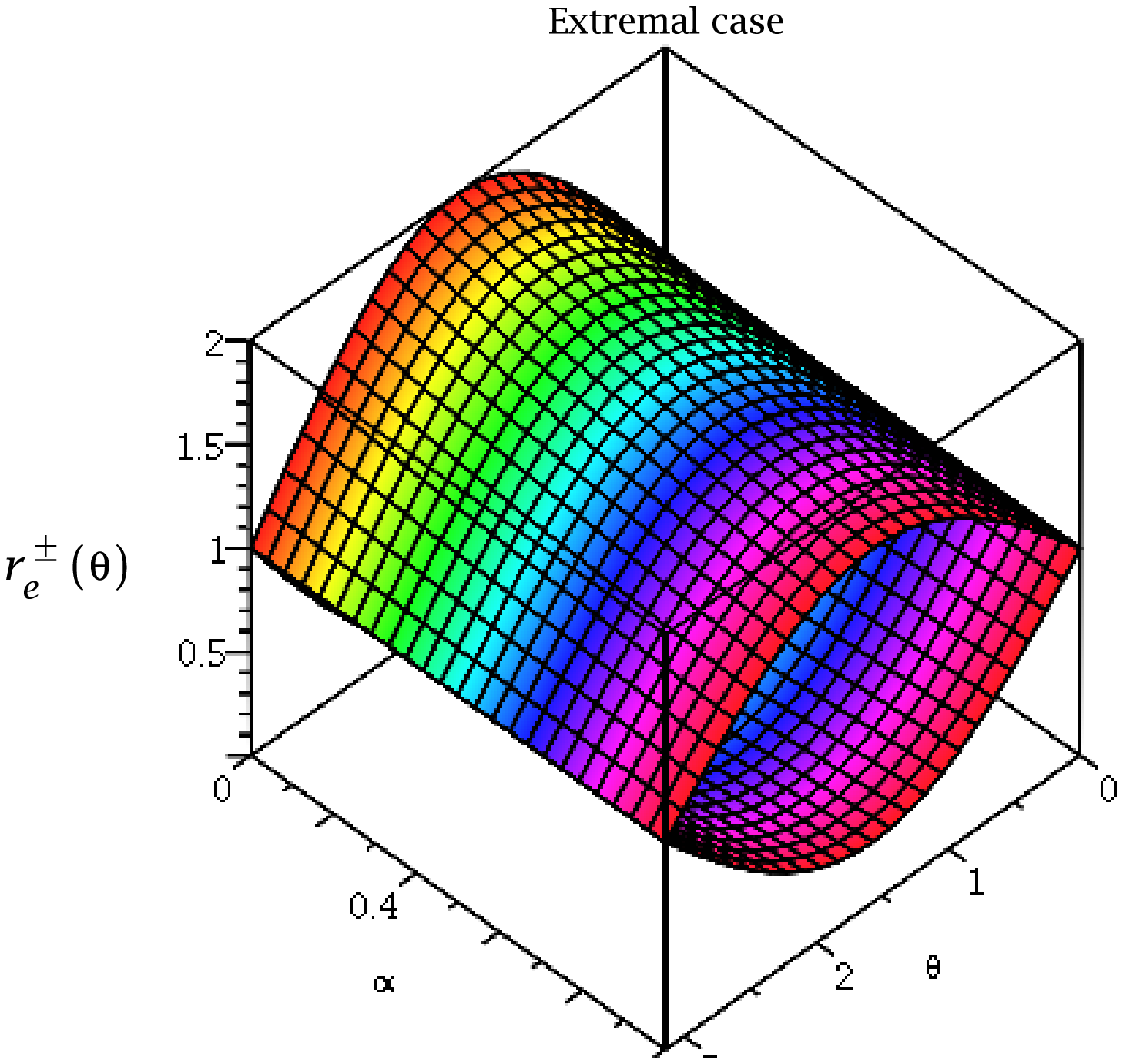}}
\end{center}
\caption{The figure implies the variation  of $r_{e}^{\pm}(\theta)$ with  $\alpha$ 
for extremal Kerr-MOG BH in equatorial plane.
\label{erx}}
\end{figure}
}
This surface is outer to the event horizon or outer horizon and it coincides with the outer horizon at 
the poles $\theta=0$ and $\theta=\pi$. The metric components of above BH  in Boyer-Lindquist 
coordinate are 
\begin{eqnarray}
g_{tt} &=& - \left(\frac{r^2-\Pi_{\alpha}+a^2 \cos^2\theta}{\rho^2}\right)\\
g_{t\phi} &=& -\left(\frac{a \, \sin^2\theta\,\Pi_{\alpha}}{\rho^2}\right)\\
g_{rr} &=& \left(\frac{\rho^2}{\Delta}\right)\\
g_{\theta\theta} &=& \rho^2 \\
g_{\phi\phi} &=& \left(r^2+a^2+\frac{a \Pi_{\alpha} \sin^2\theta}{\rho^2}\right) \sin^2\theta 
\end{eqnarray}
and
\begin{equation}
\sqrt{-g}=\rho^2 \sin\theta.
\end{equation}
where
\begin{eqnarray}
\Pi_{\alpha} &=& \left(2G_{N}{\cal M}r-\frac{\alpha}{1+\alpha} G_{N}^2{\cal M}^2\right)
\end{eqnarray}
Putting these metric components in  Eq.~(\ref{n1}), one obtains the generalized spin precession frequency 
of a test gyro for KMOG BH 
\begin{eqnarray}
\vec{\Omega}_{p} &=& \frac{\xi(r)~\sqrt{\Delta}\cos\theta~ \hat{r}+\eta(r)~\sin\theta~\hat{\theta}}{\zeta(r)},~\label{gekm}
\end{eqnarray}
where,

\begin{equation}
\begin{split}
\xi (r) & = a \Pi_{\alpha}-\frac{\Omega}{8} \left[8r^4+8a^2r^2+8 a^2\Pi_{\alpha}+3a^4+ 
4a^2\left(2\Delta-a^2\right)\cos2\theta+a^4\cos4\theta \right]\\
& +\Omega^2a^3\Pi_{\alpha}\sin^4\theta
\end{split}
\end{equation}

\begin{equation}
\begin{split}
\eta(r) &= aG_{N}{\cal M}\left(r^2-a^2 \cos^2\theta\right)-\frac{\alpha}{1+\alpha} G_{N}^2 {\cal M}^2 ar+\\ 
& \Omega\,\left(r^5-3G_{N}{\cal M}r^4+2a^2r^3\cos^2\theta-2G_{N}{\cal M}a^2r^2+a^4 \cos^4\theta r\right)\\
& +\Omega\,\left[G_{N}{\cal M}a^4\cos^2\theta\left(1+\sin^2\theta\right)+2\frac{\alpha}{1+\alpha}G_{N}^2 {\cal M}^2 
r \left(r^2+a^2\right) \right]\\
& + \Omega^2 G_{N}{\cal M} a\sin^2\theta \left[3r^4+a^2r^2+a^2\cos^2\theta\left(r^2-a^2\right)\right]\\
& - \frac{\alpha}{1+\alpha} G_{N}^2 {\cal M}^2 a\sin^2\theta \Omega^2 r \left[2r^2+a^2 \left(1+\cos^2\theta\right)\right]
\end{split}
\end{equation}

\begin{eqnarray}
\zeta(r) &=& \rho^3
\left[\left(\rho^2-\Pi_{\alpha}\right)
+2a\Omega \Pi_{\alpha} \sin^2\theta-\Omega^2\sin^2\theta
\left\{\rho^2\left(r^2+a^2\right)+a^2 \Pi_{\alpha} \sin^2\theta\right\}\right] \nonumber\\
.\label{knAB}
\end{eqnarray}
This is the expression of generalized spin frequency  which is valid both for outside and inside the 
ergosphere. Now we {could} determine the range of the angular velocity and therefore  the four velocity 
must be time-like i.e. $K^2 = g_{\phi\phi}\Omega^2+2g_{t\phi}\Omega+g_{tt}<0$. The allowed 
values of $\Omega$ at any fixed $(r,\theta)$ are 
$\Omega_-(r,\theta) < \Omega(r,\theta) < \Omega_+(r,\theta)$ where 
\begin{eqnarray}
\Omega_{\pm} =\frac{-g_{t\phi}\pm \sqrt{g_{t\phi}^2-g_{\phi\phi}g_{tt}}}{g_{\phi\phi}}. \label{okn}
\end{eqnarray}
For KMOG BH, it should be 
\begin{eqnarray}
\Omega_{\pm}&=& \frac{a \Pi_{\alpha}\sin\theta \pm \rho^2\sqrt{\Delta}}
{\sin\theta[\rho^2(r^2+a^2)+a^2\Pi_{\alpha}\sin^2\theta]}~\label{omega}
\end{eqnarray}
Now we shall calculate the values of $\Omega$ when an observer closes to the horizon. {At} the event horizon the 
angular velocity becomes $\Omega_{+} =\frac{a}{2G_{N}{\cal M}r_{+}-\frac{\alpha}{1+\alpha} G_{N}^2 {\cal M}^2}$ 
and {at} the Cauchy horizon the angular velocity becomes 
$\Omega_{-} =\frac{a}{2G_{N}{\cal M}r_{-}-\frac{\alpha}{1+\alpha} G_{N}^2 {\cal M}^2 }$. In the equatorial plane  
\begin{eqnarray}
\Omega_{\pm}|_{\theta=\pi/2}=\frac{a\Pi_{\alpha}\pm 
r^2\sqrt{\Delta}}{r^2(r^2+a^2)+a^2 \Pi_{\alpha}}. \label{oken}
\end{eqnarray}
Finally at the ring singularity $r=0$ and $\theta=\frac{\pi}{2}$
\begin{eqnarray}
\Omega_{\pm}|_{r=0, \theta=\pi/2} =1.~\label{omkn1}
\end{eqnarray}
This means that two angular frequencies coincide at the ring singularity. Variation of angular frequency~($\Omega$)
of a stationary  gyroscope could be observed from {the whale diagram}[Figure~(\ref{angf})]. The significance of
whale diagram is that the whale or a fish can move along the $\Omega_{+}$ and $\Omega_{-}$ and it is prominent 
when the MOG parameter vanishes i.e. $\alpha=0$.  {For details of whale diagram and escape cones one can see an 
interesting work by Tanatorov and Zaslavskii~\cite{zas17}. 
We set the value of 
${\cal M}=G_{N}=1$ during plot which is valid througout the work. In Fig.~\ref{angf}(a), Fig.~\ref{angf}(b) and 
Fig.~\ref{angf}(c), we  also set the MOG parameter~($\alpha=0$) and will vary the spin parameter value to differentiate 
three types of compact objects  namely non-extremal Kerr, extremal Kerr and NS. Three pictures are qualitative different 
to differentiate between non-extremal BH, extremal BH  and NS. In Fig.~\ref{angf}(d), Fig.~\ref{angf}(e) and 
Fig.~\ref{angf}(f), we set the value of MOG parameter is unity. In this case, one can see the variation of angular 
velocity with radial coordinate to differentiate three compact objects. Both $\Omega_{+}$ and $\Omega_{-}$  are 
clearly distinct in each case. The distinction is more pronounced in Fig.~\ref{angf}(g), Fig.~\ref{angf}(h) and 
Fig.~\ref{angf}(i) and so on for greater value of MOG  parameter.
\begin{figure}
\begin{center}
\subfigure[]{
\includegraphics[width=2in,angle=0]{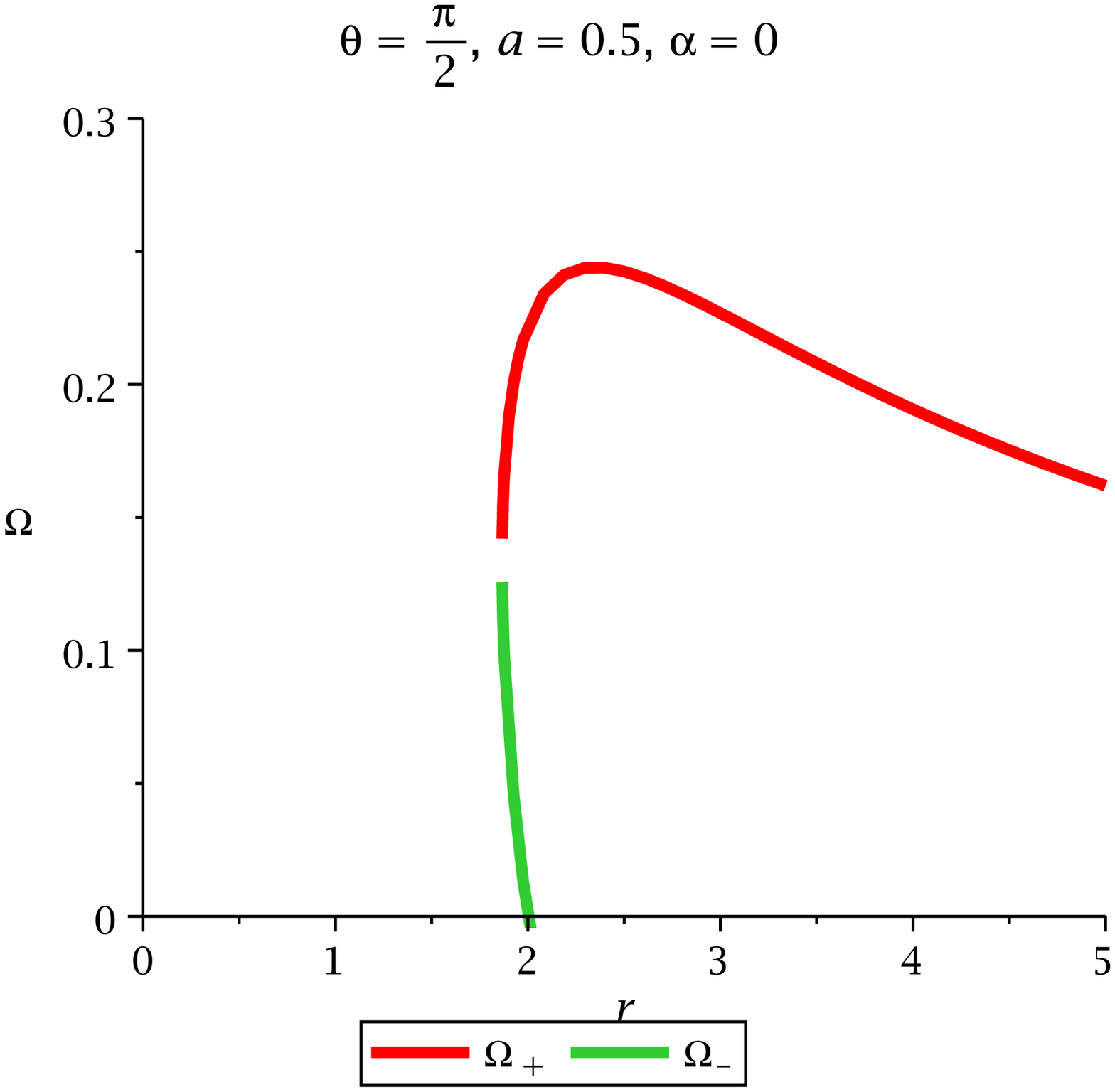}} 
\subfigure[]{
\includegraphics[width=2in,angle=0]{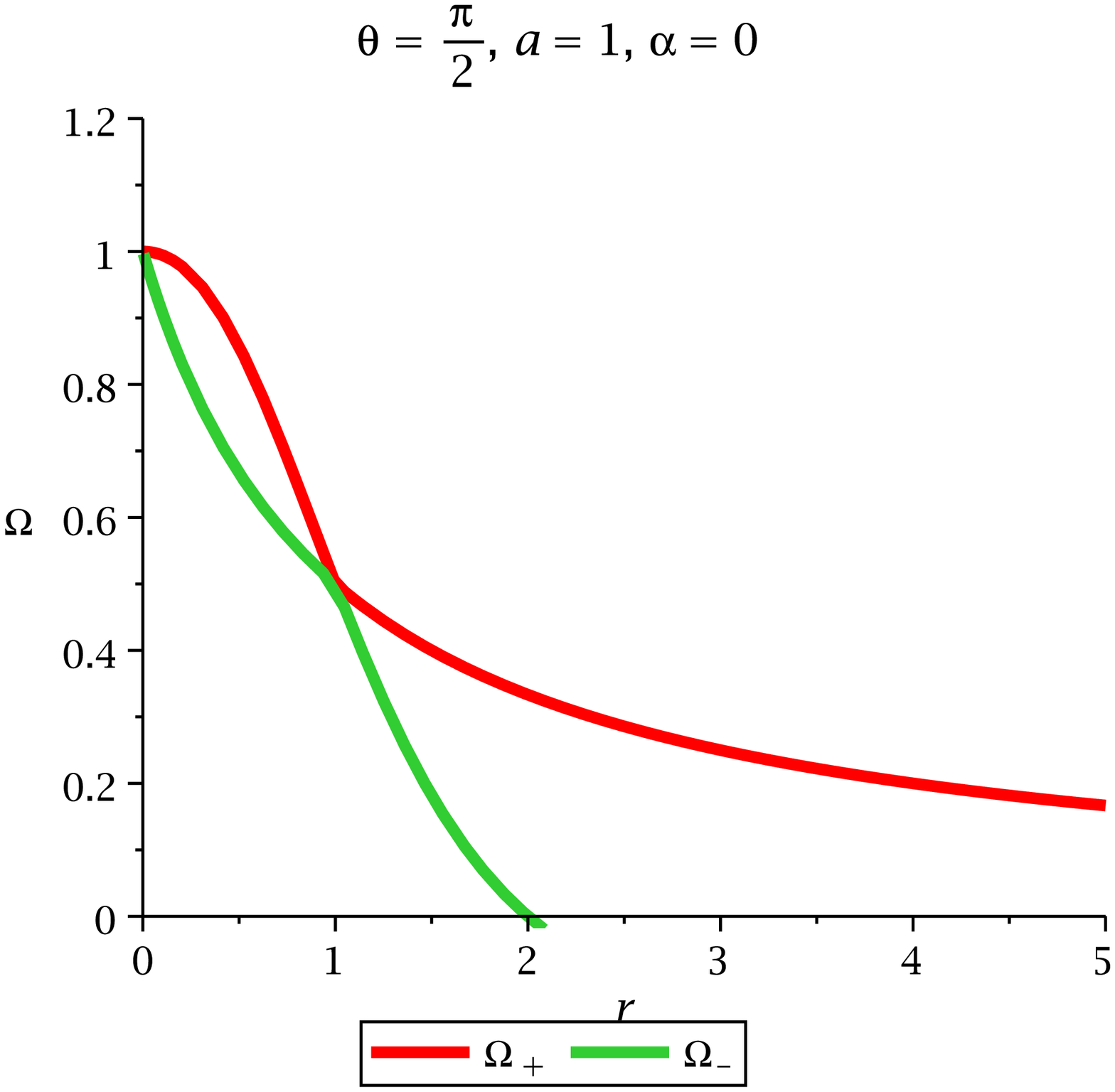}} 
\subfigure[]{
\includegraphics[width=2in,angle=0]{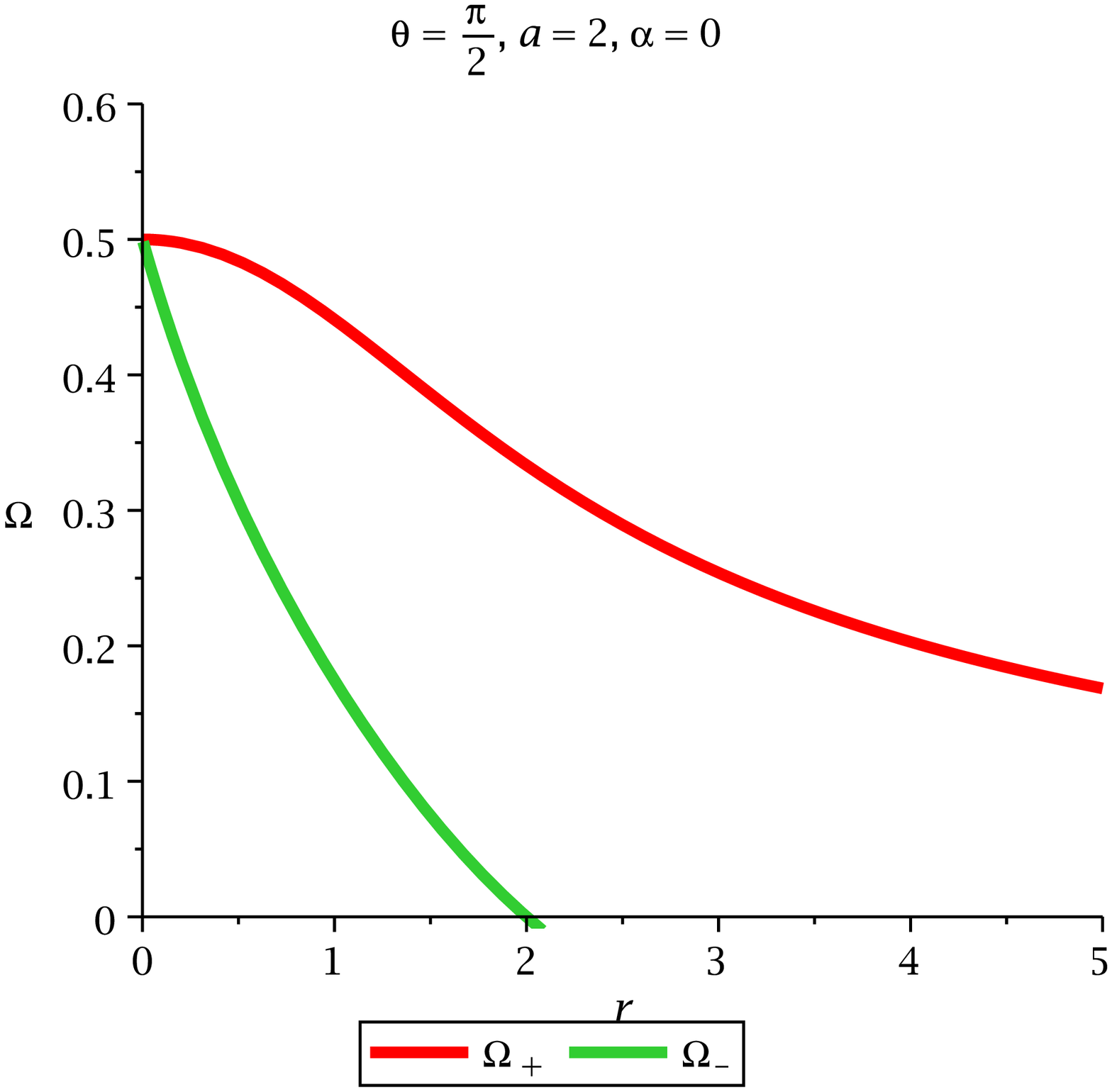}}
\subfigure[]{
\includegraphics[width=2in,angle=0]{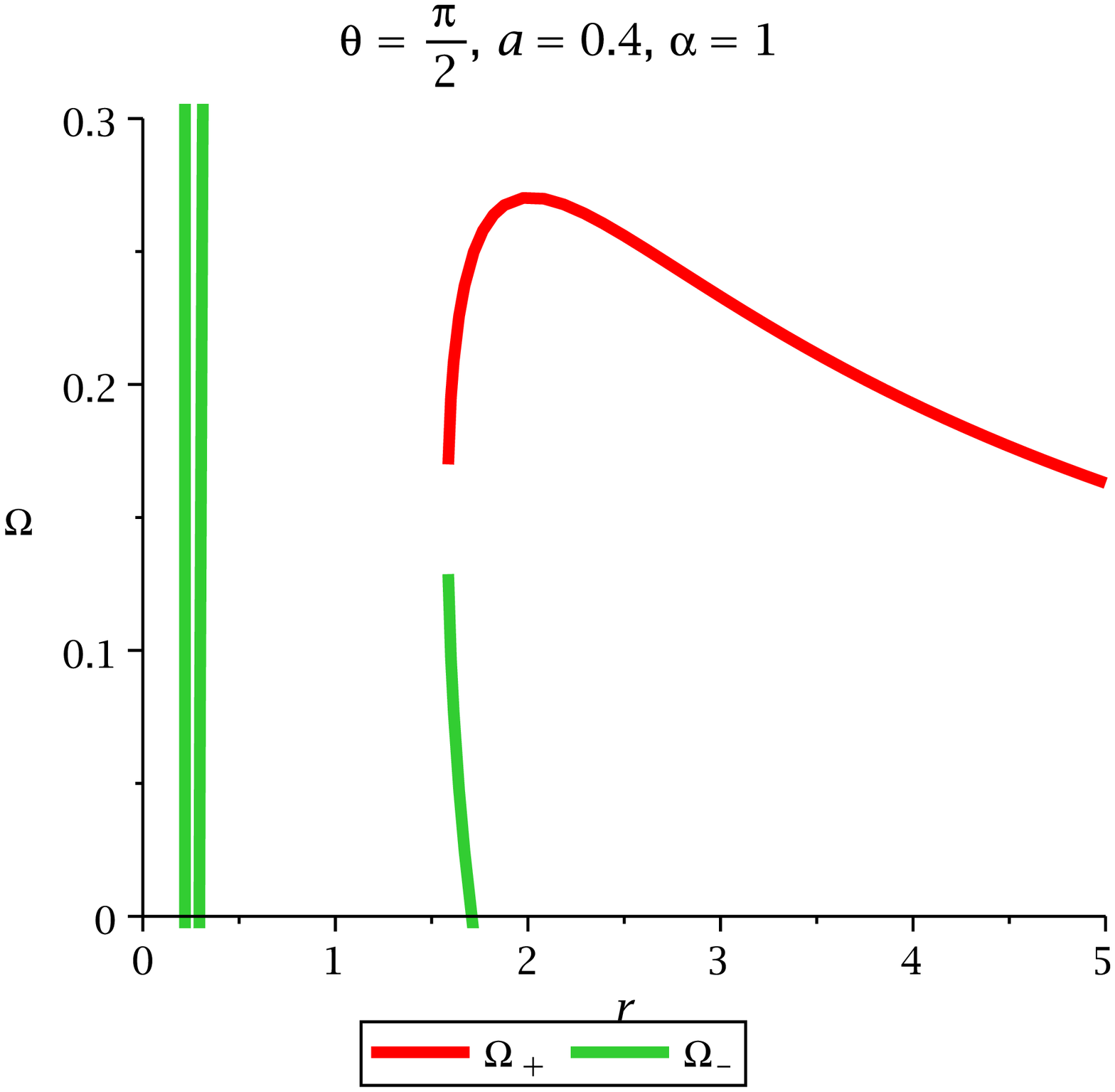}}
\subfigure[]{
\includegraphics[width=2in,angle=0]{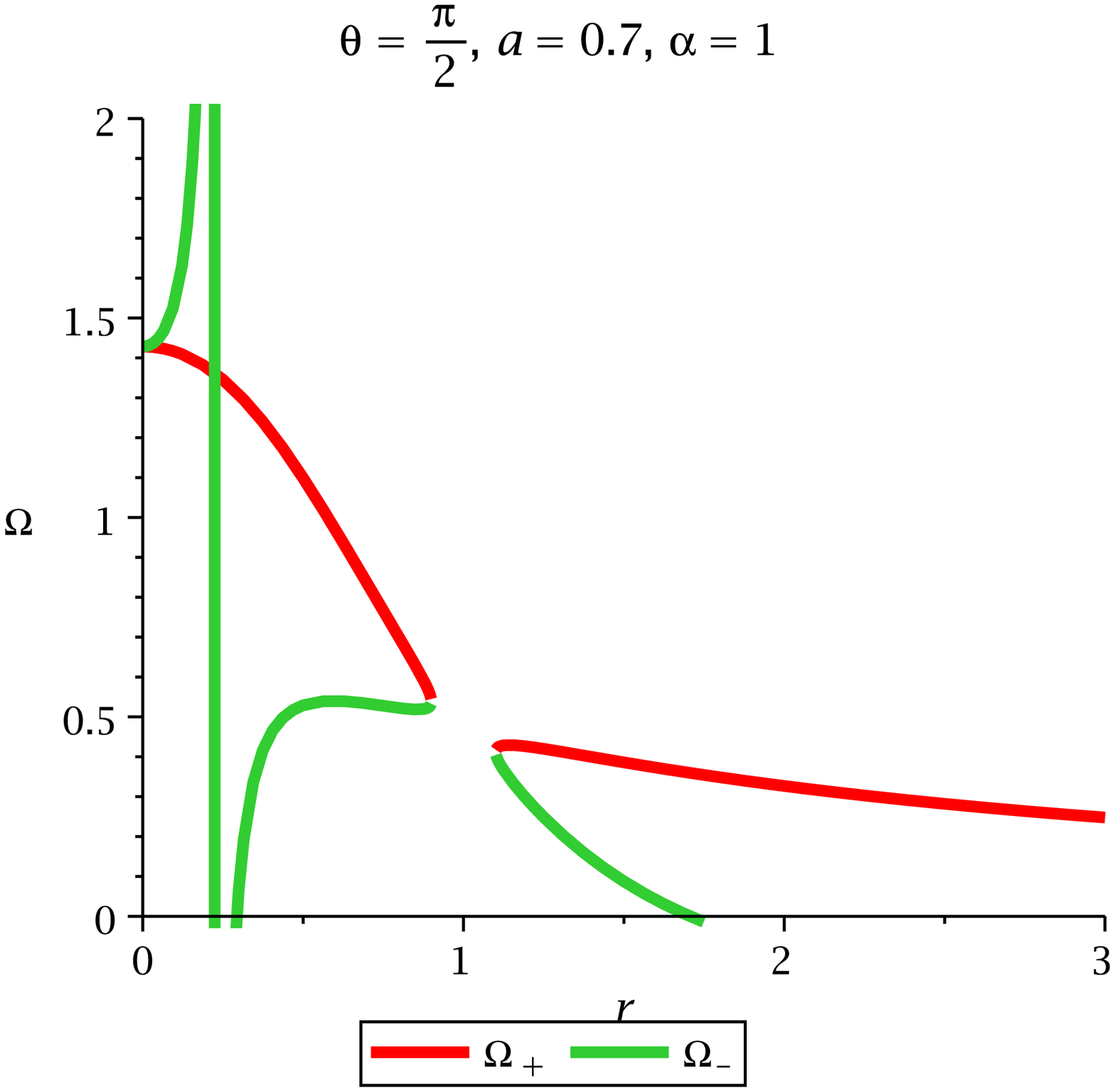}}
\subfigure[]{
\includegraphics[width=2in,angle=0]{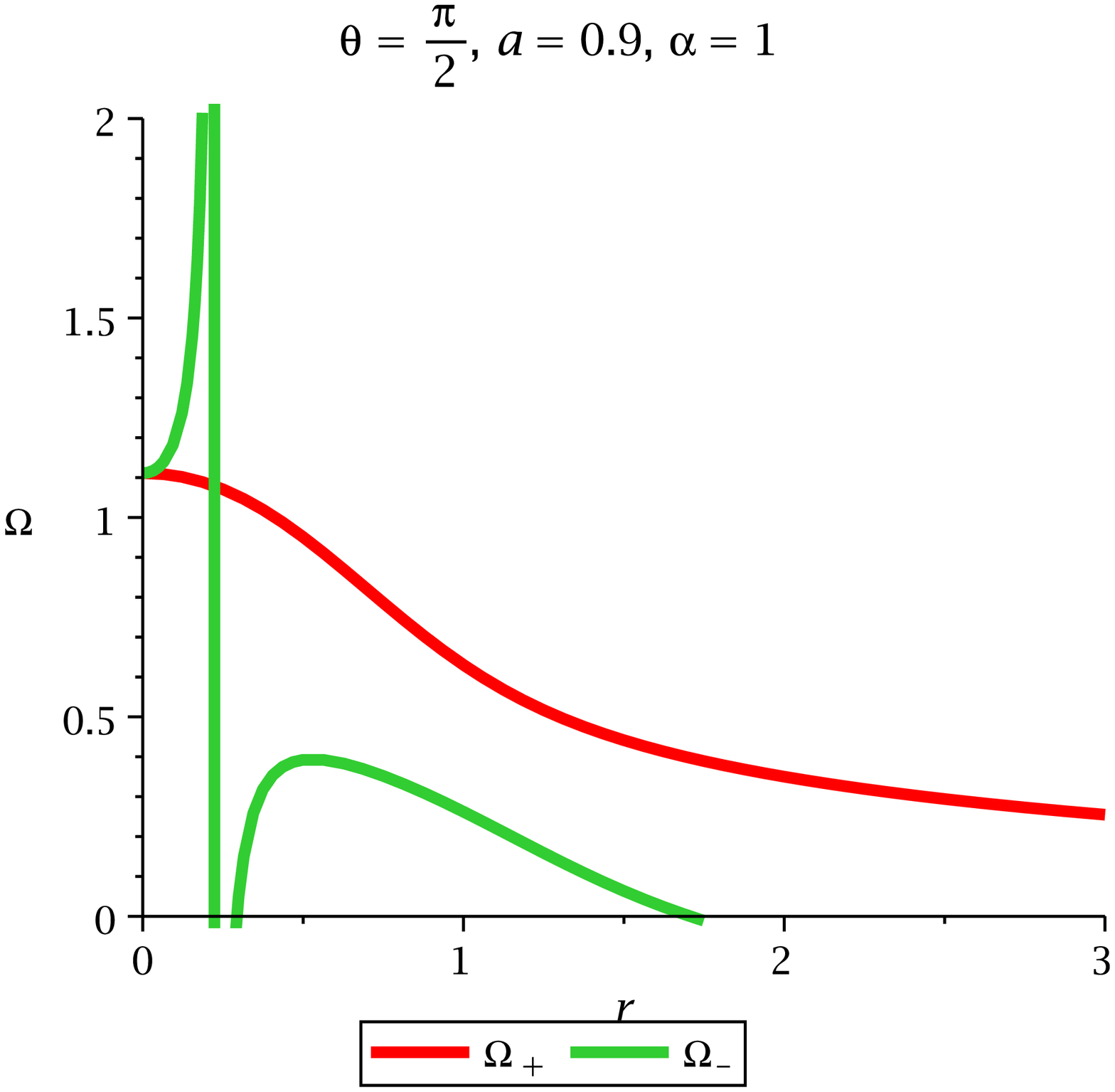}}
\subfigure[]{
\includegraphics[width=2in,angle=0]{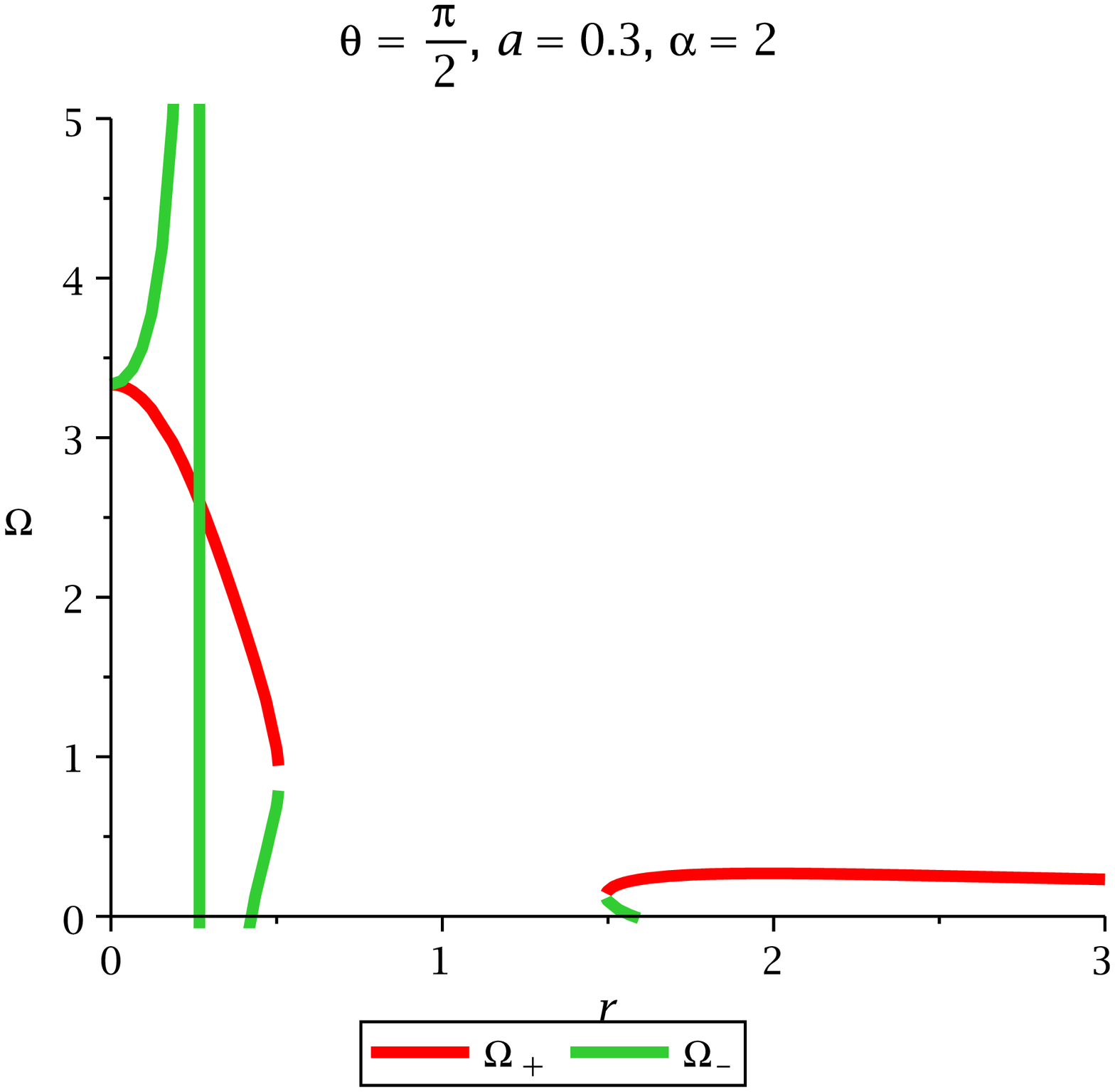}}
\subfigure[]{
\includegraphics[width=2in,angle=0]{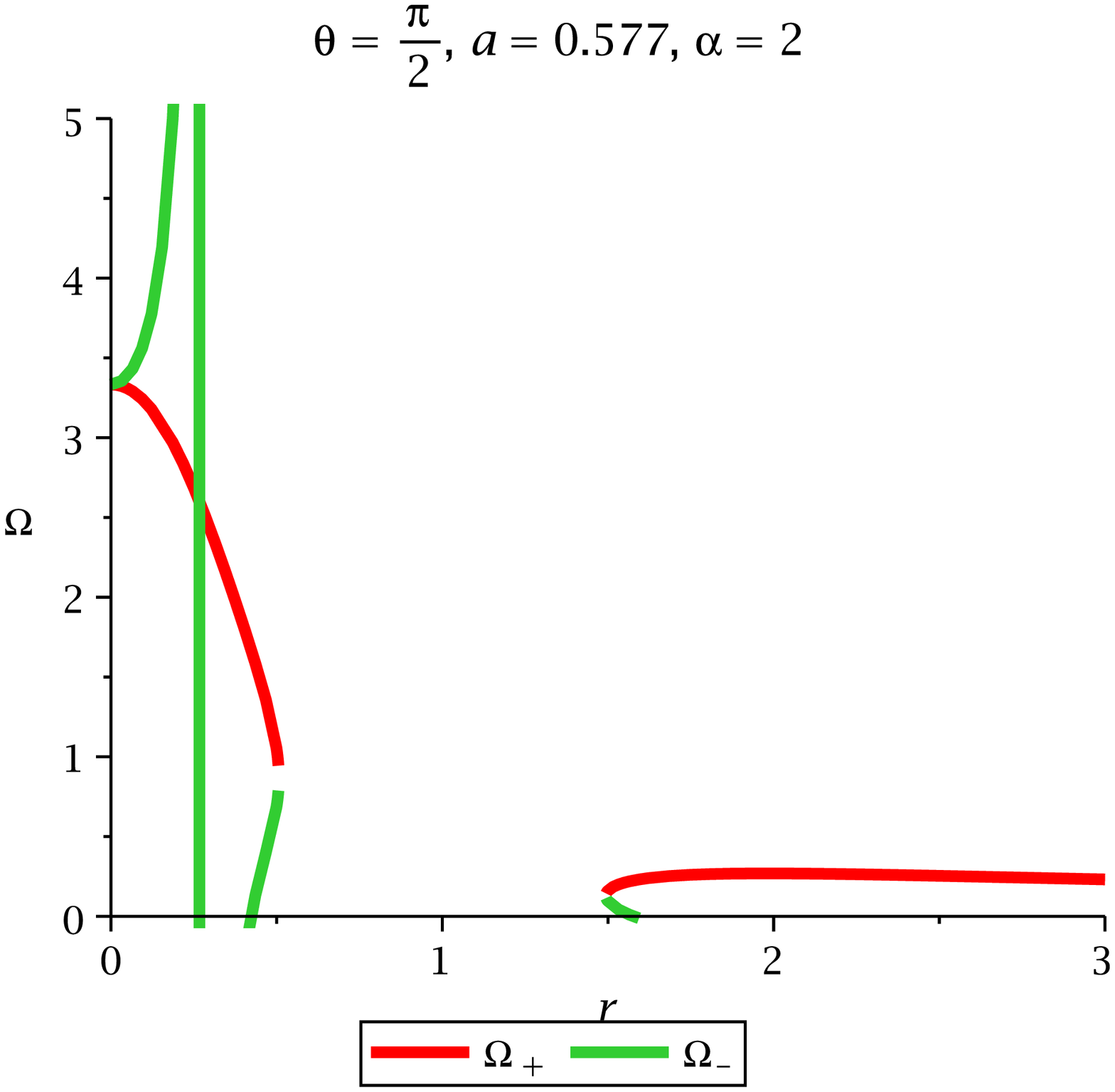}}
\subfigure[]{
\includegraphics[width=2in,angle=0]{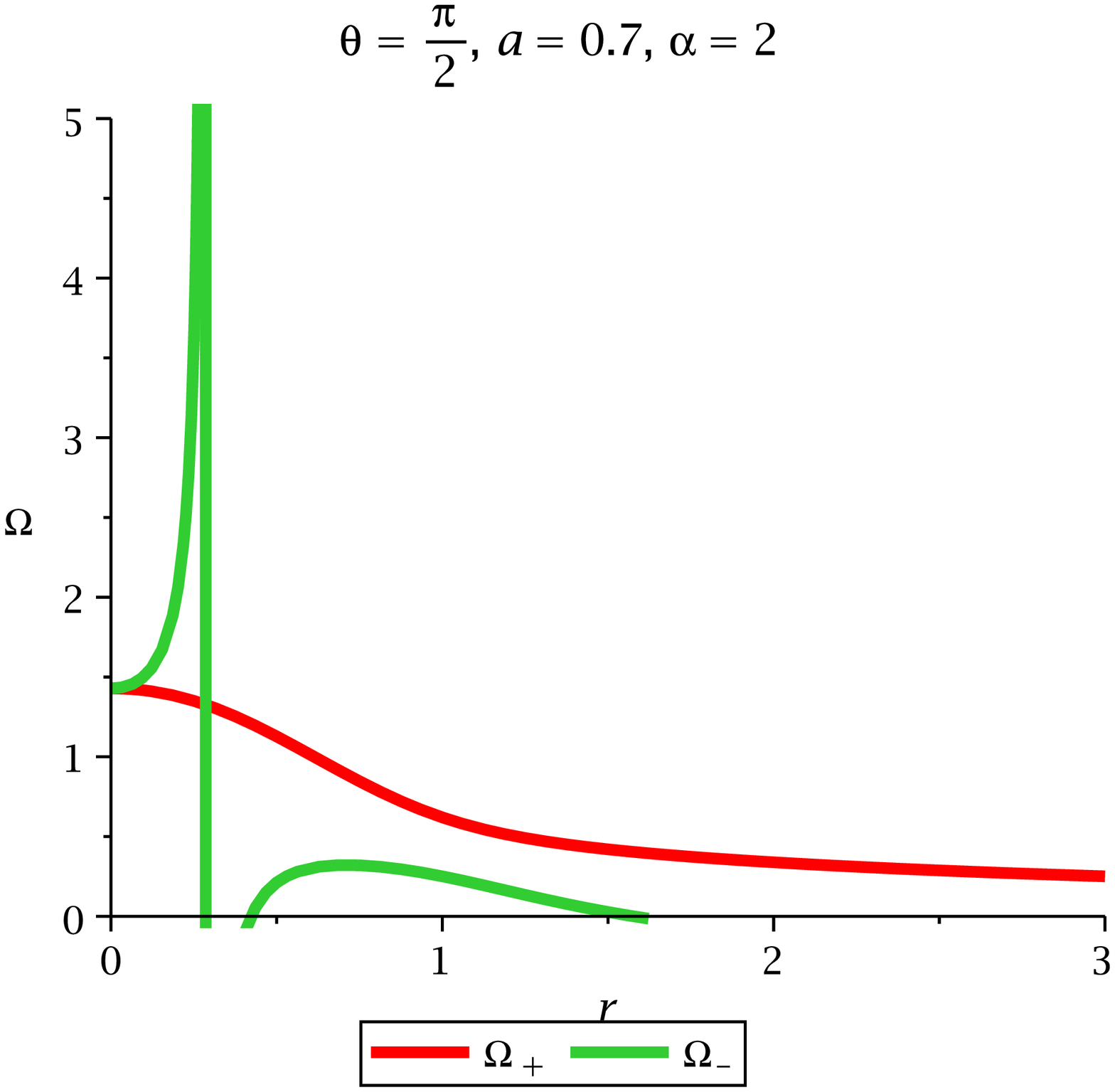}}
\subfigure[]{
\includegraphics[width=2in,angle=0]{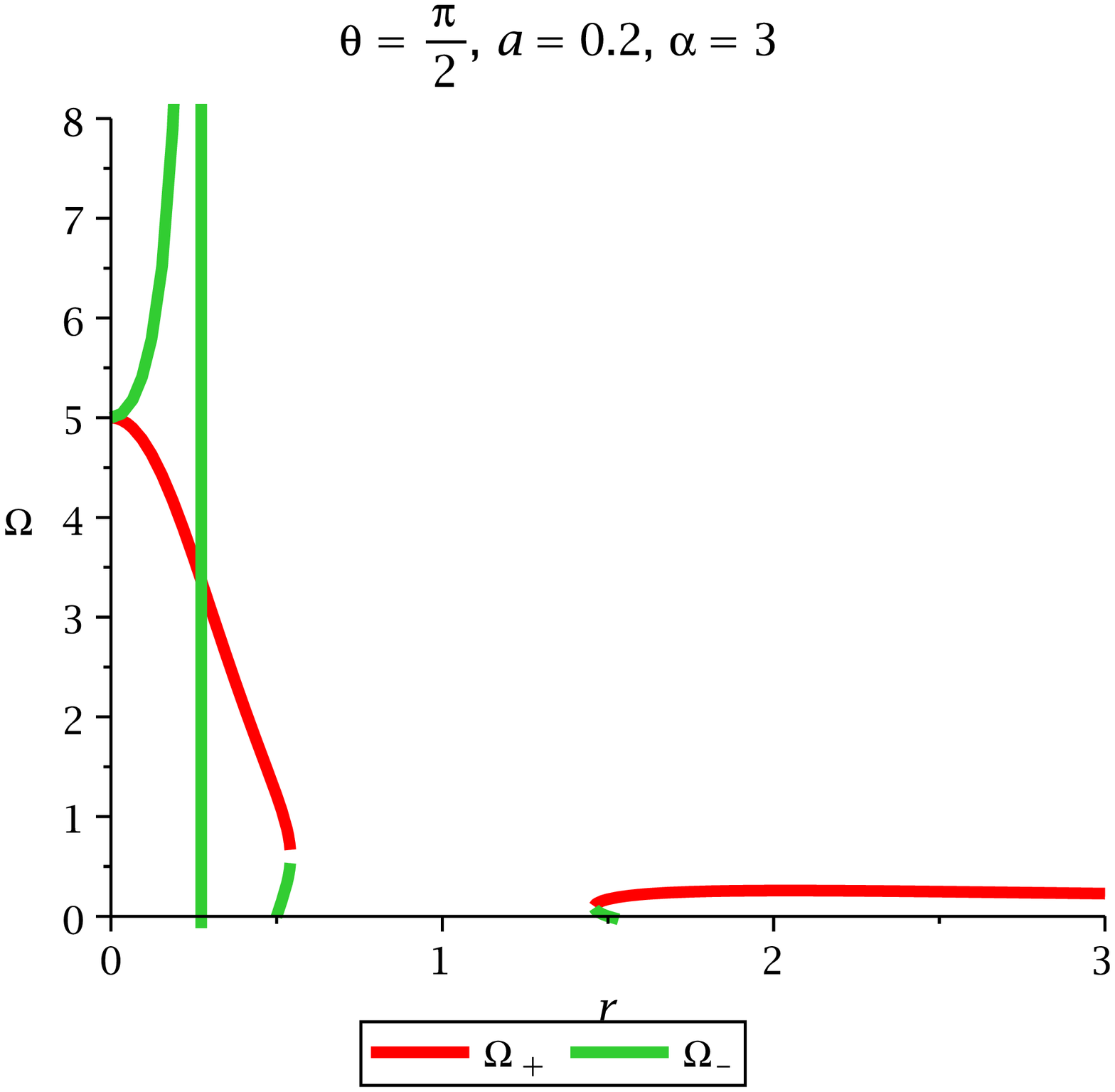}}
\subfigure[]{
\includegraphics[width=2in,angle=0]{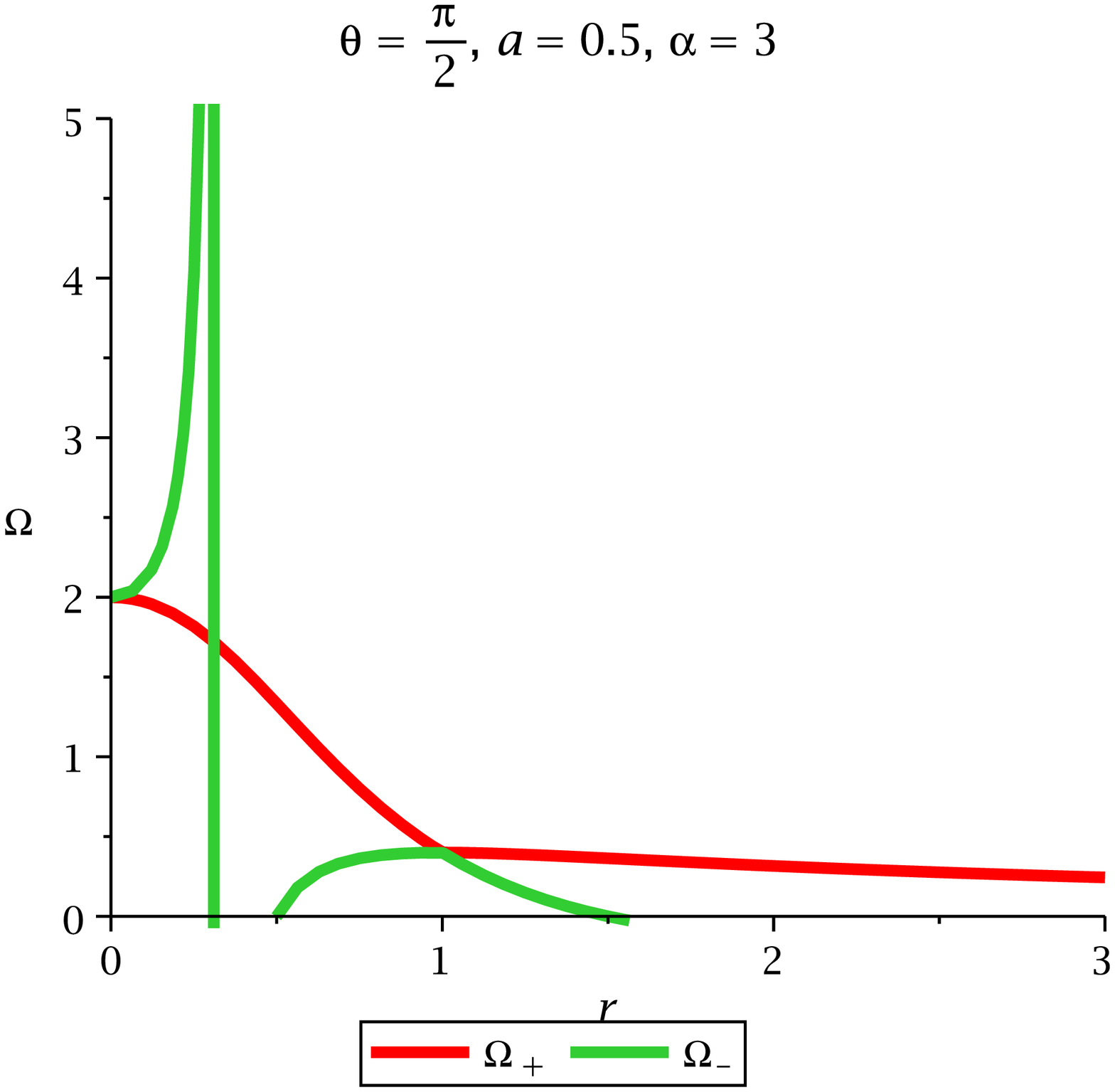}}
\subfigure[]{
\includegraphics[width=2in,angle=0]{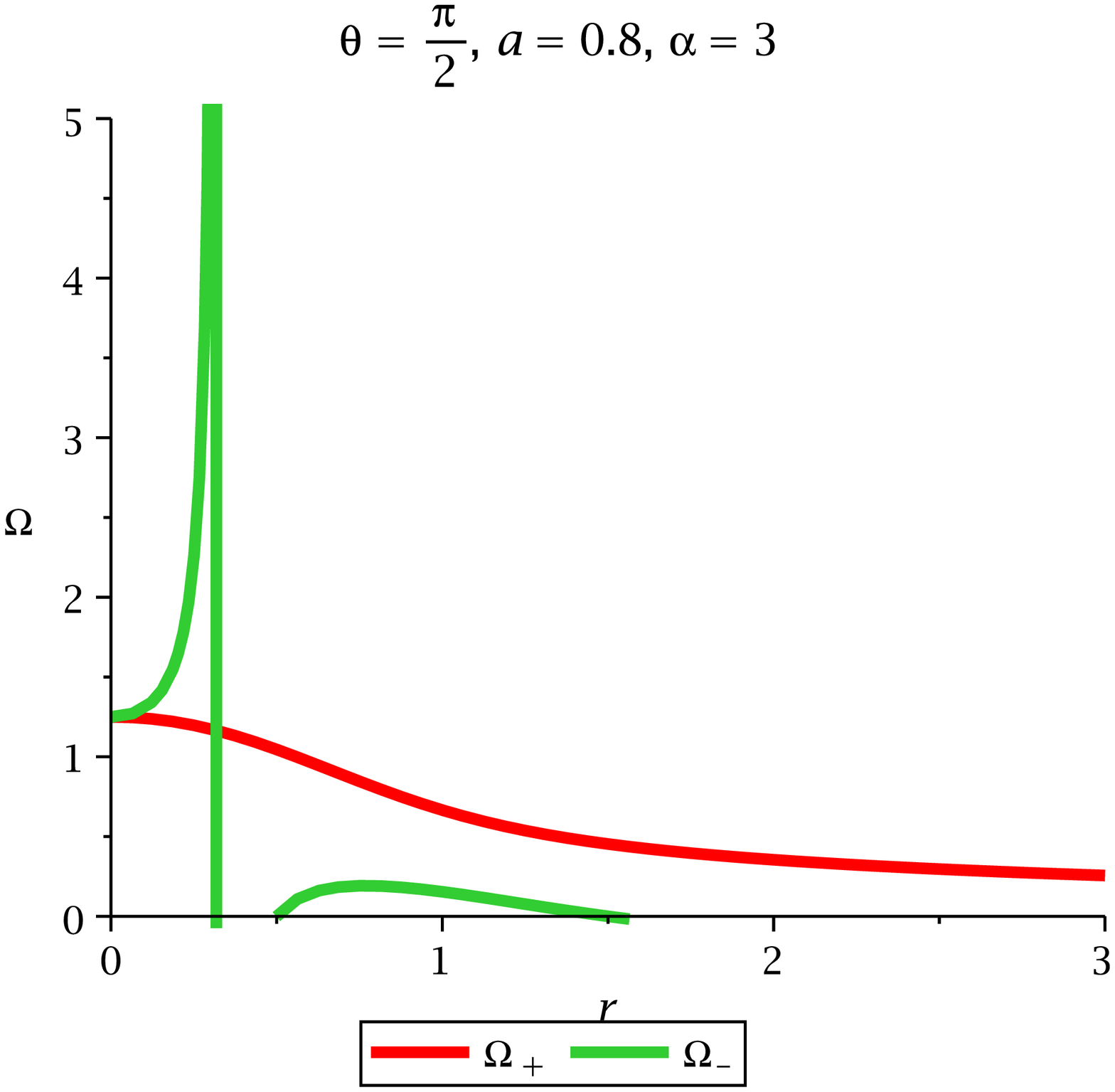}}
\caption{The figure describes the variation  of $\Omega$  with $r$ for 
different values of spin parameter, with MOG parameter and without MOG parameter. The gyroscope has 
angular frequency $\Omega$ varies with in the range $(\Omega_{+}, \Omega_{-})$. Each set of row 
depicts the variation of $\Omega$ vs. $r$ for non-extremal BH, extremal BH and NS.  
There is a qualitative difference between these plots i. e.  when we have taken into 
account the MOG parameter and without MOG parameter.}
\label{angf} 
\end{center}
\end{figure}
The angular frequency has two values i. e. in the outer horizon it is $\Omega_{+}$ and 
in the inner horizon it is $\Omega_{-}$. Now we analyze the behavior of the gyro inside 
the ergosphere of a BH and to do this job we should first compute the magnitude of the 
precession frequency for various values of $\theta$.
For this purpose, we introduce the parameter $\delta$ to scan the entire range of $\Omega$ as 
\begin{eqnarray}
\Omega &=& \delta~\Omega_{+} + (1-\delta)~\Omega_{-} = \omega -(1-2\delta)\sqrt{\omega^2-\frac{g_{tt}}{g_{\phi\phi}}}\\ 
&=& \frac{a\Pi_{\alpha}\sin\theta - (1-2\delta)~\rho^2\sqrt{\Delta}}{\sin\theta[(r^2+a^2)\rho^2+a^2\sin^2\theta \Pi_{\alpha}]}
\label{q}
\end{eqnarray}
where $0<\delta<1$ and $\omega=-g_{t\phi}/g_{\phi\phi}$. It is evident that the limiting value of 
$\delta$ gives the range of $\Omega$ from $\Omega_{+}$ to $\Omega_{-}$. Using  Eq.~(\ref{q}), one 
obtains the generalized spin precession frequency in a compact form 
\begin{eqnarray}
\vec{\Omega}_{p} &=& \Upsilon(r) 
\left[\xi(r)~\sqrt{\Delta}\cos\theta~ \hat{r}+\eta(r)~\sin\theta~\hat{\theta}\right].~\label{gekn1}
\end{eqnarray}
where
\begin{eqnarray}
\Upsilon(r) &=& \frac{(r^2+a^2)^2-a^2\Delta \sin^2\theta}{4\delta(1-\delta)\Delta~\rho^7}
\end{eqnarray}
The implication of this equation is that one could study the behaviour of generalized spin 
frequency $\vec{\Omega}_{p}$ with spin parameter $a$, $\theta$, $\delta$, $r$ and MOG parameter. 
It should be noted that the frequency vector diverges at $\delta=0, 1$, $\Delta=0$ and $\rho=0$.
This means that the generalized spin precession frequency becomes arbitrarily large at these 
values. Using this frequency one should differentiate between the BH and NS 
in a strong gravitational field.

Now consider an observer moving with a four velocity~$u$ in a stationary and axisymmetric 
spacetime. If the angular momentum is zero for a particular situation which is defined by 
\begin{eqnarray}
p_{\phi} & \equiv & \frac{\partial {\cal L}}{\partial \dot{\phi}} 
=g_{t\phi} \dot{t}+g_{\phi\phi} \dot{\phi}=\ell =0
~.\label{pps}
\end{eqnarray}
{(where ${\cal L}$ is the Lagrangian and $\ell$ is angular momentum)}
such an observer is called zero angular momentum observer~(ZAMO) which was first 
observed by Bardeen~\cite{bd,mtw}. Bardeen et al.~\cite{bpt} proved that the 
ZAMO frame is a powerful tool to analyze the physical processes near astrophysical 
object. What happens in case of Newtonian gravity? The angular momentum $\ell$ and 
angular velocity~$\Omega$ are satisfied by the relation $\ell =r^2\Omega$. It 
implies that there is no problem when we taking into consider the non-rotating 
frame i.e. $\ell=\Omega=0$. The problem should arise when we consider Einstein's 
gravity: in this case the angular momentum is satisfied the following relation
$\ell \propto (\Omega-\omega)$, where $\omega=-\frac{g_{t\phi}}{g_{\phi\phi}}$.
Here we would not obtain $\ell=0$ when $\Omega=0$. It indicates that there must 
exist two different observers i.e.  one should be zero angular momentum observer~(ZAMO) 
and the other one should be zero angular velocity observer~(ZAVO). 
In ZAMO frame, the value of angular momentum is $\ell =0$ while in ZAVO frame, the 
value of $\Omega=0$. The frame-dragging angular velocity of these two frames is 
$\omega=-\frac{g_{t\phi}}{g_{\phi\phi}}$. One should define the gravitational 
potential in the ZAMO frame as $\Phi=-\frac{1}{2} \ln|g^{tt}|$. Where $g^{tt}$ is 
contravariant component of stationary axisymmetric spacetime metric. This potential 
is again related to the  gravitational acceleration~(accelaration due to gravity)  
as felt by an observer in space $g_{\mu}=(a_{\mu})_{ZAMO}=\nabla \Phi$. The 
gravitational accelaration~$a_{\mu}$ is a kinematic invariant quantity. Using 
Eq.~(\ref{q}), one could easily see that for a particular value of 
$\delta=\frac{1}{2}$ the value of angular velocity for KMOG BH in 
ZAMO frame as 
\begin{eqnarray}
\Omega &=&  \omega  = 
\frac{a~\Pi_{\alpha}\sin\theta}{\sin\theta[(r^2+a^2)\rho^2+a^2\sin^2\theta~\Pi_{\alpha}]}
\label{q1}
\end{eqnarray}
It implies that the ZAMOs angular velocity is a function of spin parameter and 
charge parameter. In  $\theta=\frac{\pi}{2}$ plane,  the ZAMOs angular 
velocity should be 
\begin{eqnarray}
\Omega|_{\theta=\frac{\pi}{2}} &=&  \omega|_{\theta=\frac{\pi}{2}}  = 
\frac{a~\Pi_{\alpha}}{(r^2+a^2)r^2+a^2~\Pi_{\alpha}}~\label{q2}
\end{eqnarray}
The variation of $\Omega$ and $\omega$ in the equatorial plane 
could be seen from Fig.~(\ref{angf}) and Fig.~(\ref{angf1}) respectively. 
It must be noted that both these parameters are depend on MOG parameter. From 
Fig.~(\ref{angf}) and Fig.~(\ref{angf1}), it 
is observed that the three cases namely non-extremal, extremal and NS both in the presence of MOG 
parameter and in the absence of MOG parameter are qualitatively different. This means that three geometries 
are quite distinguished.
\begin{figure}
\begin{center}
\subfigure[]{
\includegraphics[width=2in,angle=0]{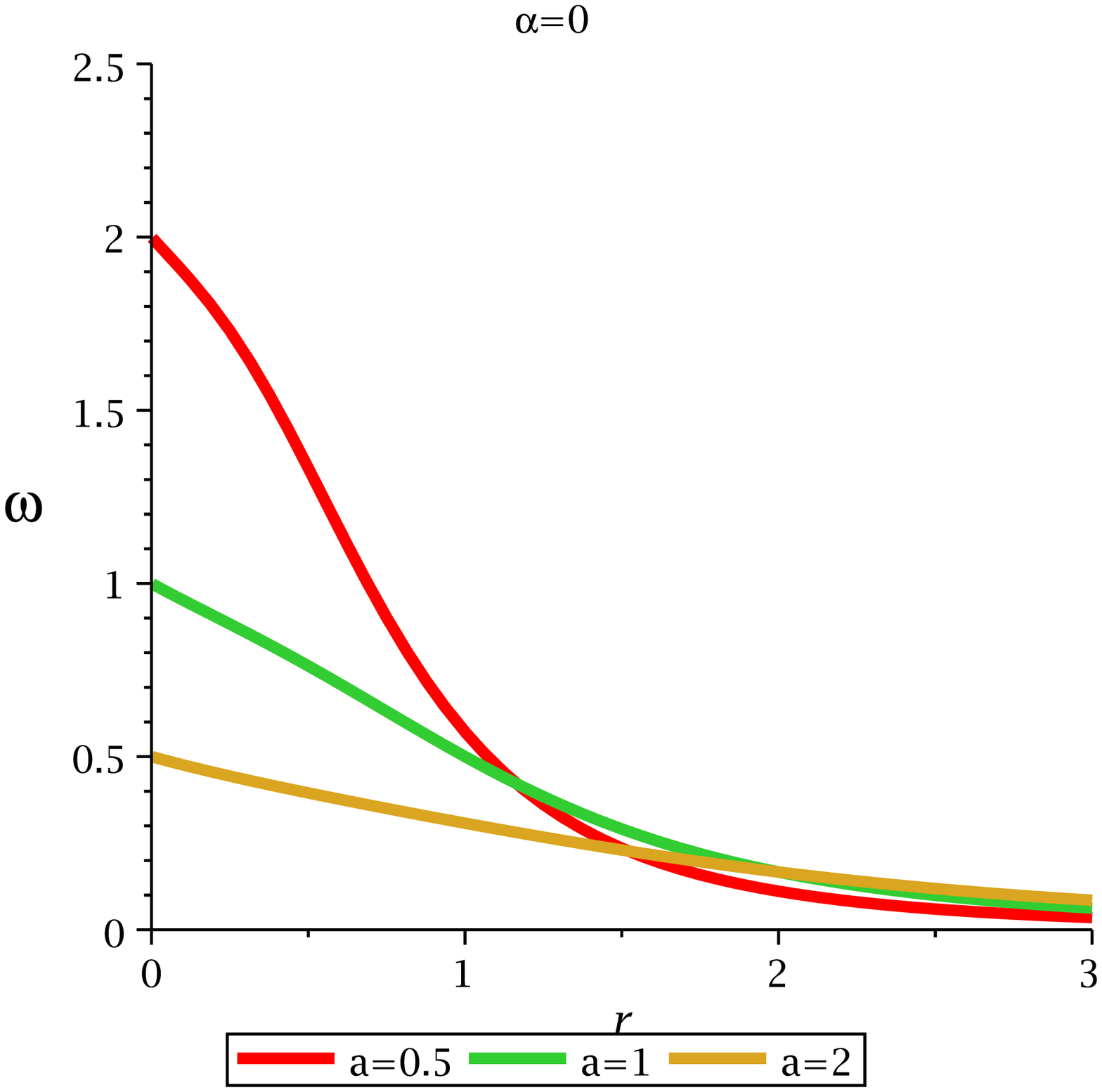}} 
\subfigure[]{
\includegraphics[width=2in,angle=0]{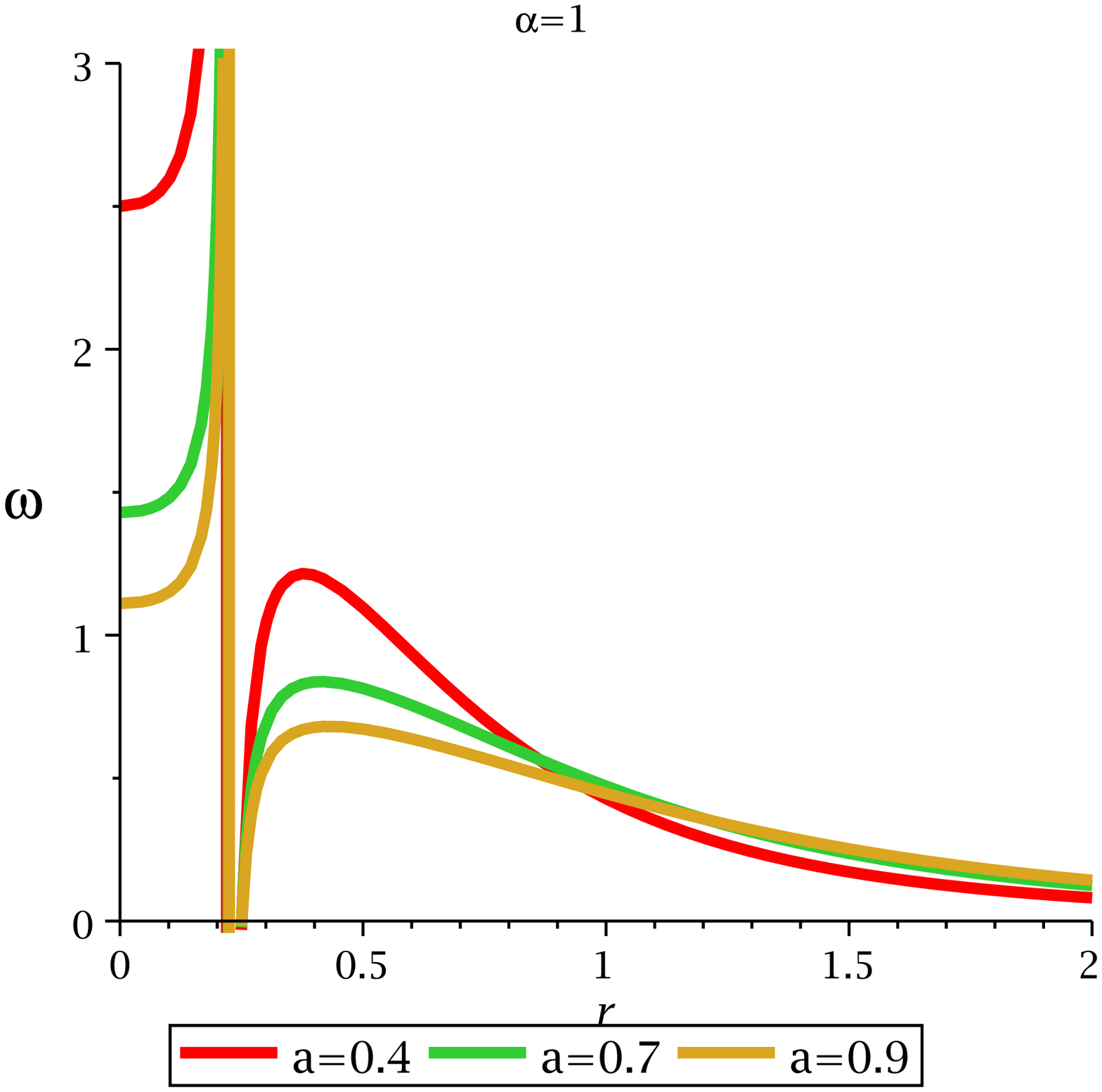}} 
\subfigure[]{
\includegraphics[width=2in,angle=0]{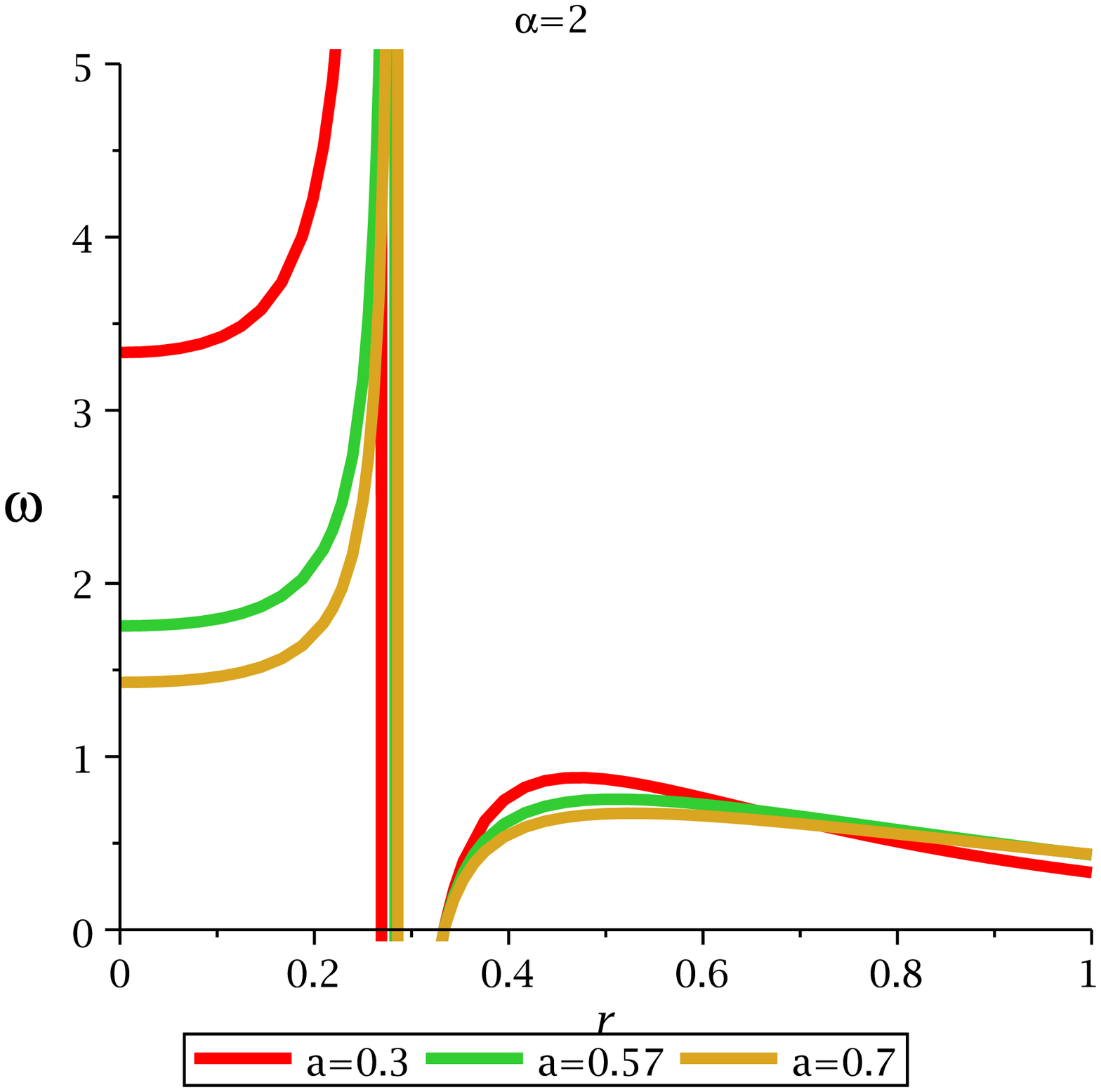}} 
%\subfigure[]{
%\includegraphics[width=2in,angle=0]{omgalpha3.eps}}
%\subfigure[]{
%\includegraphics[width=2in,angle=0]{omgs2.eps}}
%\subfigure[]{
%includegraphics[width=2in,angle=0]{omgs.eps}} 
\caption{The figure depicts the variation  of $\omega$  with $r$ 
for different values of spin parameter in  $\theta=\frac{\pi}{2}$ plane. This 
frequency is a special class of frequency called ZAMO frequency when $\delta=\frac{1}{2}$.
The first figure is considered without MOG parameter~($\alpha=0$) and  for various values of spin 
parameter to differentiate between non-extremal BH, extremal BH and NS. The second and third figure 
is considered with MOG parameter $\alpha=1$ and $\alpha=2$. In first figure, the ZAMO frequency 
decreases for each spin values~$a<1$, $a=1$ and $a>1$ for increasing the value of $r$. While in the 
second figure, the ZAMO frequency first increases then at certain point it diverges then it further 
increases to a certain peak then decreases~(for each spin values).}
\label{angf1} 
\end{center}
\end{figure}

\subsection{Behaviour of $\vec{\Omega}_{p}$ at $r=0$}
In this section we will discuss various cases of LT precession frequency. Also we will see the 
LT precession frequency structure for various useful limits. First we have taken the limit 
$r\rightarrow 0$~(i. e. at the ring singularity $r=0$, $\theta=\frac{\pi}{2}$). It should 
be noted that $\vec{\Omega}_p$ is not valid at the ring singularity~($r=0, \theta=\pi/2$). 
Yet, we should investigate its behavior in its vicinity i.e. in the region 
$r=0$, $0 \leq \theta < 90^{\circ}$. It must be noted that this region is completely outside the 
ergoregion since the ergosurface meets the ring singularity. At the $r=0$, the LT precession 
frequency becomes 
\begin{eqnarray}
\vec{\Omega}_p|_{r= 0}= \frac{\xi(\theta)~\hat{r}+\eta(\theta)~\hat{\theta}}{\zeta(\theta)},
\label{gekn2} 
\end{eqnarray}
and the magnitude of this vector is thus 
\begin{eqnarray}
{\Omega}_p|_{r= 0}= \frac{\sqrt{\xi^2(\theta)+\eta^2(\theta)}}{\zeta(\theta)}
~\label{gn2} 
\end{eqnarray}
where
$$
\xi(\theta)=\sqrt{a^2+\frac{\alpha}{1+\alpha} G_{N}^2{\cal M}^2}\times
$$
\begin{eqnarray}
\left[-\frac{\alpha}{1+\alpha} G_{N}^2{\cal M}^2
-\frac{\Omega}{8}\left\{\left(3+4\cos2\theta+\cos4\theta\right)a^3-8\frac{\alpha}{1+\alpha} G_{N}^2{\cal M}^2
~a\left(1-\cos2\theta \right)\right\}
-\frac{\alpha}{1+\alpha} G_{N}^2{\cal M}^2a^2\Omega^2\sin^4\theta\right]\nonumber\\
\end{eqnarray}
\begin{eqnarray}
\eta(\theta) &=& G_{N} {\cal M}~a^2\cos\theta \sin\theta \left[-1+a\Omega(1+\sin^2\theta)-a^2\Omega^2\sin^2\theta\right]
\end{eqnarray}
$$
\zeta(\theta) = a^2\cos^2\theta \times 
$$
\begin{eqnarray}
\left[\frac{\alpha}{1+\alpha} G_{N}^2{\cal M}^2+a^2\cos^2\theta
-2\frac{\alpha}{1+\alpha} G_{N}^2{\cal M}^2~a\Omega\sin^2\theta-a^2\Omega^2\sin^2\theta
\left(a^2\cos^2\theta-\frac{\alpha}{1+\alpha} G_{N}^2{\cal M}^2~\sin^2\theta\right)\right]\nonumber\\  
\end{eqnarray}
The valid regime of $\Omega$ is 
\begin{eqnarray}
 \Omega_{-}(\theta)< \Omega <\Omega_{+}(\theta)
\end{eqnarray}
where 
\begin{eqnarray}
\Omega_{-}(\theta) &=&-\frac{\frac{\alpha}{1+\alpha} G_{N}^2{\cal M}^2~\sin\theta+\sqrt{D}}
{a\sin\theta\left[a^2-\left(a^2+\frac{\alpha}{1+\alpha} G_{N}^2{\cal M}^2\right)\sin^2\theta\right]} \nonumber \\
\Omega_{+}(\theta) &=& \frac{-\frac{\alpha}{1+\alpha} G_{N}^2{\cal M}^2~\sin\theta
+\sqrt{D}}{a\sin\theta\left[a^2-\left(a^2+\frac{\alpha}{1+\alpha} G_{N}^2{\cal M}^2\right)\sin^2\theta\right]}\nonumber \\
\end{eqnarray}
where 
$$
D=\left(\frac{\alpha}{1+\alpha} G_{N}^2{\cal M}^2\right)^2\sin^2\theta
+\left\{a^2-\left(a^2+\frac{\alpha}{1+\alpha} G_{N}^2{\cal M}^2\right)\sin^2\theta\right\}
\left(\frac{\alpha}{1+\alpha} G_{N}^2{\cal M}^2+a^2\cos^2\theta\right)
$$
For a static observer outside the ergosphere we have to put $\Omega=0$ therefore 
we get from Eq.~(\ref{gekn2})
\begin{eqnarray}
|\vec{\Omega}_{p}|=\frac{\sqrt{\left(\frac{\alpha}{1+\alpha} G_{N}^2{\cal M}^2\right)^2
\left(a^2+\frac{\alpha}{1+\alpha} G_{N}^2{\cal M}^2\right)+{\cal M}^2a^4\sin^2\theta\cos^2\theta}}
{a^2\cos^2\theta\left(\frac{\alpha}{1+\alpha} G_{N}^2{\cal M}^2+a^2\cos^2\theta\right)}
\label{gekn3} 
\end{eqnarray}
It must be noted that $\Omega_{p}$ varies from $0 \leq \Omega_{p} < \infty$ 
for $0 \leq \theta < 90^{\circ}$ at $r=0$. This indicates that it diverges only 
at the ring singularity and in cartesian Kerr-Schild coordinates it is described 
by $x^2+y^2=a^2,~z=0$ while it is finite inside the ring singularity i. e.  
$x^2+y^2<a^2,~z=0$.

\subsection{Behaviour of $\vec{\Omega}_{p}$ at $\theta=0$}
In the limit $\theta=0$, the LT frequency is given by 
\begin{eqnarray}
\vec{\Omega}_p|_{\theta= 0} &=& 
\frac{\xi(r)|_{\theta= 0}}{\zeta(r)|_{\theta= 0}}~\sqrt{\Delta}~\hat{r},~\label{t0}
\end{eqnarray}
{where 
\begin{eqnarray}
\xi(r)|_{\theta= 0} &=&   a \Pi_{\alpha}-\frac{\Omega}{8} \left[8r^4+8a^2r^2+8 a^2\Pi_{\alpha}+3a^4+ 
4a^2(2\Delta-a^2)+a^4\right] \\
\zeta(r)|_{\theta= 0} &=& \rho^3 \left(\rho^2-\Pi_{\alpha}\right)
\end{eqnarray}
Using the identity $r^2+a^2-\Pi_{\alpha}=\Delta$, the above equations reduced to 
\begin{eqnarray}
\xi(r)|_{\theta= 0} &=& a \Pi_{\alpha}-\Omega \left(r^2+a^2\right)^2\\
\zeta(r)|_{\theta= 0} &=& \left(r^2+a^2 \right)^{3\over 2} \Delta
\end{eqnarray}
}
The magnitude of this vector is calculated to be 
\begin{eqnarray}
\Omega_p|_{\theta= 0} &=& \frac{a\Pi_{\alpha}-\Omega \left(r^2+a^2\right)^2}
{ \left(r^2+a^2\right)^\frac{3}{2}\sqrt{\Delta}}~\label{t1}
\end{eqnarray}
From the above equation one could see that the precession frequency is arbitrarily large when $\Delta=0$. The 
precession frequency is positive when $a\Pi_{\alpha}>\Omega(r^2+a^2)^2$ and negative when 
$a\Pi_{\alpha}<\Omega(r^2+a^2)^2$ and becomes zero when $a\Pi_{\alpha}=\Omega(r^2+a^2)^2$. 
From Fig.~(\ref{fgg1}),  one could see {that} the variation of precession frequency with 
radial coordinate without MOG parameter and with MOG parameter. 
From this figure, {one} can distinguish three geometrical structure of the spacetime via 
precession frequency.
\begin{figure}
\begin{center}
\subfigure[]{
\includegraphics[width=2in,angle=0]{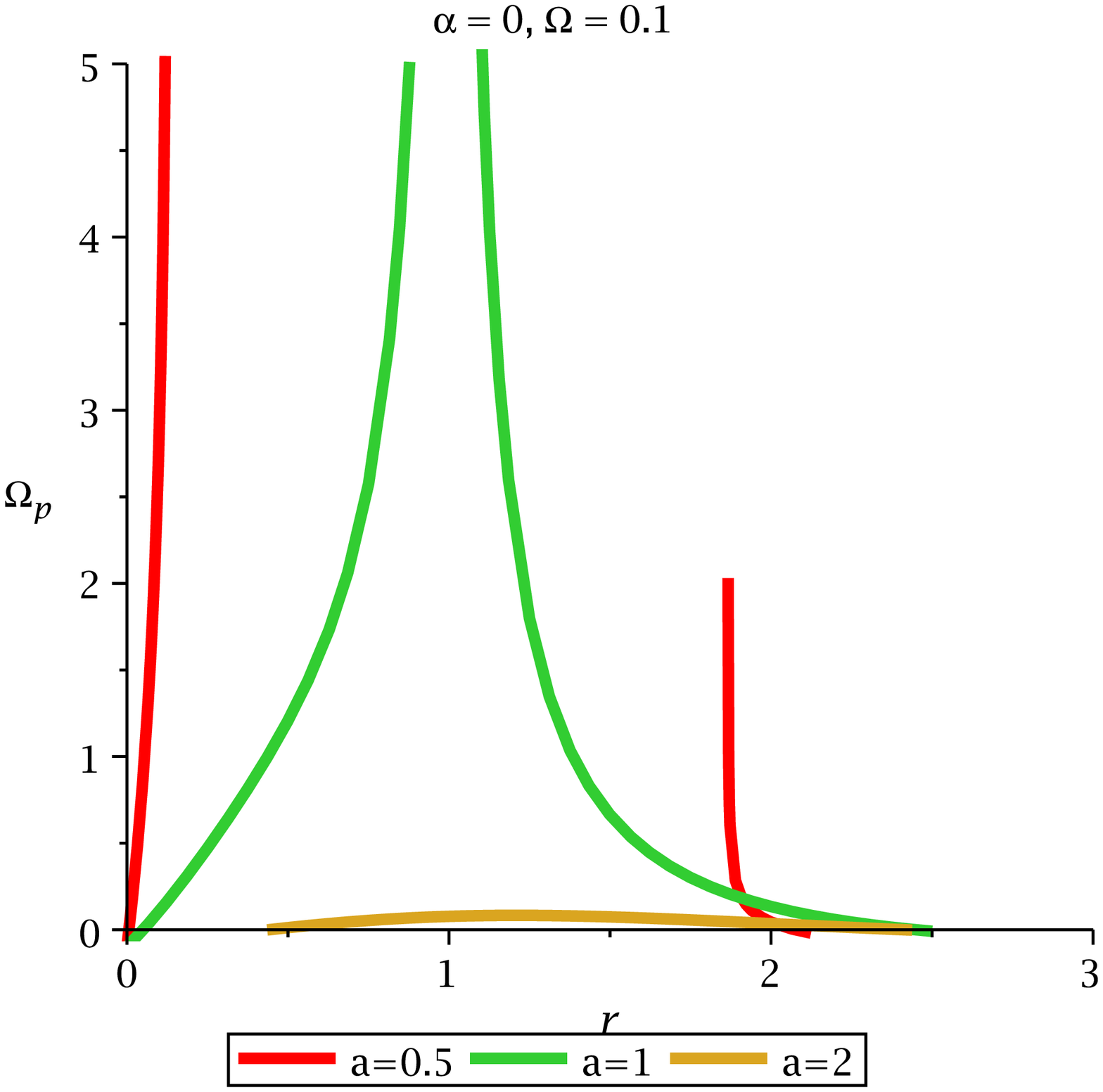}} 
\subfigure[]{
\includegraphics[width=2in,angle=0]{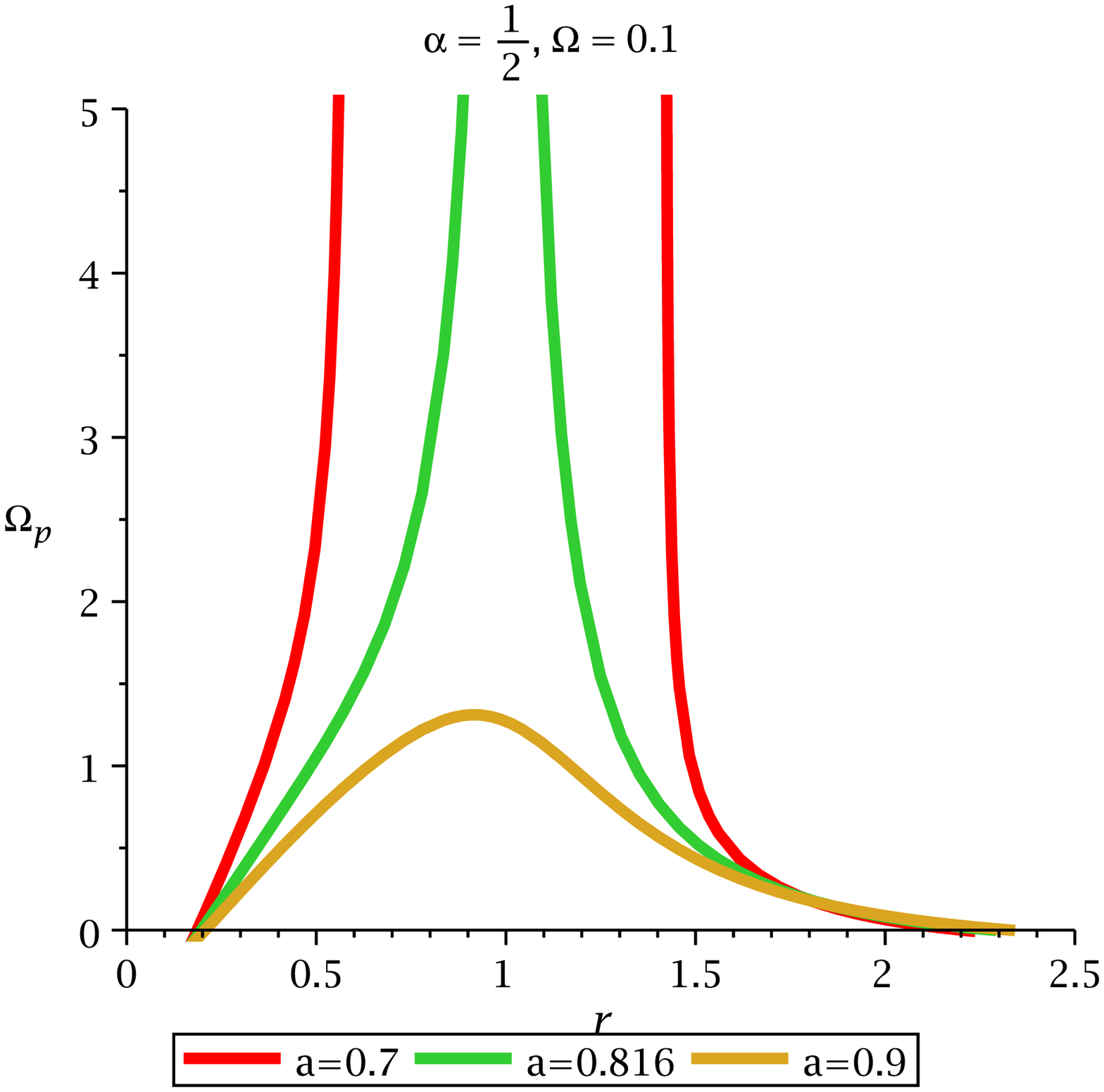}} 
\subfigure[]{
\includegraphics[width=2in,angle=0]{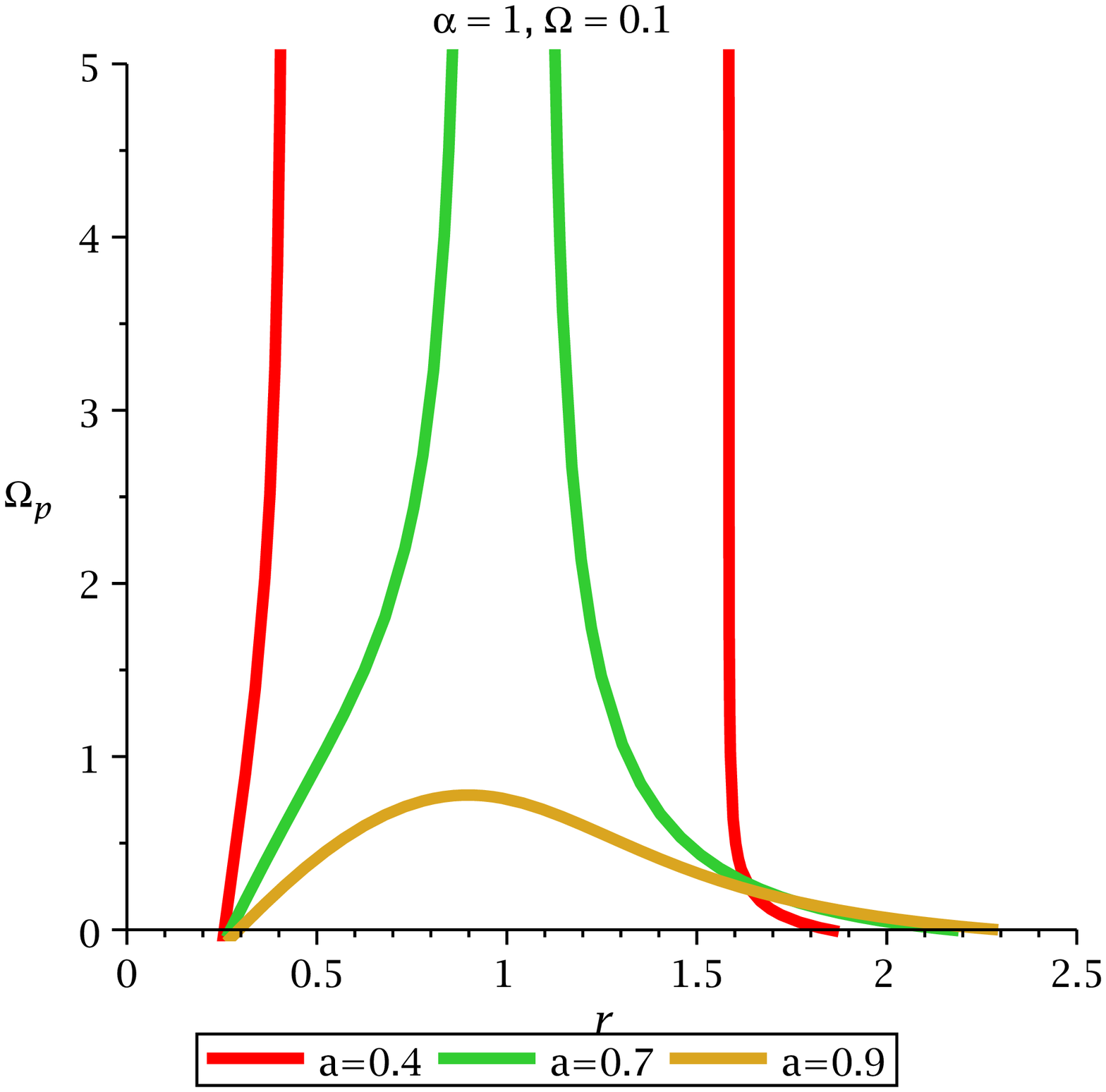}} 
\subfigure[]{
\includegraphics[width=2in,angle=0]{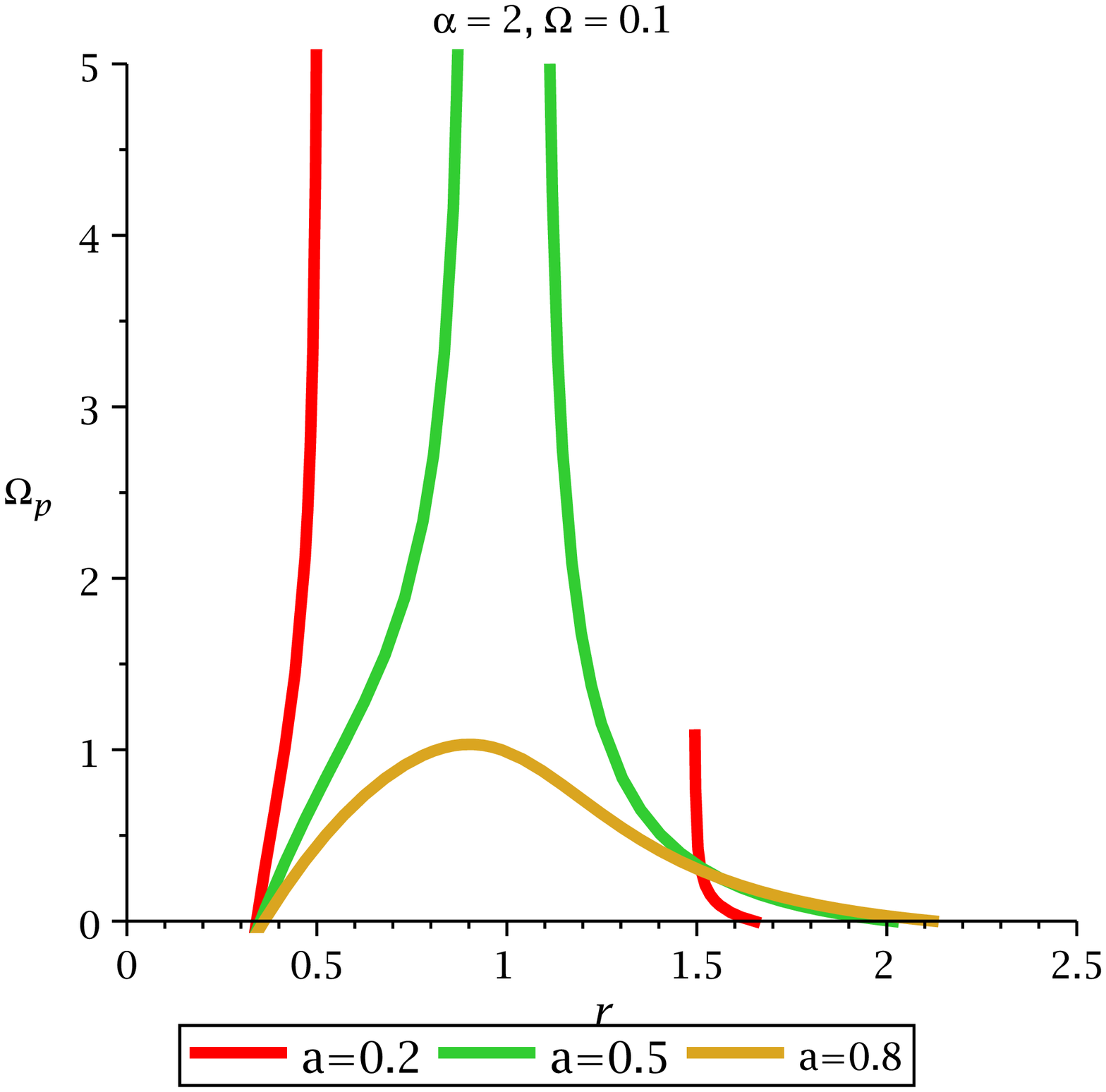}}
\subfigure[]{
\includegraphics[width=2in,angle=0]{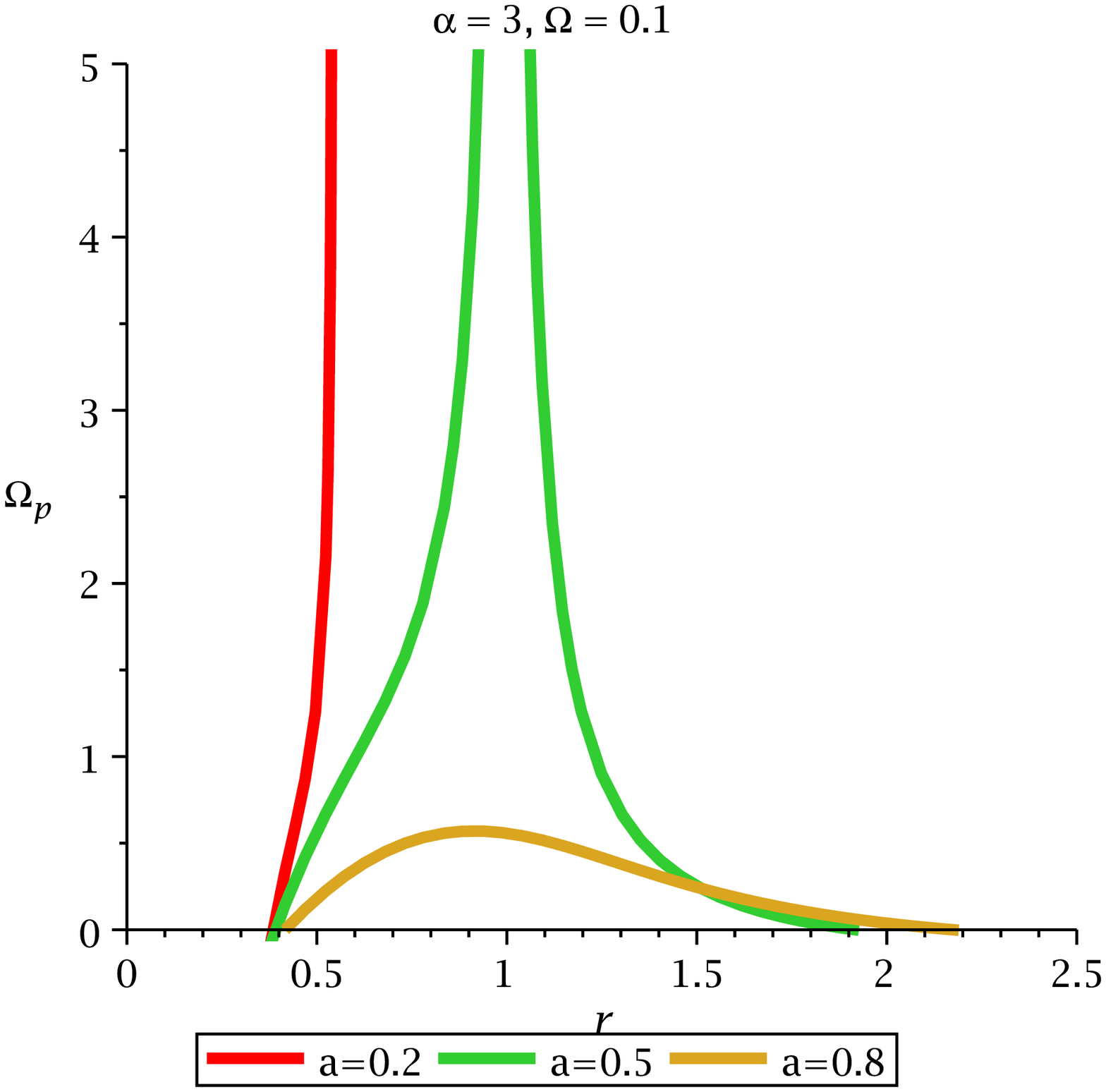}} 
\subfigure[]{
\includegraphics[width=2in,angle=0]{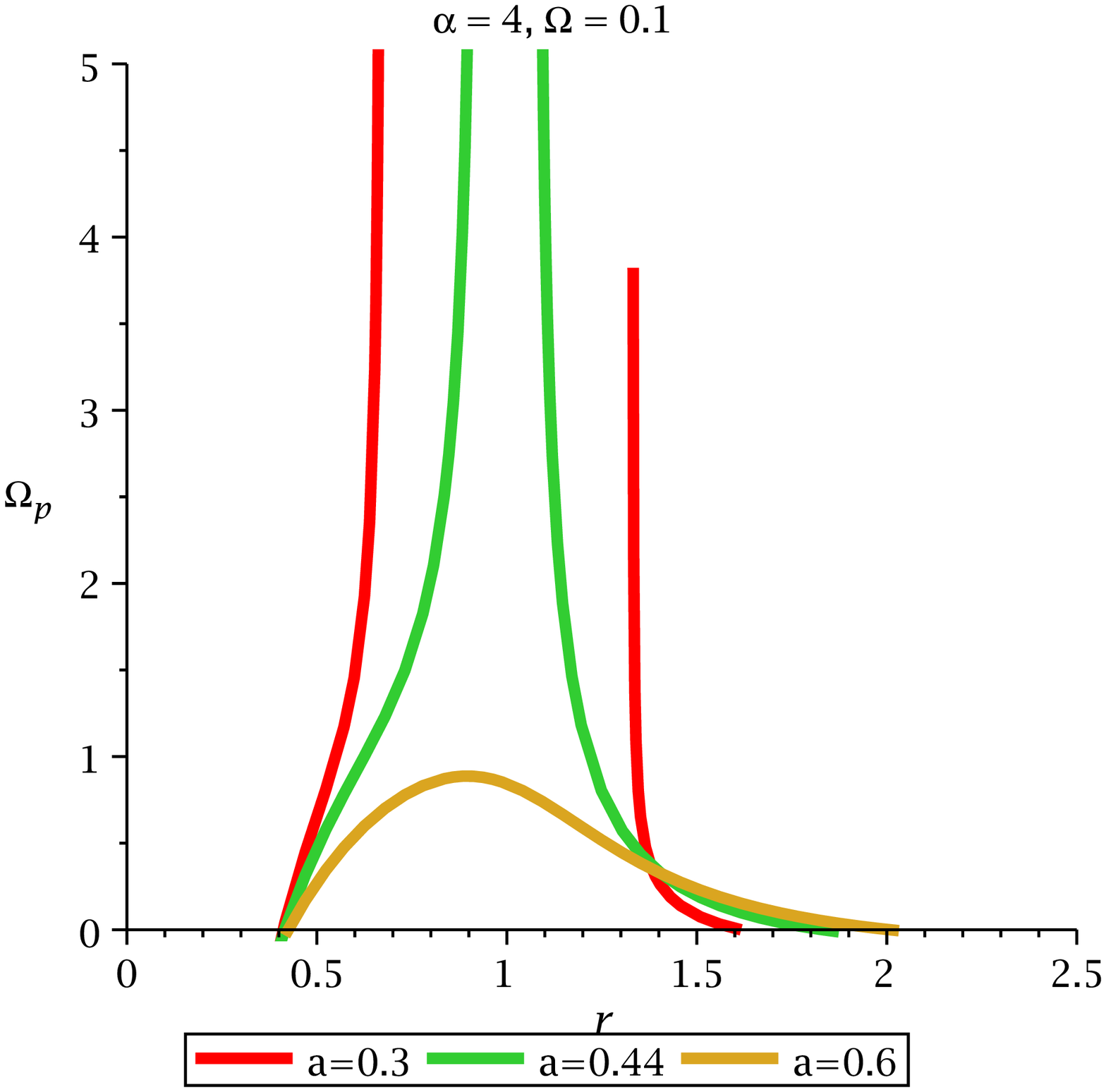}}
\caption{The figure describes the variation  of precession frequency $\Omega_{p}$  
with $r$ for $\theta=0$. In this plot we have chosen the value of $\Omega=0.1 $. We 
have varied the spin parameter values for three cases namely non-extremal BH, extremal BH and NS. 
Each plot shows the difference between precession frequency $\Omega_{p}$ having MOG parameter 
and without MOG parameter. }
\label{fgg1}
\end{center}
\end{figure}

\subsection{Behaviour of $\vec{\Omega}_{p}$ at $\theta=\frac{\pi}{6}$}
In the limit $\theta=\frac{\pi}{6}$, the LT frequency is computed as 
\begin{eqnarray}
\vec{\Omega}_{p}|_{\theta= \frac{\pi}{6}} &=& 
\frac{\sqrt{3 \Delta}\xi(r)|_{\theta=\frac{\pi}{6}}\hat{r}+\eta(r)|_{\theta=\frac{\pi}{6}} \hat{\theta}}
{2\zeta(r)|_{\theta= \frac{\pi}{6}}}~,~\label{t2}
\end{eqnarray}
The magnitude of this vector is turned out to be 
\begin{eqnarray}
\Omega_p|_{\theta=\frac{\pi}{6}} &=& \frac{\sqrt{3\Delta \xi^2 (r)|_{\theta=\frac{\pi}{6}}+\eta^2(r)|_{\theta=\frac{\pi}{6}}}}
{2\zeta(r)|_{\theta= \frac{\pi}{6}}} ~\label{t3}
\end{eqnarray}
where 
\begin{eqnarray}
\xi~(r)|_{\theta= \frac{\pi}{6}} &=& a \Pi_{\alpha}-\frac{\Omega}{16}\left(16r^4+24a^2r^2+8a^2\Pi_{\alpha}+9a^4 \right)+
\frac{\Omega^2}{16}a^3\Pi_{\alpha}
\end{eqnarray}
$$
\eta(r)|_{\theta= \frac{\pi}{6}} = aG_{N}{\cal M}\left(r^2-\frac{3}{4}a^2\right)
-\frac{\alpha}{1+\alpha} G_{N}^2 {\cal M}^2 ar+\Omega\times
$$
$$
 \left[r^5-3G_{N}{\cal M}r^4+\frac{3}{2} a^2r^3-2G_{N}{\cal M}a^2r^2+\frac{9}{16}a^4 r
+\frac{15}{16} G_{N}{\cal M}a^4+\frac{2\alpha}{1+\alpha}G_{N}^2{\cal M}^2r \left(r^2+a^2\right)\right]
$$
\begin{eqnarray}
+\frac{a\Omega^2}{4}\left[G_{N}{\cal M}\left\{3r^4+a^2r^2+\frac{3}{4}a^2\left(r^2-a^2\right)\right\}
-\frac{\alpha}{1+\alpha} G_{N}^2 {\cal M}^2 r\left(2r^2+\frac{7}{4}a^2\right)\right] 
~\label{t4}
\end{eqnarray}
$$
\zeta(r)|_{\theta= \frac{\pi}{6}} = \left(r^2+\frac{3}{4}a^2\right)^\frac{3}{2}\times
$$
\begin{eqnarray}
\left[\left(r^2+\frac{3}{4}a^2\right)-\Pi_{\alpha} +\frac{a\Omega \Pi_{\alpha}}{2}-\frac{\Omega^2}{4} 
\left\{(r^2+\frac{3}{4}a^2)(r^2+a^2)+\frac{a^2\Pi_{\alpha}}{4}\right\}\right]
.\label{t5}
\end{eqnarray}
Variation of spin precession frequency with radial coordinate with MOG parameter and without MOG parameter 
could be seen from  Fig.~(\ref{fgg2}). In Fig.~(\ref{fgg2}-a), Fig.~(\ref{fgg2}-b) and Fig.~(\ref{fgg2}-c), 
we have plotted the precession frequency with radial variable for non-extremal BH, extremal BH and NS of Kerr 
spacetime. While in Fig.~(\ref{fgg2}-d), Fig.~(\ref{fgg2}-e) and Fig.~(\ref{fgg2}-f), we have plotted the 
precession frequency for  non-extremal BH, extremal BH and NS of KMOG spacetime. 
In upper panel~[(\ref{fgg2}-a), Fig.~(\ref{fgg2}-b), Fig.~(\ref{fgg2}-c)], we can see that for BH spacetime 
the precession frequency has one minimum value while for NS it has two minima. In lower panel, 
~[(\ref{fgg2}-d), Fig.~(\ref{fgg2}-e), Fig.~(\ref{fgg2}-f)], the generalized  spin frequency has one minima for 
BH spacetime and one peak for NS.

\begin{figure}
\begin{center}
\subfigure[]{
\includegraphics[width=2in,angle=0]{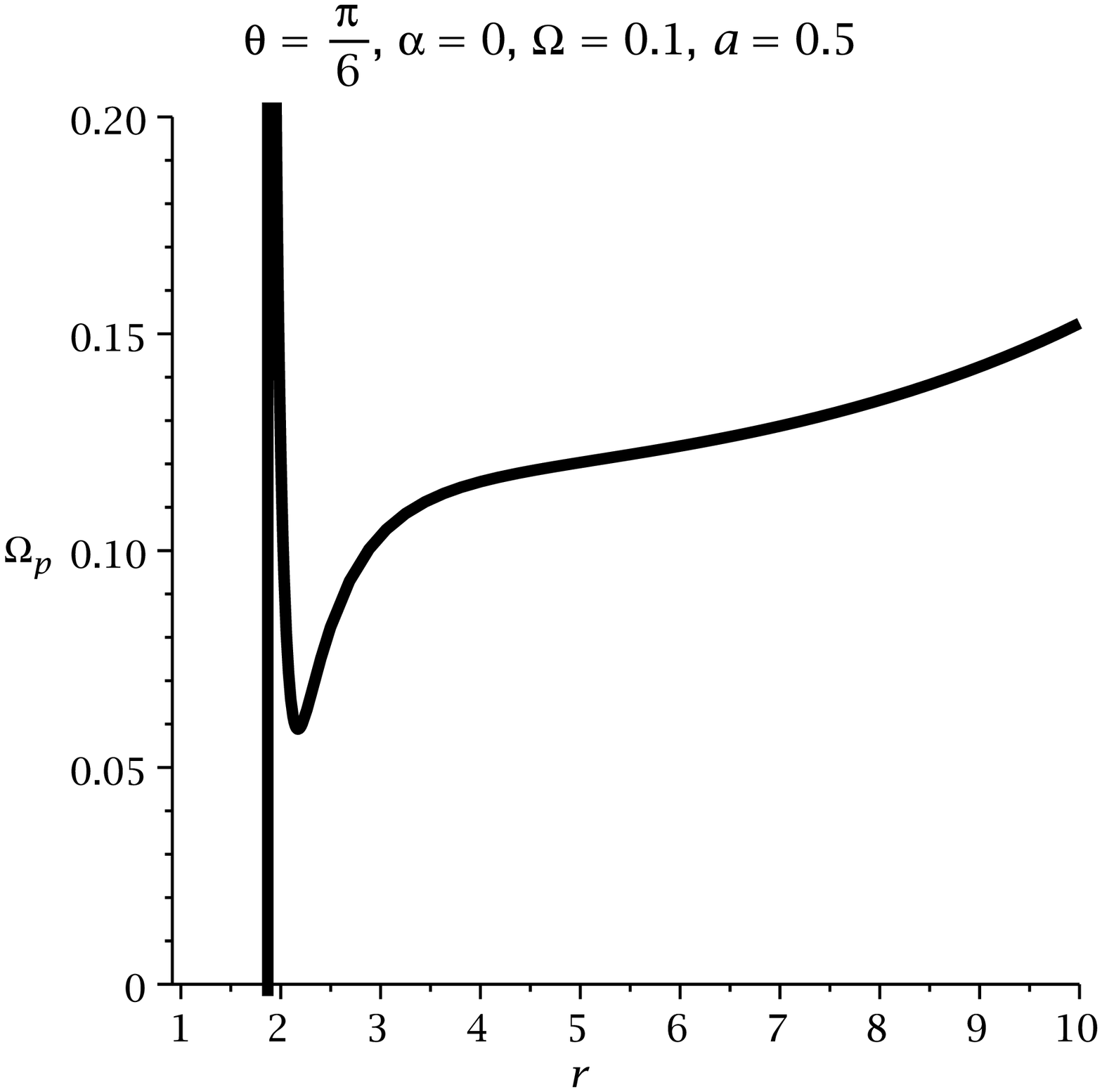}} 
\subfigure[]{
\includegraphics[width=2in,angle=0]{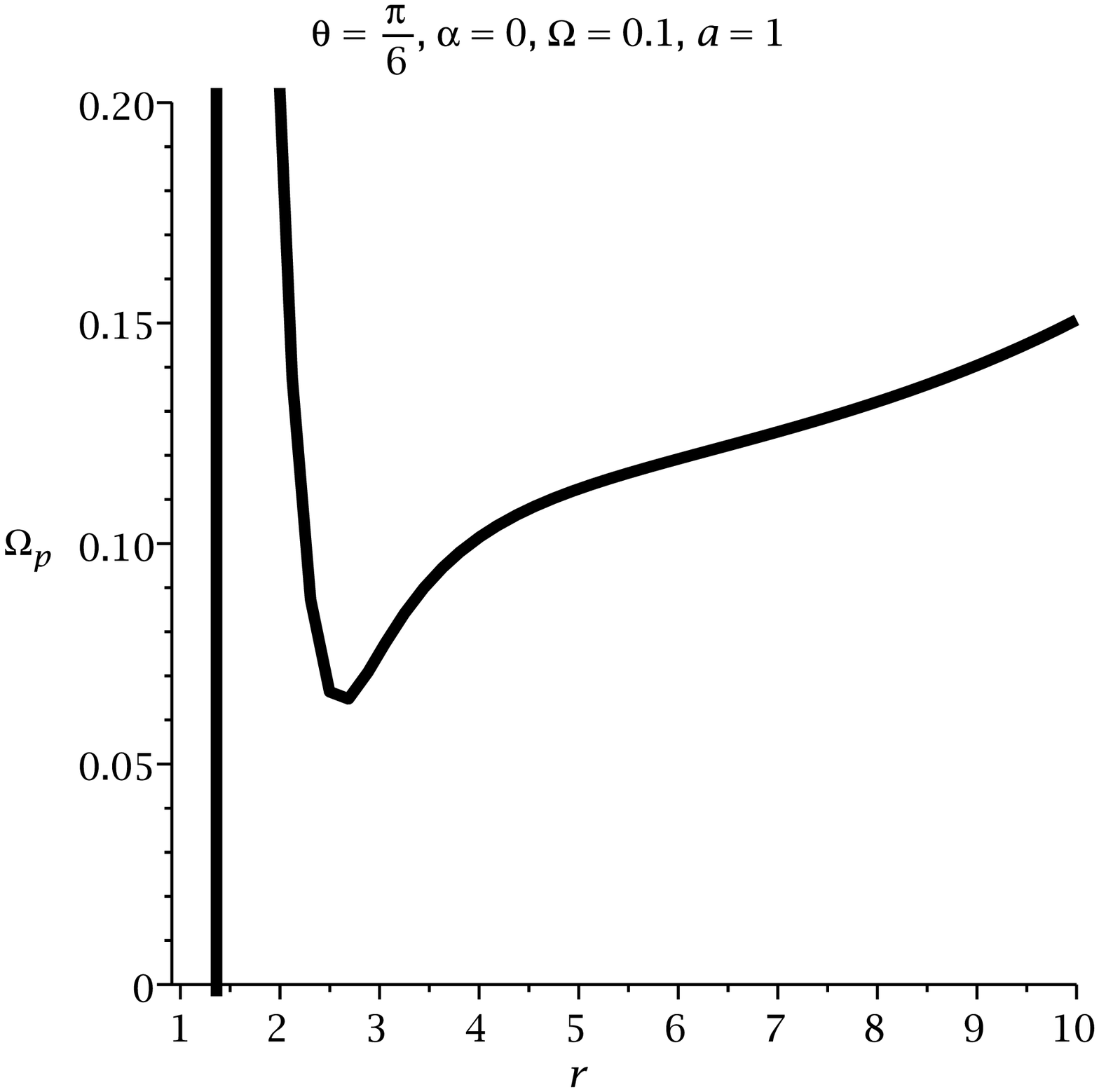}}
\subfigure[]{
\includegraphics[width=2in,angle=0]{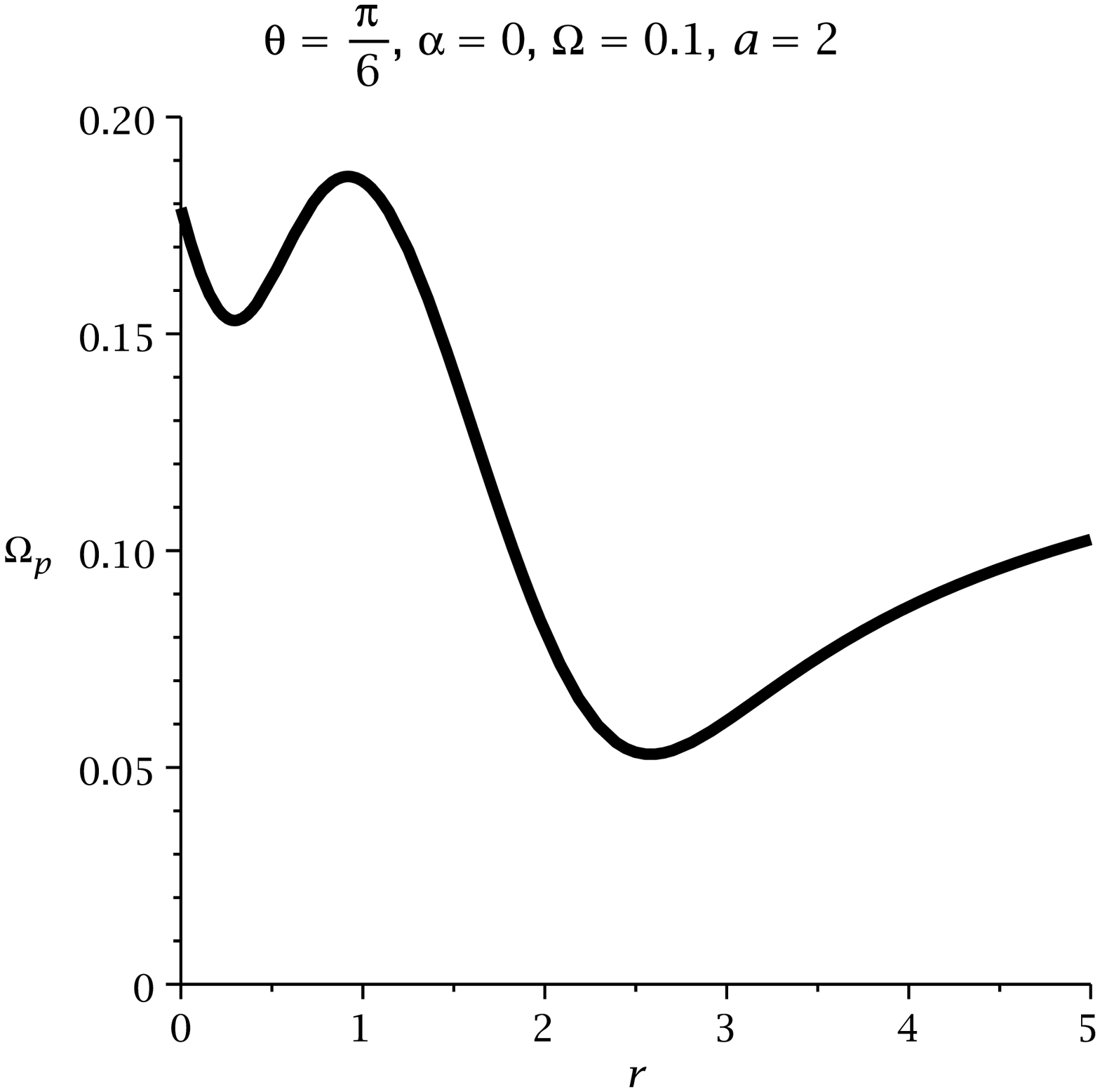}} 
\subfigure[]{
\includegraphics[width=2in,angle=0]{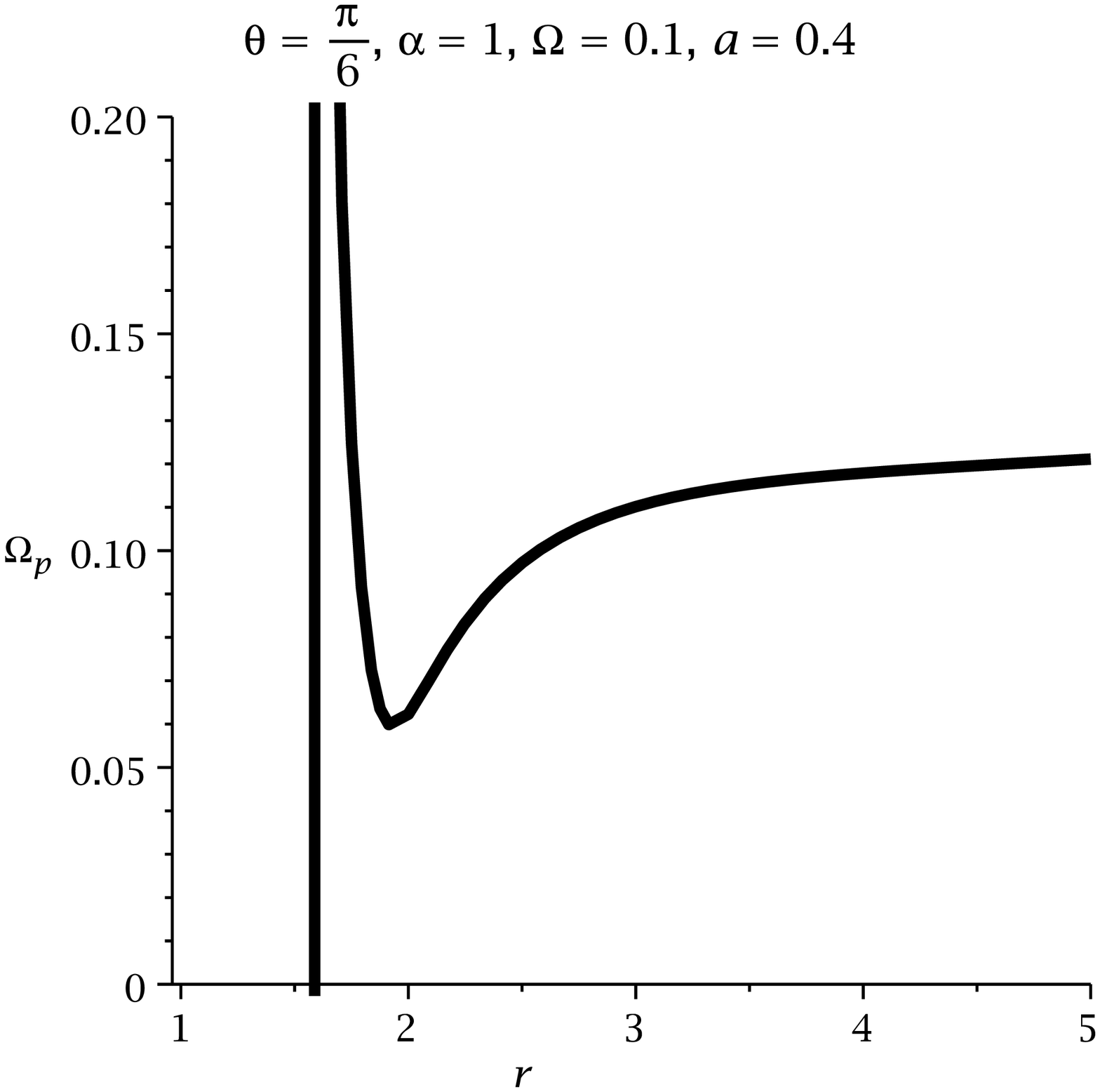}}
\subfigure[]{
\includegraphics[width=2in,angle=0]{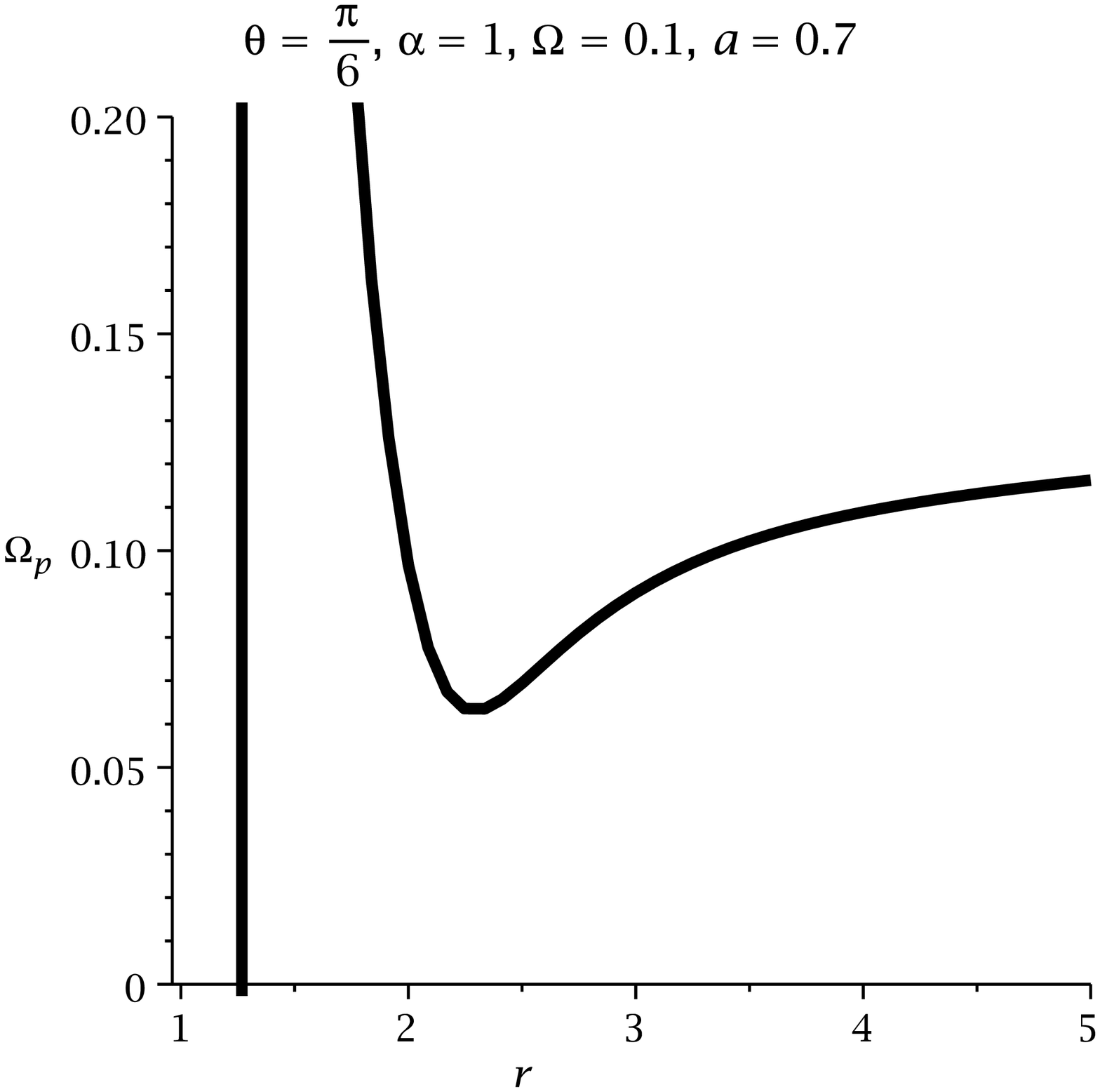}}
\subfigure[]{
\includegraphics[width=2in,angle=0]{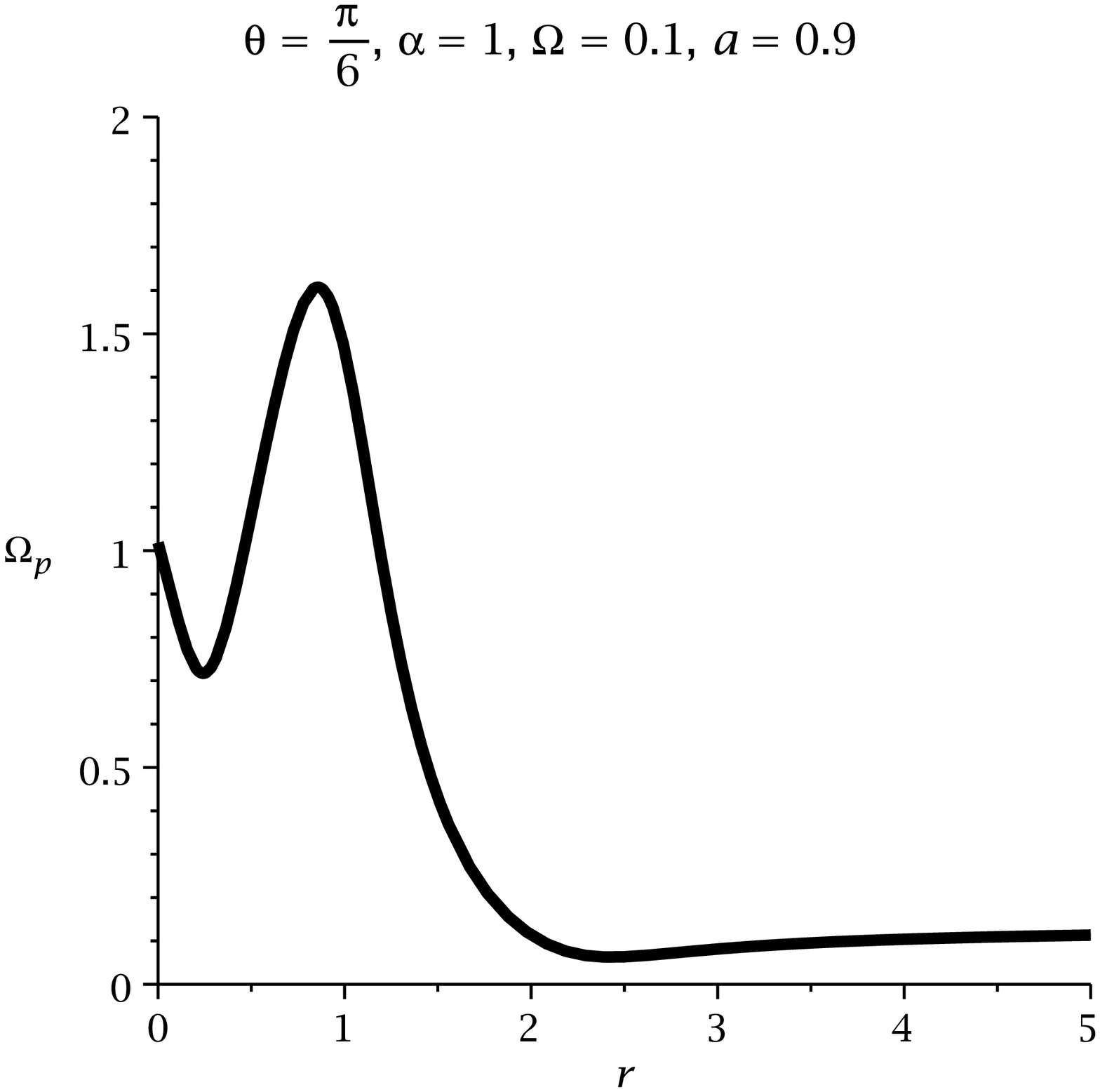}} 
\caption{Examples of the variation  of $\Omega_{p}$ with $r$ for $\theta=\frac{\pi}{6}$ in KMOG with variation 
of MOG parameter, spin parameter and $\Omega$. The upper panel describes the variation  of $\Omega_{p}$  
with $r$ for non-extremal BH, extremal BH and NS without MOG parameter. The lower panel describes the 
variation  of $\Omega_{p}$  with $r$ for non-extremal BH, extremal BH and NS with MOG parameter. }
\label{fgg2}
\end{center}
\end{figure}

\subsection{Behaviour of $\vec{\Omega}_{p}$ at $\theta=\frac{\pi}{4}$}
In the limit $\theta=\frac{\pi}{4}$, the LT frequency is derived as 
\begin{eqnarray}
\vec{\Omega}_{p}|_{\theta= \frac{\pi}{4}} &=& 
\frac{\sqrt{\Delta}\xi(r)|_{\theta=\frac{\pi}{4}}\hat{r}+\eta(r)|_{\theta=\frac{\pi}{4}} \hat{\theta}}
{\sqrt{2} \zeta(r)|_{\theta= \frac{\pi}{4}}}~,~\label{t6}
\end{eqnarray}
The magnitude of this vector is  
\begin{eqnarray}
\Omega_p|_{\theta=\frac{\pi}{4}} &=& \frac{\sqrt{\Delta \xi^2 (r)|_{\theta=\frac{\pi}{4}}
+\eta^2(r)|_{\theta=\frac{\pi}{4}}}}{\sqrt{2}\zeta(r)|_{\theta= \frac{\pi}{4}}} ~\label{t7}
\end{eqnarray}
where
\begin{eqnarray}
\xi~(r)|_{\theta= \frac{\pi}{4}} &=& a \Pi_{\alpha}-\frac{\Omega}{8}\left[8r^4+8a^2r^2+8a^2\Pi_{\alpha}+2a^4 \right]+
\frac{\Omega^2}{4}a^3\Pi_{\alpha}~\label{t8}
\end{eqnarray}
$$
\eta(r)|_{\theta= \frac{\pi}{4}} = aG_{N}{\cal M}\left(r^2-\frac{a^2}{2}\right)
-\frac{\alpha}{1+\alpha} G_{N}^2 {\cal M}^2 ar+\frac{\Omega}{4}\times
$$
$$
 \left[4r^5-12G_{N}{\cal M}r^4+4 a^2r^3-8G_{N}{\cal M}a^2r^2+a^4 r
+3 G_{N}{\cal M}a^4+\frac{8\alpha}{1+\alpha}G_{N}^2{\cal M}^2r\left(r^2+a^2\right)\right]
$$
\begin{eqnarray}
+\frac{a\Omega^2}{4}\left[G_{N}{\cal M}\left(6r^4+3a^2r^2-a^4\right)
-\frac{\alpha}{1+\alpha} G_{N}^2 {\cal M}^2 r\left(2r^2+a^2\right)\right] 
~\label{t9}
\end{eqnarray}
$$
\zeta(r)|_{\theta= \frac{\pi}{4}} = \left(r^2+\frac{a^2}{2}\right)^\frac{3}{2}\times
$$
\begin{eqnarray}
\left[\left(r^2+\frac{a^2}{2}\right)-\Pi_{\alpha} +a\Omega \Pi_{\alpha} -\frac{\Omega^2}{2} 
\left\{\left(r^2+\frac{a^2}{2}\right) \left(r^2+a^2\right)+\frac{a^2\Pi_{\alpha}}{2}\right\}\right]~.\label{t10}
\end{eqnarray}
Variation of spin precession frequency with radial coordinate with MOG parameter and without MOG parameter 
may be seen from  Fig.~(\ref{fgg3}). In Fig.~(\ref{fgg3}-a), Fig.~(\ref{fgg3}-b) and Fig.~(\ref{fgg3}-c), 
we have plotted the precession frequency with radial variable for non-extremal BH, extremal BH and NS of Kerr 
spacetime. While in Fig.~(\ref{fgg2}-d), Fig.~(\ref{fgg2}-e) and Fig.~(\ref{fgg2}-f), we have plotted the 
precession frequency for  non-extremal BH, extremal BH and NS of KMOG spacetime. 
In upper panel~[(\ref{fgg3}-a), Fig.~(\ref{fgg3}-b), Fig.~(\ref{fgg3}-c)], one can see that for BH spacetime 
the precession frequency has one minimum value while for NS it has one minimum value and one maximum value. 
In lower panel, ~[(\ref{fgg3}-d), Fig.~(\ref{fgg3}-e), Fig.~(\ref{fgg3}-f)], the generalized  spin frequency 
has one minima for BH spacetime and one peak for NS.

\begin{figure}
\begin{center}
\subfigure[]{
\includegraphics[width=2in,angle=0]{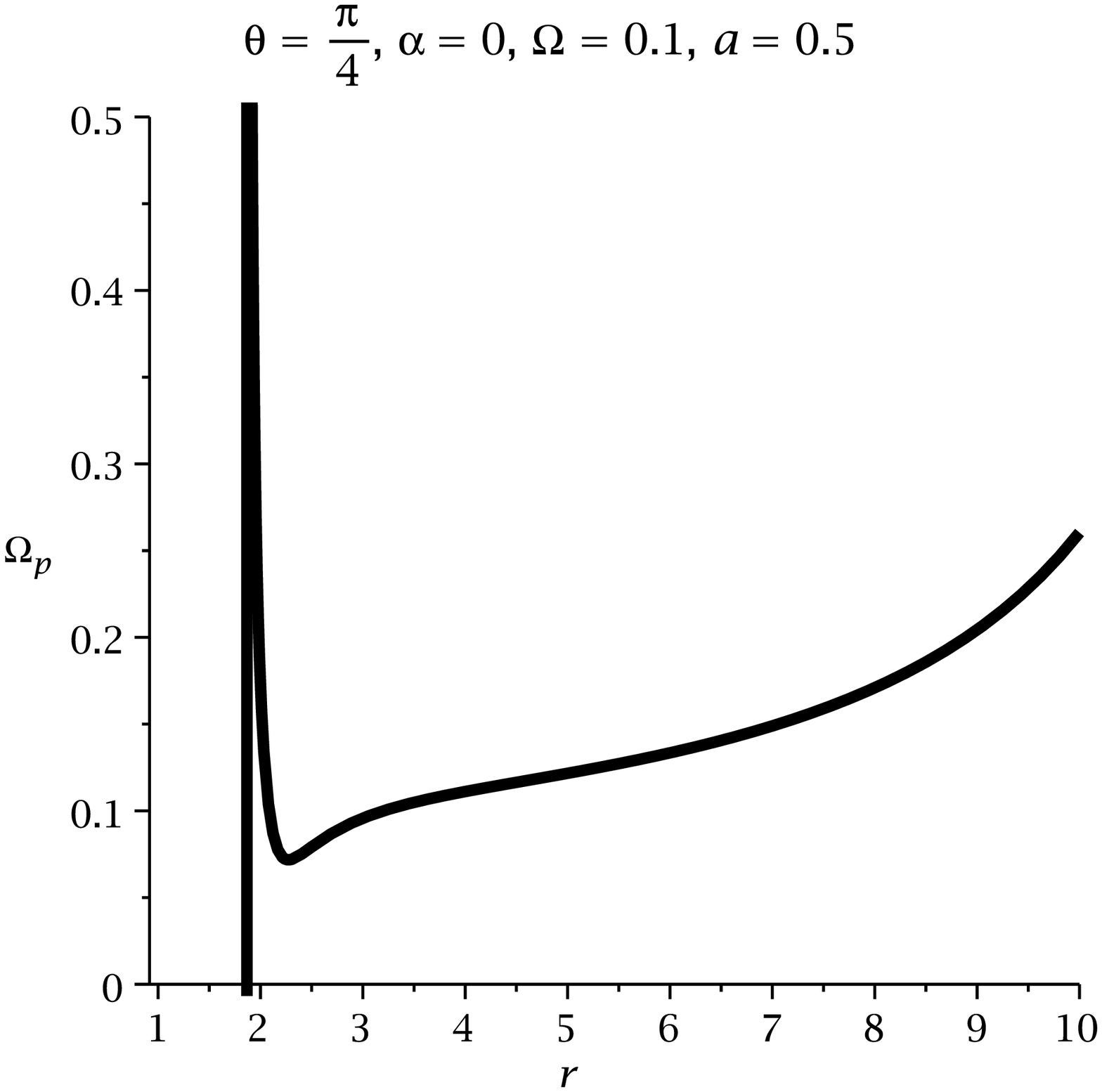}} 
\subfigure[]{
\includegraphics[width=2in,angle=0]{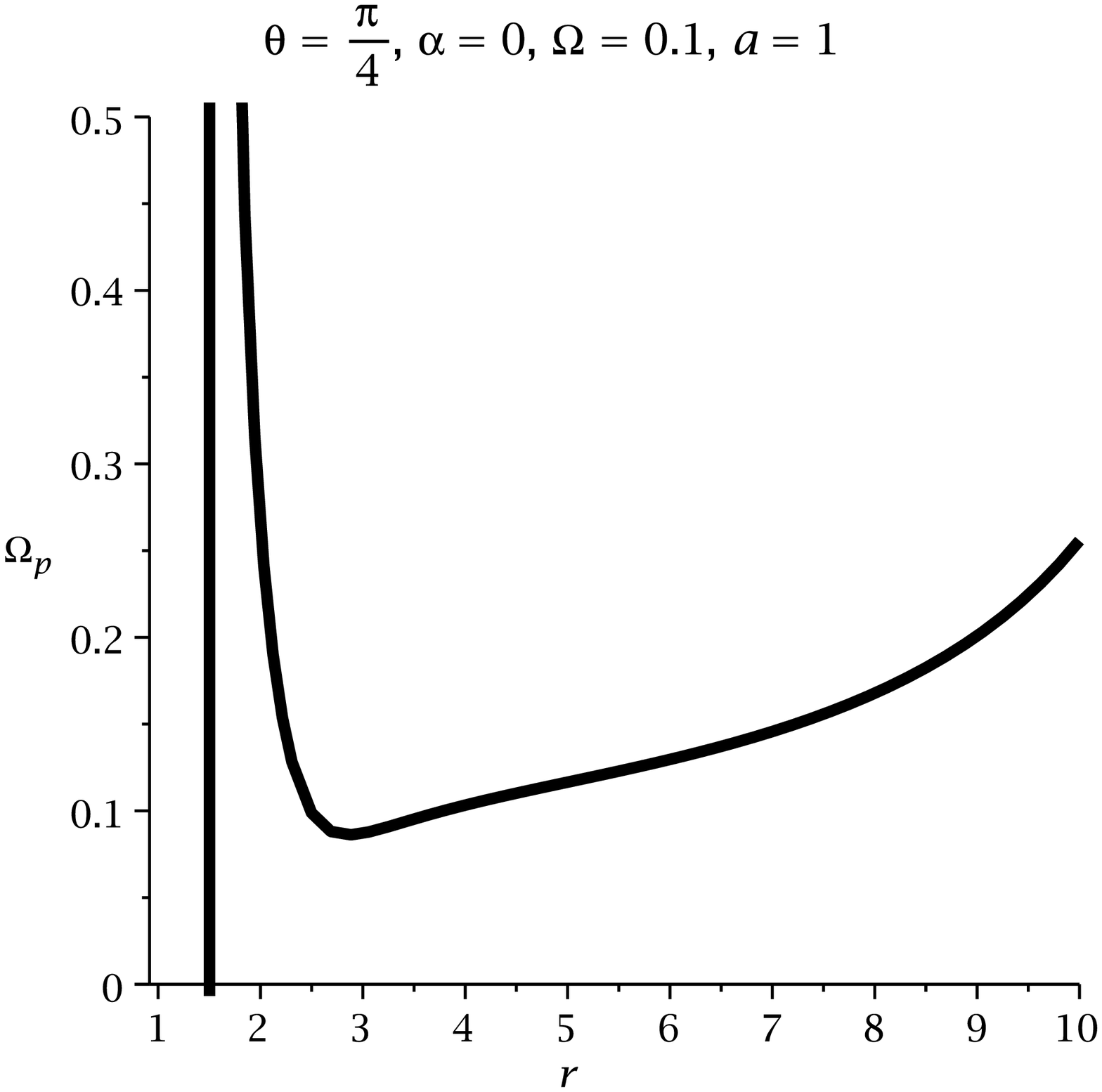}}
\subfigure[]{
\includegraphics[width=2in,angle=0]{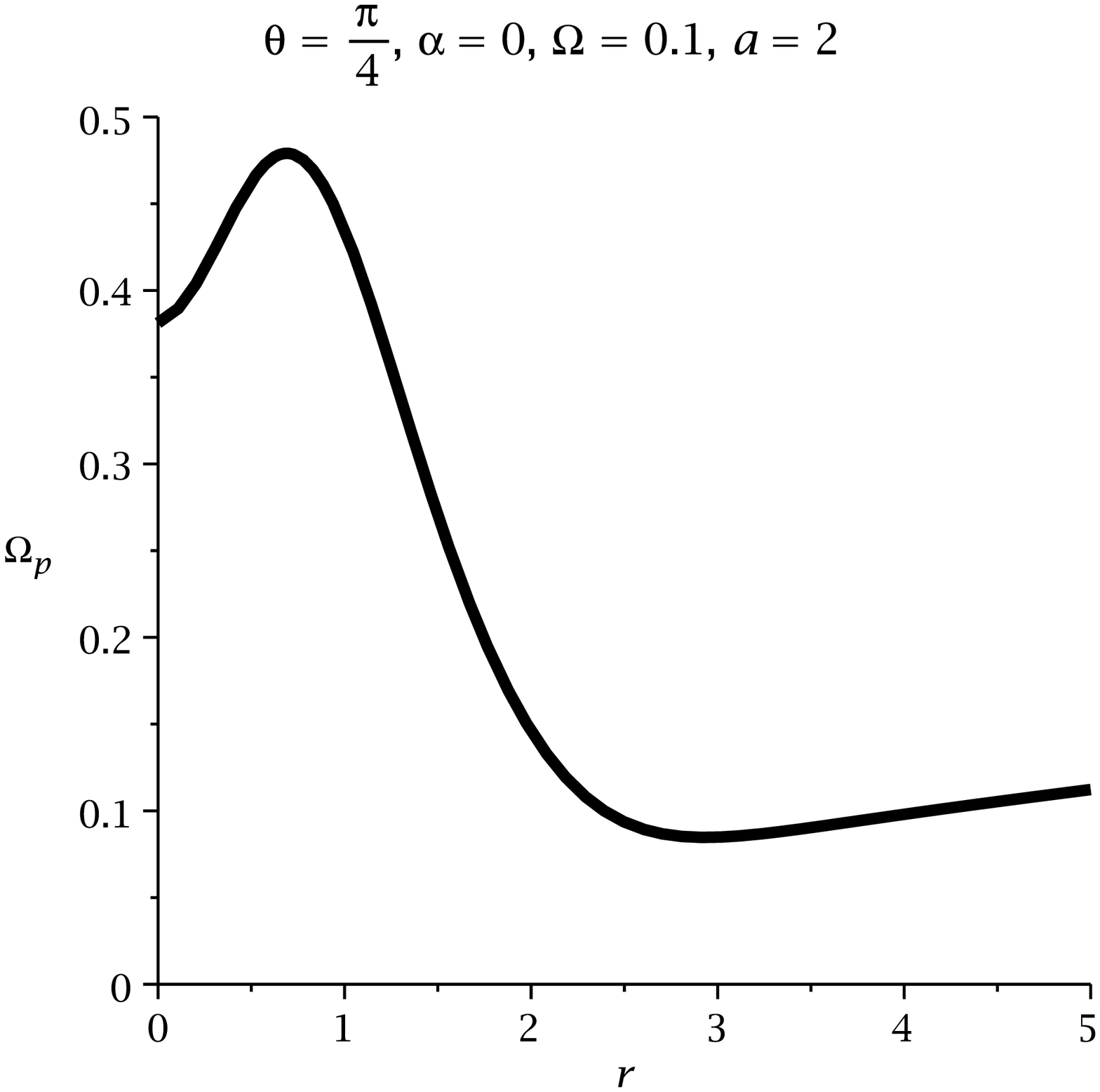}} 
\subfigure[]{
\includegraphics[width=2in,angle=0]{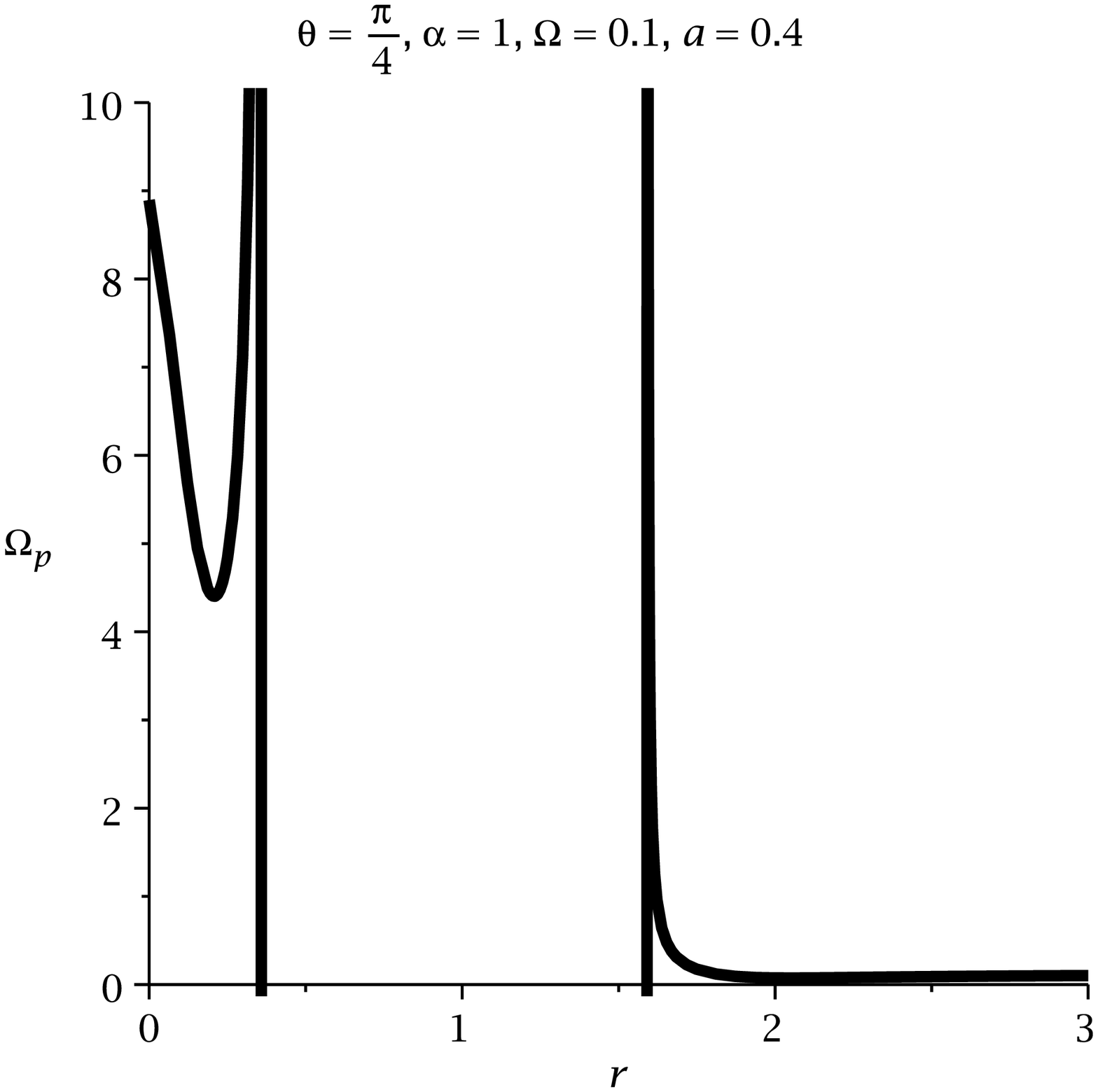}}
\subfigure[]{
\includegraphics[width=2in,angle=0]{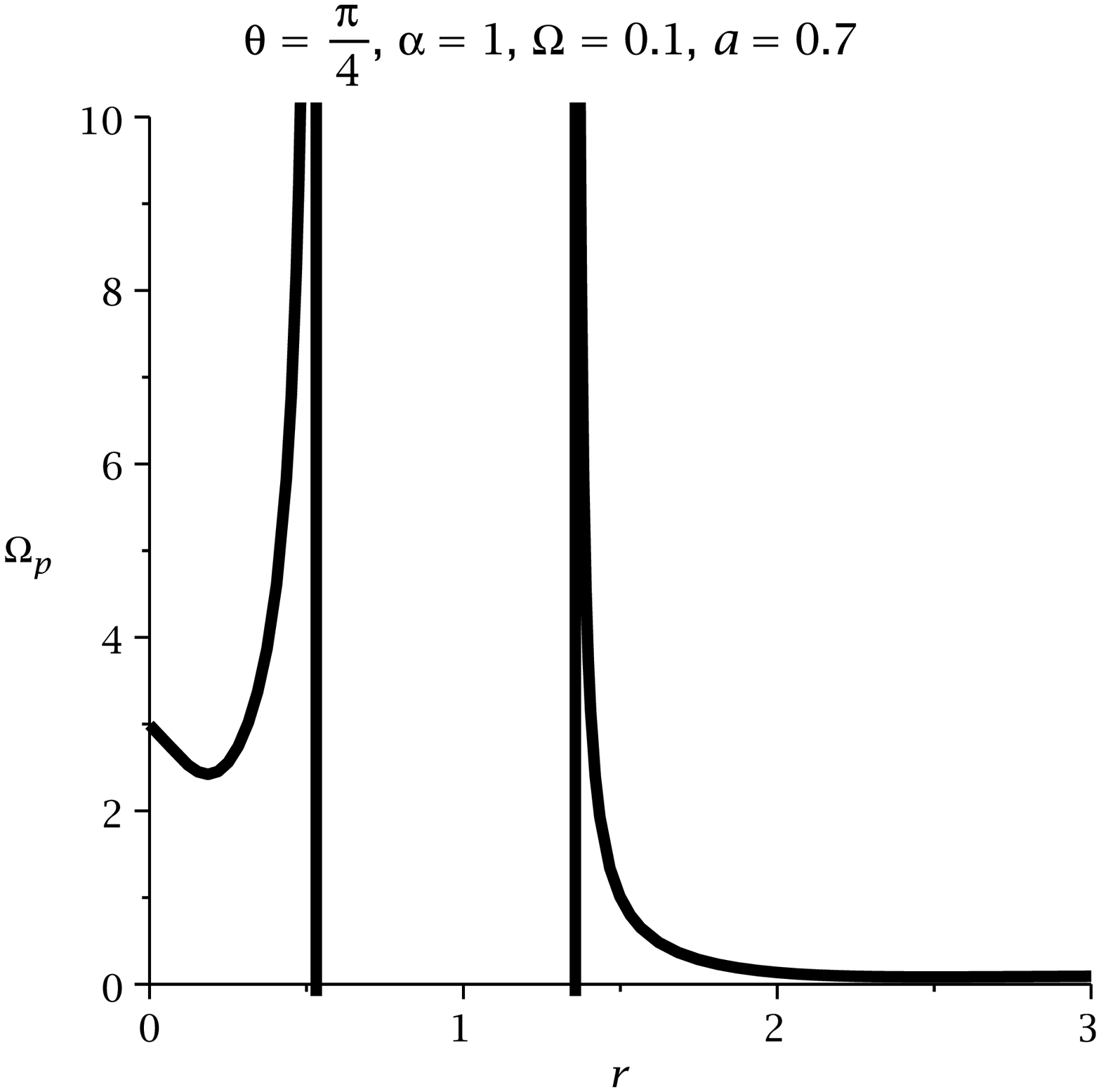}}
\subfigure[]{
\includegraphics[width=2in,angle=0]{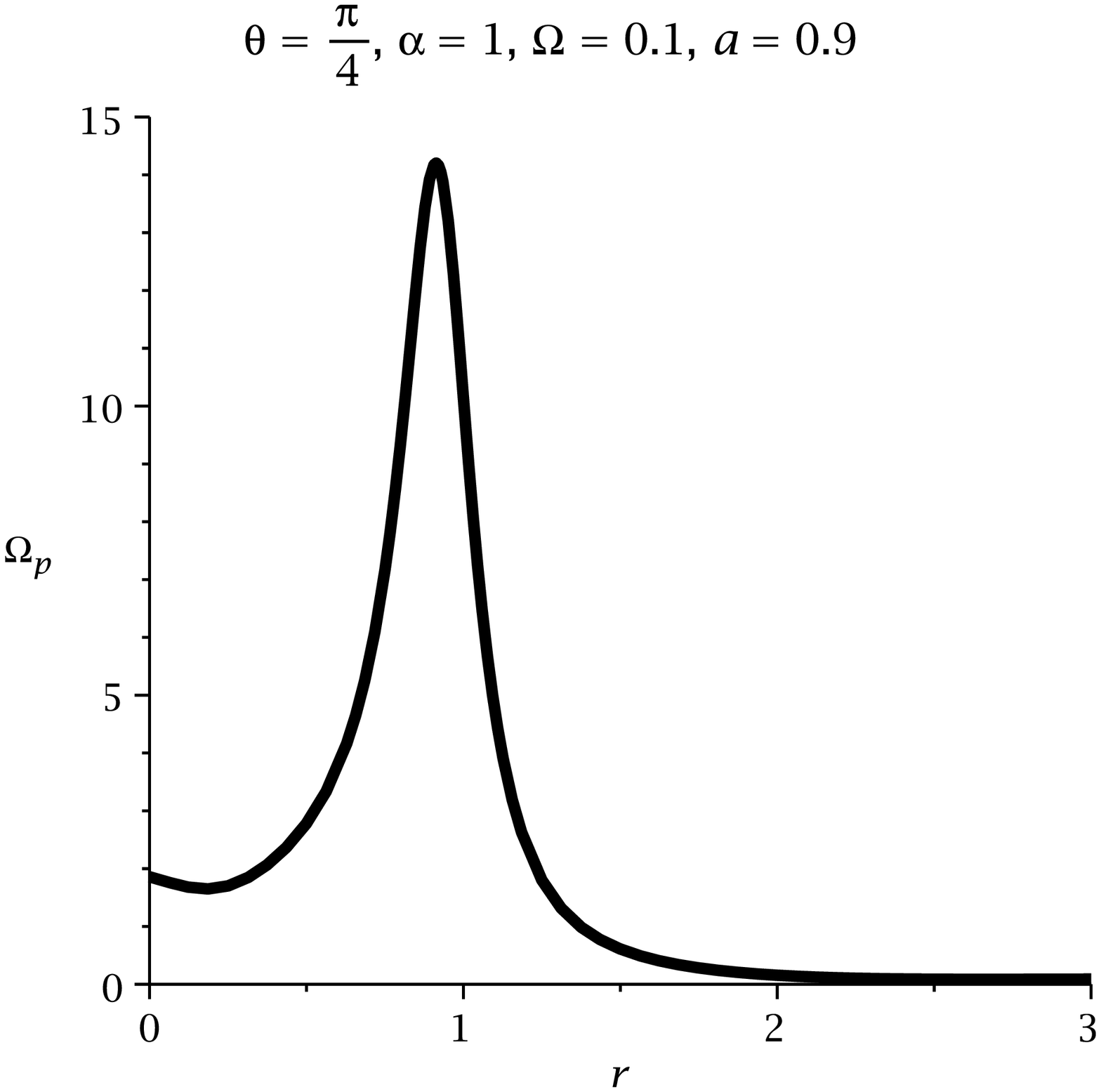}} 
\caption{ Examples of the variation  of $\Omega_{p}$ versus $r$ for $\theta=\frac{\pi}{4}$ in KMOG with variation 
of MOG parameter, spin parameter and $\Omega$. The first row describes the variation  of $\Omega_{p}$  
with $r$ for non-extremal BH, extremal BH and NS without MOG parameter. The second row describes the variation  
of $\Omega_{p}$  with $r$ for non-extremal BH, extremal BH and NS with MOG parameter. }
\label{fgg3}
\end{center}
\end{figure}

\subsection{Behaviour of $\vec{\Omega}_{p}$ at $\theta=\frac{\pi}{3}$}
In this limit $\theta=\frac{\pi}{3}$, the LT frequency is computed to be
\begin{eqnarray}
\vec{\Omega}_{p}|_{\theta= \frac{\pi}{3}} &=& 
\frac{\sqrt{\Delta}\xi(r)|_{\theta=\frac{\pi}{3}}\hat{r}+\sqrt{3}\eta(r)|_{\theta=\frac{\pi}{3}} \hat{\theta}}
{2\zeta(r)|_{\theta= \frac{\pi}{3}}}~,~\label{t12}
\end{eqnarray}
The magnitude of this vector is given by
\begin{eqnarray}
\Omega_p|_{\theta=\frac{\pi}{3}} &=& \frac{\sqrt{\Delta \xi^2 (r)|_{\theta=\frac{\pi}{3}}+3\eta^2(r)|_{\theta=\frac{\pi}{3}}}}
{2\zeta(r)|_{\theta= \frac{\pi}{3}}} ~\label{t13}
\end{eqnarray}
where 
\begin{eqnarray}
\xi~(r)|_{\theta= \frac{\pi}{3}} &=& a \Pi_{\alpha}-\frac{\Omega}{8}\left(8r^4+4a^2r^2+12a^2\Pi_{\alpha}+\frac{a^4}{2}\right)
+\frac{9}{16}a^3 \Omega^2 \Pi_{\alpha}
\end{eqnarray}
$$
\eta(r)|_{\theta= \frac{\pi}{3}} = aG_{N}{\cal M}\left(r^2-\frac{a^2}{4}\right)
-\frac{\alpha}{1+\alpha} G_{N}^2 {\cal M}^2 ar+\Omega\times
$$
$$
\left[r^5-3G_{N}{\cal M}r^4+\frac{a^2r^3}{2}-2G_{N}{\cal M}a^2r^2+\frac{a^4}{16} r
+\frac{7}{16} G_{N}{\cal M}a^4+\frac{2\alpha}{1+\alpha}G_{N}^2{\cal M}^2r(r^2+a^2)\right]
$$
\begin{eqnarray}
+\frac{3}{4}a\Omega^2\left[G_{N}{\cal M}\left(3r^4+a^2r^2+\frac{a^2}{4}(r^2-a^2)\right)
-\frac{\alpha}{1+\alpha} G_{N}^2 {\cal M}^2 r\left(2r^2+\frac{5}{4}a^2\right)\right]
~\label{t14}
\end{eqnarray}
$$
\zeta(r)|_{\theta= \frac{\pi}{3}} = \left(r^2+\frac{a^2}{4}\right)^\frac{3}{2}\times
$$
\begin{eqnarray}
\left[\left(r^2+\frac{a^2}{4}\right)-\Pi_{\alpha} +\frac{3}{2}a\Omega \Pi_{\alpha}-\frac{3}{4}\Omega^2 
\left\{\left(r^2+\frac{a^2}{4}\right) \left(r^2+a^2\right)+\frac{3a^2\Pi_{\alpha}}{4}\right\}\right]
.\label{t15}
\end{eqnarray}
Analogously, the variation of spin precession frequency with radial coordinate with MOG parameter 
and without MOG parameter may be seen from  Fig.~(\ref{nnn3}).
\begin{figure}
\begin{center}
\subfigure[]{
\includegraphics[width=2in,angle=0]{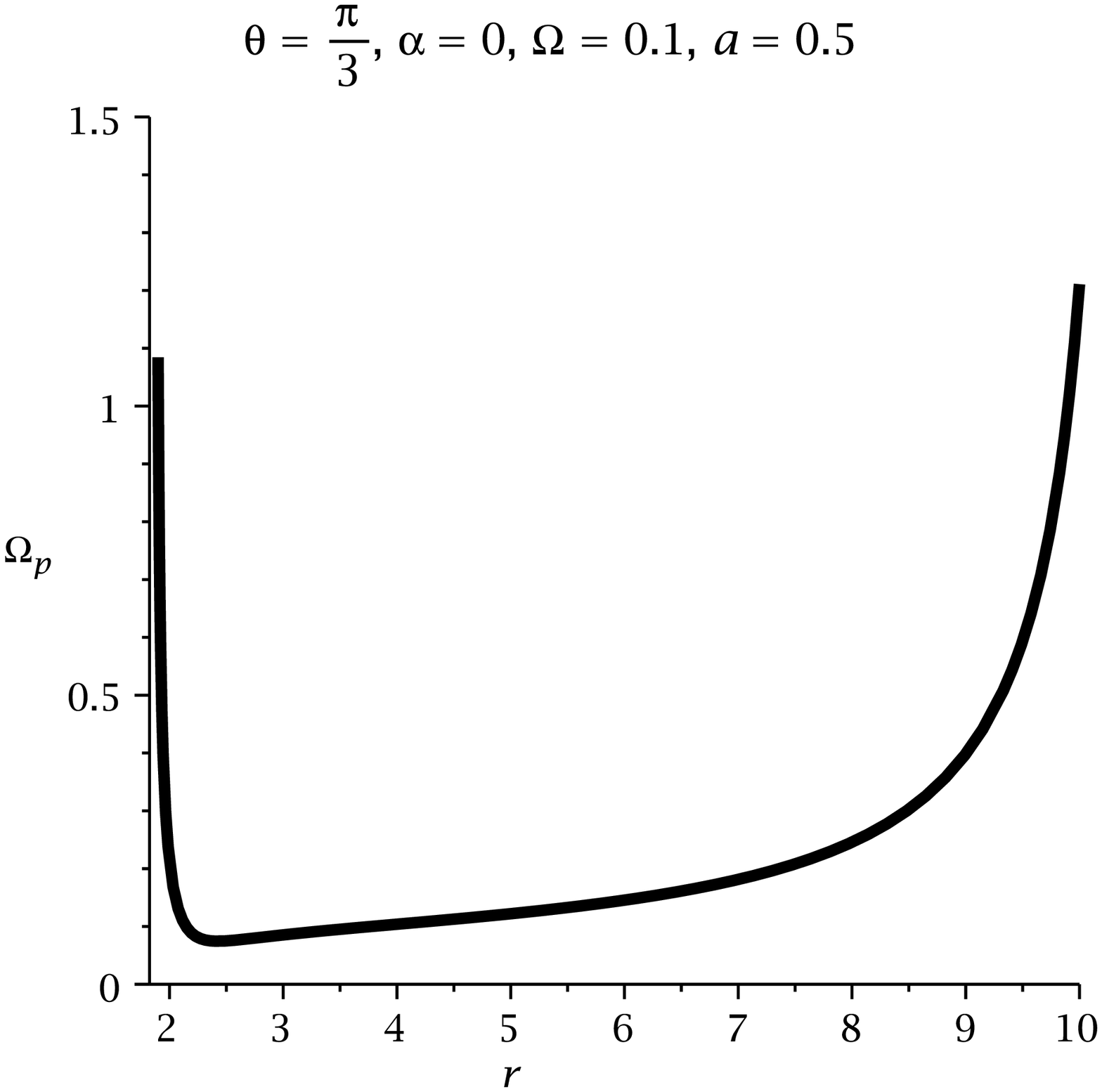}} 
\subfigure[]{
\includegraphics[width=2in,angle=0]{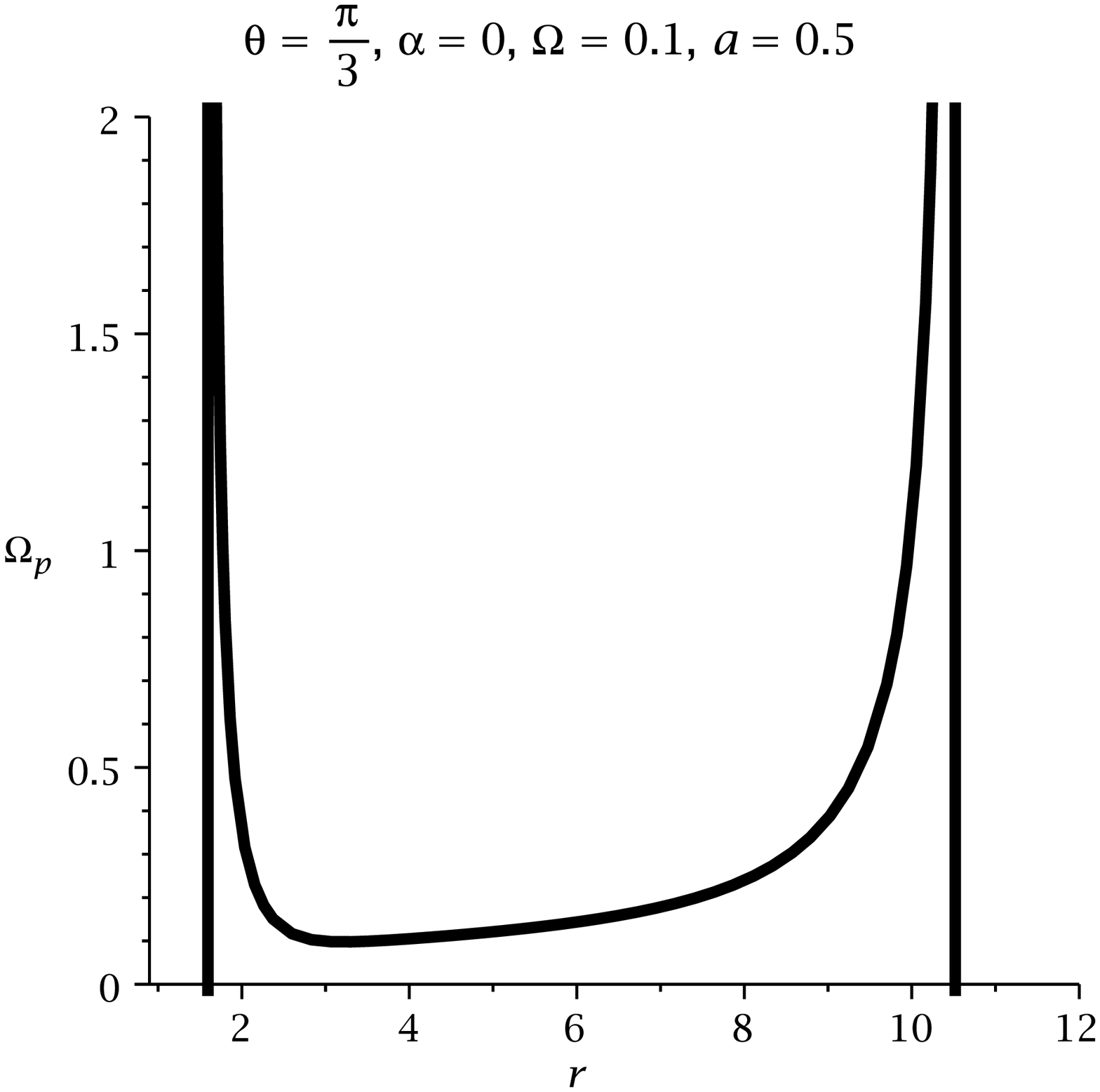}}
\subfigure[]{
\includegraphics[width=2in,angle=0]{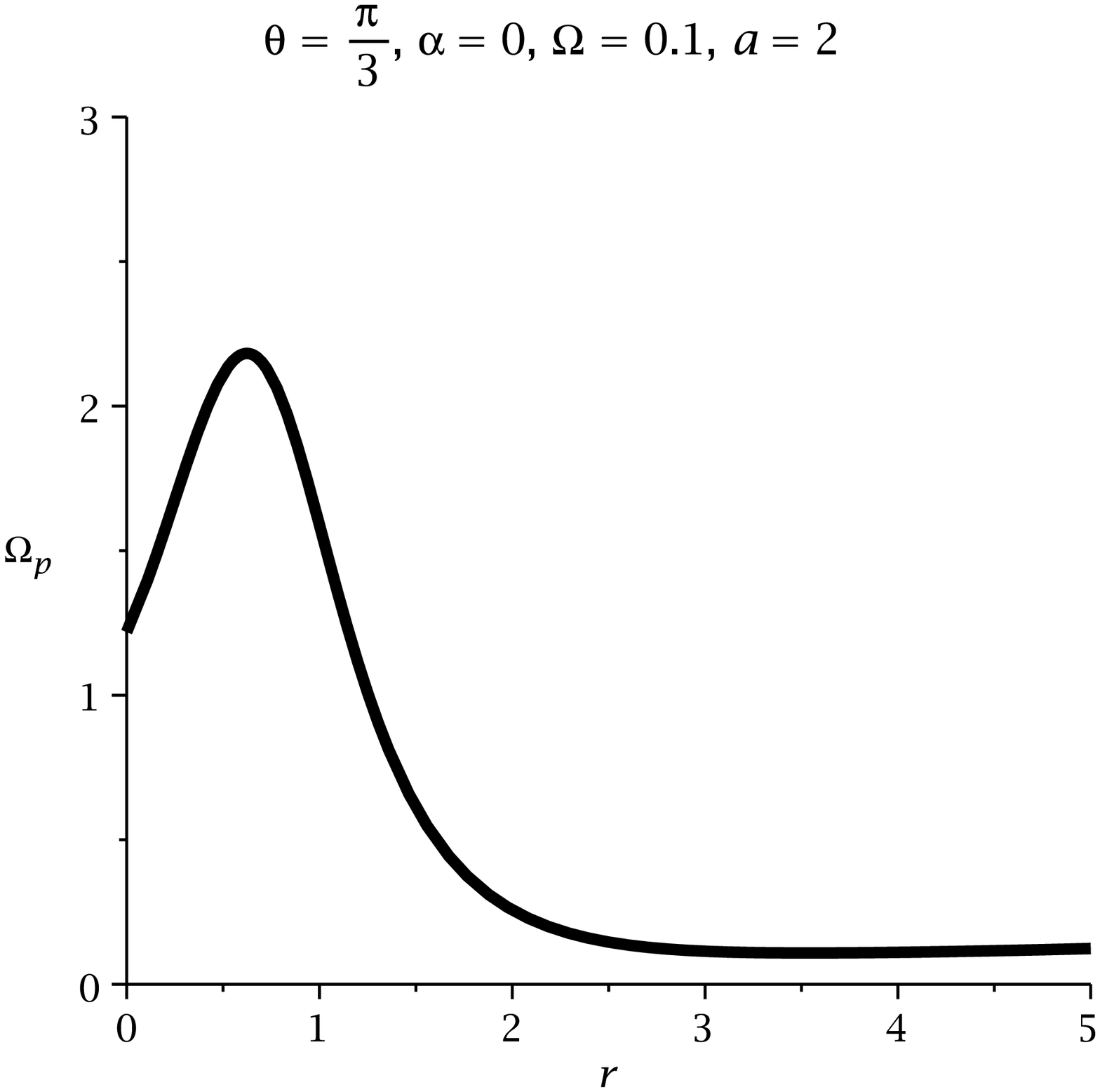}}
\subfigure[]{
\includegraphics[width=2in,angle=0]{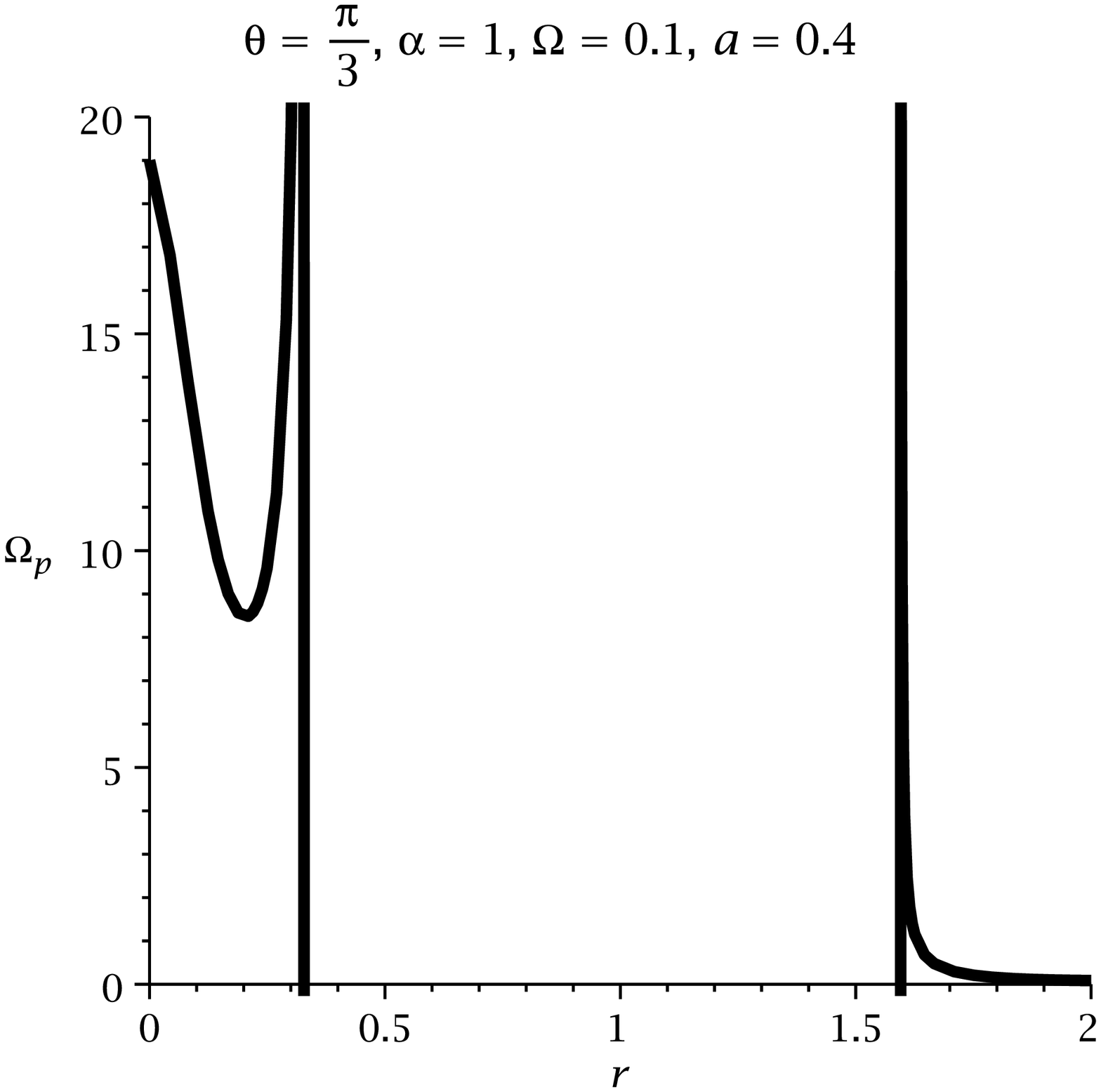}}
\subfigure[]{
\includegraphics[width=2in,angle=0]{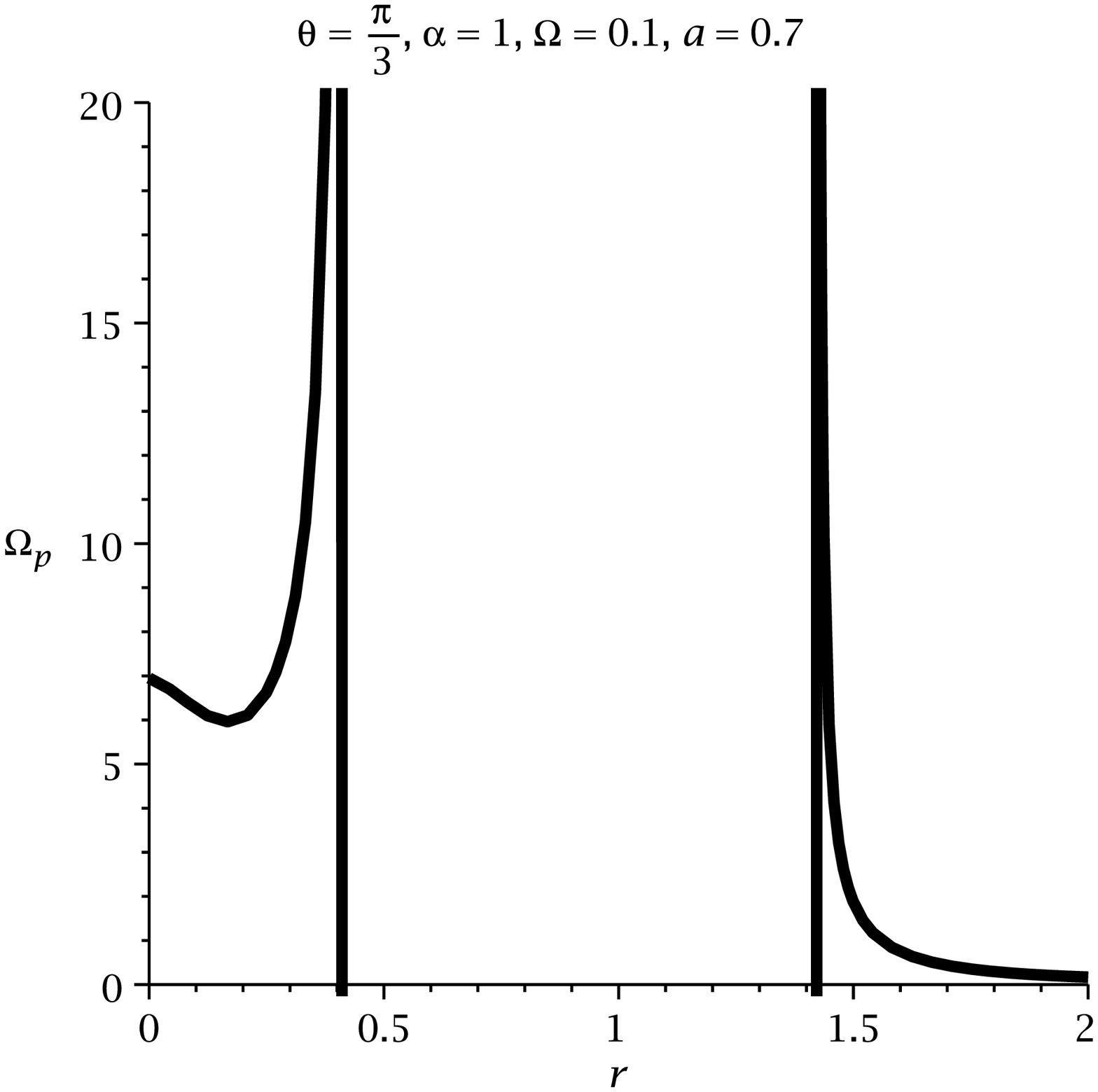}}
\subfigure[]{
\includegraphics[width=2in,angle=0]{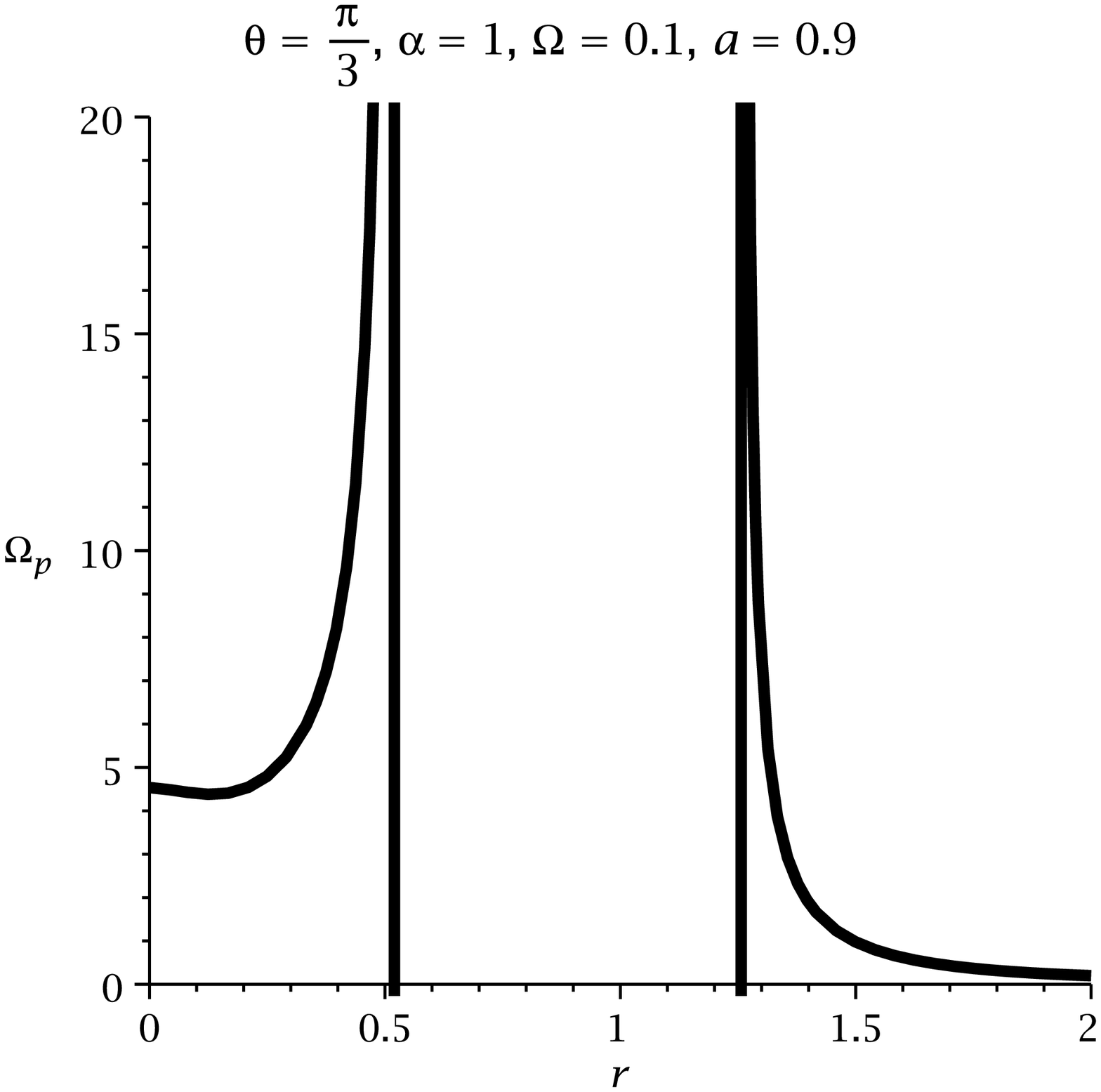}}
\caption{ Examples of the variation  of $\Omega_{p}$ vs. $r$ for $\theta=\frac{\pi}{3}$ in KMOG 
with variation of MOG parameter, spin parameter and $\Omega$. The first row describes the variation  of $\Omega_{p}$  
with $r$ for non-extremal BH, extremal BH and NS without MOG parameter. The second row describes the variation  
of $\Omega_{p}$  with $r$ for non-extremal BH, extremal BH and NS with MOG parameter.}
\label{nnn3}
\end{center}
\end{figure}

\subsection{Behaviour of $\vec{\Omega}_{p}$ at $\theta=\frac{\pi}{2}$}
In the equatorial plane the precession frequency vector is given by 
\begin{eqnarray} 
\vec{\Omega}_{p}|_{\theta=\frac{\pi}{2}} &=& 
\frac{\eta(r)|_{\theta=\frac{\pi}{2}}}{\zeta(r)|_{\theta=\frac{\pi}{2}}}\hat{\theta}
~\label{t16t}
\end{eqnarray}
The magnitude of this vector is thus
\begin{eqnarray}
\Omega_p|_{\theta=\frac{\pi}{2}} &=& \frac{\eta(r)|_{\theta=\frac{\pi}{2}}}{\zeta(r)|_{\theta=\frac{\pi}{2}}}  ~\label{t17t}
\end{eqnarray}
where 
\begin{eqnarray}
\eta(r)_{\theta=\frac{\pi}{2}} &=& ar\left(G_{N}{\cal M} r-\frac{\alpha}{1+\alpha} G_{N}^2 {\cal M}^2\right)\nonumber\\
&& 
+\Omega~\left[r^5-3G_{N} {\cal M}r^4-2G_{N} {\cal M}a^2r^2
+2\frac{\alpha}{1+\alpha} G_{N}^2{\cal M}^2r\left(r^2+a^2\right)\right]\nonumber\\
&& 
+a\Omega^2\left[G_{N} {\cal M}r^2\left(3r^2+a^2 \right)-\frac{\alpha}{1+\alpha} G_{N}^2{\cal M}^2r \left(2r^2+a^2\right)\right] \\ 
\zeta(r)_{\theta=\frac{\pi}{2}} &=& r^3\,\left[r^2-\Pi_{\alpha}+2a\Omega \Pi_{\alpha}-\Omega^2\left\{r^2\left(r^2+a^2\right)
+a^2\Pi_{\alpha}\right\}\right]
\end{eqnarray}
Variation of spin precession frequency with radial coordinate with 
MOG parameter and without MOG parameter may be seen from  Fig.~(\ref{nnn4}).
\begin{figure}
\begin{center}
\subfigure[]{
\includegraphics[width=2in,angle=0]{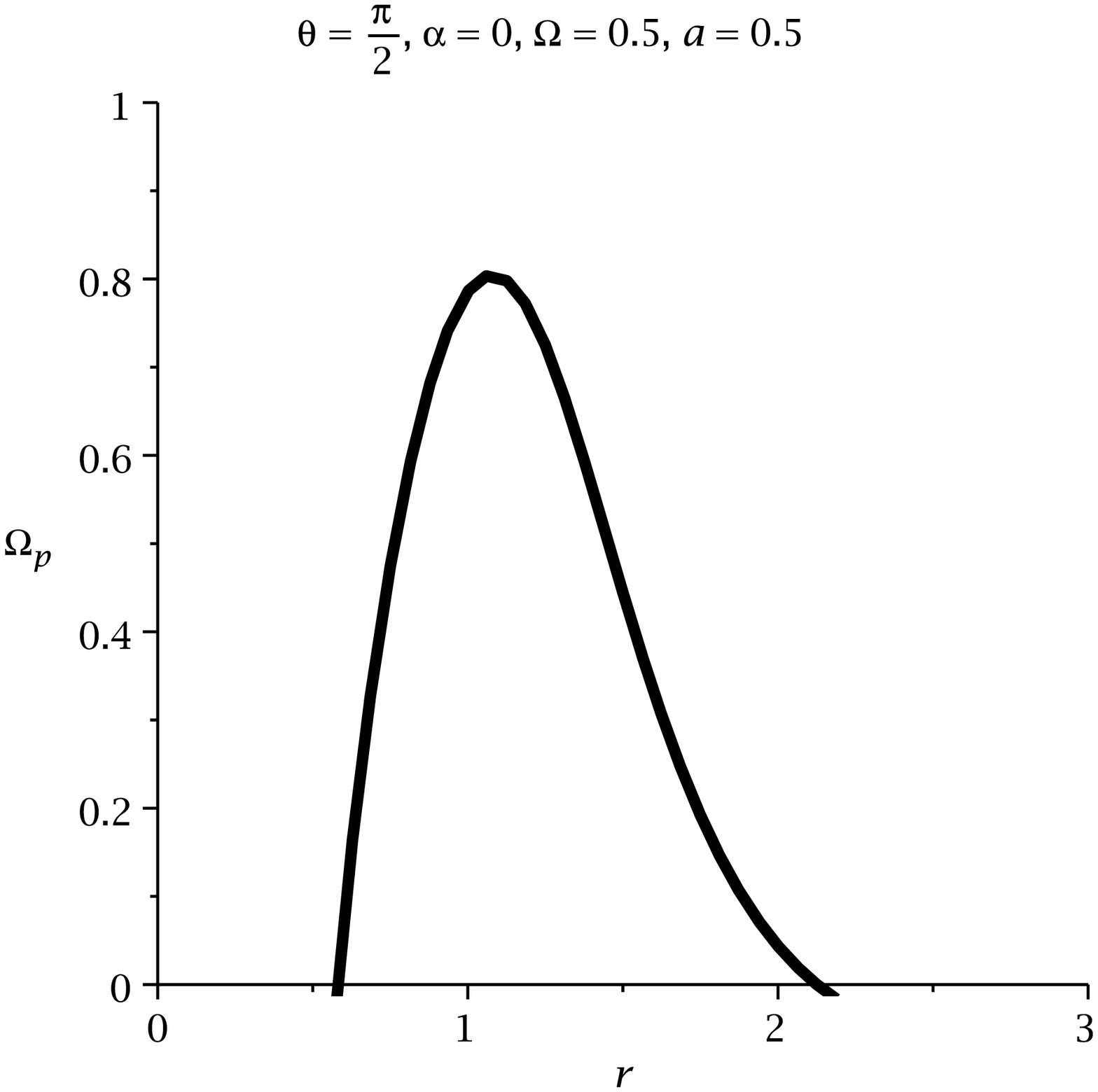}} 
\subfigure[]{
\includegraphics[width=2in,angle=0]{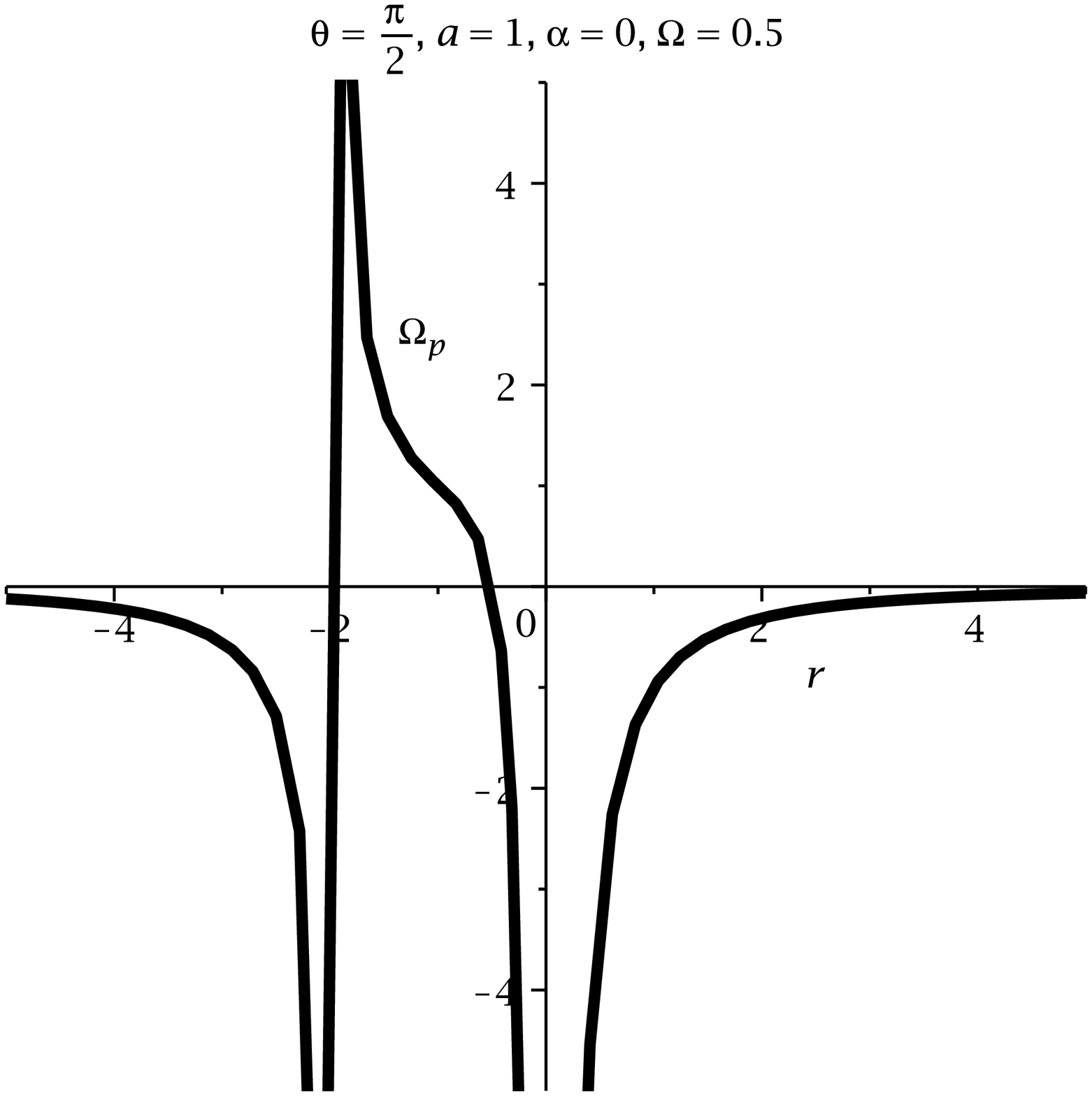}}
\subfigure[]{
\includegraphics[width=2in,angle=0]{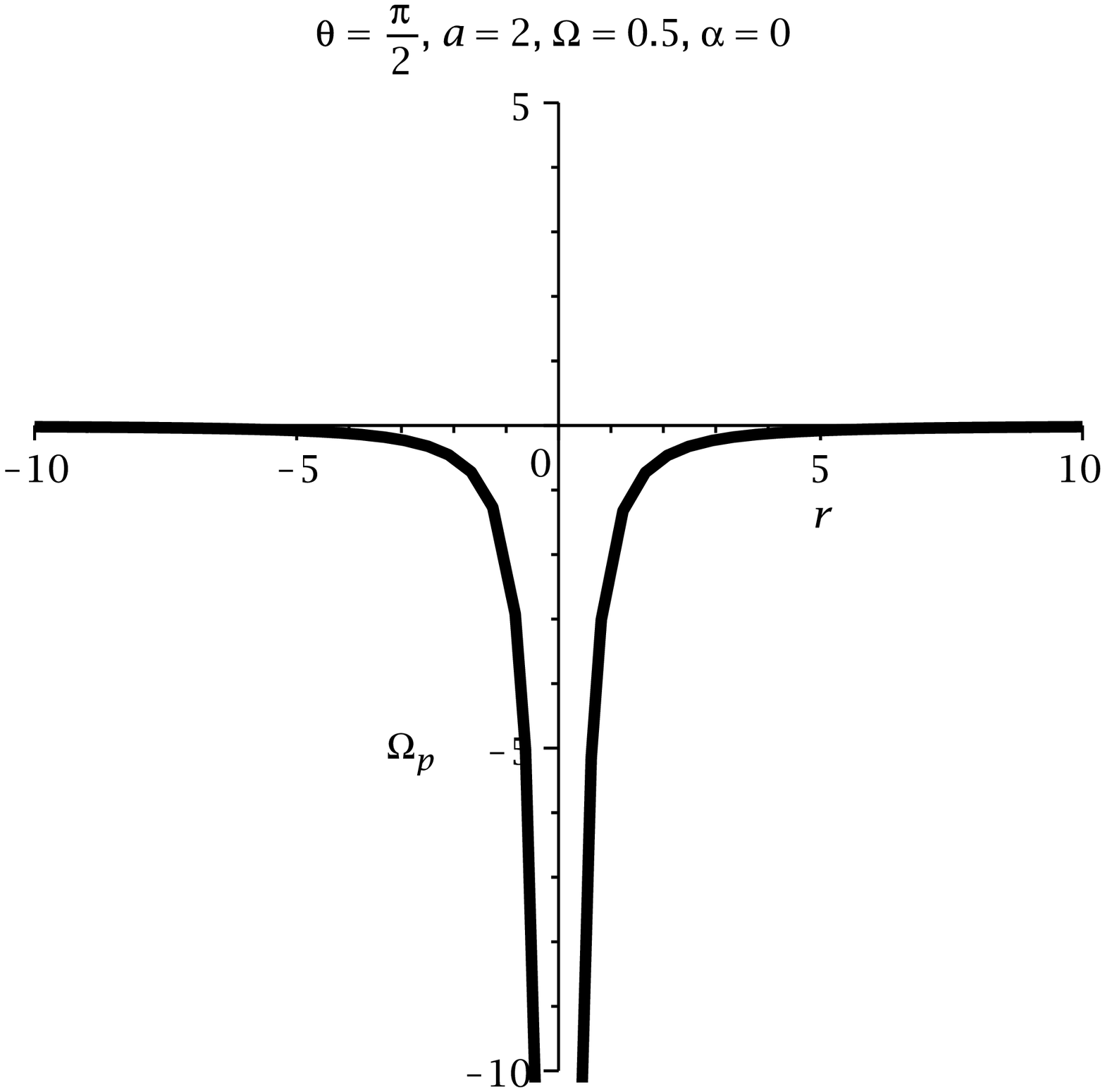}}
\subfigure[]{
\includegraphics[width=2in,angle=0]{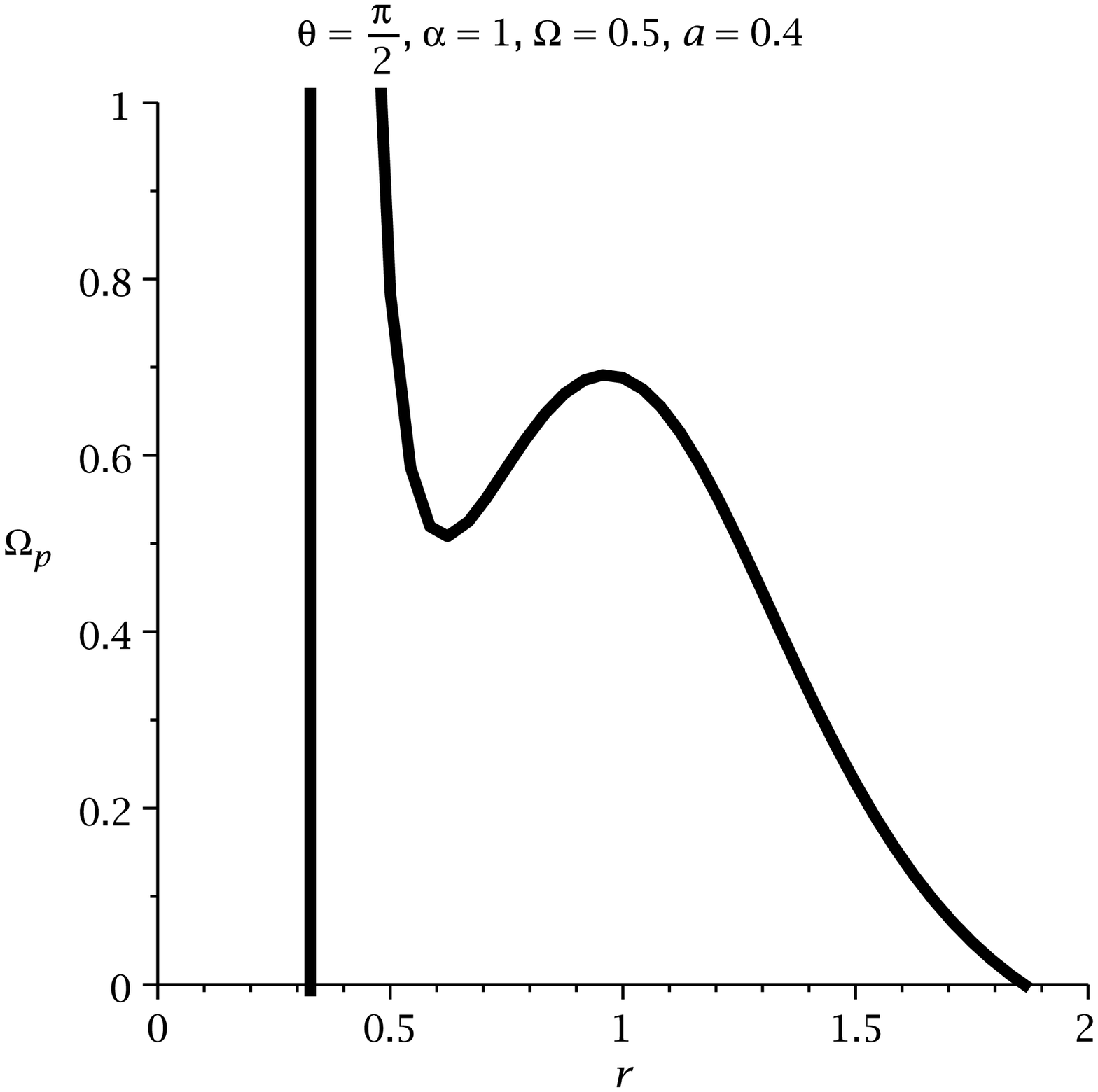}}
\subfigure[]{
\includegraphics[width=2in,angle=0]{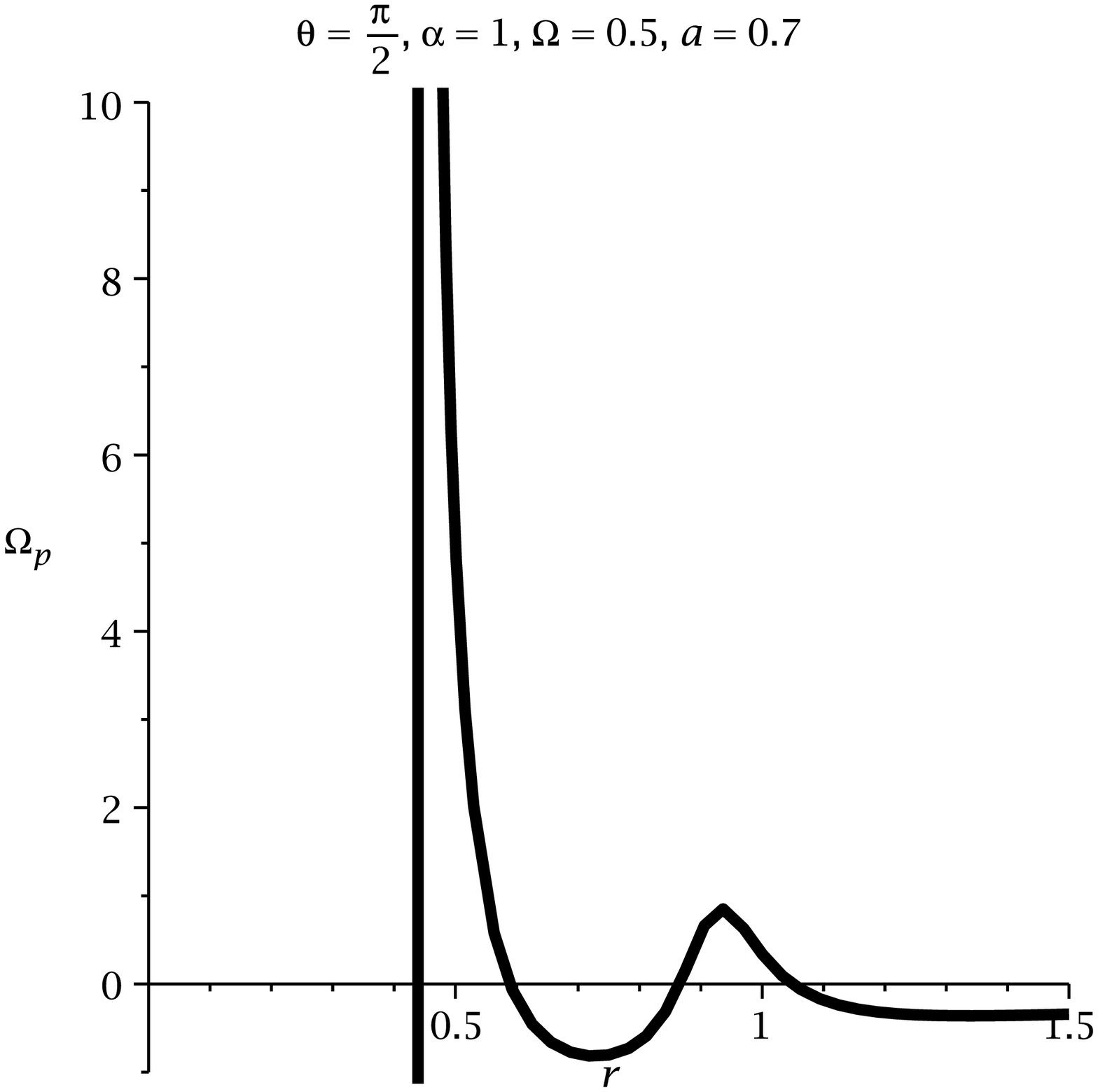}}
\subfigure[]{
\includegraphics[width=2in,angle=0]{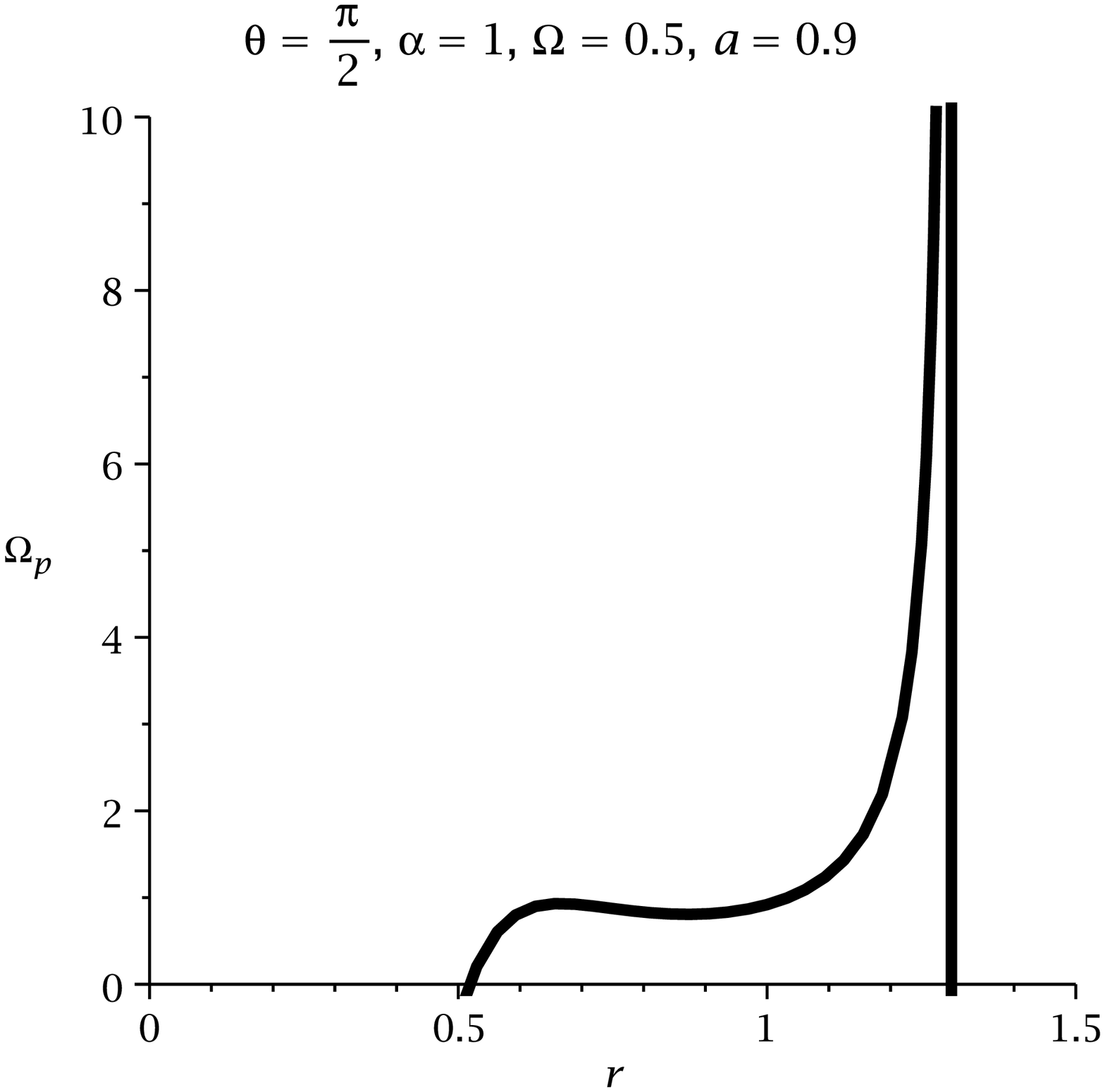}}
\caption{ Examples of the variation  of $\Omega_{p}$ versus $r$ for $\theta=\frac{\pi}{2}$ in KMOG with variation 
of MOG parameter, spin parameter and $\Omega$. The first row describes the variation  of $\Omega_{p}$  
with $r$ for non-extremal BH, extremal BH and NS without MOG parameter. The second row describes the variation  
of $\Omega_{p}$  with $r$ for non-extremal BH, extremal BH and NS with MOG parameter.}
\label{nnn4}
\end{center}
\end{figure}
The range of $\Omega$ could be determined by using Eq.~(\ref{oken}). As we said earlier 
that the outer horizon and the outer ergosurface are located at 
$r_{+} = G_{N}{\cal M} + \sqrt{\frac{G_{N}^2{\cal M}^2}{1+\alpha}-a^2}$ and 
$r_{e}^{+} = G_{N}{\cal M} + \sqrt{\frac{G_{N}^2{\cal M}^2}{1+\alpha}-a^2 \cos^2~\theta}$. While 
in the equatorial plane, the ergosphere is located at $r_{e}^{+} =m+\epsilon$ where $m=G_{N}{\cal M}$
and $\epsilon=\frac{m}{\sqrt{1+\alpha}}$.
Therefore one could obtain the precession frequency at the equatorial
ergosurface $r=r_{e}^{+}=m+\epsilon$ 
\begin{eqnarray}
{\Omega}_{p}|_{r=r_{e}^{+}} &=& \frac{\eta(r)|_{r=r_{e}^{+}}}{\zeta(r)|_{r=r_{e}^{+}}} 
~\label{t19}
\end{eqnarray}
where 
\begin{eqnarray}
\eta(r)|_{r=r_{e}^{+}} &=& (m+\epsilon)\left[a m_{\epsilon}+\Omega m_{\epsilon\epsilon}
+a\Omega^2m_{\epsilon\epsilon\epsilon}\right] ~\label{t20}
\end{eqnarray}
where 
$$
m_{\epsilon}=m^2+m\epsilon-\alpha\epsilon^2
$$
$$
m_{\epsilon\epsilon}= (m+\epsilon)^2(\epsilon^2-m\epsilon-2m^2+2\alpha\epsilon^2)-2ma^2(m+\epsilon)+2\alpha\epsilon^2a^2
$$
$$
m_{\epsilon\epsilon\epsilon}=3m(m+\epsilon)^3-2\alpha\epsilon^2(m+\epsilon)^2+ma^2(m+\epsilon)-\alpha\epsilon^2a^2
$$
and 
\begin{eqnarray}
\zeta(r)|_{r=r_{e}^{+}} &=& (m+\epsilon)^3\left[ m_{\chi}+2a\Omega m_{\chi\chi}
-\Omega^2m_{\chi\chi\chi}\right] ~\label{t21}
\end{eqnarray}
where 
$$
m_{\chi}=(1+\alpha)\epsilon^2-m^2
$$
$$
m_{\chi\chi}=2m(m+\epsilon)-\alpha\epsilon^2
$$
$$
m_{\chi\chi\chi}=(m+\epsilon)^4+a^2(m+\epsilon)^2+2ma^2(m+\epsilon)-\alpha\epsilon^2a^2
$$
In the extremal limit $a=\epsilon$, the precession frequency becomes 
\begin{eqnarray}
{\Omega}_{p}|_{\theta=\frac{\pi}{2}}^{r=r_{e}^{+}} 
&=& \frac{\epsilon \left[\epsilon-\Omega (3\epsilon^2+2\epsilon m+m^2)
+\Omega^2(\epsilon^3+m^3+4\epsilon m^2+7m \epsilon^2)\right]}
{(\epsilon+m)^3\left[2\epsilon\Omega-\Omega^2(3\epsilon^2+m^2+2\epsilon m)\right]} 
~\label{t22}
\end{eqnarray}
The value of $\Omega$ lies in the range $0 < \Omega < \frac{2\epsilon}{3\epsilon^2+m^2+2\epsilon m}$.
It indicates that if the mass or angular momentum of the central object increases then the 
value of $\Omega$ both at the ergosurface and in the $\theta=\frac{\pi}{2}$ plane becomes 
decreases. It may be written as in terms of MOG parameter as 
$$
{\Omega}_{p}|_{\theta=\frac{\pi}{2}}^{r=r_{e}^{+}} 
= \frac{1+\alpha}{\left(1+\sqrt{1+\alpha} \right)^3}\times
$$
\begin{eqnarray}
\frac{1- G_{N}{\cal M}\Omega\left(2+\sqrt{1+\alpha}+\frac{3}{\sqrt{1+\alpha}} \right)+ G_{N}^2{\cal M}^2\Omega^2 
\left(4+\sqrt{1+\alpha}+\frac{7}{\sqrt{1+\alpha}}+\frac{1}{1+\alpha}\right)}
{ G_{N}^2{\cal M}^2\Omega\left[2- G_{N}{\cal M}\Omega\left(2+\sqrt{1+\alpha}+\frac{3}{\sqrt{1+\alpha}}
\right) \right]}
~\label{t23}
\end{eqnarray}
when the MOG parameter reduces to zero value one gets the result of extremal Kerr BH as 
${\Omega}_{p}|_{\theta=\frac{\pi}{2}}^{r=r_{e}^{+}}=\frac{1-6 G_{N}{\cal M}\Omega+13 G_{N}^2{\cal M}^2\Omega^2}
{16  G_{N}^2{\cal M}^2\Omega(1-3 G_{N}{\cal M}\Omega)}$.
One could observe the variation of precession frequency with MOG parameter in the equatorial ergosphere  
from the Fig.~(\ref{nn3}).
\begin{figure}
\begin{center}
\subfigure[]{
\includegraphics[width=2in,angle=0]{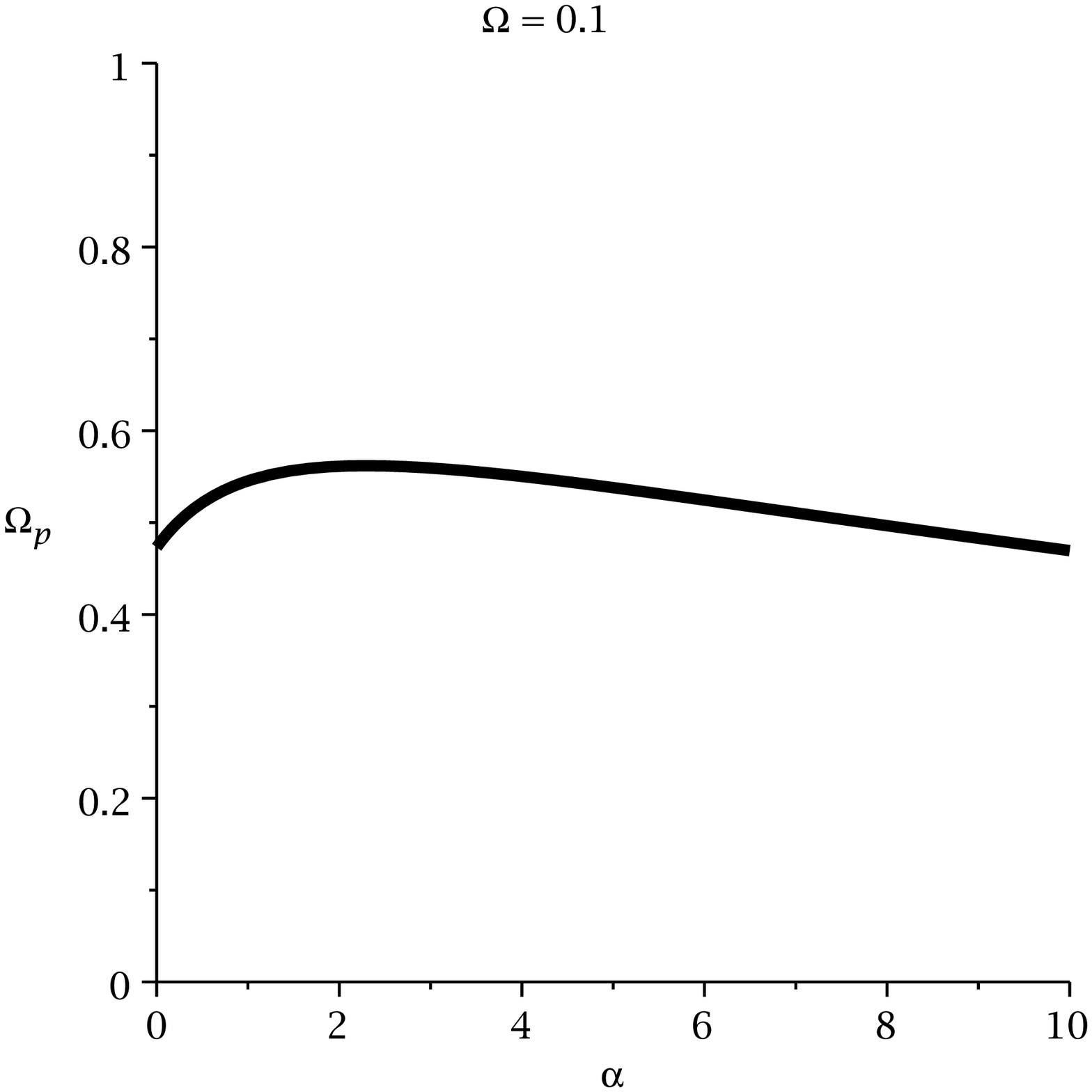}}
\subfigure[]{
\includegraphics[width=2in,angle=0]{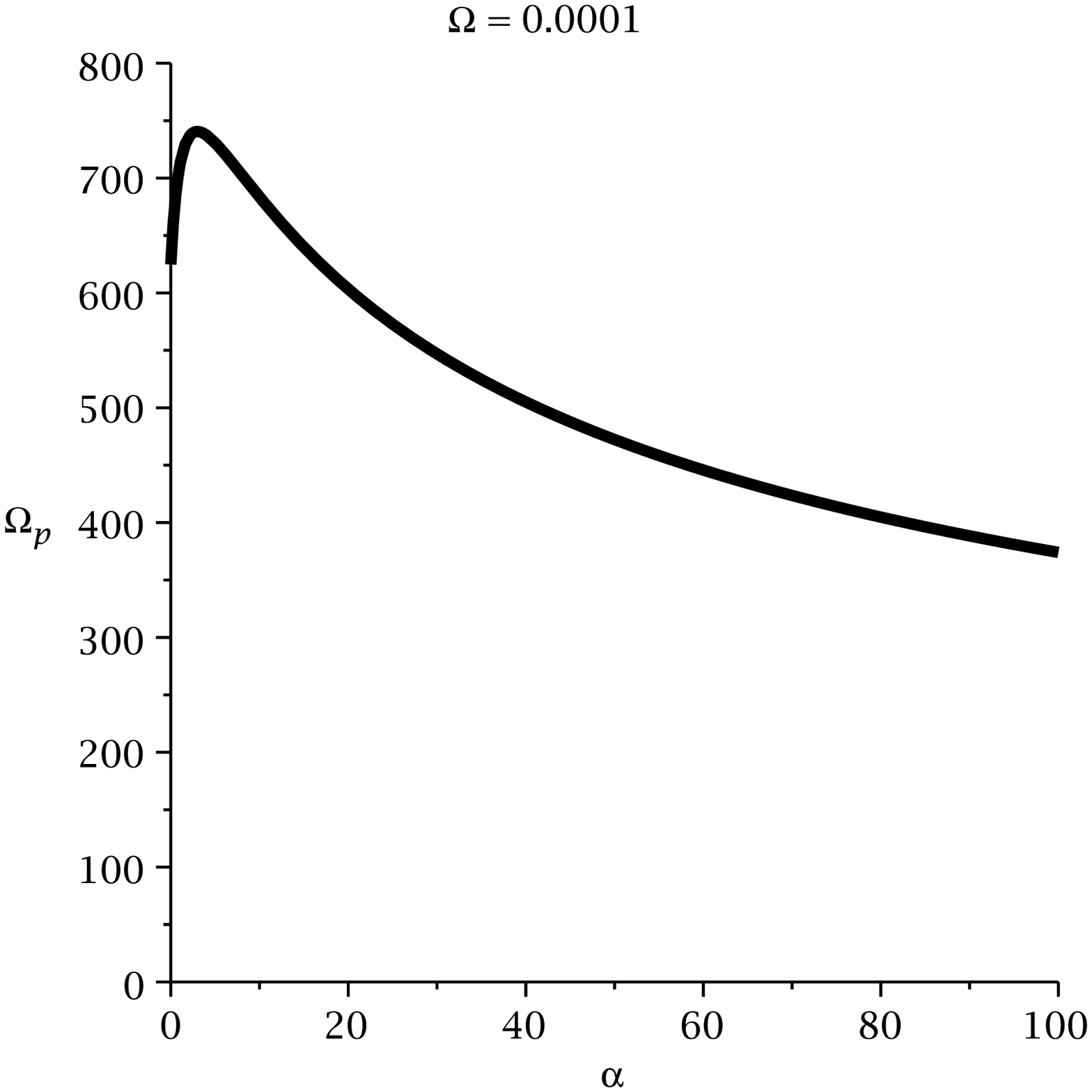}} 
\subfigure[]{
\includegraphics[width=2in,angle=0]{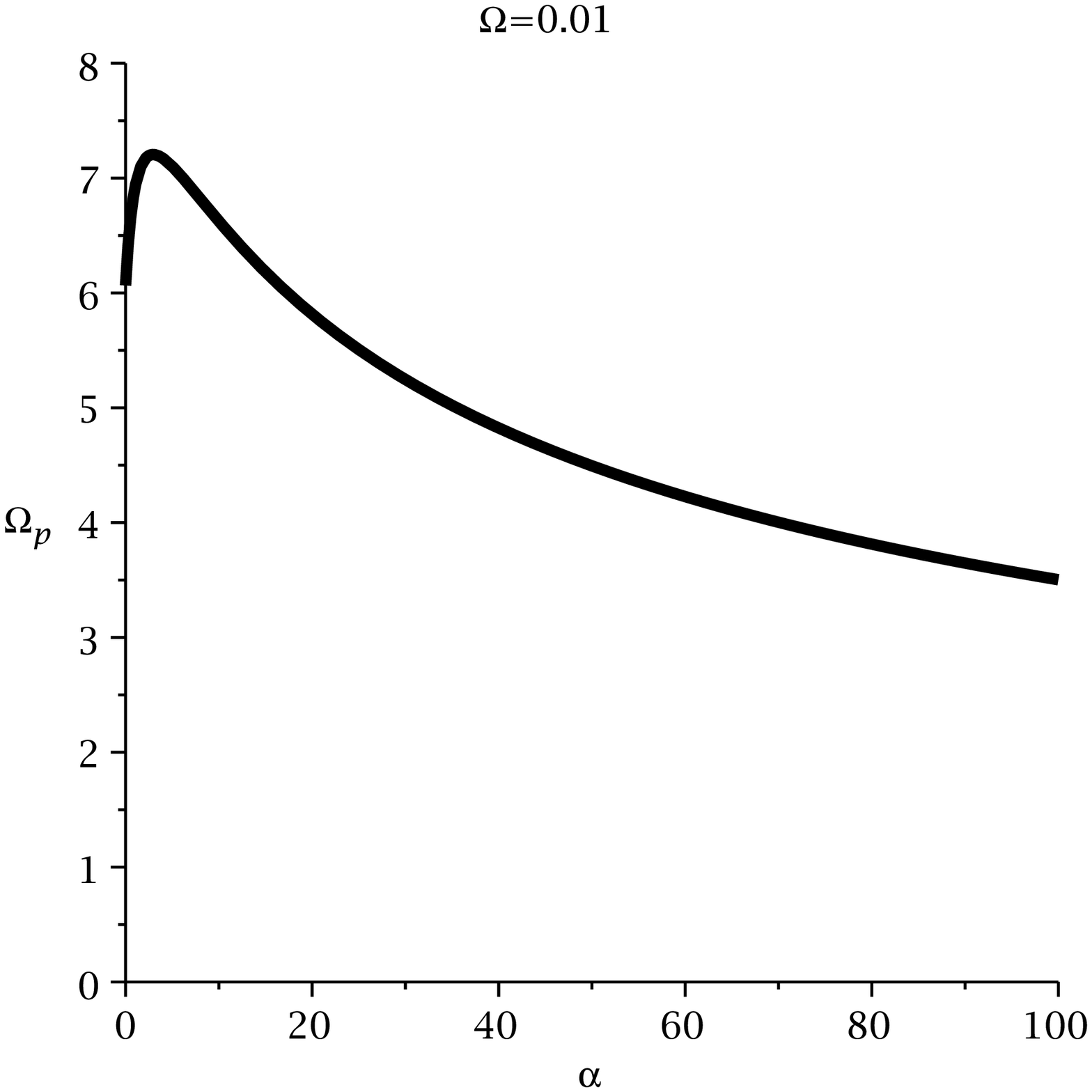}} 
\subfigure[]{
\includegraphics[width=2in,angle=0]{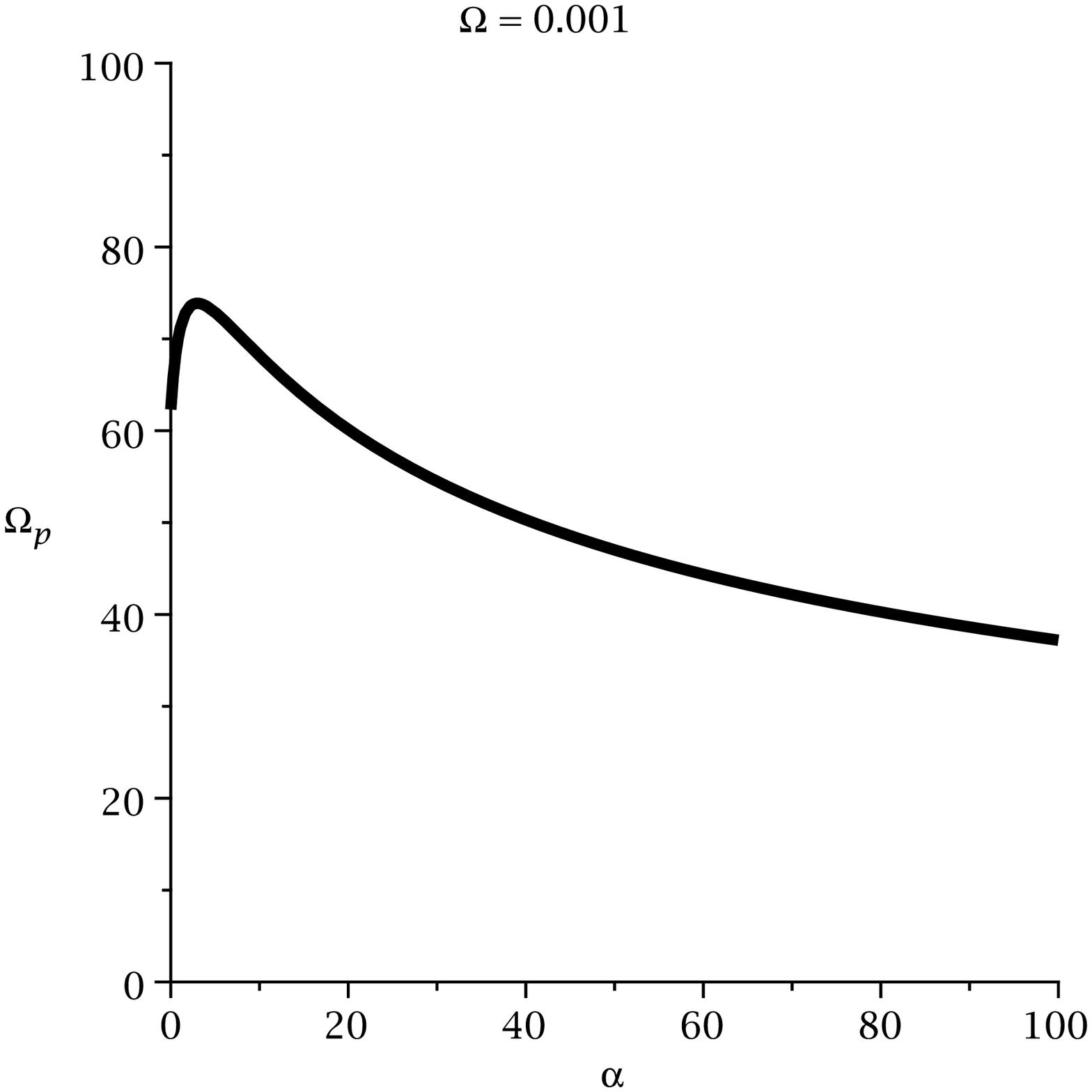}}
\subfigure[]{
\includegraphics[width=2in,angle=0]{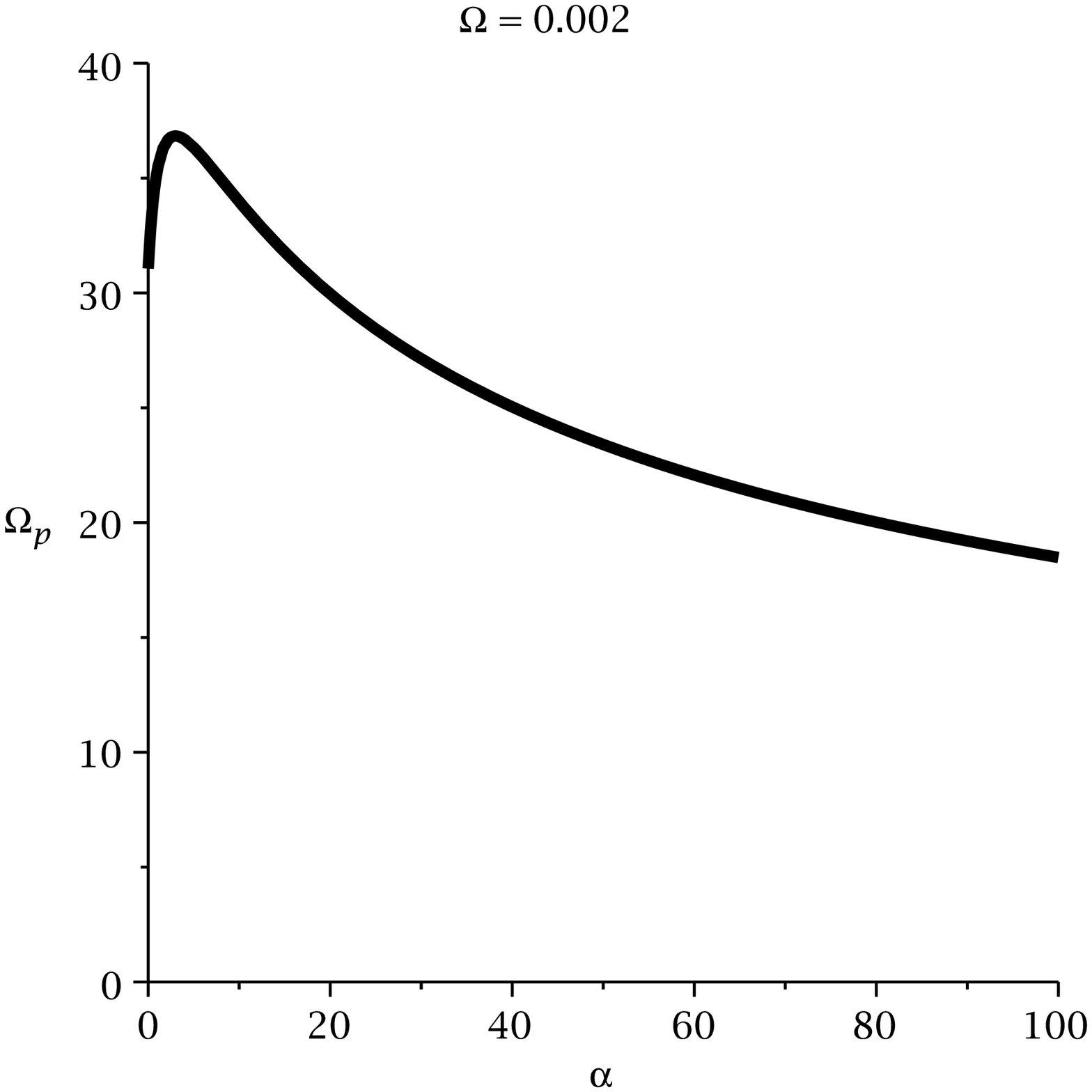}}
\subfigure[]{
\includegraphics[width=2in,angle=0]{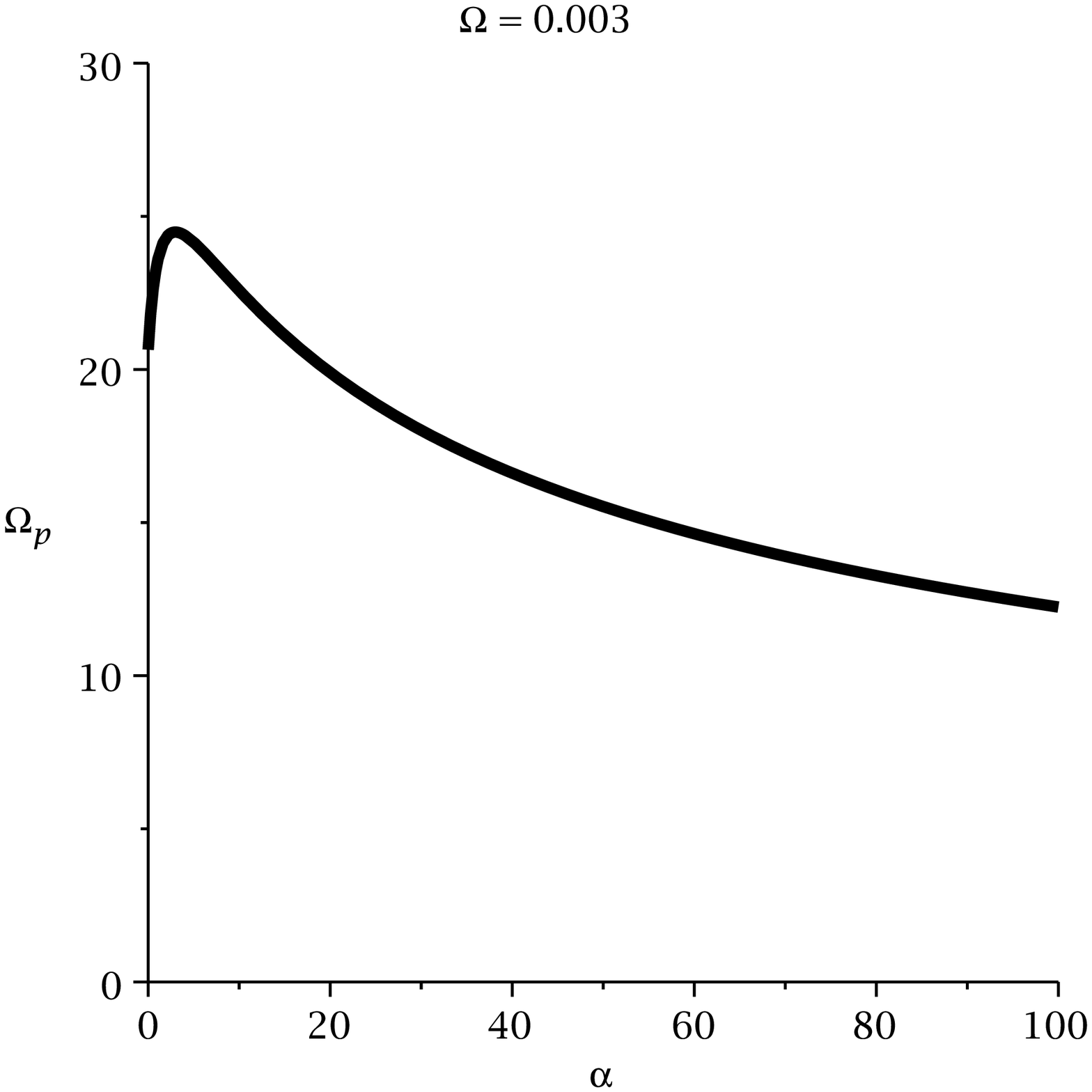}}
%\subfigure[]{
%\includegraphics[width=2in,angle=0]{pi2erg2.eps}}
%\subfigure[]{
%\includegraphics[width=2in,angle=0]{pi2ergo2.eps}} 
%\subfigure[]{
%\includegraphics[width=2in,angle=0]{pi2ergo3.eps}}
%\subfigure[]{
%\includegraphics[width=2in,angle=0]{pi2ergo4.eps}}
%\subfigure[]{
%\includegraphics[width=2in,angle=0]{pi2ergo5.eps}}
%\subfigure[]{
%\includegraphics[width=2in,angle=0]{pi2ergo6.eps}}
\caption{The figure describes the variation  of $\Omega_{p}$  
with MOG parameter~($\alpha$) for various values of $\Omega$.}
\label{nn3}
\end{center}
\end{figure}
It should be noted that the precession frequency at the outer 
horizon $r=r_{+}=m$ of extremal KMOG BH is computed to be 
\begin{eqnarray}
{\Omega}_{p}|_{\theta=\frac{\pi}{2}}^{r=r_{+}}
=-\frac{\epsilon }{m^2}~\label{t24}
\end{eqnarray}
In terms of MOG parameter the above expression can be written as 
\begin{eqnarray}
{\Omega}_{p}|_{\theta=\frac{\pi}{2}}^{r=r_{+}}=-\frac{1}{\sqrt{1+\alpha} G_{N}{\cal M}} ~\label{t25}
\end{eqnarray}
when $\alpha=0$, the precession frequency reduces to extremal BH, 
${\Omega}_{p}|_{\theta=\frac{\pi}{2}}^{r=r_{+}}=-\frac{1}{G_{N}{\cal M}}$.

\subsection{ $\vec{\Omega}_{p}$ in the Schwarzschild-MOG  spacetime}
In the limit $a=0$, the KMOG BH reduces to Schwarzschild-MOG~(SMOG) spacetime. 
In this case, the LT precession frequency becomes
\begin{eqnarray}
 \vec{\Omega}_{p}|_{a=0}=\Omega ~ \frac{-(r^2-\Pi_{\alpha})^{\frac{1}{2}}\cos\theta ~\hat{r}+
(r^2-\Pi_{\alpha}-G_{N} {\cal M}r+\frac{\alpha}{1+\alpha} G_{N}^2{\cal M}^2)\sin\theta ~\hat{\theta}}
{r^2-\Pi_{\alpha}-r^4\Omega^2\sin^2\theta}~\label{rn}
\end{eqnarray}
where $\Omega$ might have any value so that $u$ should be timelike. 

It must be noted that SMOG spacetime is a spherically symmetric 
solution of the  Einstein equation. It denotes a BH of {mass ${\cal M}$}. 
In the equatorial plane $\Omega_{p}|_{\theta=\frac{\pi}{2}}$ 
becomes 
\begin{eqnarray}
\Omega_{p} &=& \Omega ~ \frac{r^2-\Pi_{\alpha}-G_{N} {\cal M}r+
\frac{\alpha}{1+\alpha} G_{N}^2{\cal M}^2}{r^2-\Pi_{\alpha}-\Omega^2 r^4}.~\label{rn1}
\end{eqnarray}
It implies that when a gyroscope moving in the equatorial plane of the 
Reissner-Nordstr\"{o}m spacetime~(static spacetime) it becomes precess. 
Now if the gyro moves along circular orbits then $\Omega_{p}$ becomes 
Keplerian frequency i.e. 
\begin{eqnarray}
\Omega_{p}=\Omega=\Omega_{K}= \sqrt{\frac{G_{N} {\cal M}r-\frac{\alpha}{1+\alpha} G_{N}^2{\cal M}^2}{r^{4}}}.
\end{eqnarray}
This equation indicates the precession frequency in the Copernican 
frame which is derived with respect to the proper time $\tau$. The 
proper time $\tau$  is computed in the Copernican frame is related to the 
coordinate frame $t$ via the relation 
$d\tau=\sqrt{1-\frac{3G_{N} {\cal M}}{r}+2\left(\frac{\alpha}{1+\alpha}\right)\frac{ G_{N}^2{\cal M}^2}{r^2}} dt$. Thus 
one obtains the precession frequency in the coordinate basis $\Omega^{'}$ as,
\begin{equation}
 \Omega^{'} = \left(\frac{G_{N} {\cal M}r-\frac{\alpha}{1+\alpha} G_{N}^2{\cal M}^2}{r^4}\right)^{\frac{1}{2}}
 \sqrt{1-\frac{3G_{N} {\cal M}}{r}+2\left(\frac{\alpha}{1+\alpha}\right)\frac{ G_{N}^2{\cal M}^2}{r^2}}.
\end{equation}
Thus one can compute the geodetic precession frequency which is the difference
between $\Omega^{'}$ and $\Omega$, and one obtains 
\begin{eqnarray}
\Omega_{\rm geodetic} &=& \left(\frac{G_{N} {\cal M}r-\frac{\alpha}{1+\alpha} G_{N}^2{\cal M}^2}{r^4}\right)^{\frac{1}{2}} 
\left(1- \sqrt{1-\frac{3G_{N} {\cal M}}{r}+2\left(\frac{\alpha}{1+\alpha}\right)\frac{ G_{N}^2{\cal M}^2}{r^2}}\right)
\end{eqnarray}
This is called the geodetic precession of a test gyroscope around a nonrotating 
spherical object of mass $G_{N} {\cal M}$. In the limit when $\alpha=0$, one gets the 
geodetic precession of Schwarzschild BH~\cite{chiba}. One could see the geodetic precession of a 
nonrotating spherical object from the Fig.~(\ref{nn4}) for different values of MOG parameter. 
\begin{figure}
\begin{center}
\subfigure[]{
\includegraphics[width=2in,angle=0]{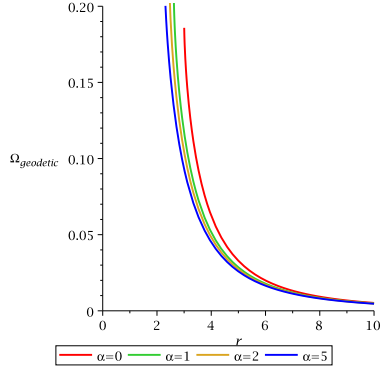}} 
\caption{\label{angf2} The figure describes the variation  of geodetic precession  with $r$ 
for different values of spin parameter in the $\theta=\frac{\pi}{2}$ plane.}
\label{nn4}
\end{center}
\end{figure}

\section{Frame-Dragging effect of KMOG spacetime with $\Omega=0$}
For strengthening our work, in this section now we will focus on particularly LT effect by 
calculating the frequency $\Omega_{LT}$ and considering the angular velocity of test gyro 
is equal to zero i.e. $\Omega=0$~(We ommit other precession 
e.g. geodetic precession etc.). The theoretical prescription of this formalism for $\Omega=0$ was 
first introduced by Chakraborty \& Majumdar~\cite{chp} in the strong gravity regime and for arbitrary value 
of $\Omega$. The result of the weak field limit of the LT frequency can be easily obtained from the 
general formalism. Therefore, one can derive the LT precession frequency for KMOG BH in MOG by 
using Eq.~(\ref{n2}). 

As we have said earlier when $\Omega=0$ then $\vec{\Omega_{p}}=\vec{\Omega_{LT}}$. Thus one can compute 
for the metric~(\ref{mg2.1}) the LT frequency vector is
\begin{eqnarray}
\vec{\Omega}_{LT} &=& \frac{\chi(r)~\sqrt{\Delta}\cos\theta~\hat{r}+\mu(r)~\sin\theta~\hat{\theta}}{\sigma(r)}
,~\label{gekm3}
\end{eqnarray}
where,
$$
\chi (r)= a \Pi_{\alpha}
$$
$$
\mu(r) = aG_{N}{\cal M}(r^2-a^2\cos^2\theta)-\frac{\alpha}{1+\alpha} G_{N}^2 {\cal M}^2 ar
$$
$$
\sigma(r) = \rho^3(\rho^2-\Pi_{\alpha})
$$
The magnitude of this vector is computed to be 
\begin{eqnarray}
\Omega_{LT}(r,\theta) &=& \frac{\sqrt{\Delta~\chi^2(r)~\cos^2\theta+\mu^2(r)~\sin^2\theta}}
{\sigma(r)} ~\label{ttl3}
\end{eqnarray}
It clearly shows that the LT frequency is affected by the MOG parameter~$\alpha$. This is very interesting in a sense 
that the deformation parameter changes the geometry of the spacetime and it also changes the 
LT frequency. This could be seen from the graphical plot. Without  MOG parameter the LT frequency 
reduces to Kerr BH. The presence of the 
MOG parameter decreases the LT frequency. Now we would derive the LT frequency for various values of $\theta$.
One could see the variation of Lense-Thirring frequency with the radial coordinate from the subsequent 
figures. From this diagram one can observed how much amount of LT frequency is changed due to the 
deformation parameter.
%%%%%%%%%%%%%%%%%%%%%%%%%%%%%%%%%%%%%%%%%%%%%%%%
\subsection{Behaviour of $\vec{\Omega}_{LT}$ at $\theta=0$}
First we consider the case $\theta=0$.  In this case the LT frequency vector becomes
\begin{eqnarray}
\vec{\Omega}_{LT}|_{\theta= 0} &=& 
\frac{\chi(r)|_{\theta= 0}}{\sigma(r)|_{\theta= 0}}~\sqrt{\Delta}~\hat{r},~\label{tl0}
\end{eqnarray}
The magnitude of this vector is calculated to be 
\begin{eqnarray}
\Omega_{LT}|_{\theta= 0} &=&  \frac{a\Pi_{\alpha}}{(r^2+a^2)^\frac{3}{2}\sqrt{\Delta}}~\label{tl1}
\end{eqnarray}
It follows that the LT frequency depends on the MOG parameter. Using this equation one can differentiate 
between BHs and NS in MOG theory. It could be observed from the  Fig.~\ref{gm}. In this diagram, 
we have plotted the LT frequency $\Omega_{LT}$ with respect 
to the radial coordinate for angular coordinate value $\theta=0$. We have speculated that the LT 
frequency is diminished due to the presence of the MOG parameter in compared to Kerr BH.
\begin{figure}
\begin{center}
\subfigure[]{
\includegraphics[width=2in,angle=0]{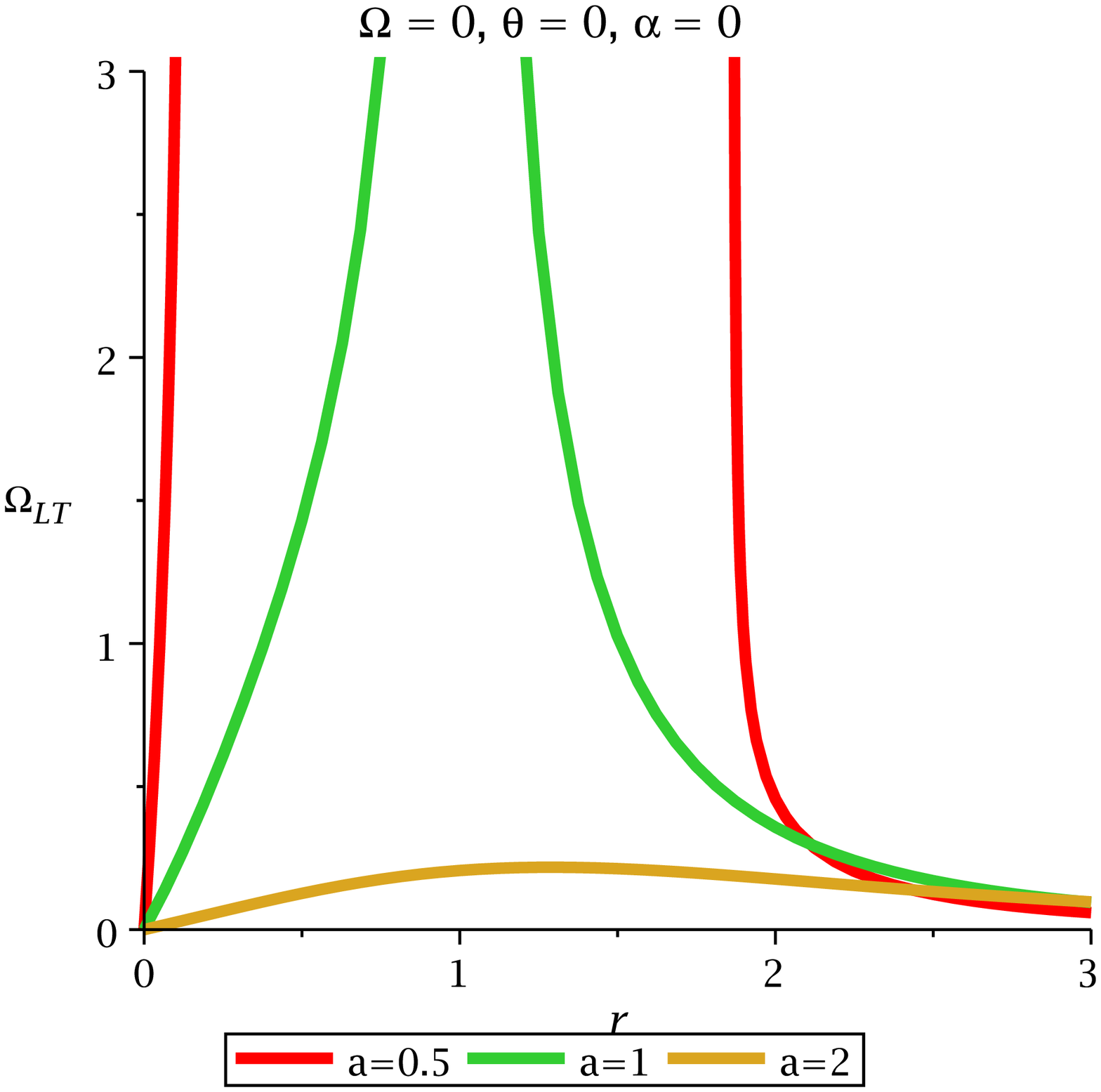}} 
\subfigure[]{
\includegraphics[width=2in,angle=0]{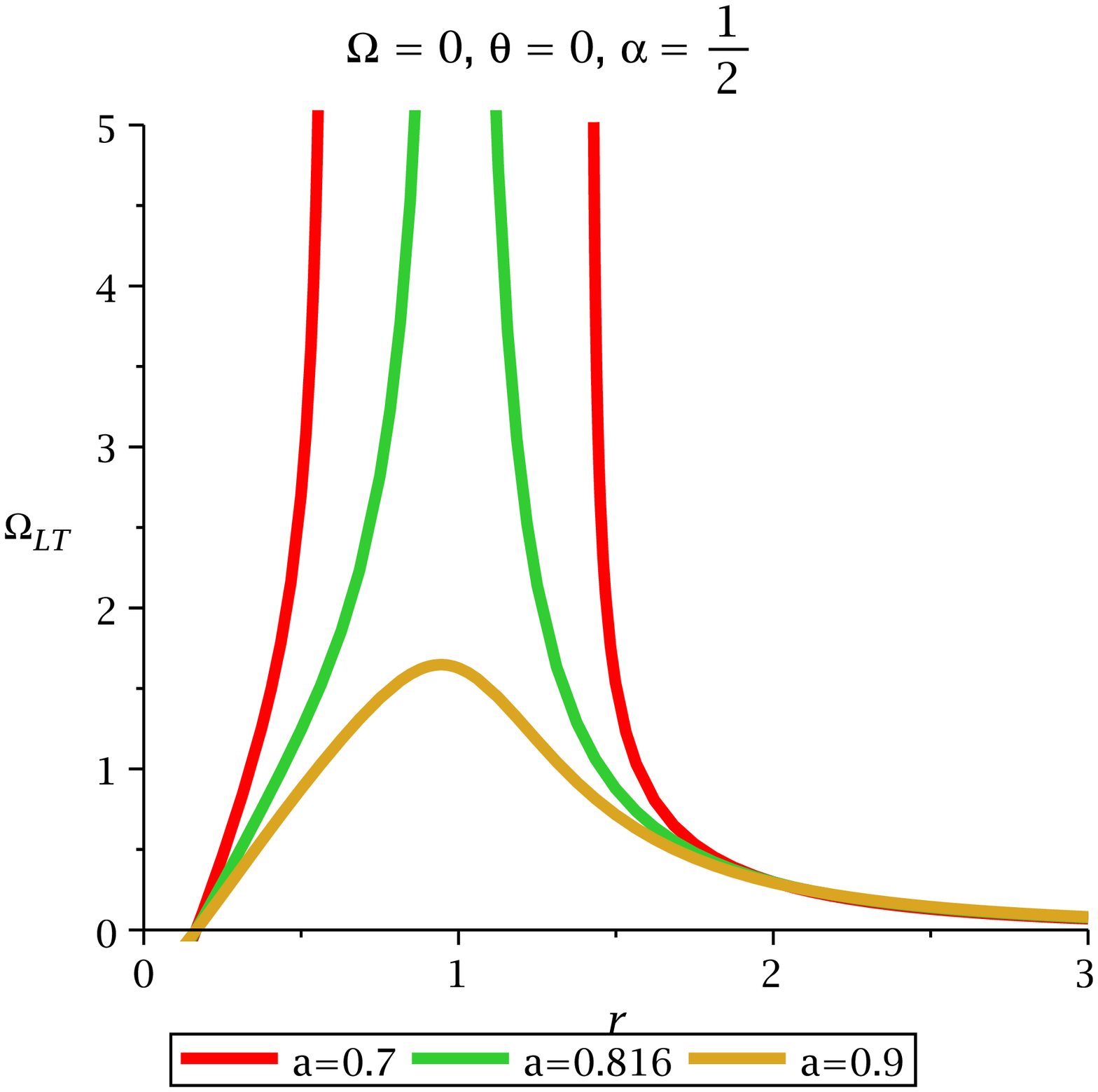}} 
\subfigure[]{
\includegraphics[width=2in,angle=0]{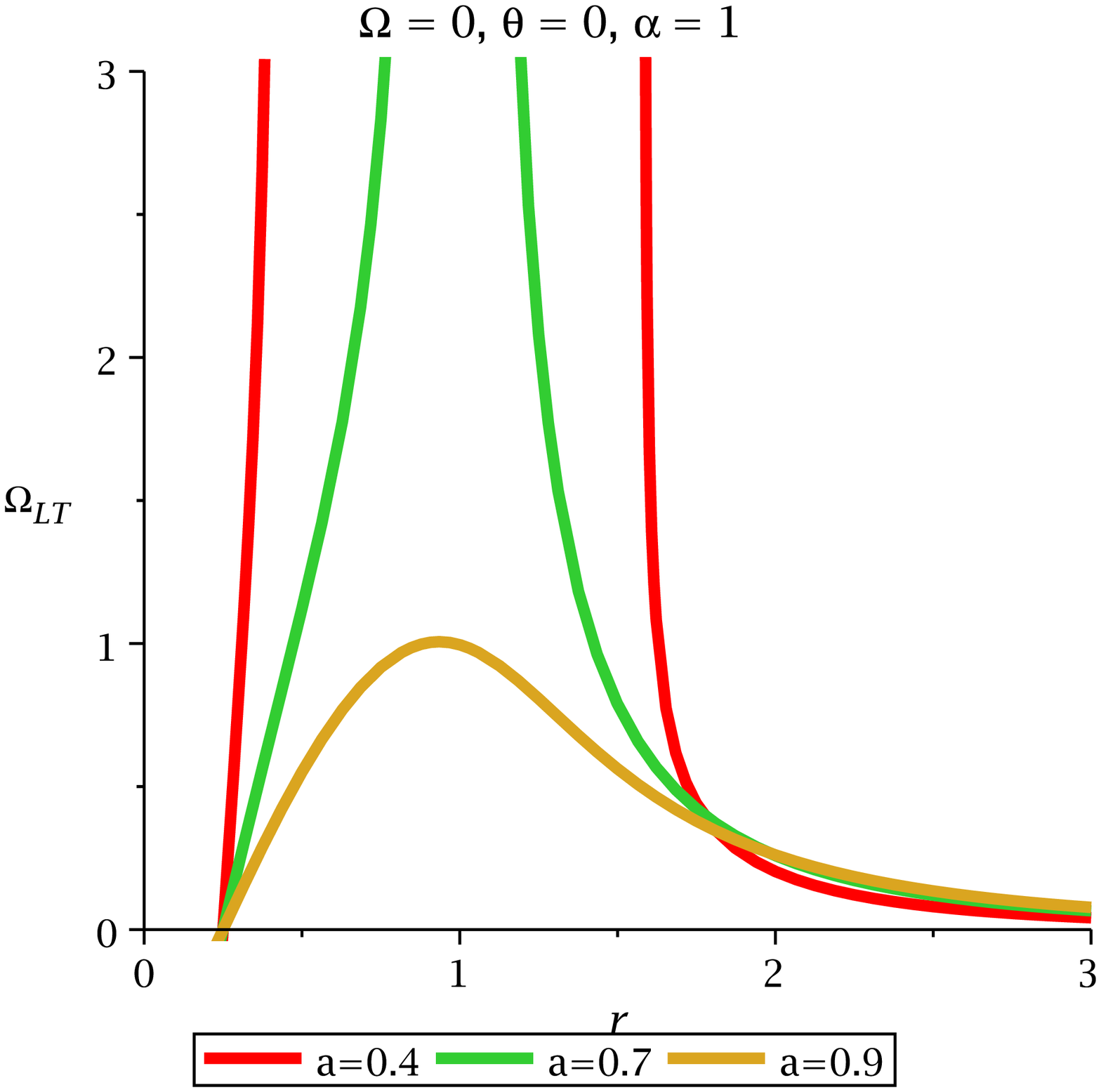}} 
\subfigure[]{
\includegraphics[width=2in,angle=0]{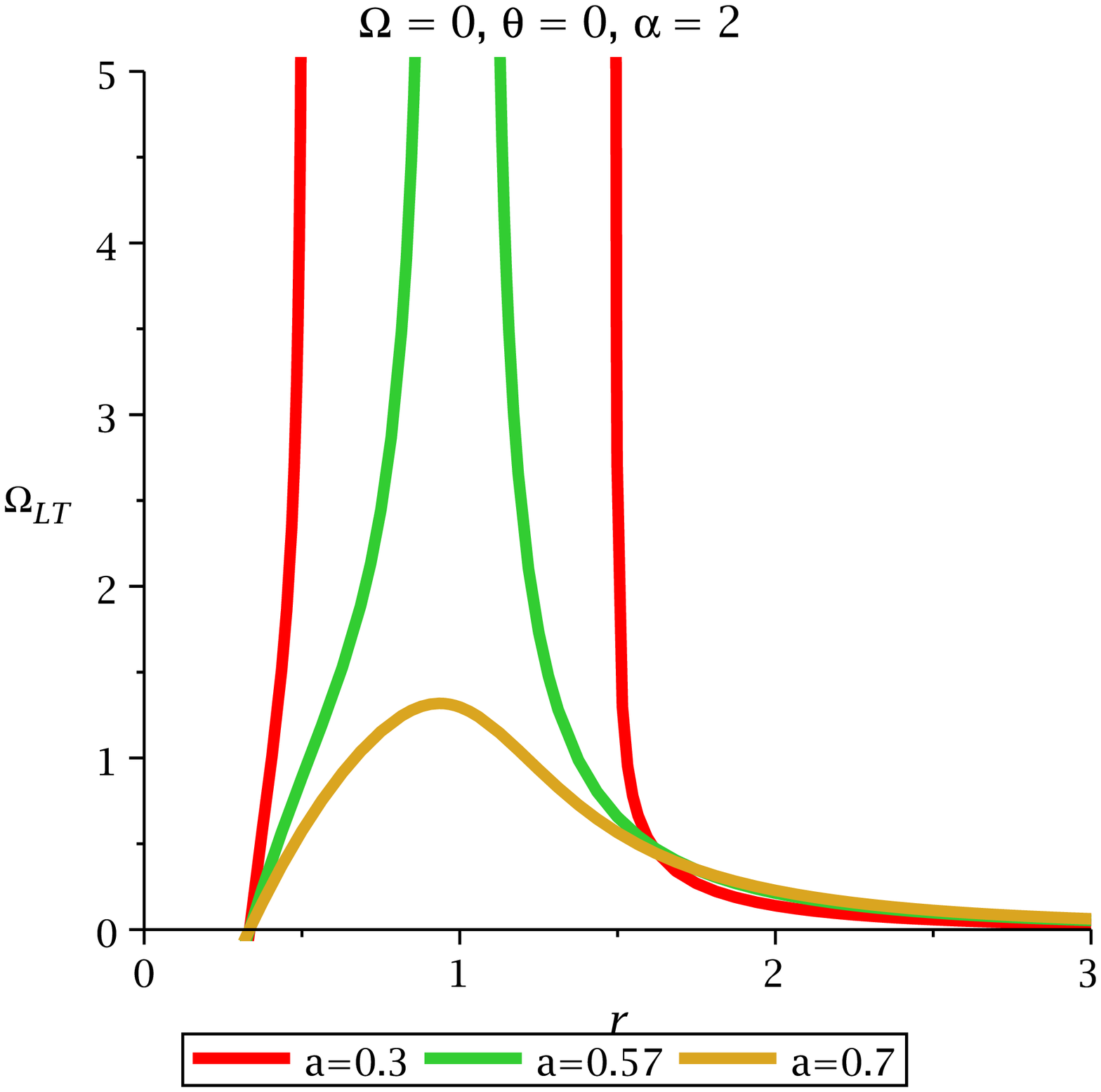}} 
\subfigure[]{
\includegraphics[width=2in,angle=0]{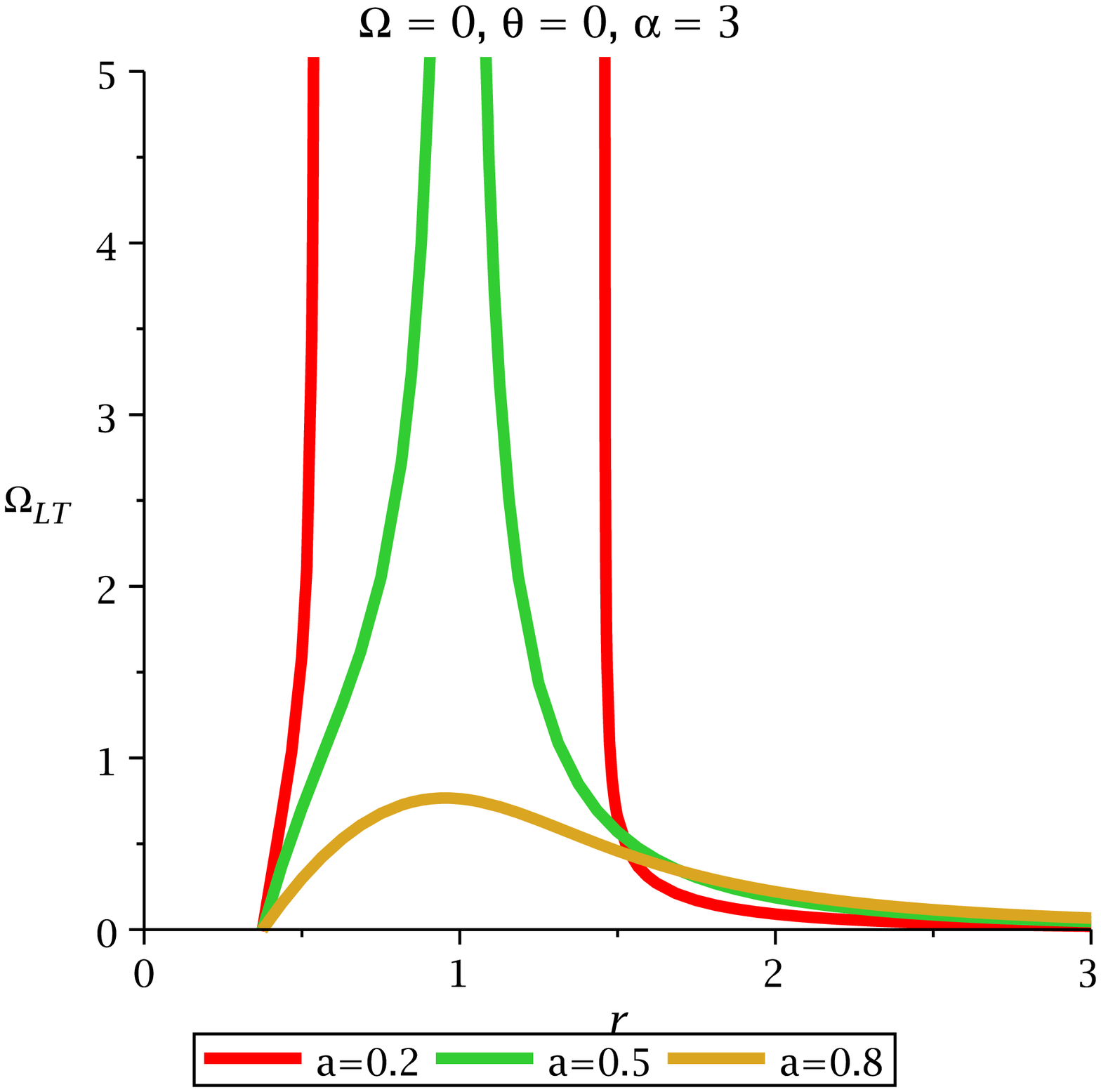}} 
\subfigure[]{
\includegraphics[width=2in,angle=0]{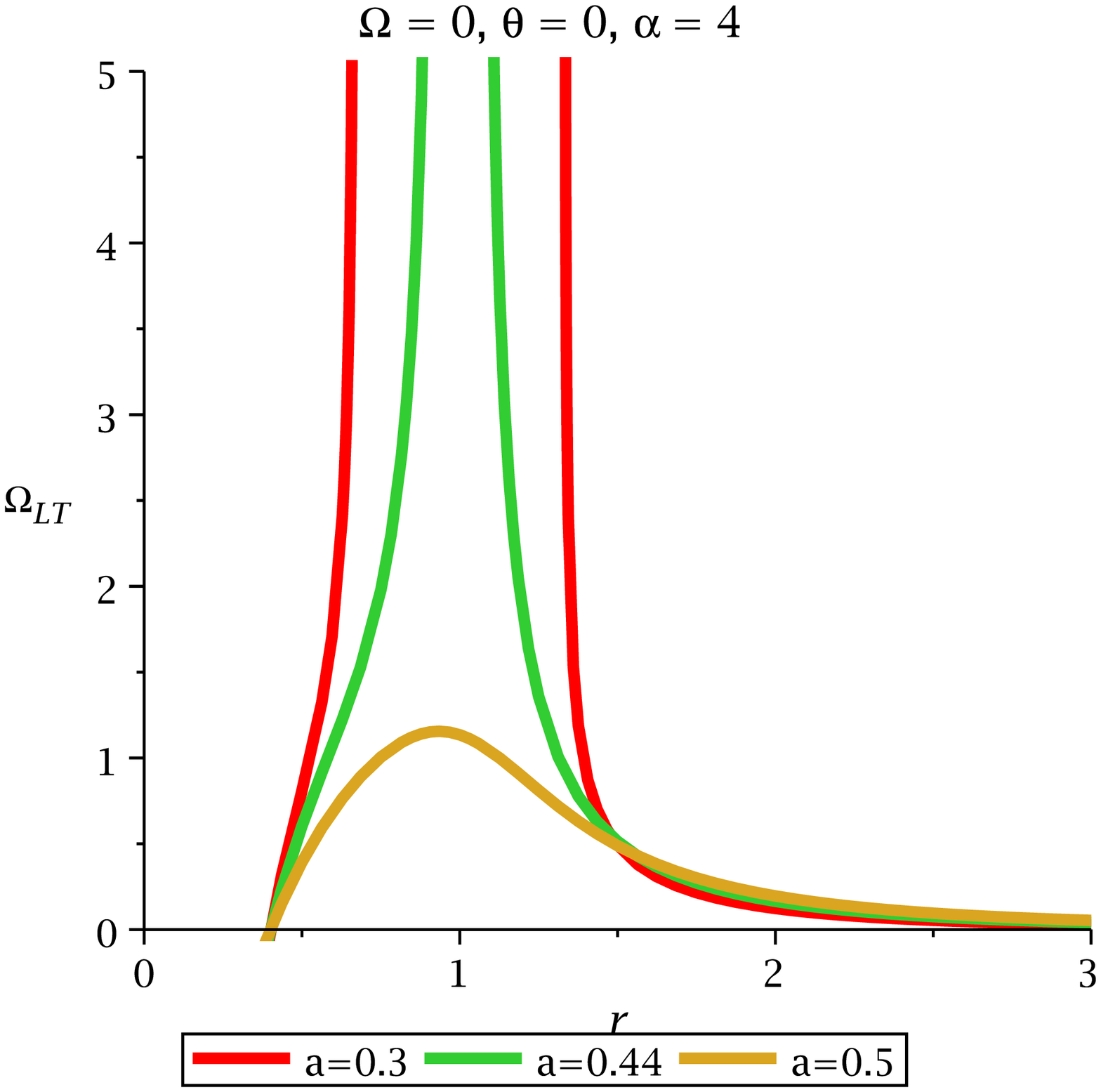}} 
\end{center}
\caption{ Examples of the variation  of $\Omega_{LT}$ versus $r$ for $\theta=0$ in KMOG with variation 
of MOG parameter and  spin parameter. Here  $\Omega=0$. The first figure describes the variation  
of $\Omega_{LT}$  with $r$ for non-extremal BH, extremal BH and NS without MOG parameter. 
The rest of the figure describes the variation  of $\Omega_{LT}$  with $r$ for non-extremal BH, 
extremal BH and NS with MOG parameter.  Using these plots one can easily distinguish 
between three compact objects.} \label{gm}
\end{figure}
%%%%%%%%%%%%%%%%%%%%%%%%%%%%%%%%%%%%%%%%%%%%%%
\subsection{Behaviour of $\vec{\Omega}_{LT}$ at $\theta=\frac{\pi}{6}$}
In this limit $\theta=\frac{\pi}{6}$, the LT frequency is computed as 
\begin{eqnarray}
\vec{\Omega}_{LT}|_{\theta= \frac{\pi}{6}} &=& 
\frac{\sqrt{3 \Delta}\chi(r)|_{\theta=\frac{\pi}{6}}\hat{r}+\mu(r)|_{\theta=\frac{\pi}{6}} \hat{\theta}}
{2\sigma(r)|_{\theta= \frac{\pi}{6}}}~,~\label{tl2}
\end{eqnarray}
The magnitude of this vector is calculated to be 
\begin{eqnarray}
\Omega_{LT}|_{\theta=\frac{\pi}{6}} &=& \frac{\sqrt{3\Delta \chi^2 (r)|_{\theta=\frac{\pi}{6}}
+\mu^2(r)|_{\theta=\frac{\pi}{6}}}}{2\sigma(r)|_{\theta= \frac{\pi}{6}}} ~\label{tl3}
\end{eqnarray}
where 
\begin{eqnarray}
\chi~(r)|_{\theta= \frac{\pi}{6}} &=& a \Pi_{\alpha}
\end{eqnarray}
$$
\mu(r)|_{\theta= \frac{\pi}{6}} = aG_{N}{\cal M}\left(r^2-\frac{3}{4}a^2\right)
-\frac{\alpha}{1+\alpha} G_{N}^2 {\cal M}^2 ar
$$
and
$$
\sigma(r)|_{\theta= \frac{\pi}{6}} = \left(r^2+\frac{3}{4}a^2\right)^\frac{3}{2}
\left[\left(r^2+\frac{3}{4}a^2\right)-\Pi_{\alpha}\right]
$$
From the above equation one can easily seen that the LT frequency is dependent on the MOG parameter
and spin parameter. It could seen from the diagram~(\ref{nl}) by plotting the LT precession frequency 
with the radial coordinate.
\begin{figure}
\begin{center}
\subfigure[]{
\includegraphics[width=2in,angle=0]{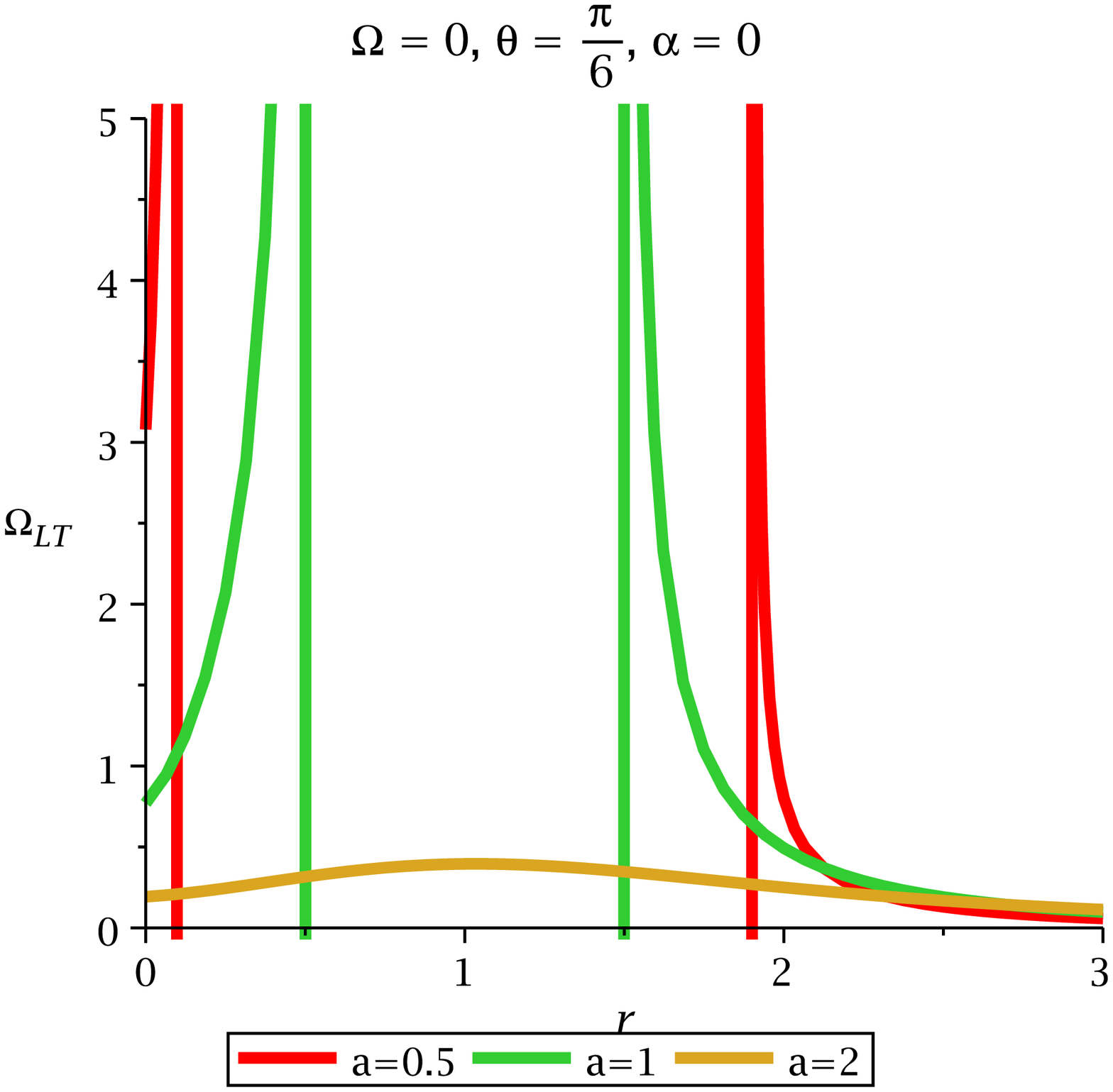}} 
\subfigure[]{
\includegraphics[width=2in,angle=0]{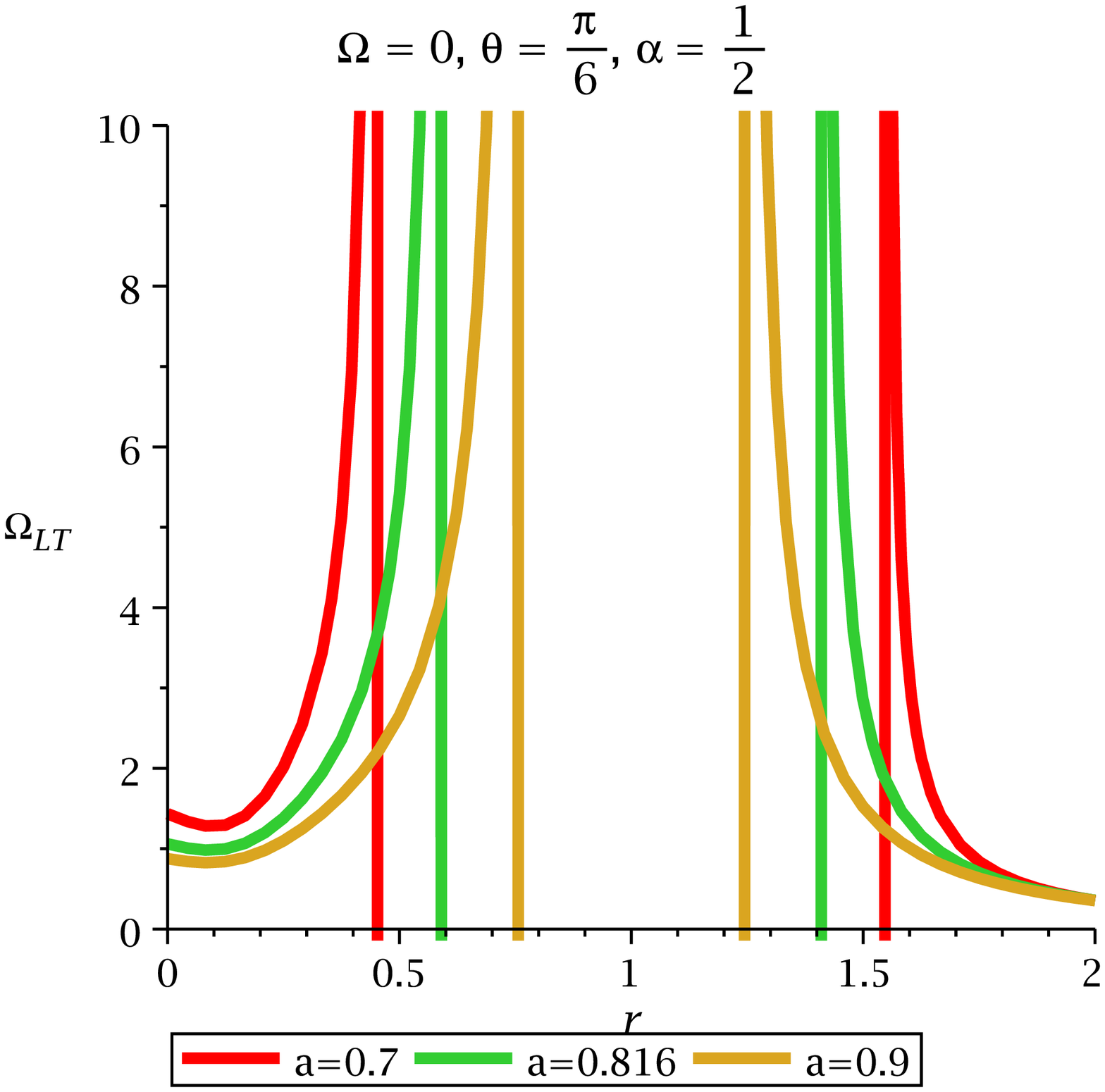}}
\subfigure[]{
\includegraphics[width=2in,angle=0]{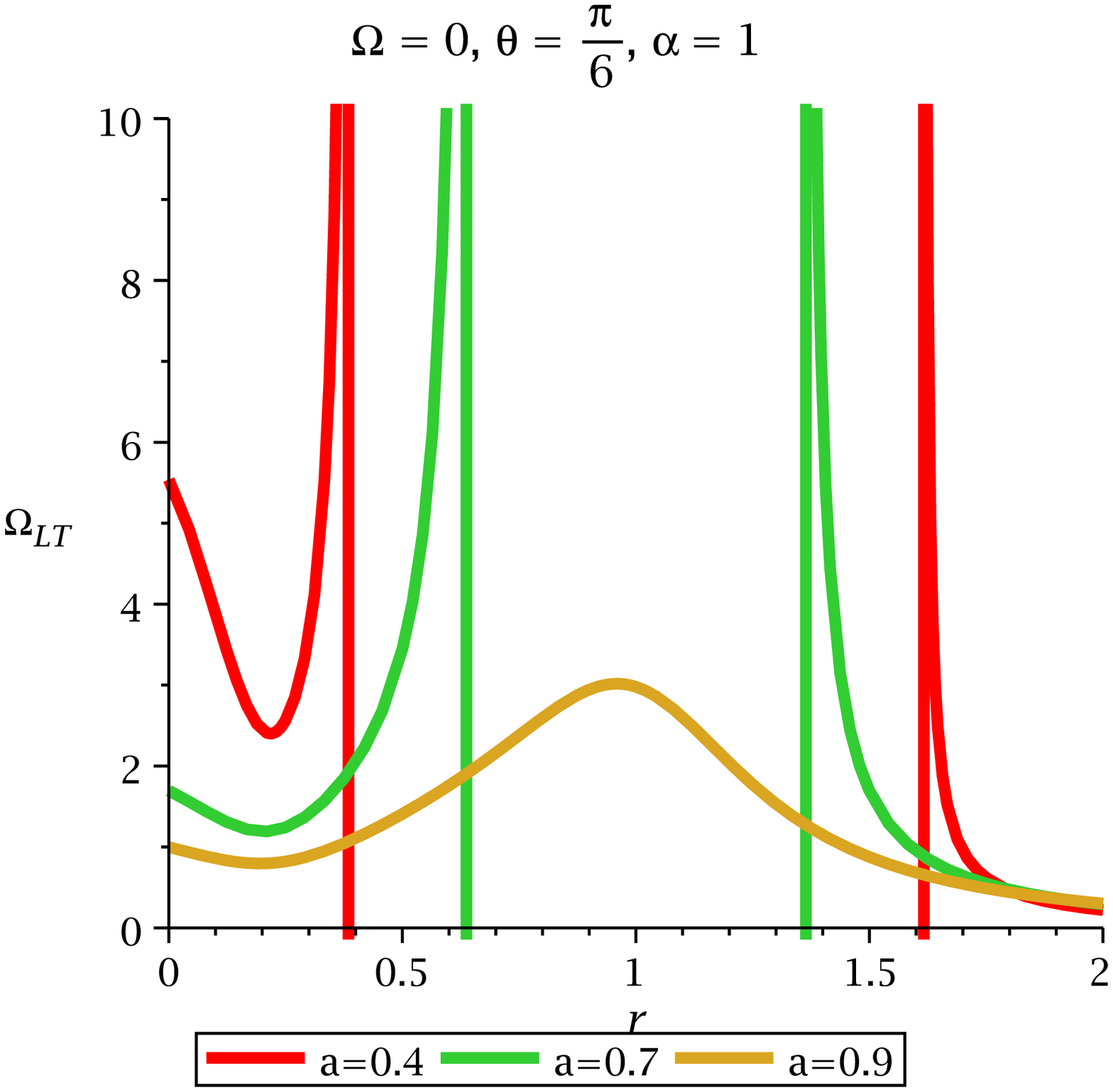}} 
\subfigure[]{
\includegraphics[width=2in,angle=0]{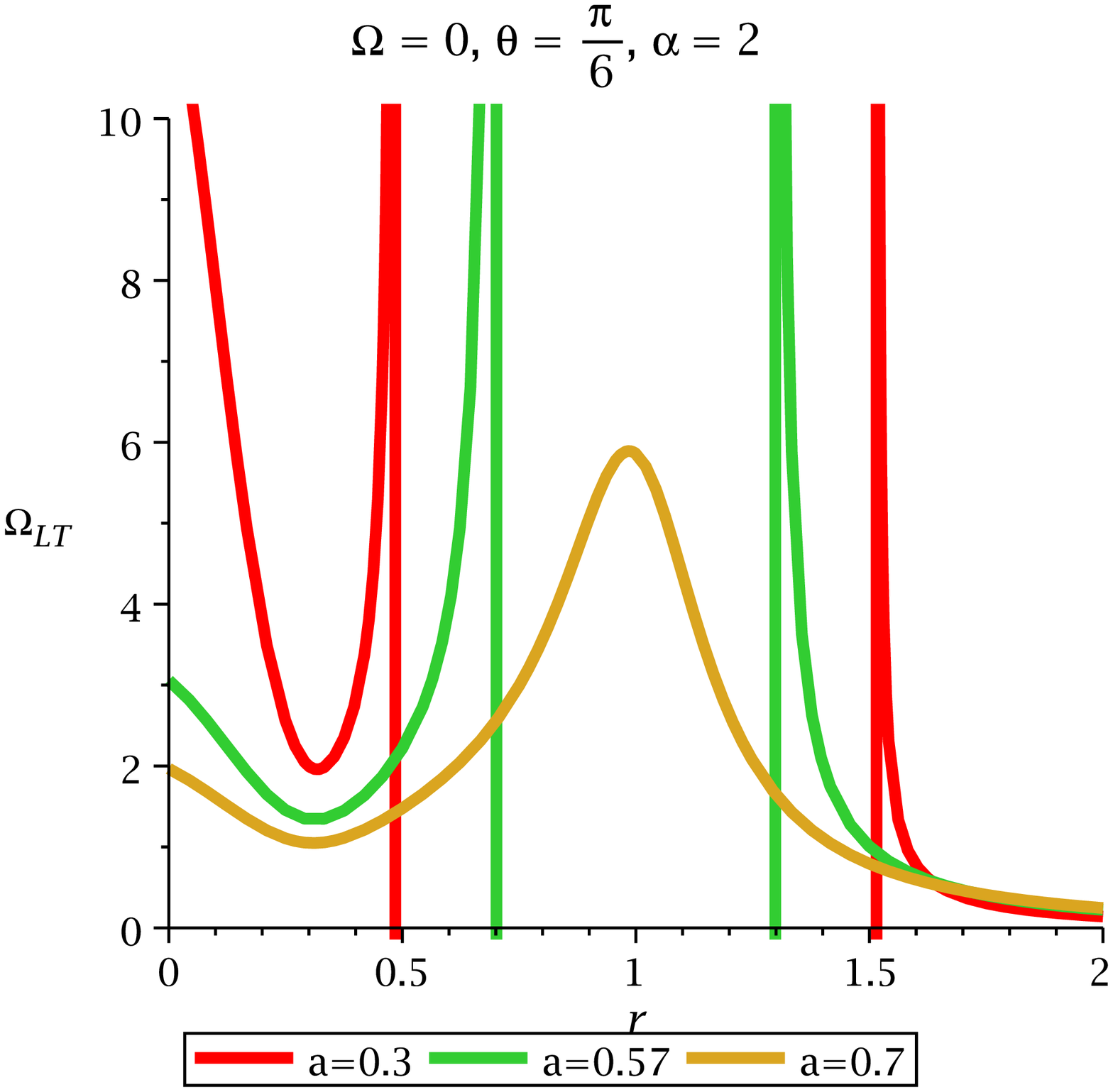}} 
\subfigure[]{
\includegraphics[width=2in,angle=0]{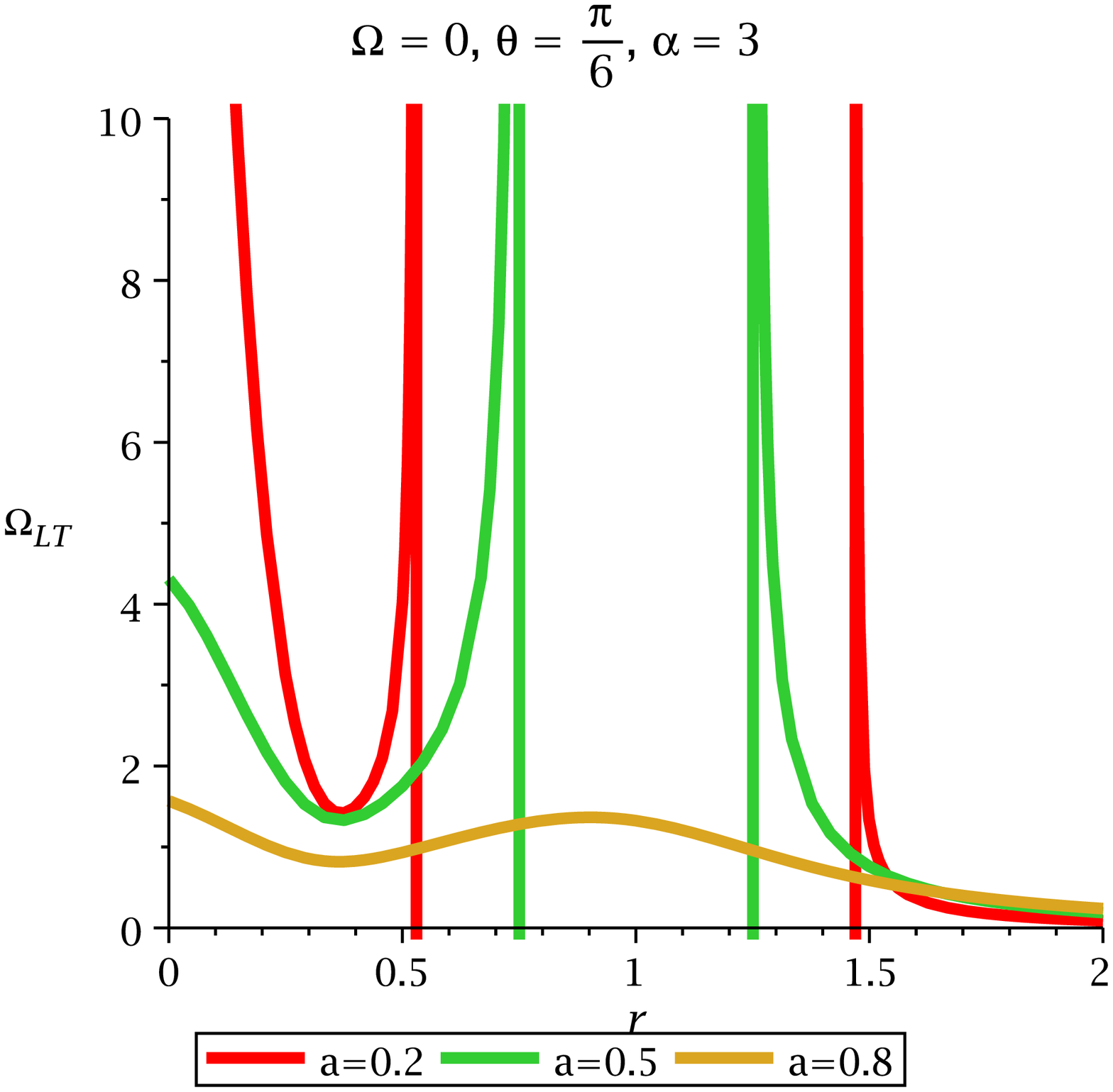}}
\subfigure[]{
\includegraphics[width=2in,angle=0]{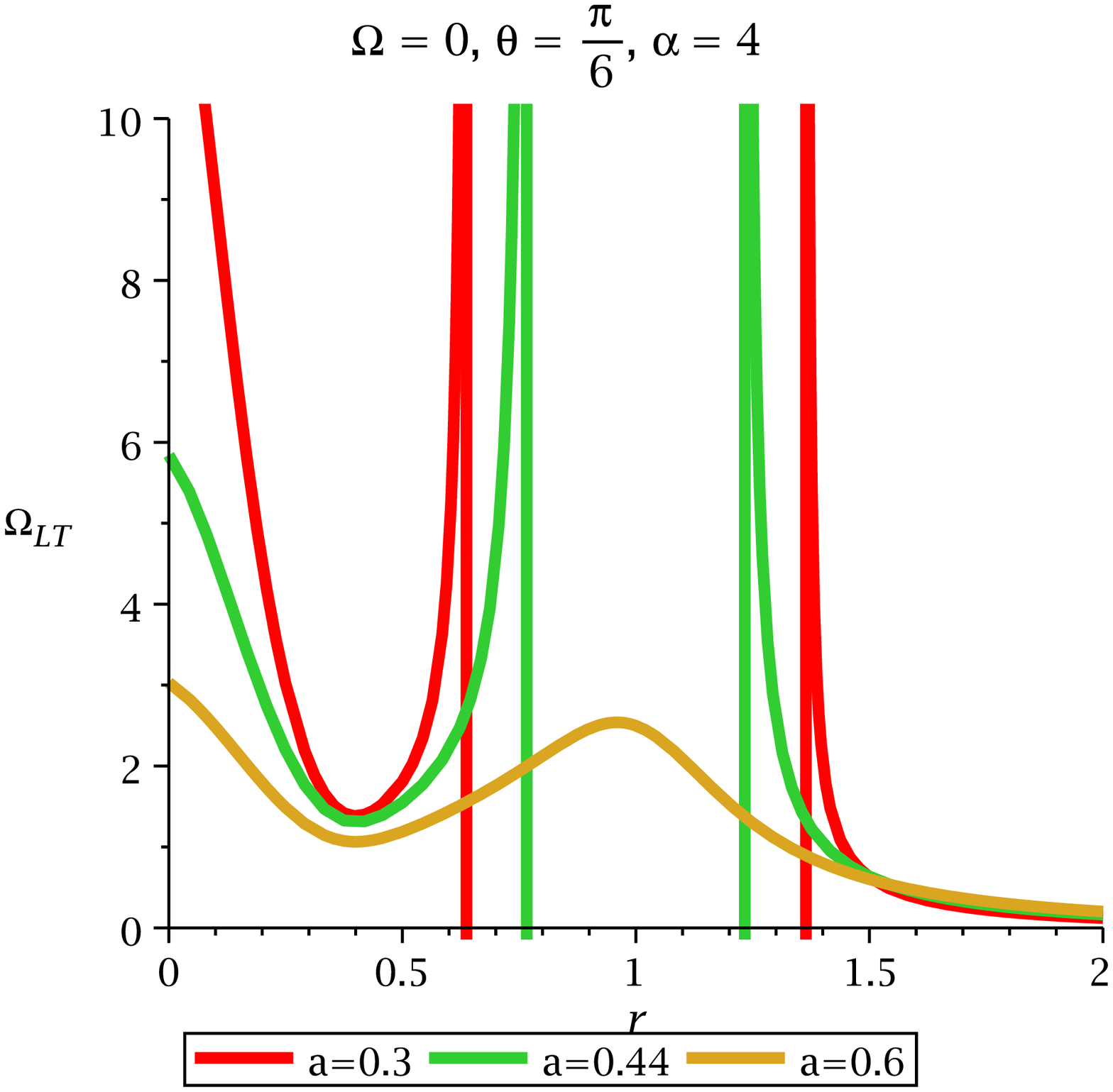}} 
\caption{Examples of the variation  of $\Omega_{LT}$ versus $r$ for $\theta=\frac{\pi}{6}$ in 
KMOG with variation of MOG parameter and  spin parameter. Here  $\Omega=0$. The first figure 
describes the variation  of $\Omega_{LT}$  with $r$ for non-extremal BH, extremal BH and NS 
without MOG parameter. 
The rest of the figure describes the variation  of $\Omega_{LT}$  with $r$ for 
non-extremal BH, extremal BH and NS with MOG parameter. }
\label{nl}
\end{center}
\end{figure}
%%%%%%%%%%%%%%%%%%%%%%%%%%%%%%%%%%%%%%%%%%%%%%%%%

\subsection{Behaviour of $\vec{\Omega}_{LT}$ at $\theta=\frac{\pi}{4}$}
In the limit $\theta=\frac{\pi}{4}$, the LT frequency is computed to be 
\begin{eqnarray}
\vec{\Omega}_{LT}|_{\theta= \frac{\pi}{4}} &=& 
\frac{\sqrt{\Delta}\chi(r)|_{\theta=\frac{\pi}{4}}\hat{r}+\mu(r)|_{\theta=\frac{\pi}{4}}\hat{\theta}}
{\sqrt{2}\sigma(r)|_{\theta= \frac{\pi}{4}}}~,~\label{ttl6}
\end{eqnarray}
The magnitude of this vector is thus 
\begin{eqnarray}
\Omega_{LT}|_{\theta=\frac{\pi}{4}} &=& \frac{\sqrt{\Delta \chi^2 (r)|_{\theta=\frac{\pi}{4}}
+\mu^2(r)|_{\theta=\frac{\pi}{4}}}}{\sqrt{2}\sigma(r)|_{\theta= \frac{\pi}{4}}} ~\label{ttl7}
\end{eqnarray}
where
\begin{eqnarray}
\chi~(r)|_{\theta= \frac{\pi}{4}} &=& a \Pi_{\alpha}~\label{tl8}
\end{eqnarray}
$$
\mu(r)|_{\theta= \frac{\pi}{4}} = aG_{N}{\cal M}\left(r^2-\frac{a^2}{2}\right)
-\frac{\alpha}{1+\alpha} G_{N}^2 {\cal M}^2 ar
$$
$$
\sigma(r)|_{\theta= \frac{\pi}{4}} = \left(r^2+\frac{a^2}{2}\right)^\frac{3}{2}
\left[\left(r^2+\frac{a^2}{2}\right)-\Pi_{\alpha} \right]
$$
One could differentiate the non-extremal BH, extremal BH and NS from the 
diagram~(\ref{pi4q})
\begin{figure}
\begin{center}
\subfigure[]{
\includegraphics[width=2in,angle=0]{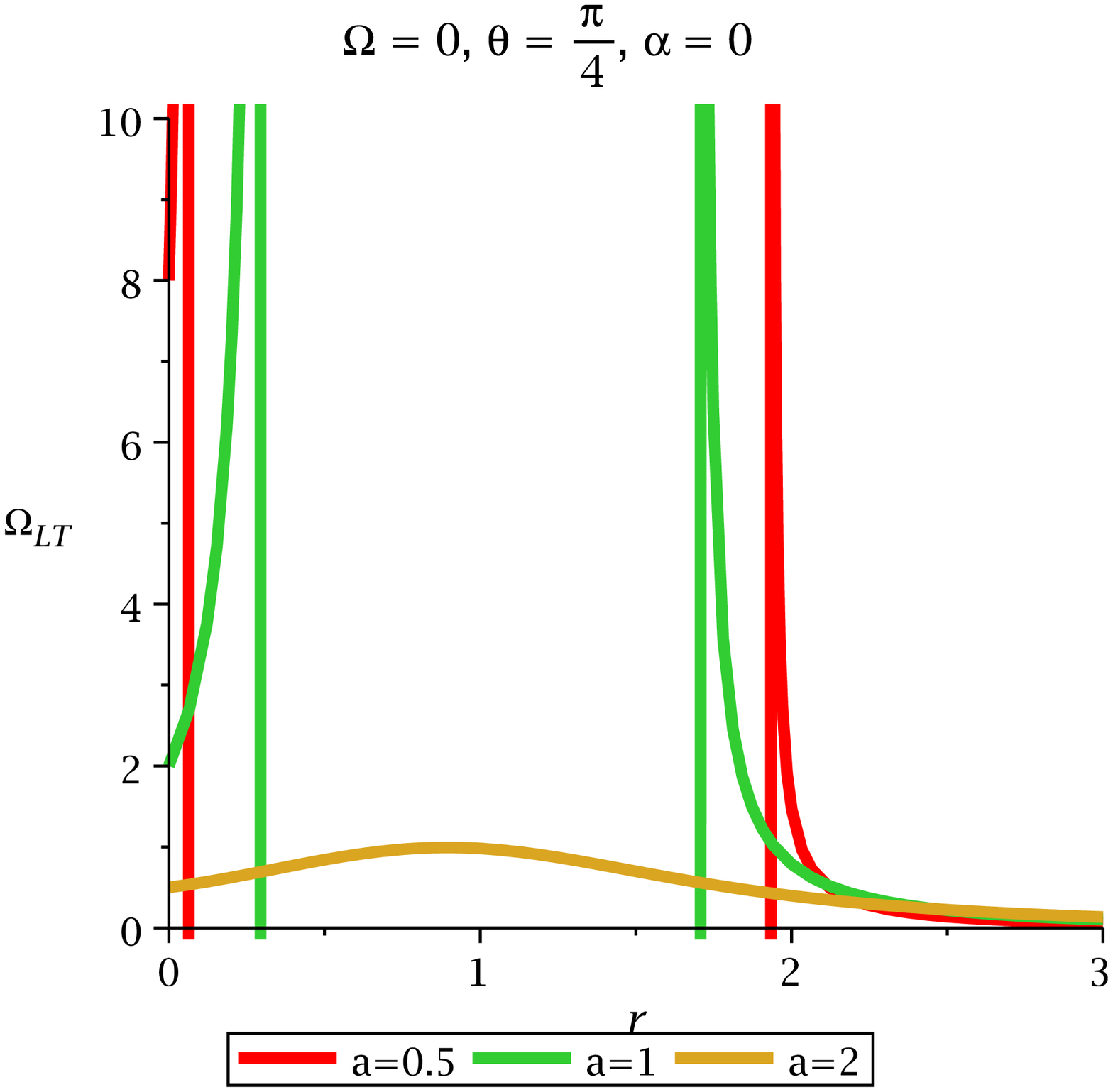}} 
\subfigure[]{
\includegraphics[width=2in,angle=0]{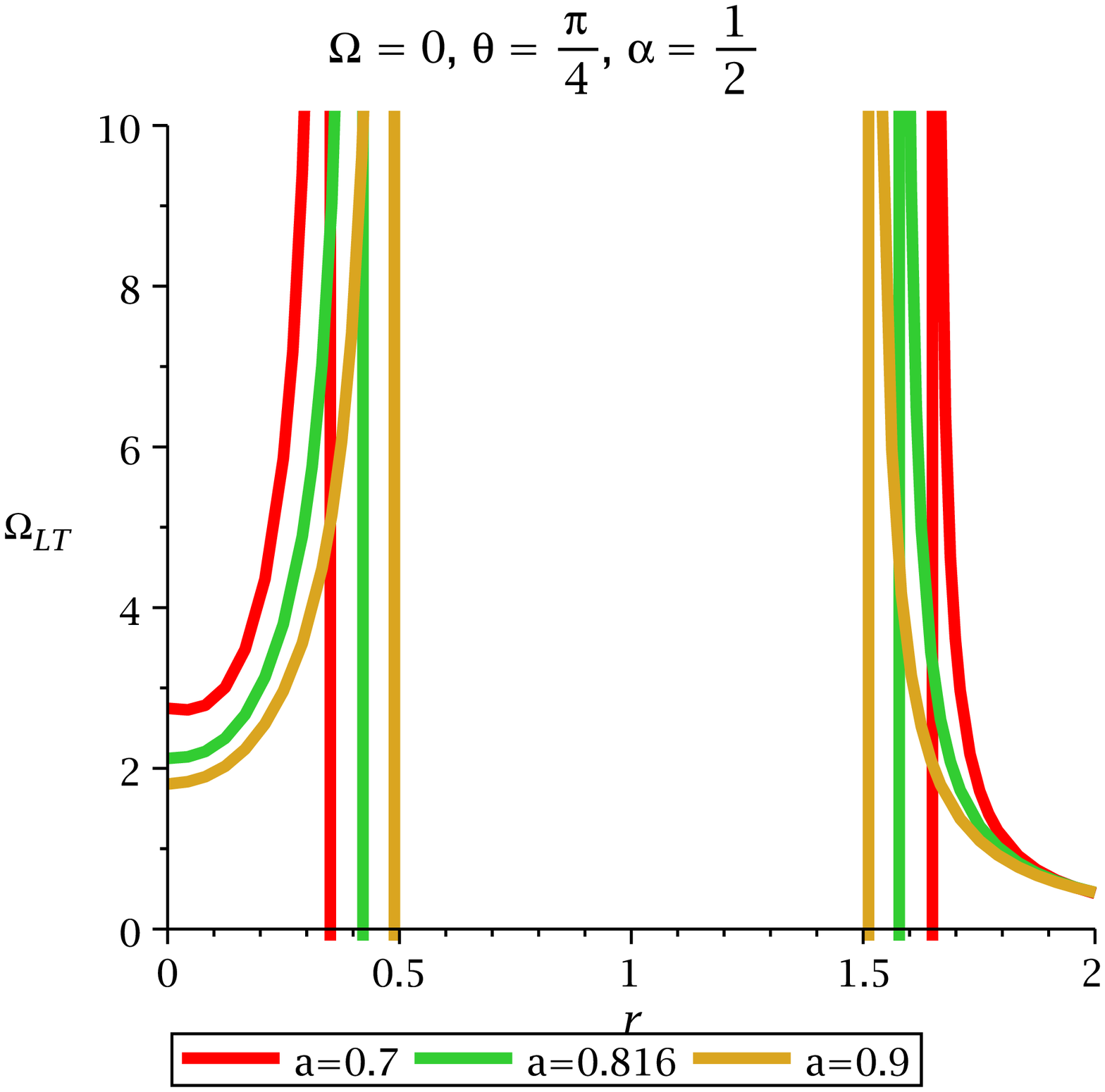}}
\subfigure[]{
\includegraphics[width=2in,angle=0]{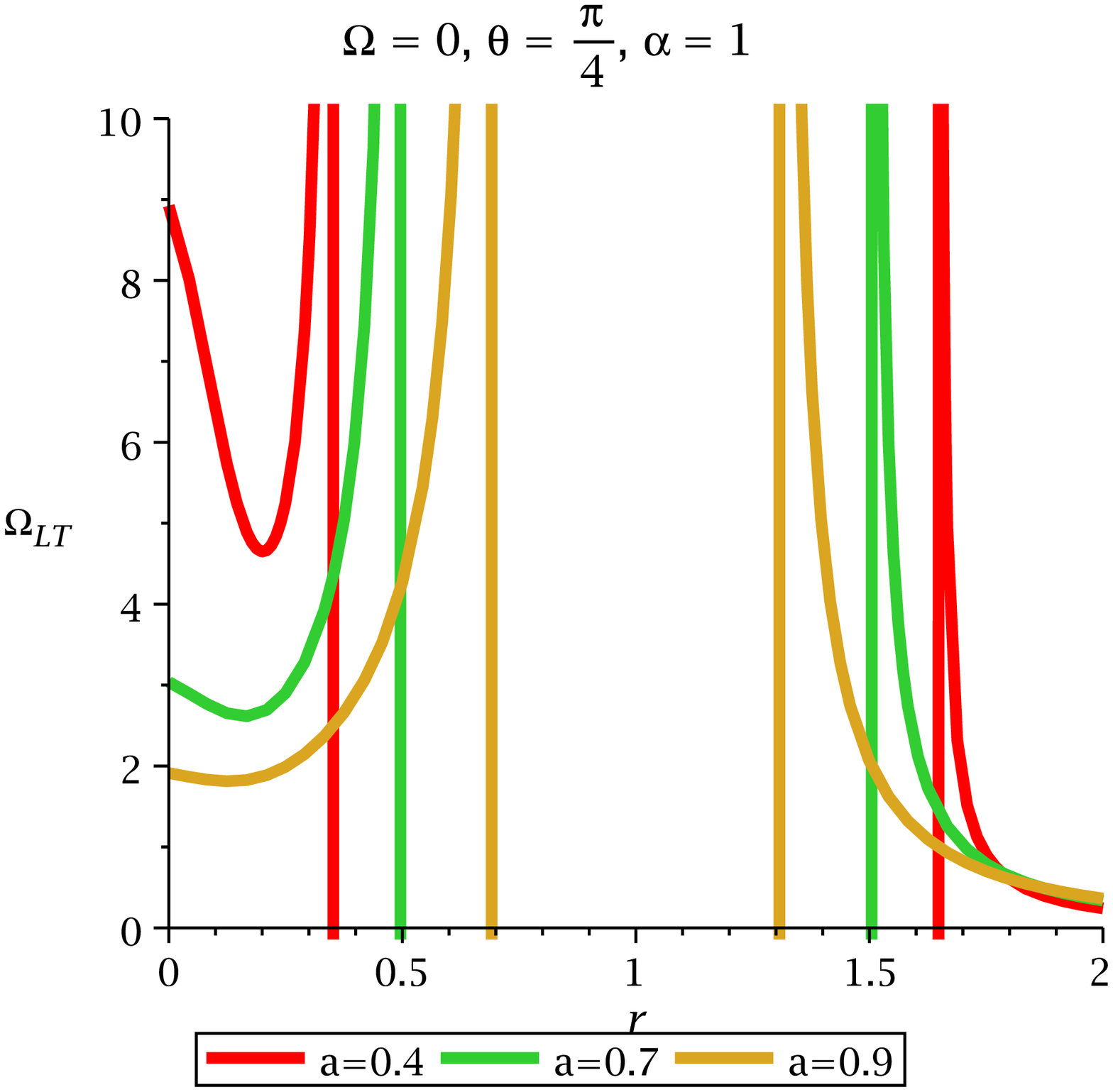}} 
\subfigure[]{
\includegraphics[width=2in,angle=0]{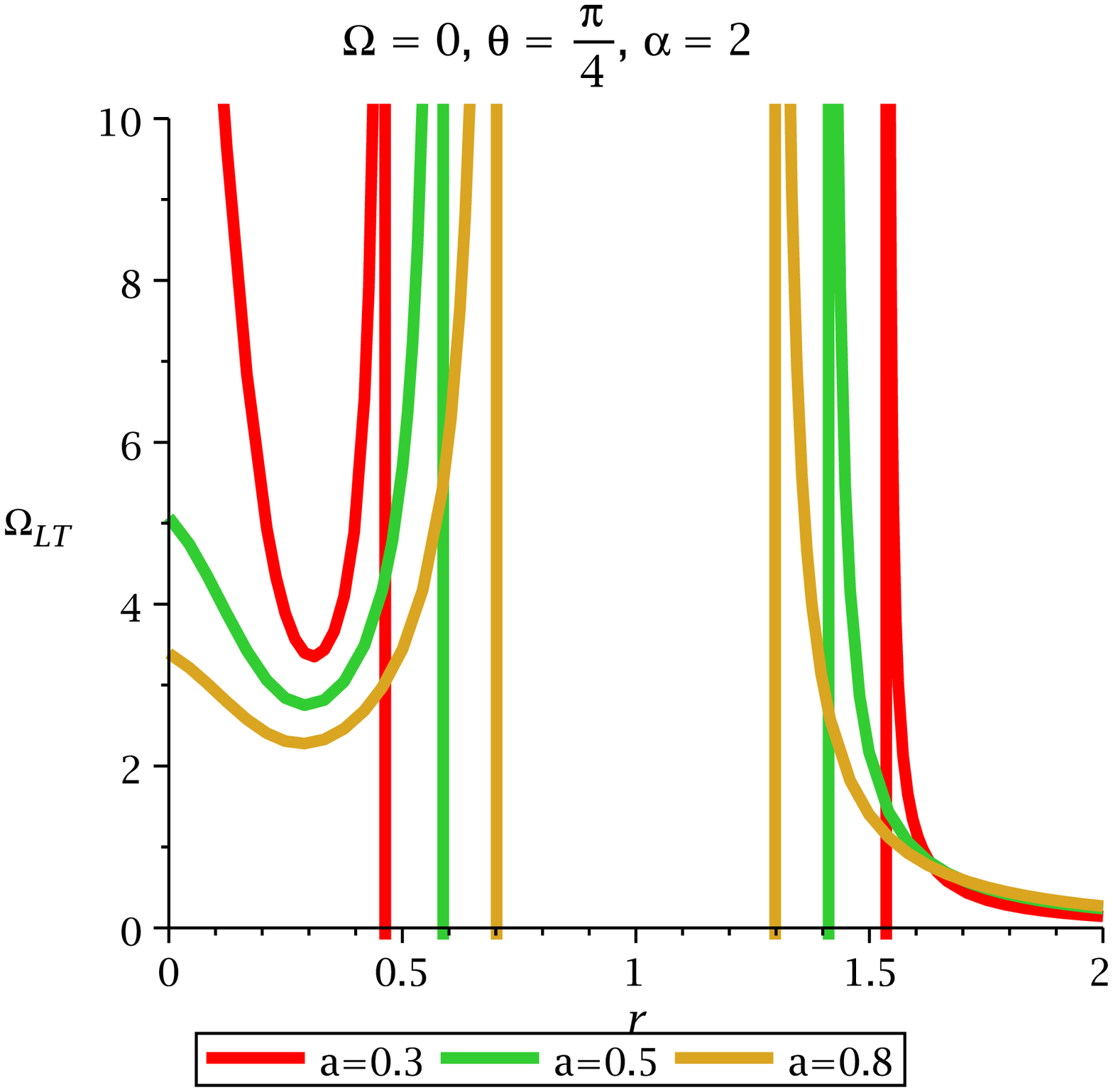}} 
\subfigure[]{
\includegraphics[width=2in,angle=0]{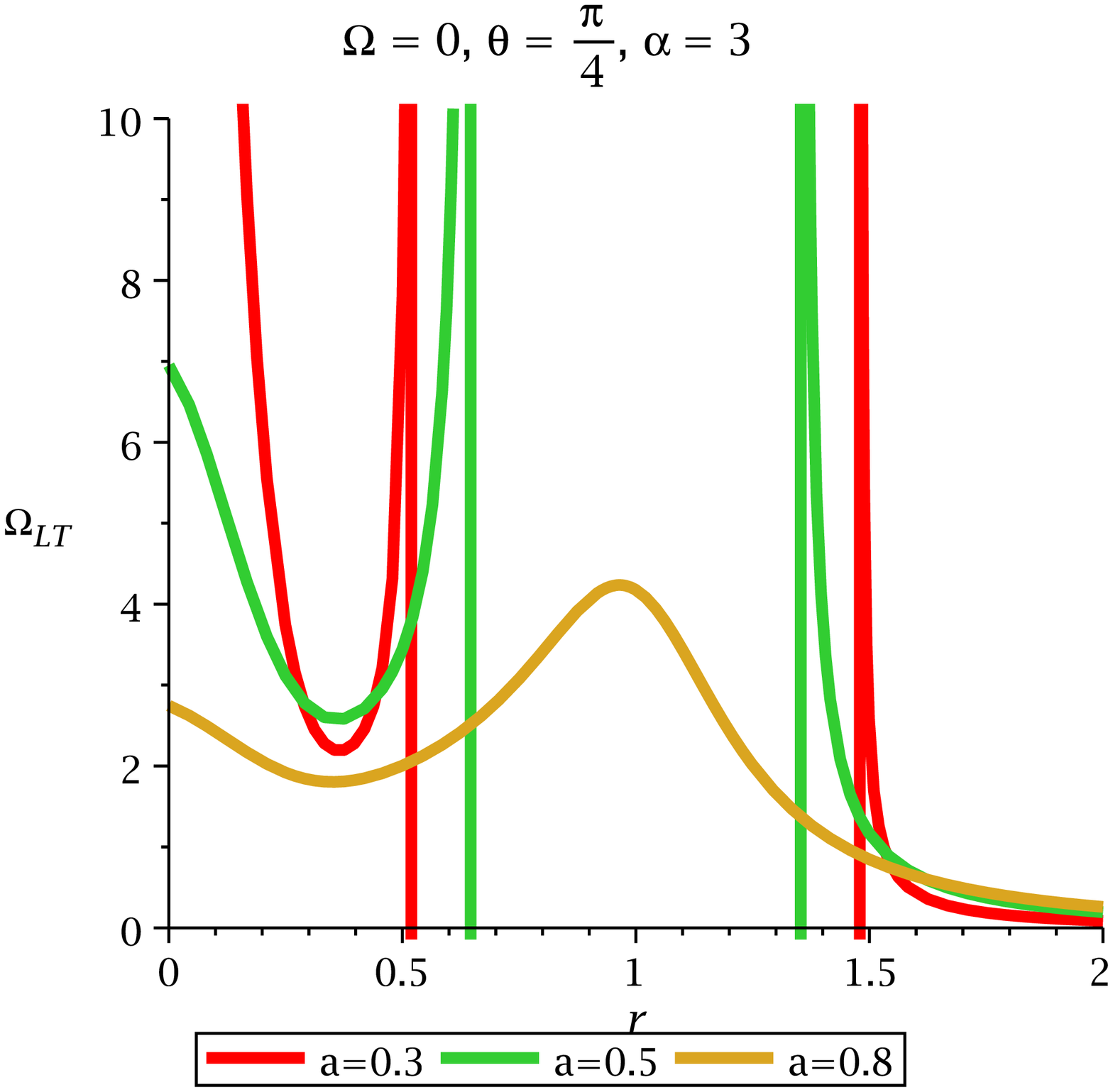}}
\subfigure[]{
\includegraphics[width=2in,angle=0]{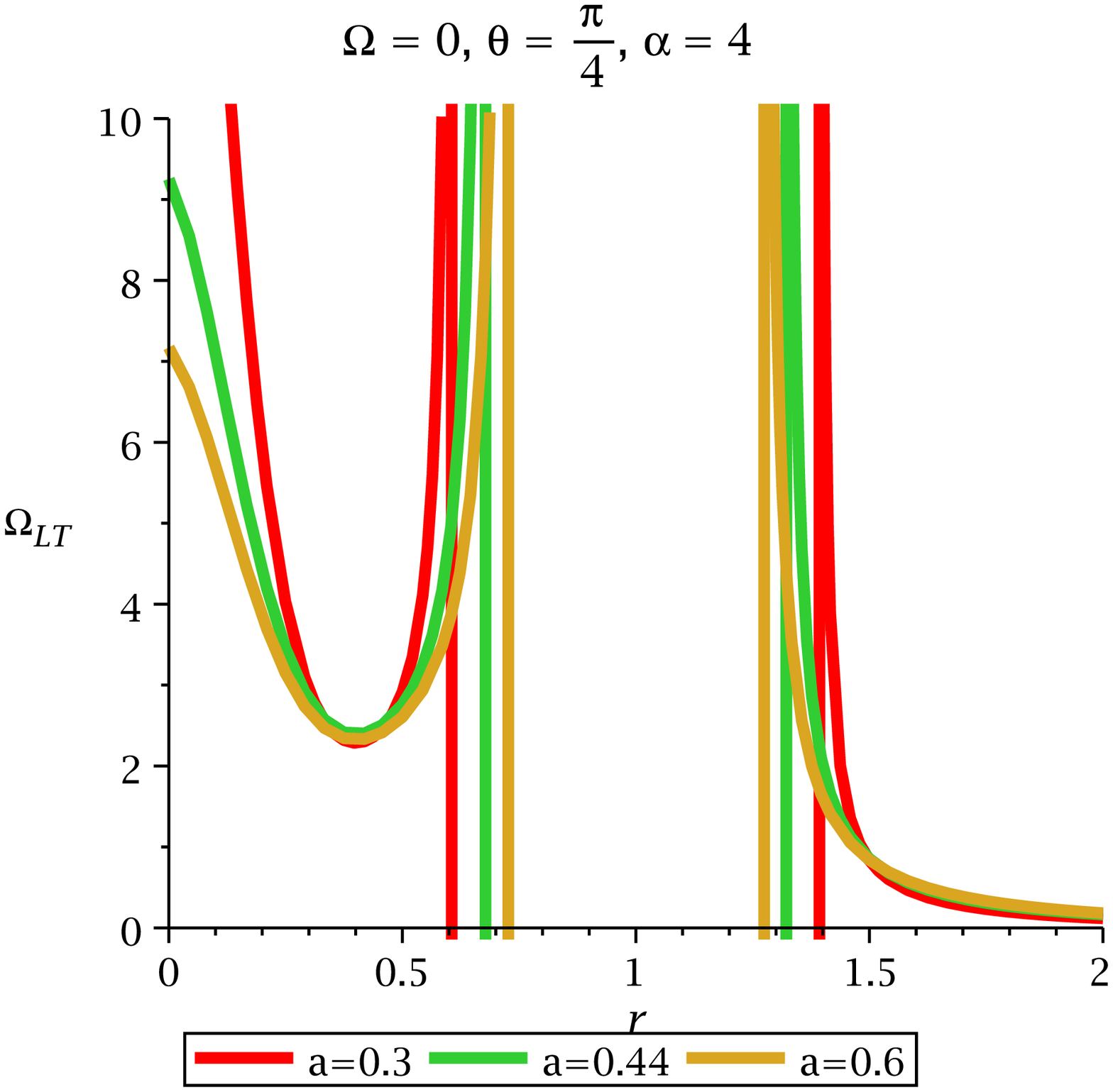}} 
\caption{Examples of the variation  of $\Omega_{LT}$ versus $r$ for $\theta=\frac{\pi}{4}$ in 
KMOG with variation of MOG parameter and  spin parameter. Here  $\Omega=0$. The first figure 
describes the variation  of $\Omega_{LT}$  with $r$ for non-extremal BH, extremal BH and NS 
without MOG parameter. The rest of the figure describes the variation  of $\Omega_{LT}$  
with $r$ for non-extremal BH, extremal BH and NS with MOG parameter.}
\label{pi4q}
\end{center}
\end{figure}

\subsection{Behaviour of $\vec{\Omega}_{LT}$ at $\theta=\frac{\pi}{3}$}
Similarly for $\theta=\frac{\pi}{3}$, the LT frequency is 
\begin{eqnarray}
\vec{\Omega}_{LT}|_{\theta= \frac{\pi}{3}} &=& 
\frac{\sqrt{\Delta}\chi(r)|_{\theta=\frac{\pi}{3}}\hat{r}+\sqrt{3}\mu(r)|_{\theta=\frac{\pi}{3}} \hat{\theta}}
{2\sigma(r)|_{\theta= \frac{\pi}{3}}}~,~\label{tl5}
\end{eqnarray}
The magnitude of this vector is given by
\begin{eqnarray}
\Omega_{LT}|_{\theta=\frac{\pi}{3}} &=& 
\frac{\sqrt{\Delta \chi^2 (r)|_{\theta=\frac{\pi}{3}}+3\mu^2(r)|_{\theta=\frac{\pi}{3}}}}
{2\sigma(r)|_{\theta= \frac{\pi}{3}}} ~\label{vt16}
\end{eqnarray}
where 
\begin{eqnarray}
\chi~(r)|_{\theta= \frac{\pi}{3}} &=& a \Pi_{\alpha}
\end{eqnarray}
$$
\mu(r)|_{\theta= \frac{\pi}{3}} = aG_{N}{\cal M}\left(r^2-\frac{a^2}{4}\right)
-\frac{\alpha}{1+\alpha} G_{N}^2 {\cal M}^2 ar
$$
and
$$
\sigma(r)|_{\theta= \frac{\pi}{3}} = \left(r^2+\frac{a^2}{4}\right)^\frac{3}{2}
\left[\left(r^2+\frac{a^2}{4}\right)-\Pi_{\alpha}\right]
$$
Variation of spin precession frequency with radial coordinates for various values 
of spin parameter could be seen from the Fig.~(\ref{pi3q}).
\begin{figure}
\begin{center}
\subfigure[]{
\includegraphics[width=2in,angle=0]{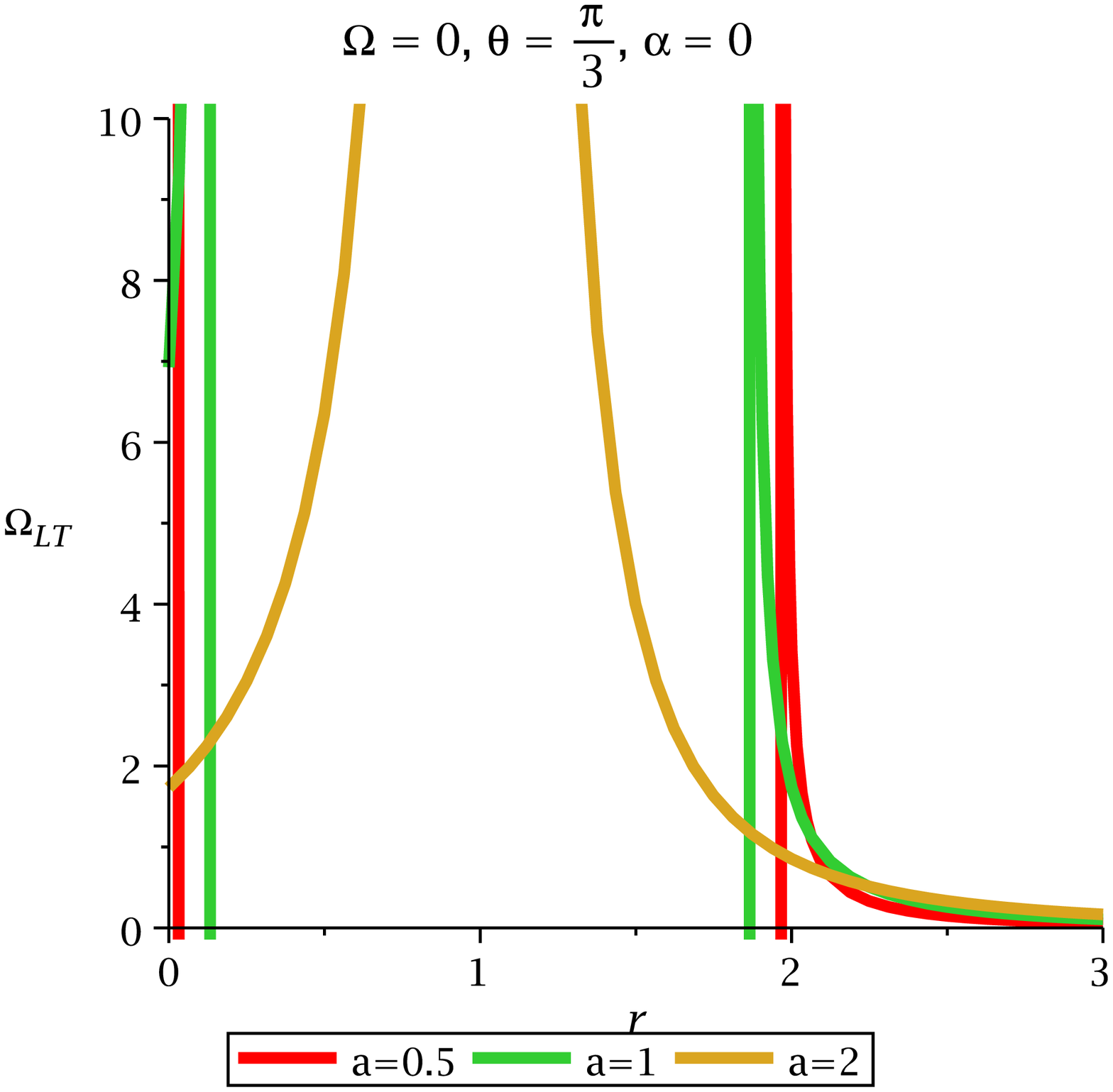}} 
\subfigure[]{
\includegraphics[width=2in,angle=0]{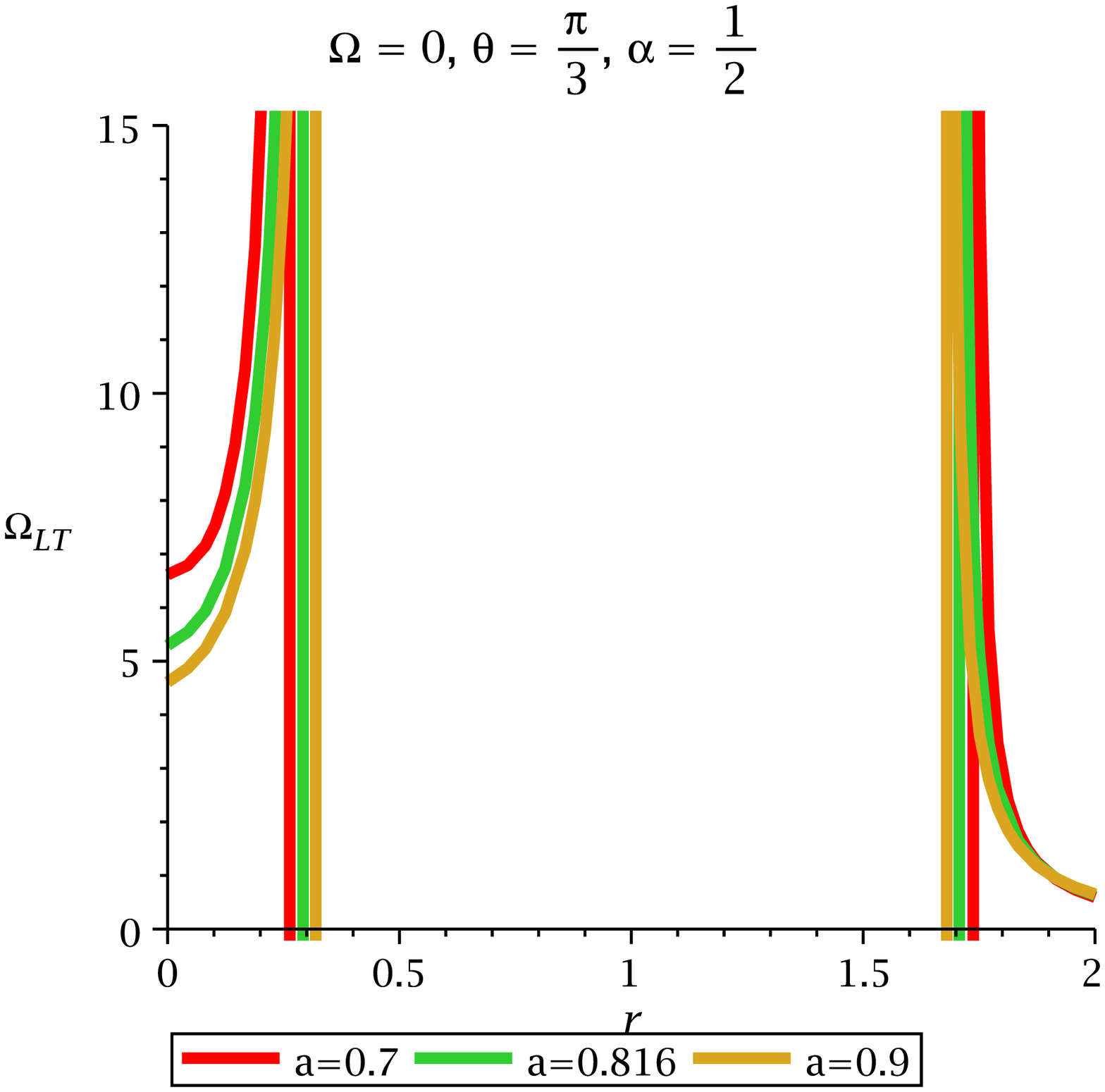}}
\subfigure[]{
\includegraphics[width=2in,angle=0]{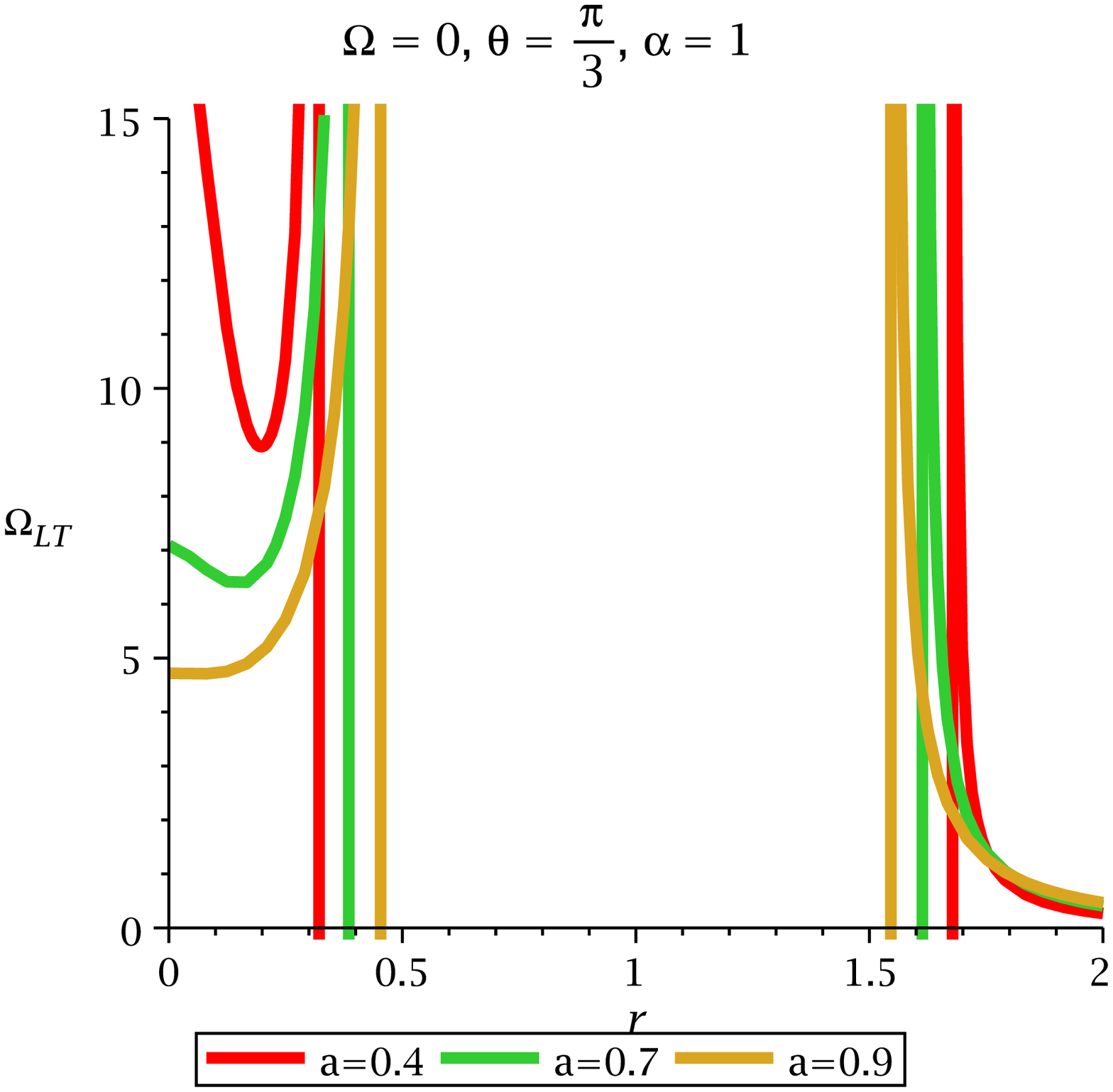}} 
\subfigure[]{
\includegraphics[width=2in,angle=0]{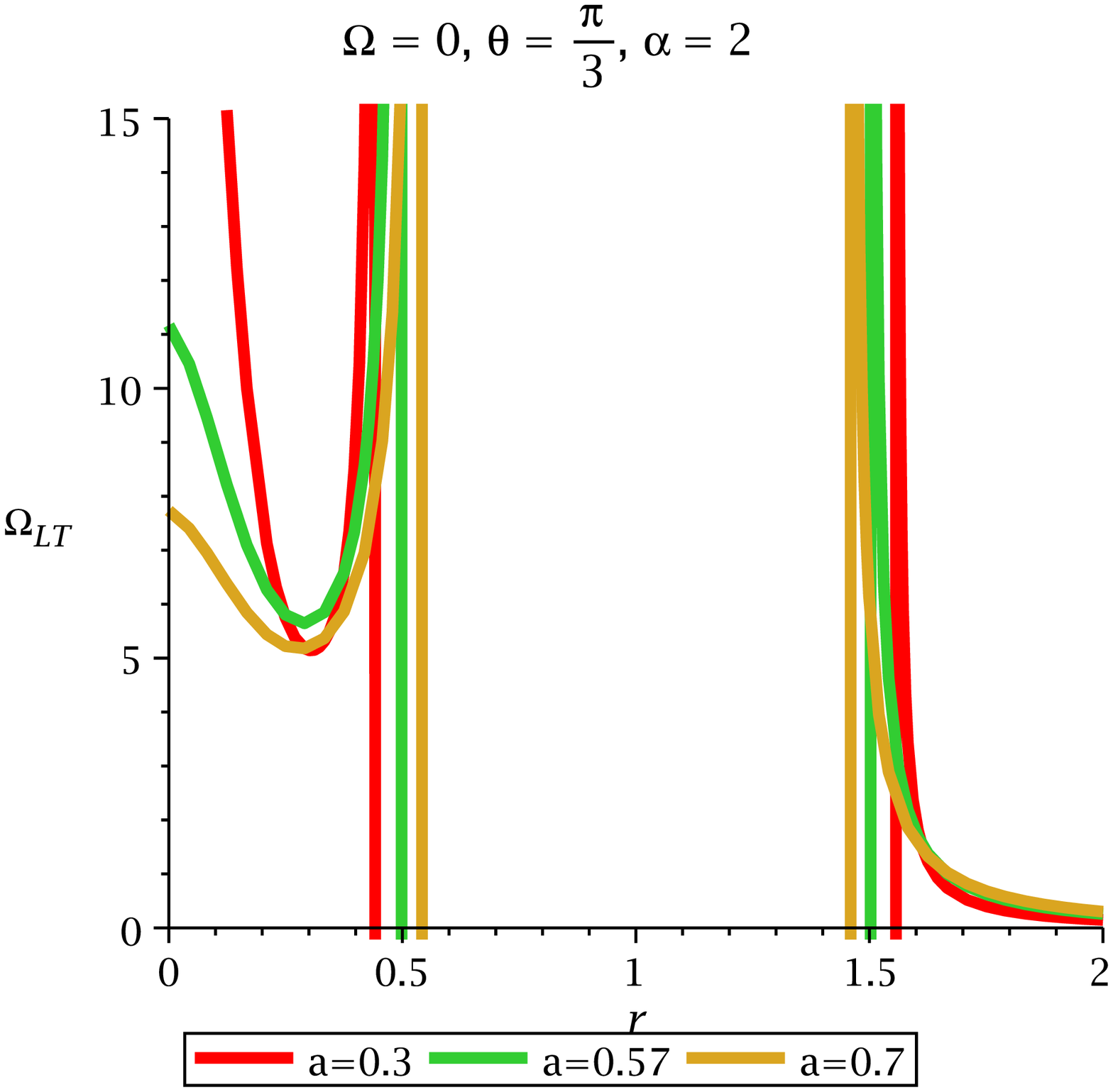}} 
\subfigure[]{
\includegraphics[width=2in,angle=0]{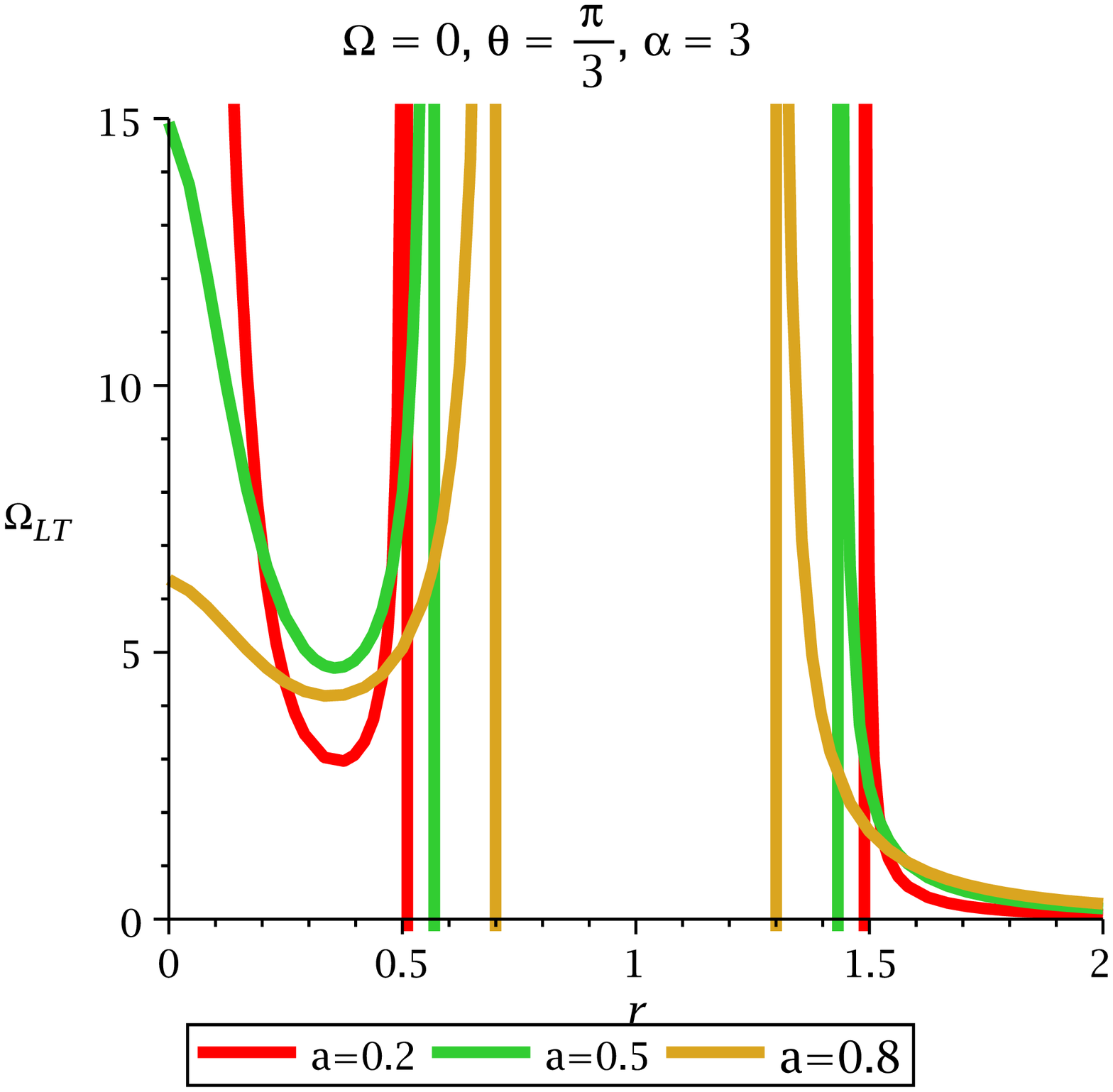}}
\subfigure[]{
\includegraphics[width=2in,angle=0]{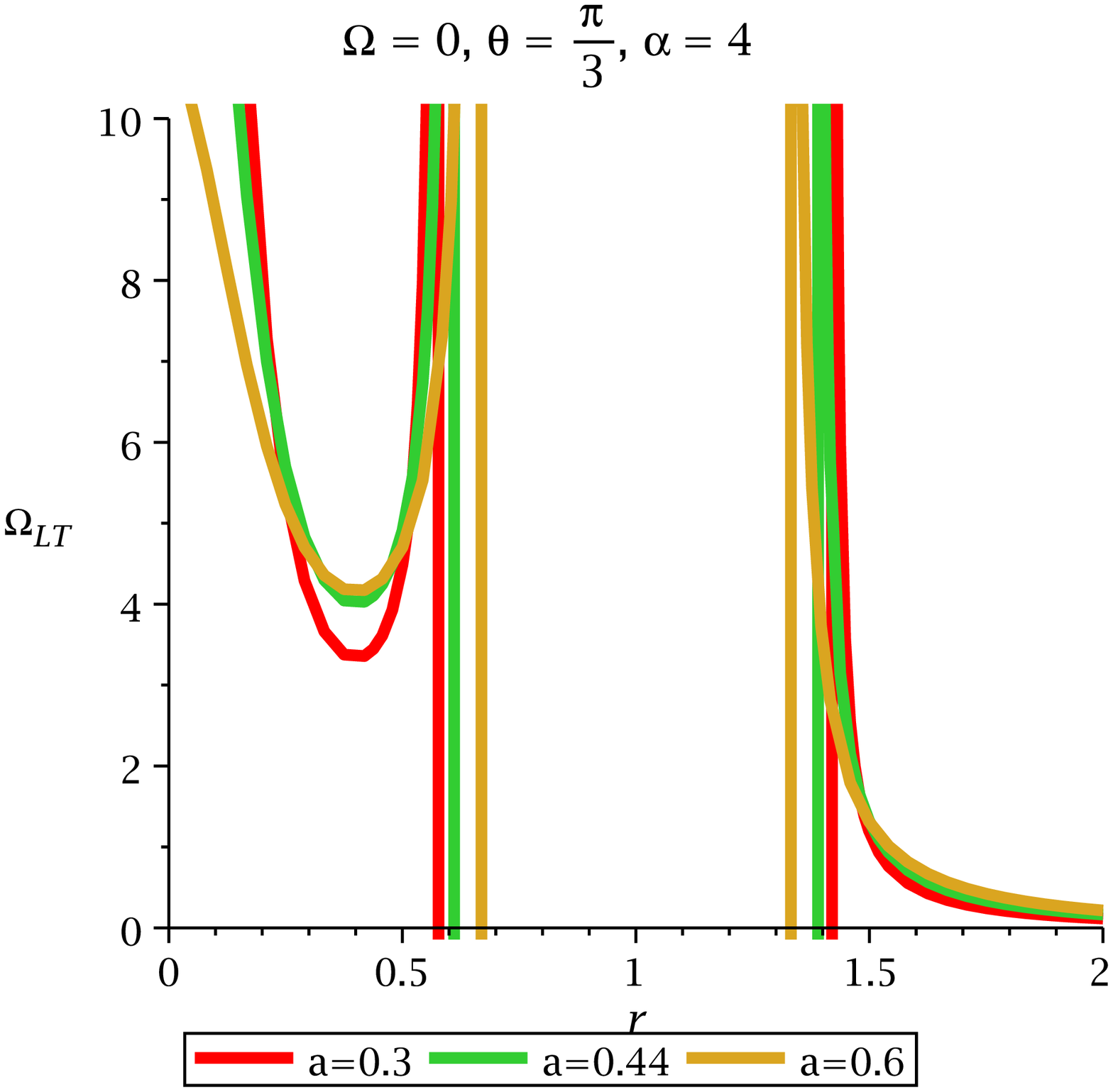}} 
\caption{Examples of the variation  of $\Omega_{LT}$ versus $r$ for $\theta=\frac{\pi}{3}$ in 
KMOG with variation of MOG parameter and  spin parameter. Here  $\Omega=0$. The first figure describes the variation  
of $\Omega_{LT}$  with $r$ for non-extremal BH, extremal BH and NS without MOG parameter. 
The rest of the figure describes the variation  of $\Omega_{LT}$  with $r$ for non-extremal BH, 
extremal BH and NS with MOG parameter.}
\label{pi3q}
\end{center}
\end{figure}
%%%%%%%%%%%%%%%%%%%%%%%%%%%%%%%%%%%%%%%%%%%

\subsection{Behaviour of $\vec{\Omega}_{LT}$ at $\theta=\frac{\pi}{2}$}
On the equatorial plane the precession frequency vector is given by 
\begin{eqnarray} 
\vec{\Omega}_{LT}|_{\theta=\frac{\pi}{2}} &=& 
\frac{\mu(r)|_{\theta=\frac{\pi}{2}}}{\sigma(r)|_{\theta=\frac{\pi}{2}}}\hat{\theta}
~\label{tl16}
\end{eqnarray}
The magnitude of this vector is then
\begin{eqnarray}
\Omega_{LT}|_{\theta=\frac{\pi}{2}} &=& \frac{\mu(r)|_{\theta=\frac{\pi}{2}}}{\sigma(r)|_{\theta=\frac{\pi}{2}}} 
~\label{vt17}
\end{eqnarray}
where 
\begin{eqnarray}
\mu(r)|_{\theta=\frac{\pi}{2}} &=& ar\left(G_{N}{\cal M} r-\frac{\alpha}{1+\alpha} G_{N}^2 {\cal M}^2\right)\\
\sigma(r)|_{\theta=\frac{\pi}{2}} &=& r^3\left(r^2-\Pi_{\alpha}\right)
\end{eqnarray}
Variation of spin precession frequency with radial coordinates for various values of spin parameter 
could be seen from the Fig.~(\ref{pi2q}).
\begin{figure}
\begin{center}
\subfigure[]{
\includegraphics[width=2in,angle=0]{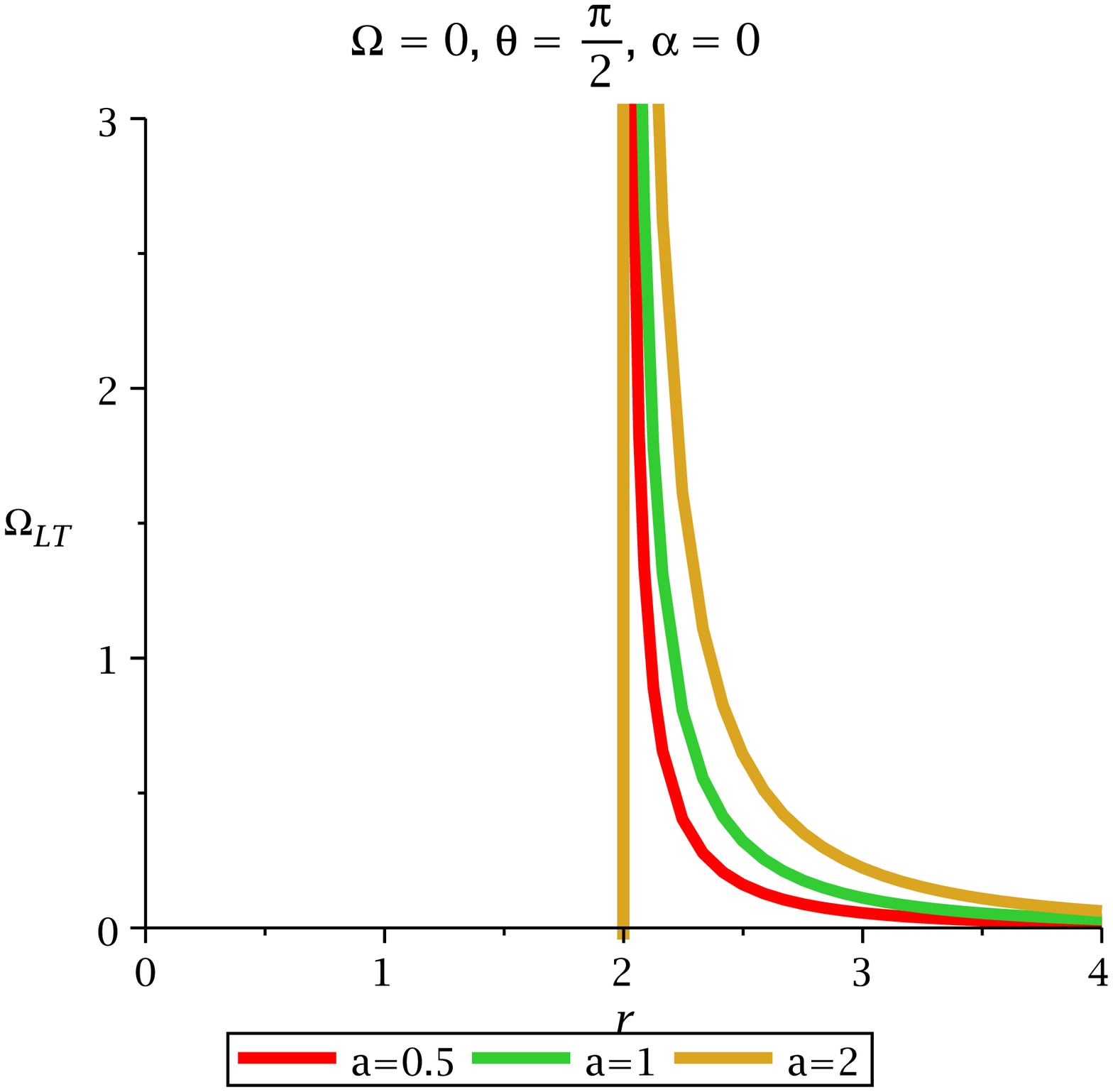}} 
\subfigure[]{
\includegraphics[width=2in,angle=0]{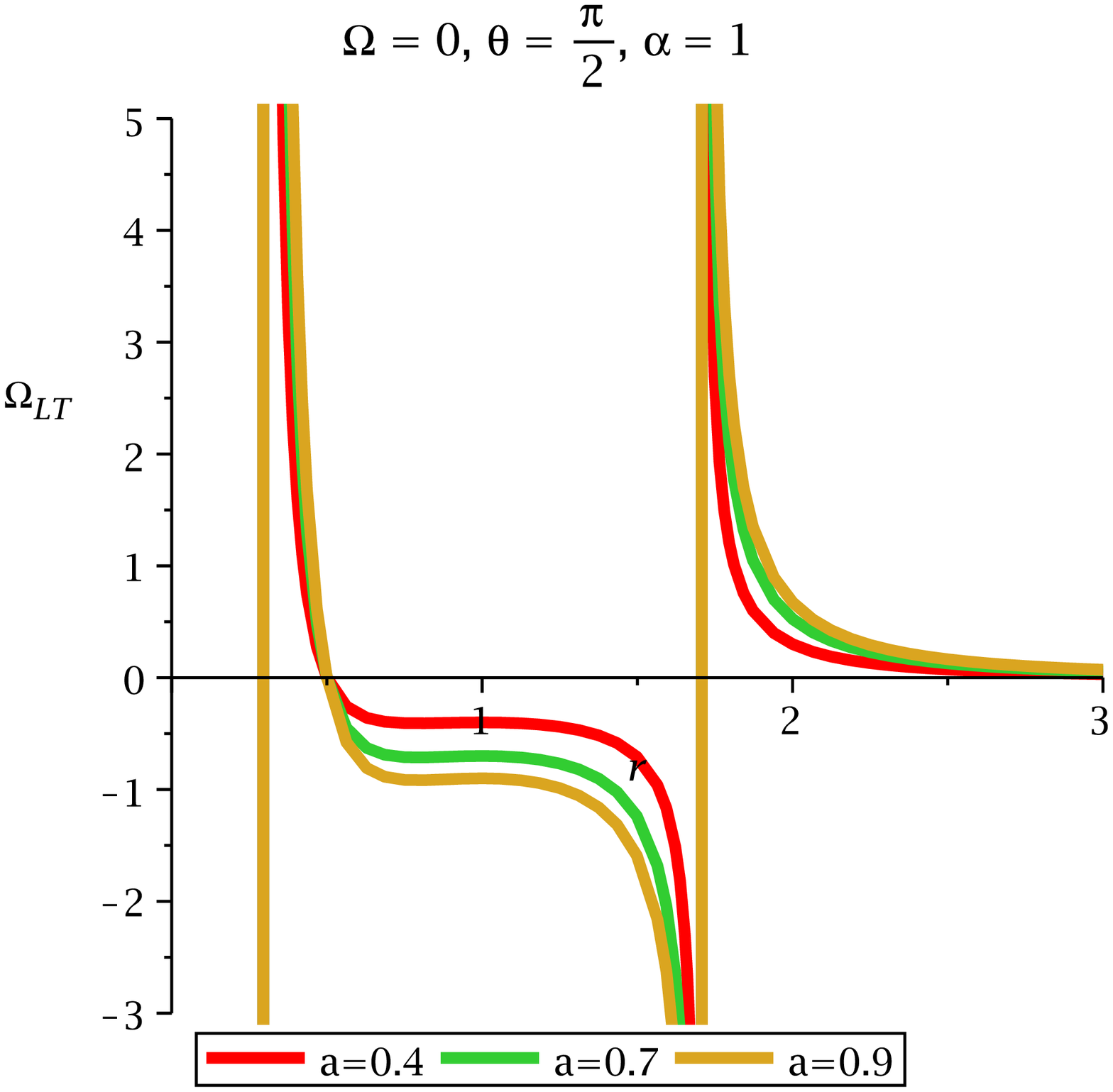}} 
\subfigure[]{
\includegraphics[width=2in,angle=0]{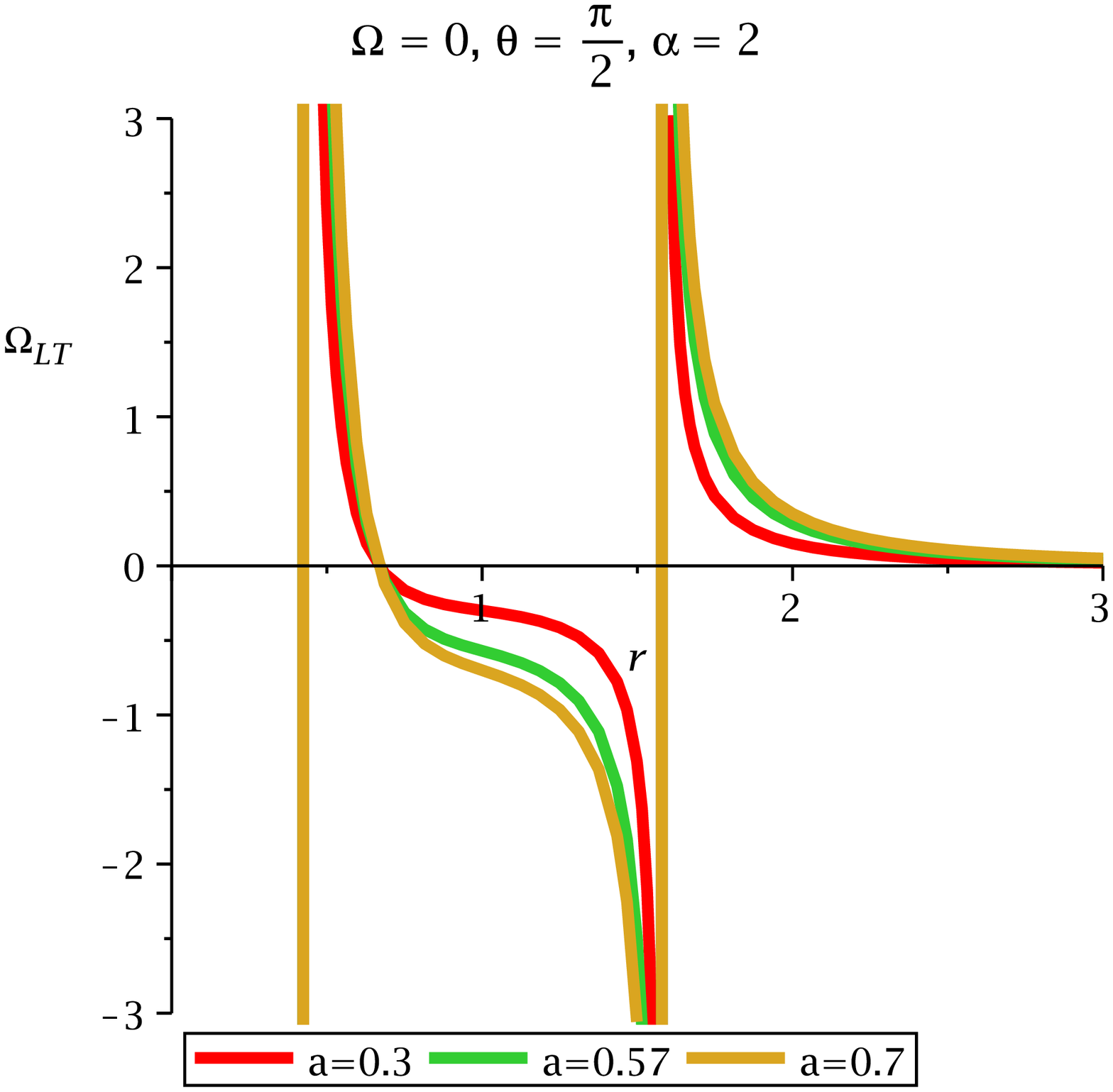}} 
\caption{ Examples of the variation  of $\Omega_{LT}$ versus $r$ for $\theta=\frac{\pi}{2}$ in 
KMOG with variation of MOG parameter and  spin parameter. Here  $\Omega=0$. The first figure 
describes the variation  of $\Omega_{LT}$  with $r$ for non-extremal BH, extremal BH and NS 
without MOG parameter. The second and third figure describes the variation  of $\Omega_{LT}$  
with $r$ for non-extremal BH, extremal BH and NS with MOG parameter.}
\label{pi2q}
\end{center}
\end{figure}
%%%%%%%%%%%%%%%%%%%%%%%%%%%%%%%%%%%%%%%%%%%%%%%%%%%%%%%%%%%%%%%

Using these plots, one can differentiate between non-extremal BH, extremal BH and NS in KMOG spacetime 
and Kerr spacetime.  It could be easily say that the presence of the MOG parameter drastically  changes 
the geometry of the BH spacetime. More appropriately, one could say that 
the presence of the MOG parameter enoromously changes the shape of the LT frequency diagram in 
contrast with the zero MOG parameter  LT frequency diagram. Futhermore, it is observed that in the 
NS case the geometric structure drastically different from BH spacetime.

\subsection{Lense-Thirring Precession in  Extremal KMOG spacetime}
It is very crucial to study the LT precession in case of extremal KMOG BH 
in comparison to extremal Kerr BH.  This is because the extremal BH has several important features. 
One crucial feature is that it has no Hawking temperature i.e. $T_{H}=0$. It has also no bifurcaton 
2-sphere. Moreover, it has no trapped surface. Whereas its near-extremal counterpart possecess all 
the said features. Extremal BHs also playing  a major role both in  string theory and quantum gravity. 
They have used to count the string states in string theory while in quantum gravity they have used 
as a theoretical toy. In supersymmetric theory, extremal BHs satisfied the BPS~(Bogomolnyi-Prasad-Sommerfield) 
bound and they are invariant under several super charges. They are also stable and do not radiate 
Hawking radiation. Their entropy was calculated in string theory~\cite{vafa}.  
The other definition of extremal BH is that it  is a BH when two horizons are coincident. 
The extremal limit of KMOG BH is defined by $a=\frac{G_{N}{\cal M}}{\sqrt{1+\alpha}}$ or 
$J=\frac{G_{N}^2{\cal M}^2}{\sqrt{1+\alpha}}$. Thus one gets the extremal horizon is at 
$r_{ex}=G_{N}{\cal M}=a\sqrt{1+\alpha}$. So far we have not written the exact expression of 
LT precession frequency vector for extremal KMOG BH now we should write this expression as 
$$
\vec \Omega_{LT} = \frac{G_{N}{\cal M}}{\sqrt{1+\alpha}} \times
$$
\begin{eqnarray}
\frac{\left[ \Pi_{\alpha}(r-G_{N}{\cal M})\cos\theta~\hat{r}
+G_{N}{\cal M} \left(r^2-\frac{\alpha}{1+\alpha} G_{N}{\cal M} r-\frac{G_{N}^2{\cal M}^2}{1+\alpha}\cos^2\theta \right)
\sin\theta~\hat{\theta}\right]}
{\left(r^2+\frac{G_{N}^2{\cal M}^2}{1+\alpha}\cos^2\theta\right)^\frac{3}{2}  
\left(r^2-\Pi_{\alpha}+\frac{G_{N}^2{\cal M}^2}{1+\alpha}\cos^2\theta\right)}
\end{eqnarray}
Taking magnitude of this vector one obtains the LT frequency for extremal KMOG BH as
$$
\Omega_{LT} (r,\theta,\alpha) = \frac{G_{N}{\cal M}}{\sqrt{1+\alpha}}
\times
$$
\begin{eqnarray}
\frac{\left[\Pi_{\alpha}^2(r-G_{N}{\cal M})^2 \cos^2\theta +
G_{N}^2{\cal M}^2\left(r^2-\frac{\alpha}{1+\alpha} G_{N}{\cal M} r-\frac{G_{N}^2{\cal M}^2}{1+\alpha}\cos^2\theta\right)^2
\sin^2\theta\right]^{\frac{1}{2}}}
{\left(r^2+\frac{G_{N}^2{\cal M}^2}{1+\alpha}\cos^2\theta\right)^\frac{3}{2}  
\left(r^2-\Pi_{\alpha}+\frac{G_{N}^2{\cal M}^2}{1+\alpha}\cos^2\theta\right) }
\end{eqnarray}
Now we will compute the LT frequency of extremal KMOG BH for various angles starting from polar region to 
equatorial plane and the variation of the said frequency could be observed from the plot. 

Case I:\\ 
First we take the value of $\theta=0$, then one gets the LT frequency as
\begin{eqnarray}
\Omega_{LT}  =  \frac{G_{N} {\cal M}}{\sqrt{1+\alpha}} \frac{\Pi_{\alpha}}
{\left(r^2+\frac{G_{N}^2{\cal M}^2}{1+\alpha}\right)^\frac{3}{2}\left(r-G_{N}{\cal M}\right)}
= f(r, {\cal M}, \alpha)
\end{eqnarray}
It follows that the frequency is a function of  $f(r, {\cal M}, \alpha)$ while for Kerr BH it is a function 
of $f(r, {\cal M})$ only. It also should be noted that at the extremal horizon the LT frequency diverges both 
for extremal KMOG and extremal Kerr BH. Variation of Lense-Thirring frequency of extremal KMOG BH could be 
observed from  Fig.~(\ref{xq}) for different spin values. 
\begin{figure}
\begin{center}
\subfigure[]{
\includegraphics[width=2in,angle=0]{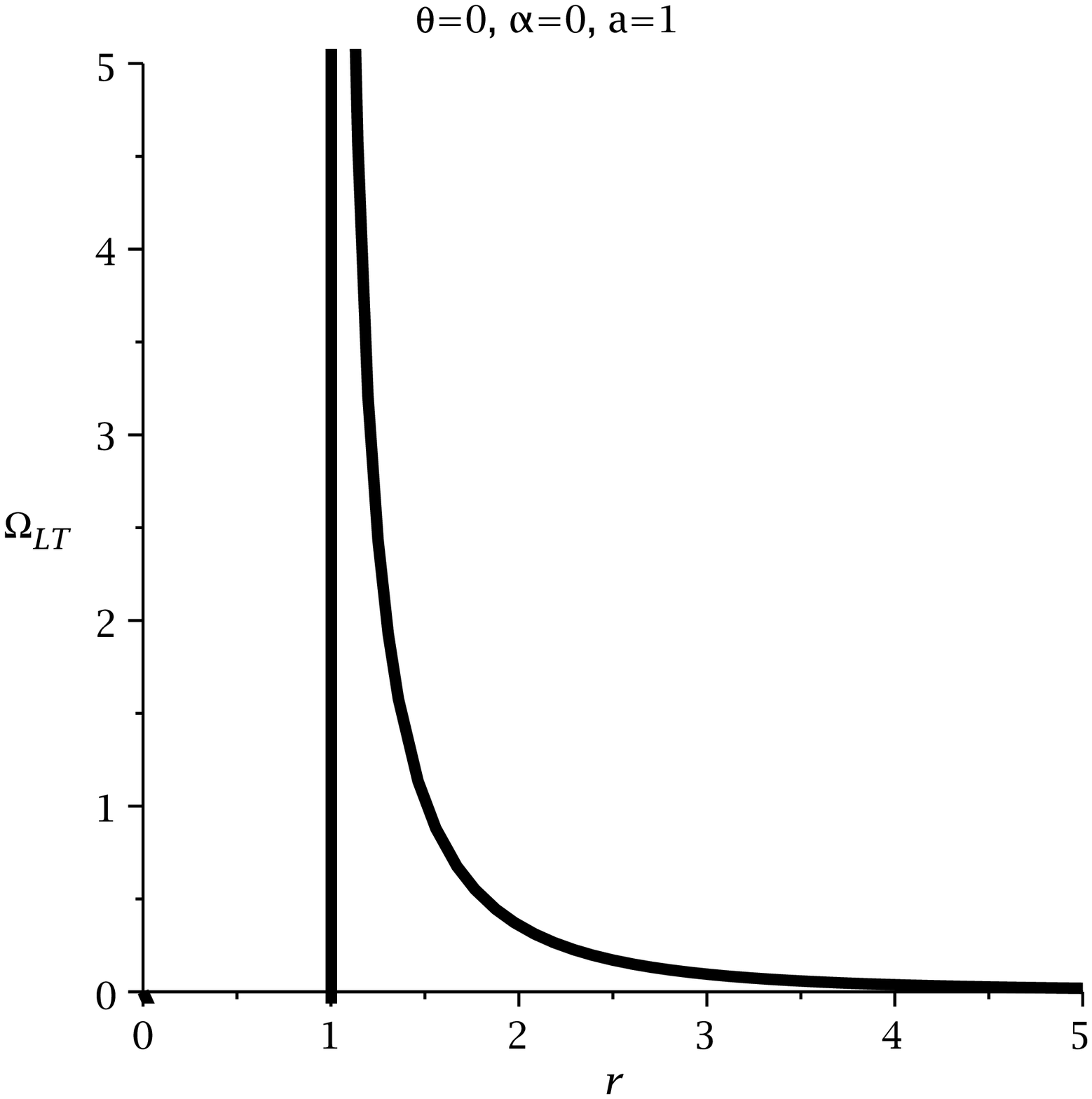}} 
\subfigure[]{
\includegraphics[width=2in,angle=0]{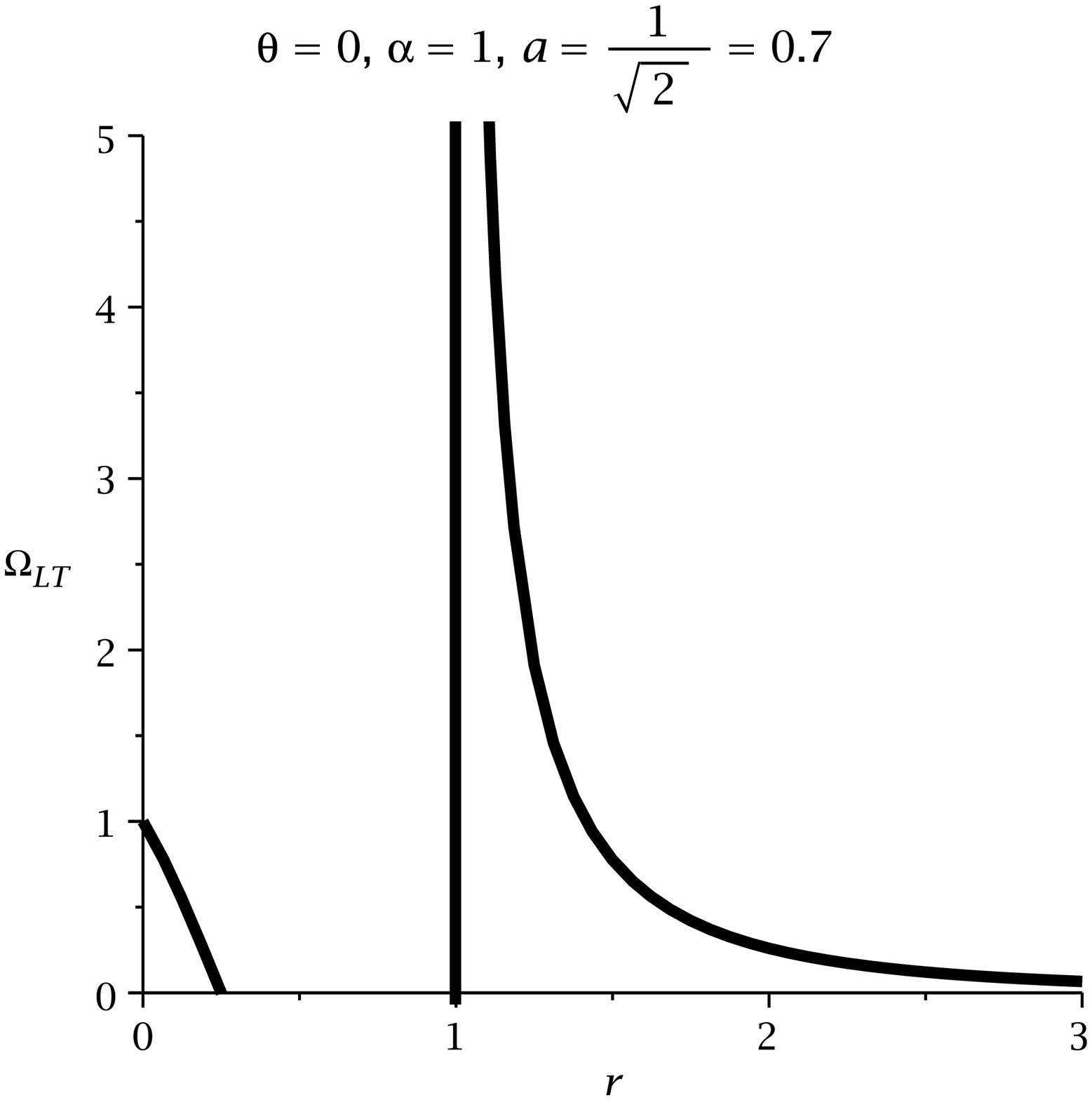}} 
\subfigure[]{
\includegraphics[width=2in,angle=0]{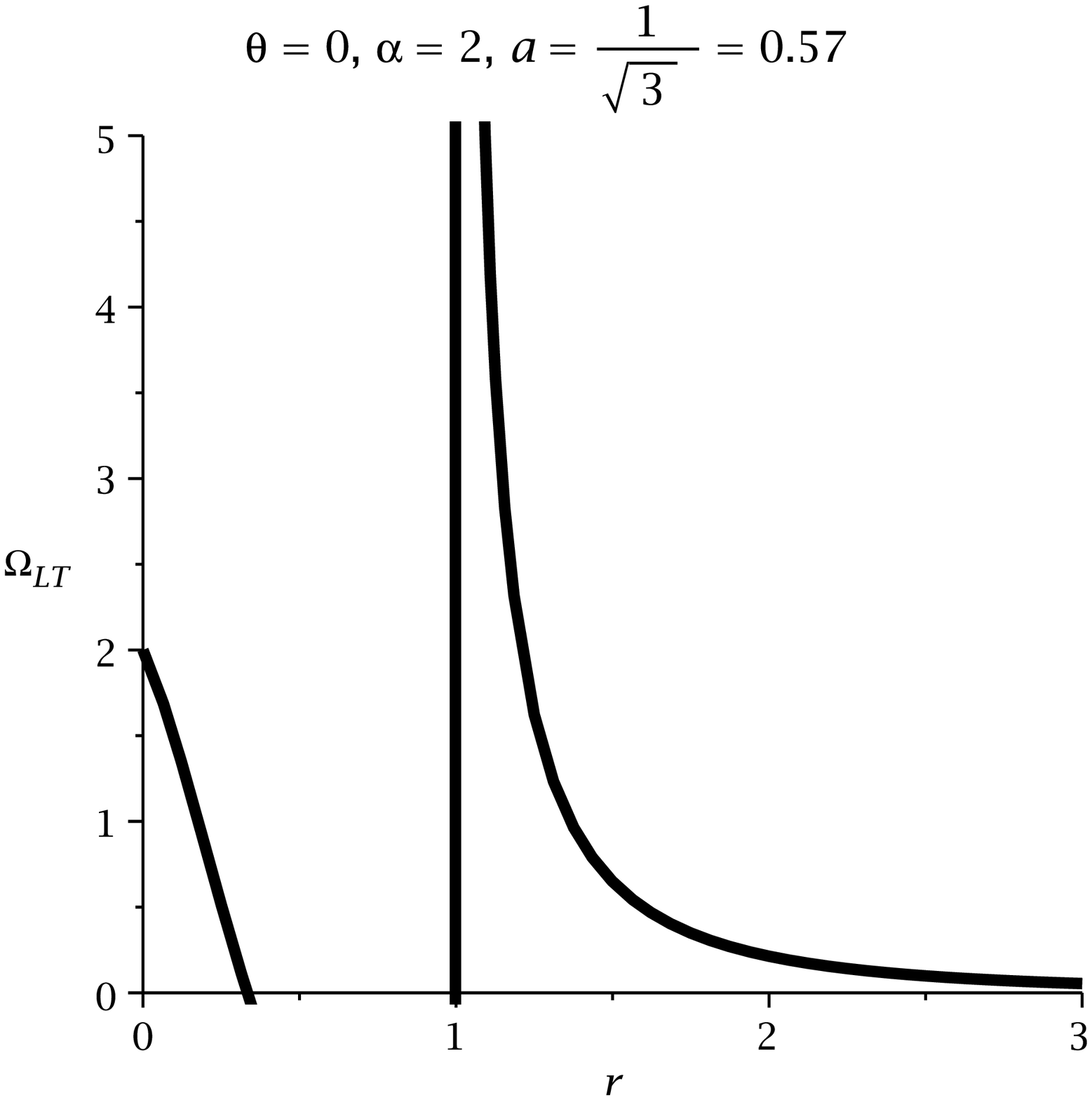}} 
\caption{The first figure 
describes the variation  of $\Omega_{LT}$  with $r$ for extremal BH in KMOG 
without MOG parameter. The second and third figure describes the variation  
of $\Omega_{LT}$  with $r$ for extremal BH with MOG parameter. Here we set $\theta=0$.}
\label{xq}
\end{center}
\end{figure}

Case II: \\
For $\theta=\frac{\pi}{6}$, the LT frequency is derived to be
$$
\Omega_{LT} (r,\frac{\pi}{6}) = \frac{G_{N} {\cal M}}{2\sqrt{1+\alpha}}  \times
$$
\begin{eqnarray}
\frac{\left[3 \Pi_{\alpha}^2(r-G_{N}{\cal M})^2 
+ G_{N}^2{\cal M}^2\left(r^2-\frac{\alpha}{1+\alpha} G_{N}{\cal M} r-\frac{3 G_{N}^2{\cal M}^2}{4(1+\alpha)}\right)^2
\right]^{\frac{1}{2}}}{\left[r^2+\frac{3G_{N}^2{\cal M}^2}{4(1+\alpha)}\right]^\frac{3}{2} 
\left[r^2-\Pi_{\alpha}+\frac{3G_{N}^2{\cal M}^2}{4(1+\alpha)}\right]}
\end{eqnarray}

\begin{figure}
\begin{center}
\subfigure[]{
\includegraphics[width=2in,angle=0]{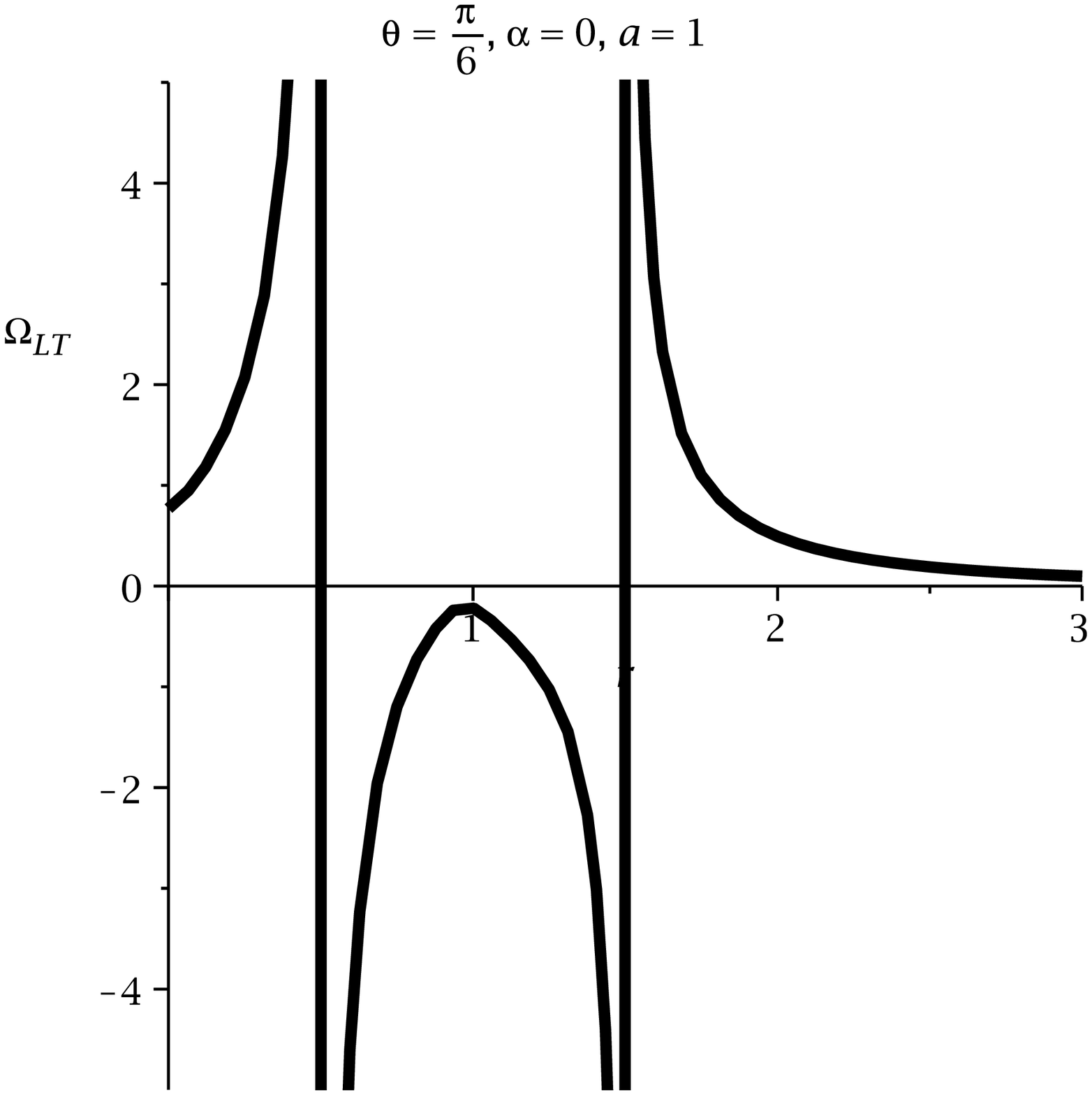}} 
\subfigure[]{
\includegraphics[width=2in,angle=0]{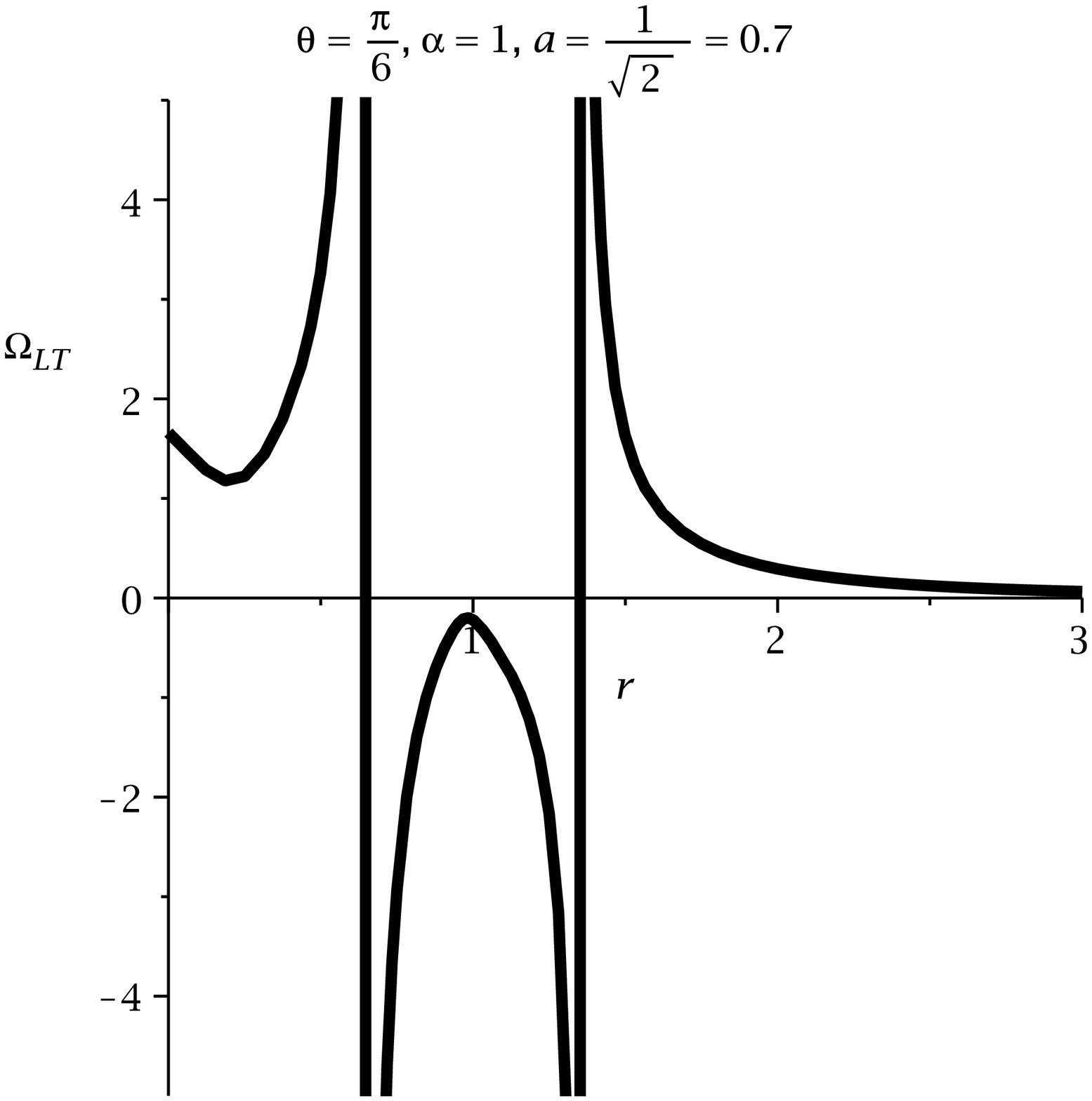}} 
\subfigure[]{
\includegraphics[width=2in,angle=0]{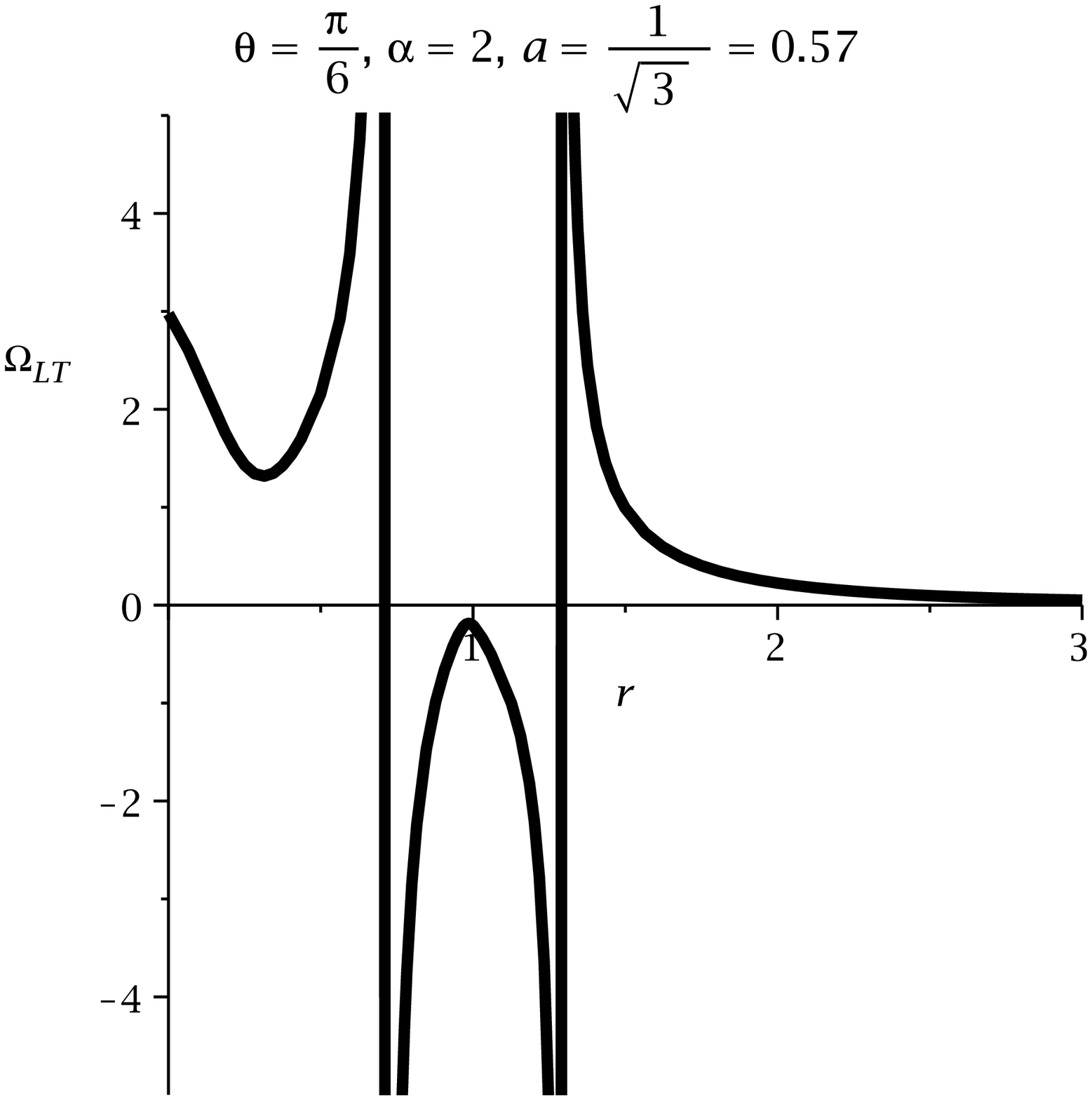}} 
\caption{ The first figure 
describes the variation  of $\Omega_{LT}$  with $r$ for extremal BH in KMOG 
without MOG parameter. The second and third figure describes the variation  
of $\Omega_{LT}$  with $r$ for extremal BH in KMOG with MOG parameter. 
Here we set $\theta=\frac{\pi}{6}$.}
\label{xq1}
\end{center}
\end{figure}

Case III: \\
For $\theta=\frac{\pi}{4}$, the LT frequency is computed to be
$$
\Omega_{LT} (r, \frac{\pi}{4}) = \frac{G_{N} {\cal M}}{\sqrt{2(1+\alpha)}}   \times
$$
\begin{eqnarray}
\frac{\left[ \Pi_{\alpha}^2(r-G_{N}{\cal M})^2 
+ G_{N}^2{\cal M}^2\left(r^2-\frac{\alpha}{1+\alpha} G_{N}{\cal M} r-\frac{ G_{N}^2{\cal M}^2}{2(1+\alpha)}\right)^2
\right]^{\frac{1}{2}}}{\left[r^2+\frac{G_{N}^2{\cal M}^2}{2(1+\alpha)}\right]^\frac{3}{2} 
\left[r^2-\Pi_{\alpha}+\frac{G_{N}^2{\cal M}^2}{2(1+\alpha)}\right]}
\end{eqnarray}

\begin{figure}
\begin{center}
\subfigure[]{
\includegraphics[width=2in,angle=0]{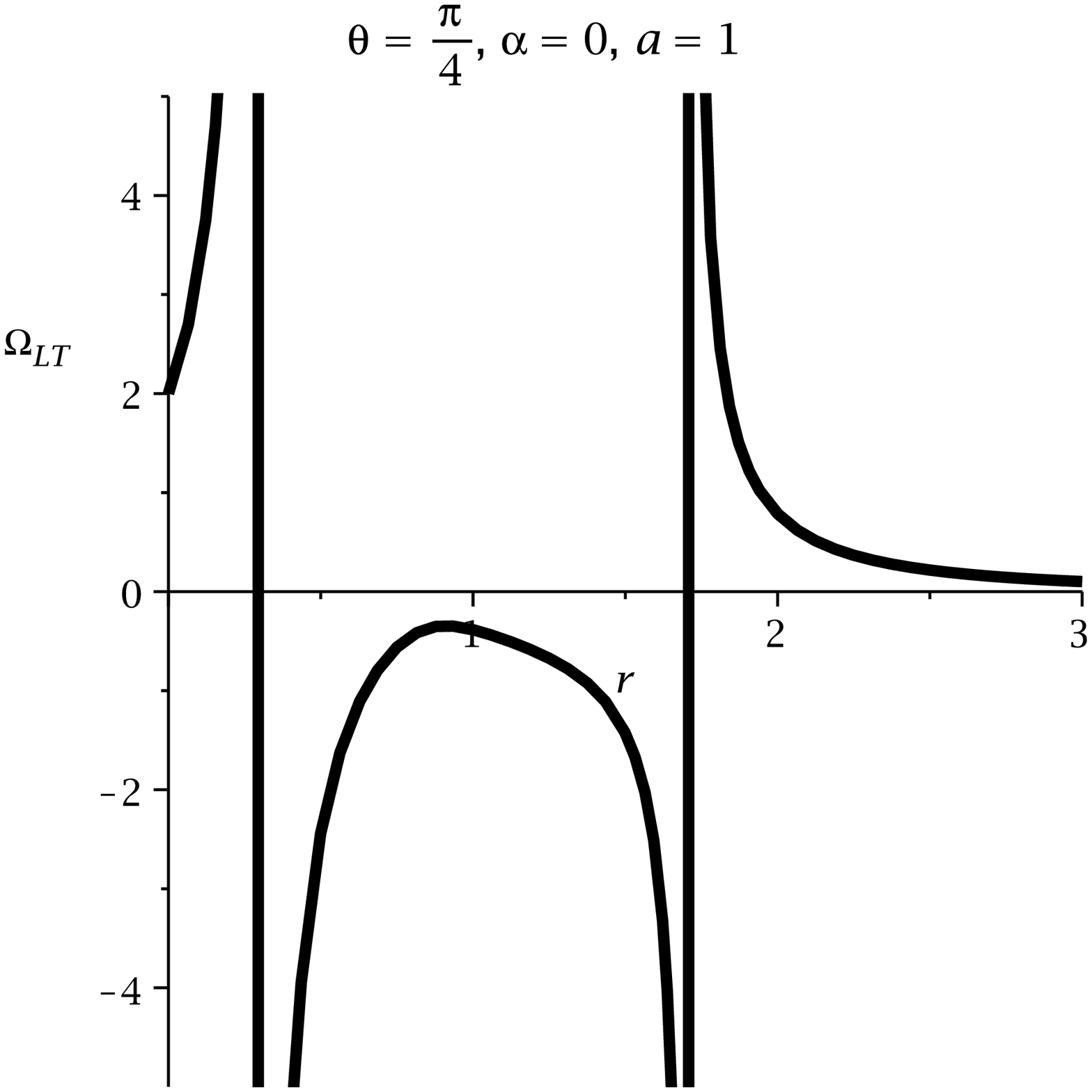}} 
\subfigure[]{
\includegraphics[width=2in,angle=0]{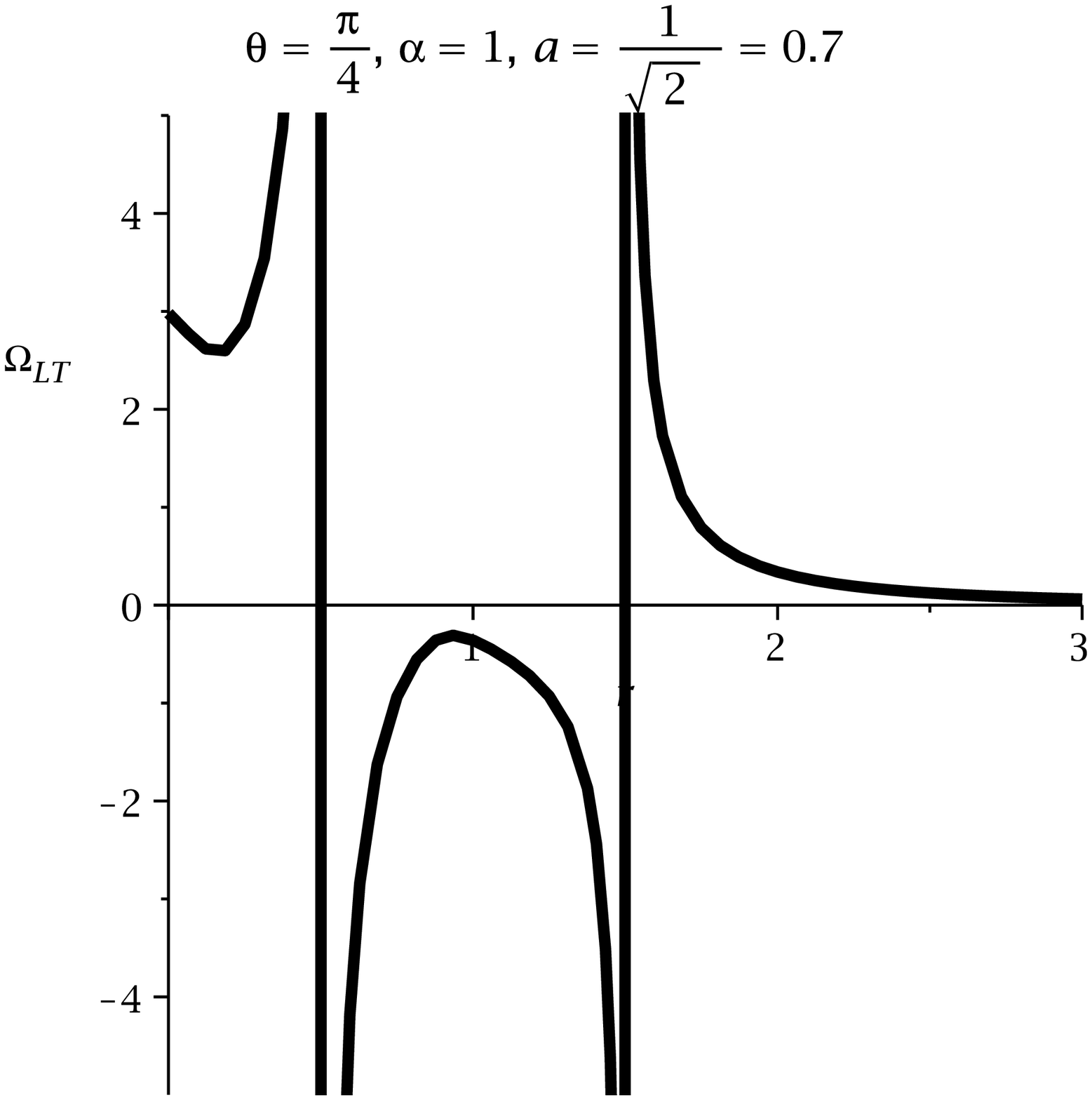}} 
\subfigure[]{
\includegraphics[width=2in,angle=0]{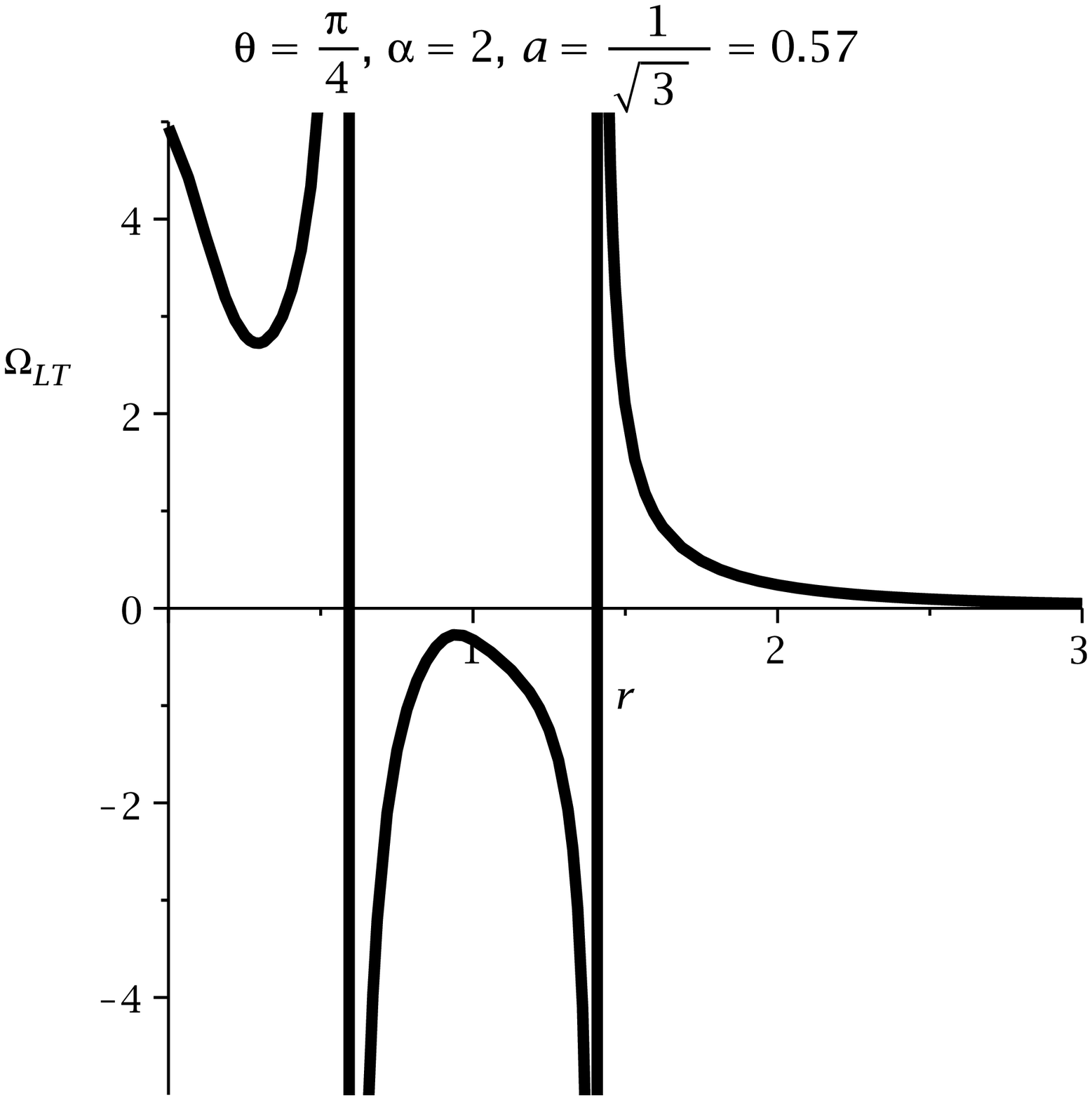}} 
\caption{The first figure describes the variation  of $\Omega_{LT}$  with $r$ 
for extremal BH in KMOG without MOG parameter. The second and third figure describes the variation  
of $\Omega_{LT}$  with $r$ for extremal BH in KMOG with MOG parameter. Here we set $\theta=\frac{\pi}{4}$.}
\label{xq2}
\end{center}
\end{figure}

Case IV: \\
For $\theta=\frac{\pi}{3}$, the LT frequency is derived to be
$$
\Omega_{LT} (r, \frac{\pi}{3}) = \frac{G_{N} {\cal M}}{2\sqrt{1+\alpha}} \times
$$
\begin{eqnarray}
\frac{\left[\Pi_{\alpha}^2 (r-G_{N}{\cal M})^2 
+ 3G_{N}^2{\cal M}^2\left(r^2-\frac{\alpha}{1+\alpha} G_{N}{\cal M} r-\frac{ G_{N}^2{\cal M}^2}{4(1+\alpha)}\right)^2
\right]^{\frac{1}{2}}}{\left[r^2+\frac{G_{N}^2{\cal M}^2}{4(1+\alpha)}\right]^\frac{3}{2} 
\left[r^2-\Pi_{\alpha}+\frac{G_{N}^2{\cal M}^2}{4(1+\alpha)}\right]}
\end{eqnarray}
\begin{figure}
\begin{center}
\subfigure[]{
\includegraphics[width=2in,angle=0]{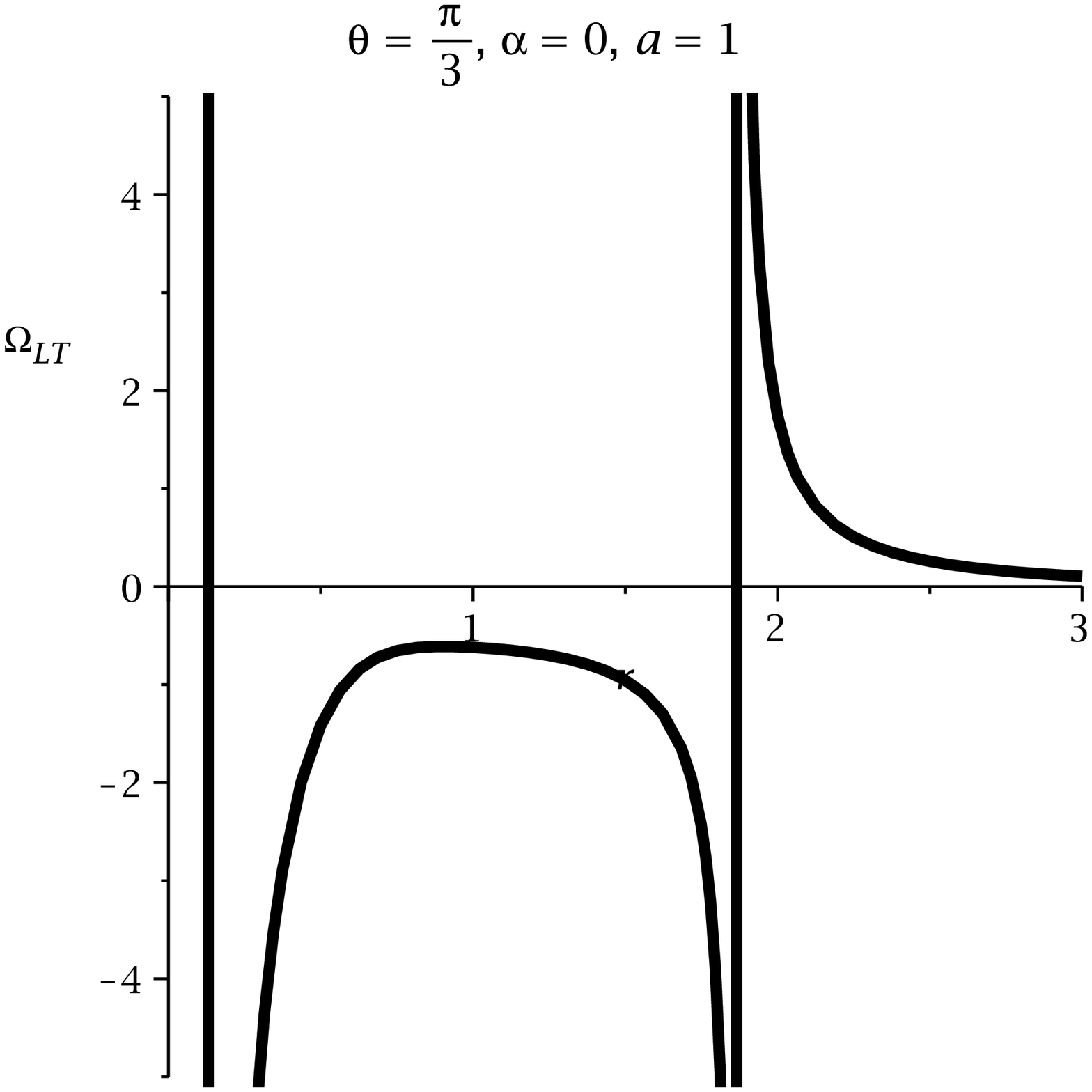}} 
\subfigure[]{
\includegraphics[width=2in,angle=0]{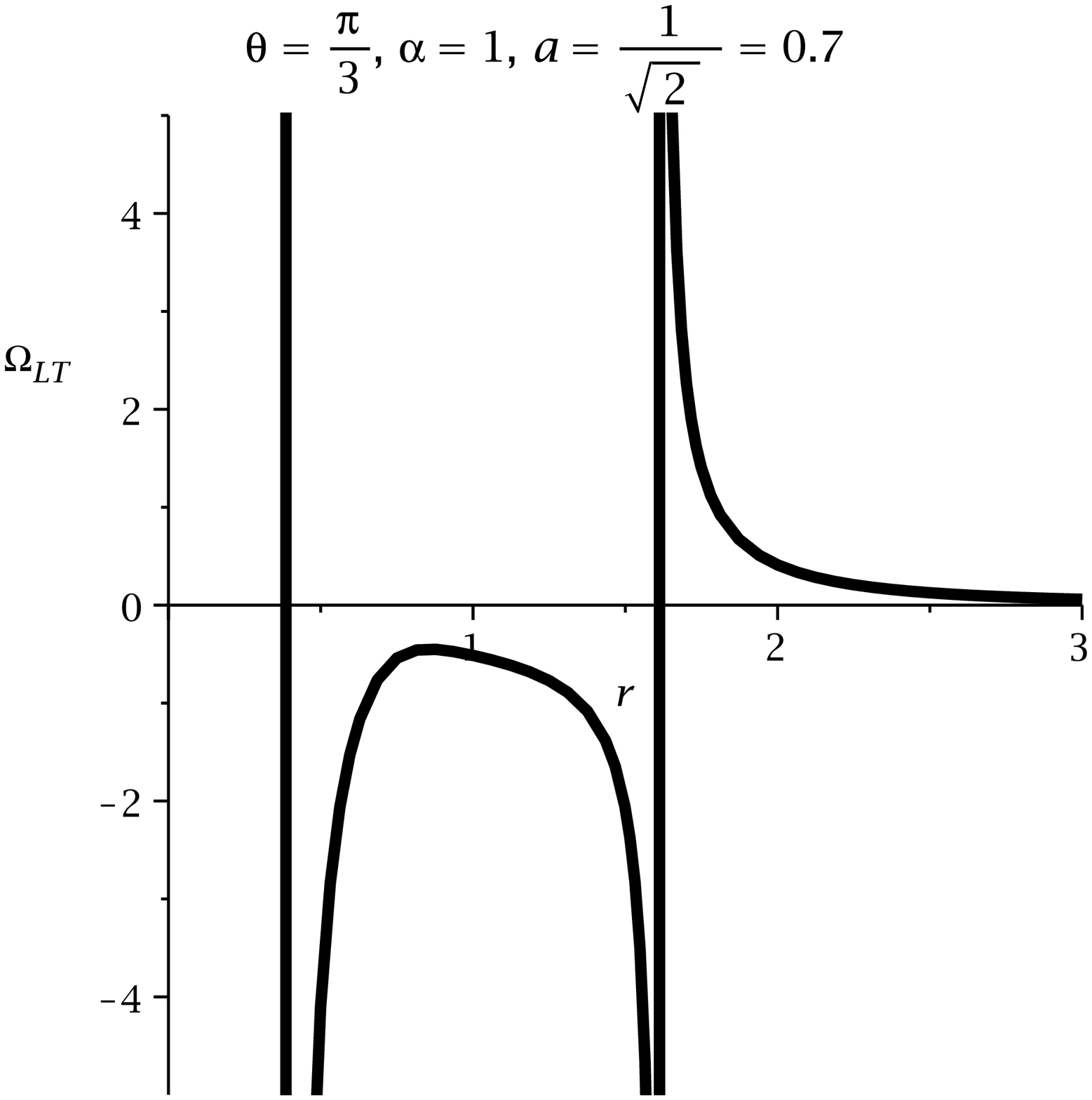}} 
\subfigure[]{
\includegraphics[width=2in,angle=0]{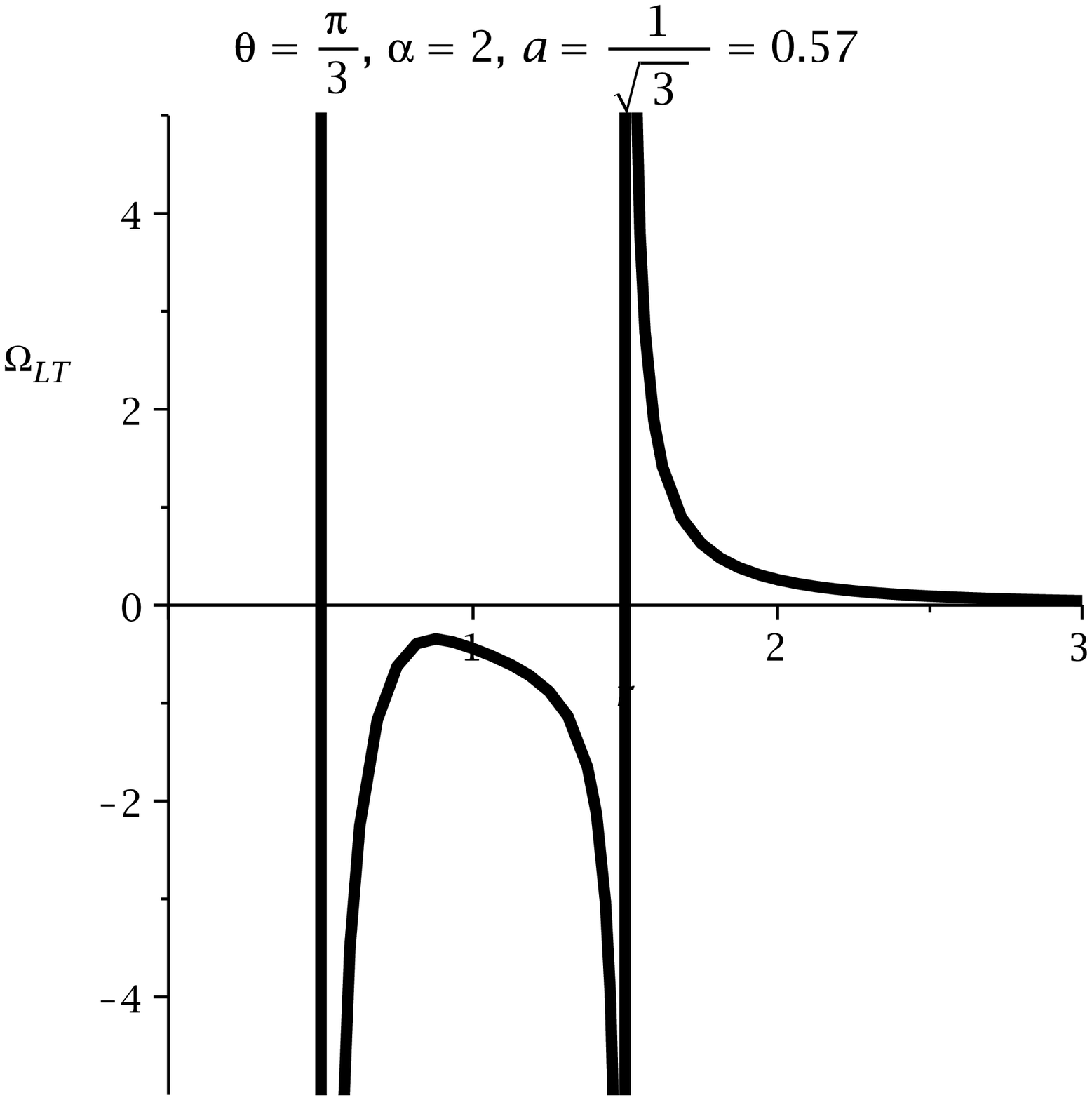}} 
\caption{The first figure describes the variation  of $\Omega_{LT}$  with $r$ 
for extremal BH in KMOG without MOG parameter. The second and third figure describes the variation  
of $\Omega_{LT}$  with $r$ for extremal BH in KMOG with MOG parameter. 
Here we set $\theta=\frac{\pi}{3}$.}
\label{xq3}
\end{center}
\end{figure}

Case V: \\
Finally, for $\theta=\frac{\pi}{2}$, the LT frequency is 
\begin{eqnarray}
\Omega_{LT}  =  \frac{G_{N}^2 {\cal M}^2}{\sqrt{1+\alpha}} \frac{\left(r-\frac{\alpha}{1+\alpha} G_{N}{\cal M} \right)}
{r^2(r^2-\Pi_{\alpha})} 
\end{eqnarray}
It should be noted that in each cases the LT frequency reduces to Kerr BH for $\alpha=0$. The variation 
of these frequencies could be observed from following Fig. ~(\ref{xq5}). 

\begin{figure}
\begin{center}
\subfigure[]{
\includegraphics[width=2in,angle=0]{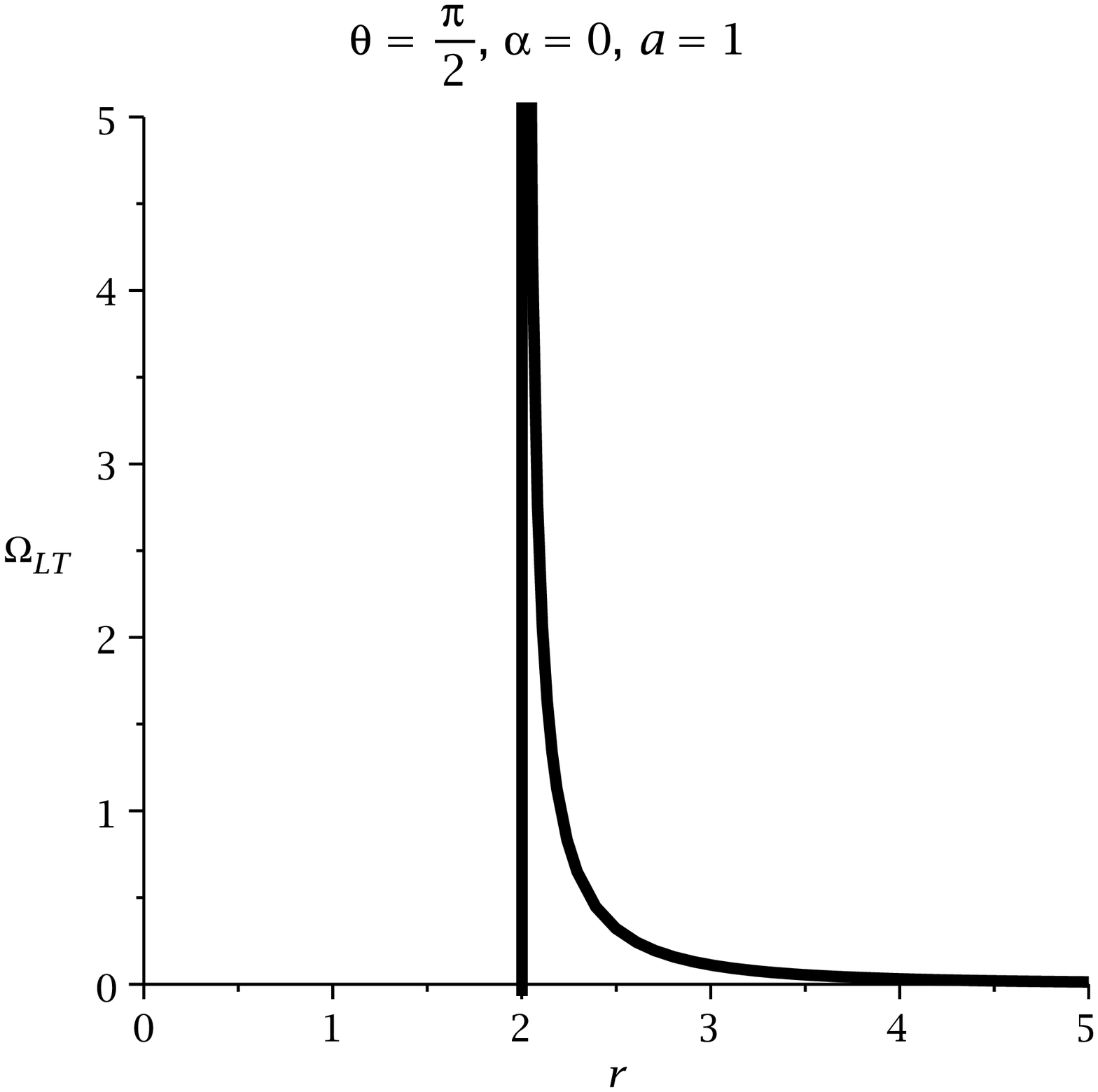}} 
\subfigure[]{
\includegraphics[width=2in,angle=0]{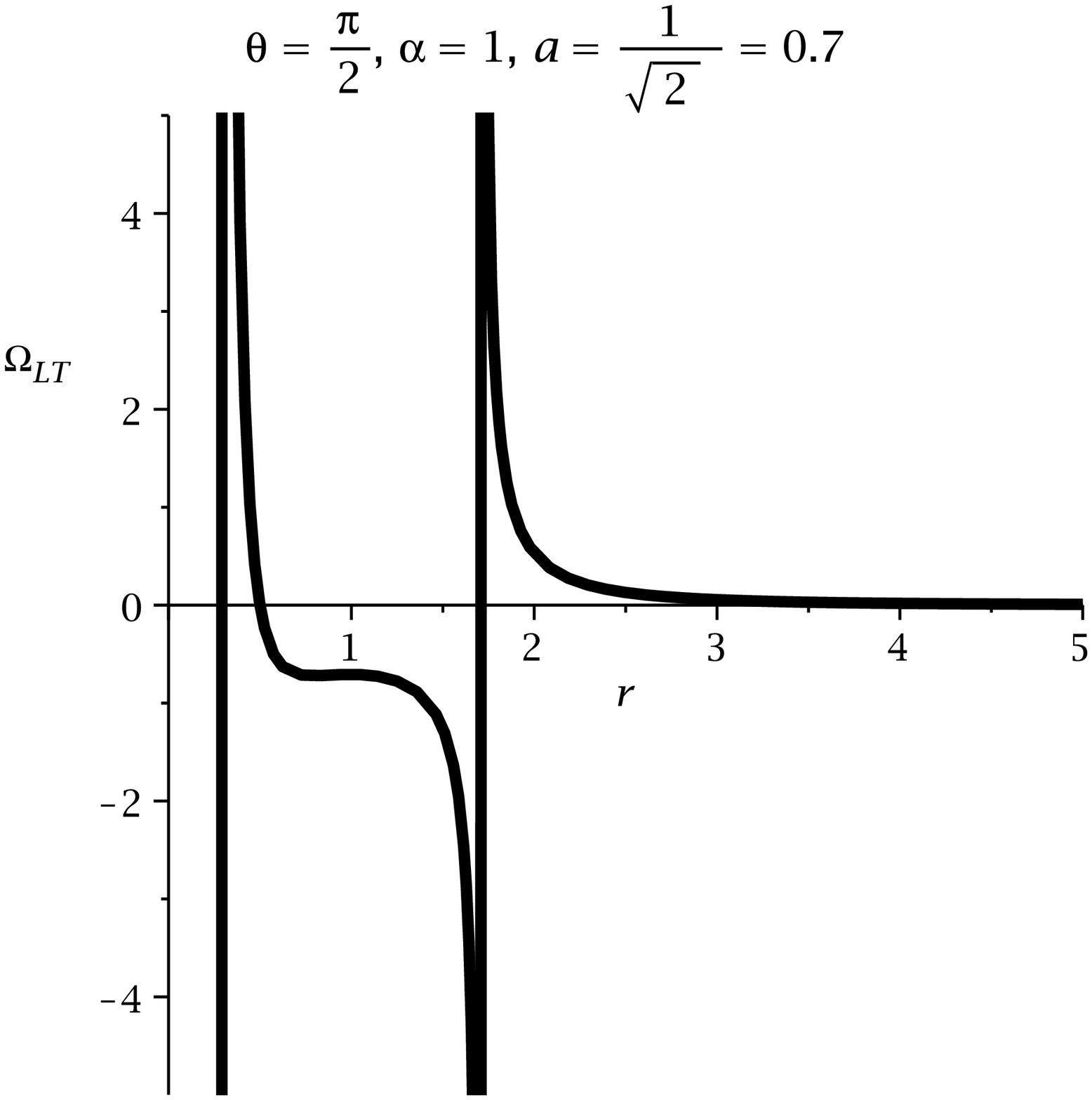}} 
\subfigure[]{
\includegraphics[width=2in,angle=0]{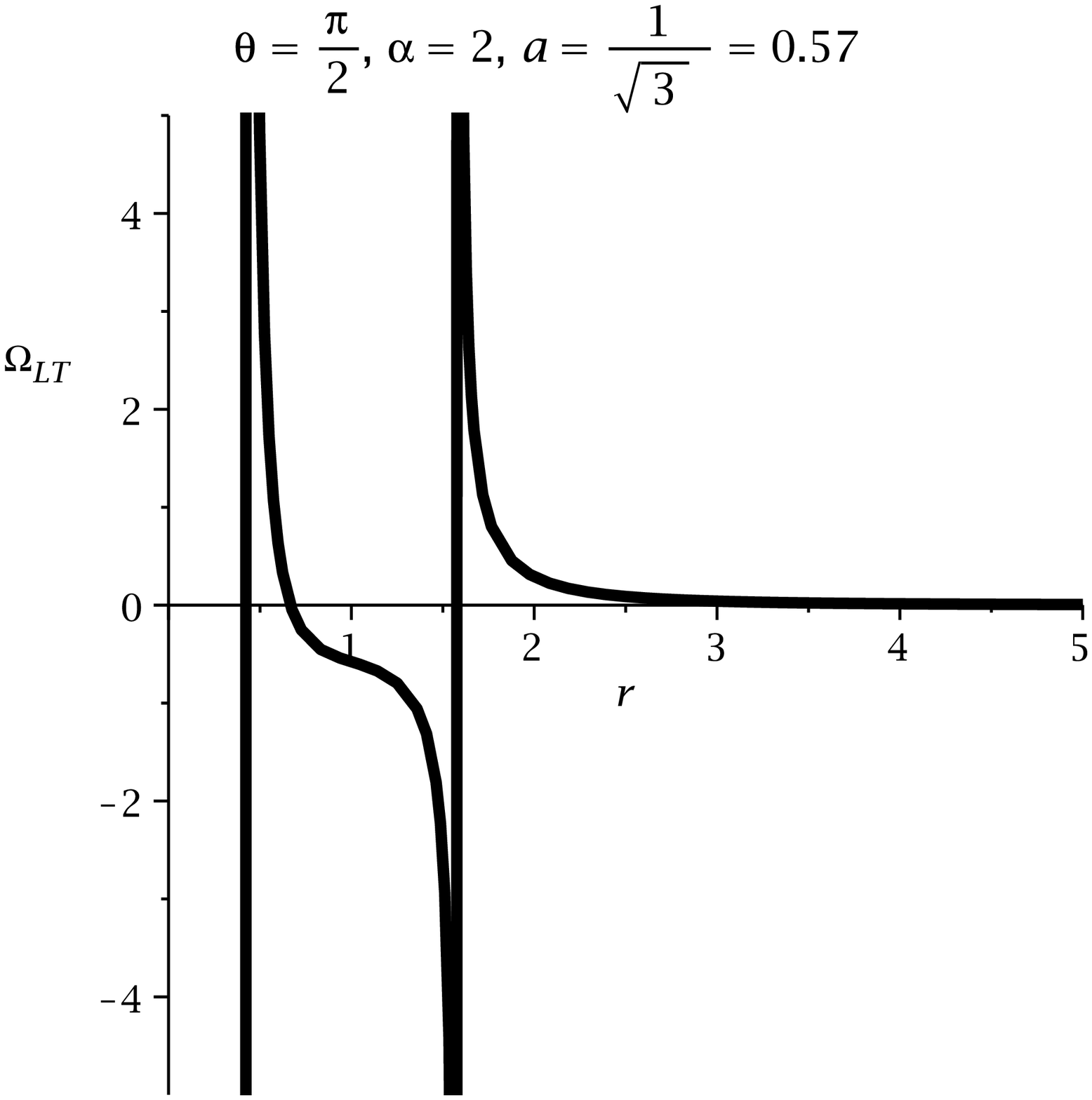}} 
\caption{ The first figure describes the variation  of $\Omega_{LT}$  with $r$ 
for extremal BH in KMOG without MOG parameter. The second and third figure describes the variation  
of $\Omega_{LT}$  with $r$ for extremal BH in KMOG with MOG parameter. Here we set $\theta=\frac{\pi}{2}$.}
\label{xq5}
\end{center}
\end{figure}

\section{Accretion Disk properties in KMOG spacetime}
To differentiate non-extremal BH spacetime, extremal BH spacetime and NS, it is essential to study 
accretion disk physics  around a stationary, axisymmetric spacetime. To distinguish three geometries of  
{the} compact object, {also} it is  essential to study three fundamental frequencies of accretion disk. This could 
be possible only by {studying} the geodesic motion of test particle around the compact objects. Again to study 
the physics of accretion disk, one should  compute the stable circular orbit of the said compact object. The main 
stable orbit is inner-most stable circular orbit~(ISCO) or last stable circular orbit~(LSCO). This ISCO can 
help us to distinguish three geometries. In Kerr geometry, the ISCO radii depends on spin parameter ~($a$) while 
in KMOG geometry, it depends on both spin parameter~($a_{\alpha}$) and MOG parameter~($\alpha$). We previously 
noticed that in KMOG geometry, the spin parameter decreases as MOG value increases~(See Fig.~(\ref{spn})). 

Three fundamenta frequencies which are very important 
for accretion disk physics of KMOG BH, namely the Keplerian frequency~($\Omega_{\phi}$), the radial 
epicyclic frequency~($\Omega_{r}$) and the vertical epicyclic frequency~($\Omega_{\theta}$).  They 
can easily be derived using the formulae~(\ref{a2},\ref{a3},\ref{a4}) given in 
Appendix:~[See also~\cite{epjc19}]
\footnote{Note that in Ref.~\cite{epjc19} the term $G_{N}$ is missing, here we have included it in 
each frequencies formulae} 
\begin{eqnarray}
\Omega_{\phi} &=& \pm \frac{\sqrt{\Pi_{\alpha}-G_{N}{\cal M}r}}{r^2\pm a \sqrt{\Pi_{\alpha}-G_{N}{\cal M}r}} \label{ekm}\\
\Omega_{r} &=& \frac{\sqrt{G_{N}{\cal M}r \Delta-4\left(\Pi_{\alpha}-G_{N}{\cal M}r\right) 
\left(\sqrt{\Pi_{\alpha}-G_{N}{\cal M}r} \mp a \right)^2}} 
{r \left(r^2 \pm a\sqrt{\Pi_{\alpha}-G_{N}{\cal M}r} \right)}~\label{ekm1}\\
\Omega_{\theta} &=& \frac{\sqrt{r^2\left(\Pi_{\alpha}-G_{N}{\cal M}r\right) \mp 2 a \Pi_{\alpha}\sqrt{\Pi_{\alpha}-G_{N}{\cal M}r}
+a^2\left(2\Pi_{\alpha}-G_{N}{\cal M}r \right)}}{r \left(r^2 \pm a\sqrt{\Pi_{\alpha}-G_{N}{\cal M}r} \right)}~\label{ekm2}\\
\end{eqnarray}
It is important to note that the upper sign for corotating~(direct) orbit and lower sign for 
counter-rotating~(retrograde) orbit. When MOG value vanishes, one obtains the epicyclic frequencies of 
Kerr BH.

The variation of these frequencies for three compact objects namely 
non-extremal BH, extremal BH and NS could be seen from the following diagram~
[Fig. ~(\ref{xq6}), Fig. ~(\ref{xq7}), Fig. ~(\ref{xq8})]. 
From first figure when $\alpha=0$ means that for Kerr BH, it is observed that the  Keplerian frequency 
decreases with increasing the radial value starting from maximum value for non-extremal BH~($a=0.5$). 
As spin value increases the value of Keplerian frequency decreases. While in second figure where the 
MOG parameter is present then the picture is quite different from the former. Here the Keplerian 
frequency has gone through a peak for different spin values.
%%%%%%%%%%%%%%%%%%%%%%%%%%%%%%%%Keplerian Frequency%%%%%%%%%%%%%%%%%%%%%
\begin{figure}
\begin{center}
\subfigure[]{
\includegraphics[width=2in,angle=0]{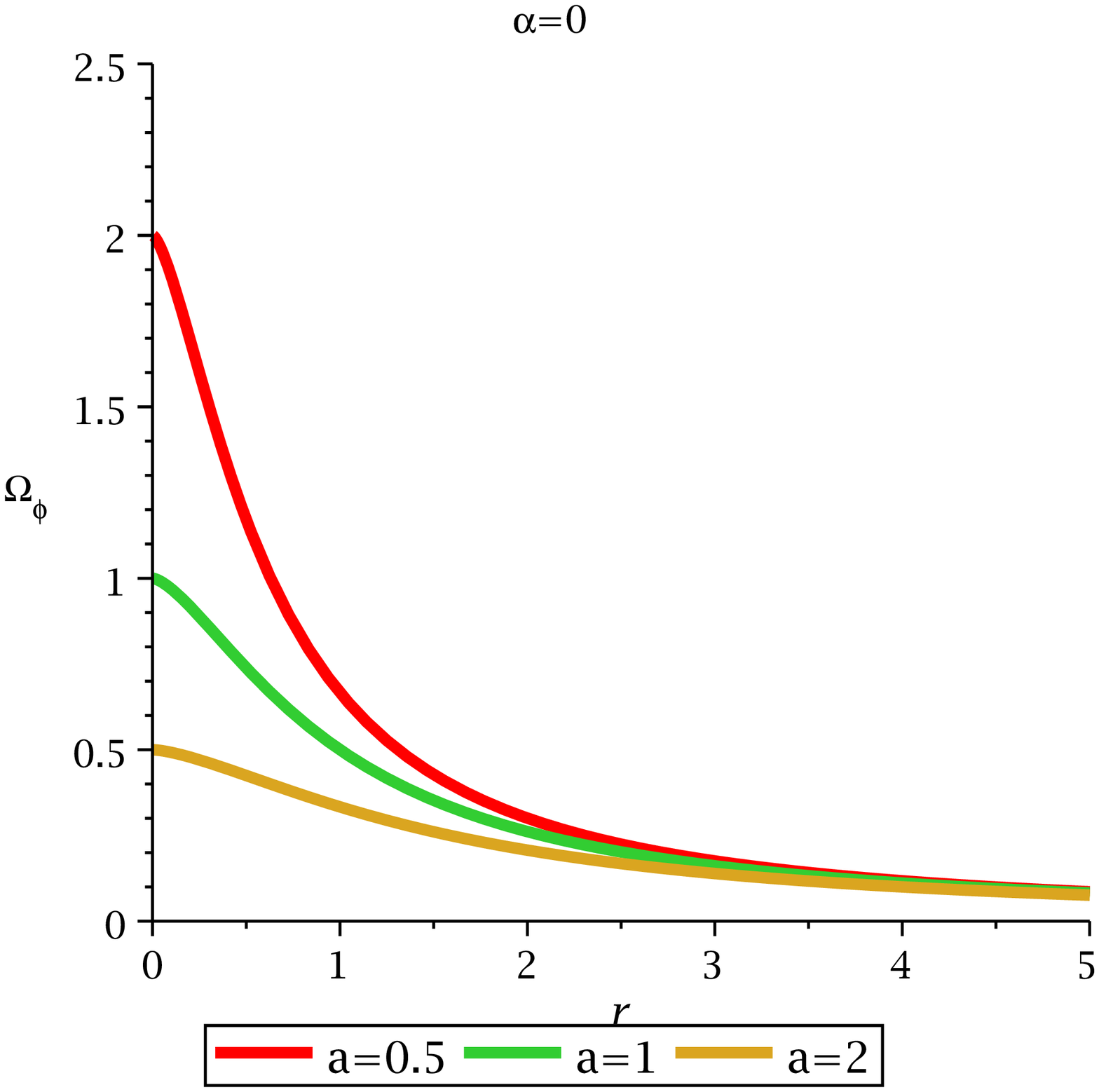}} 
\subfigure[]{
\includegraphics[width=2in,angle=0]{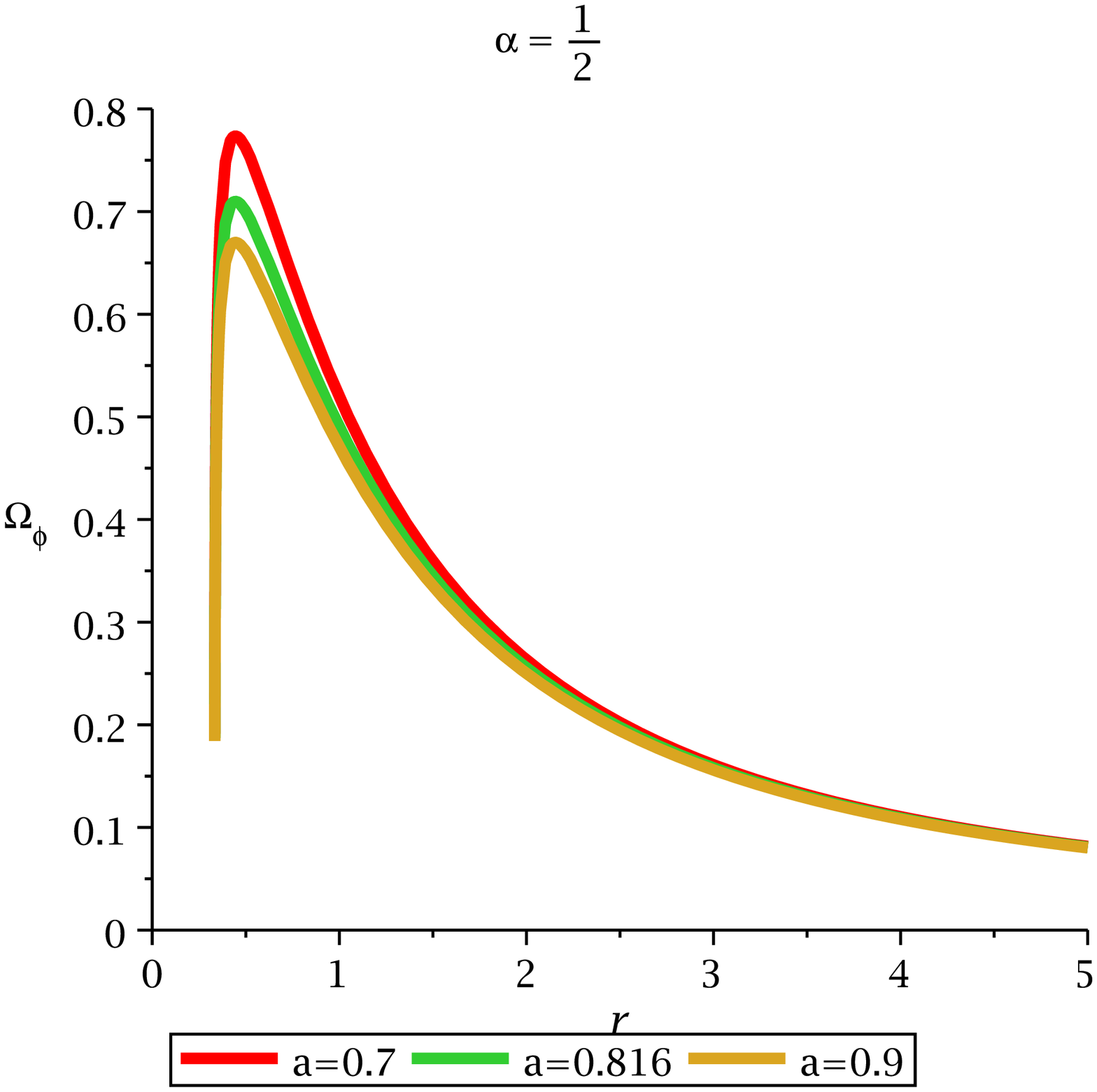}} 
\subfigure[]{
\includegraphics[width=2in,angle=0]{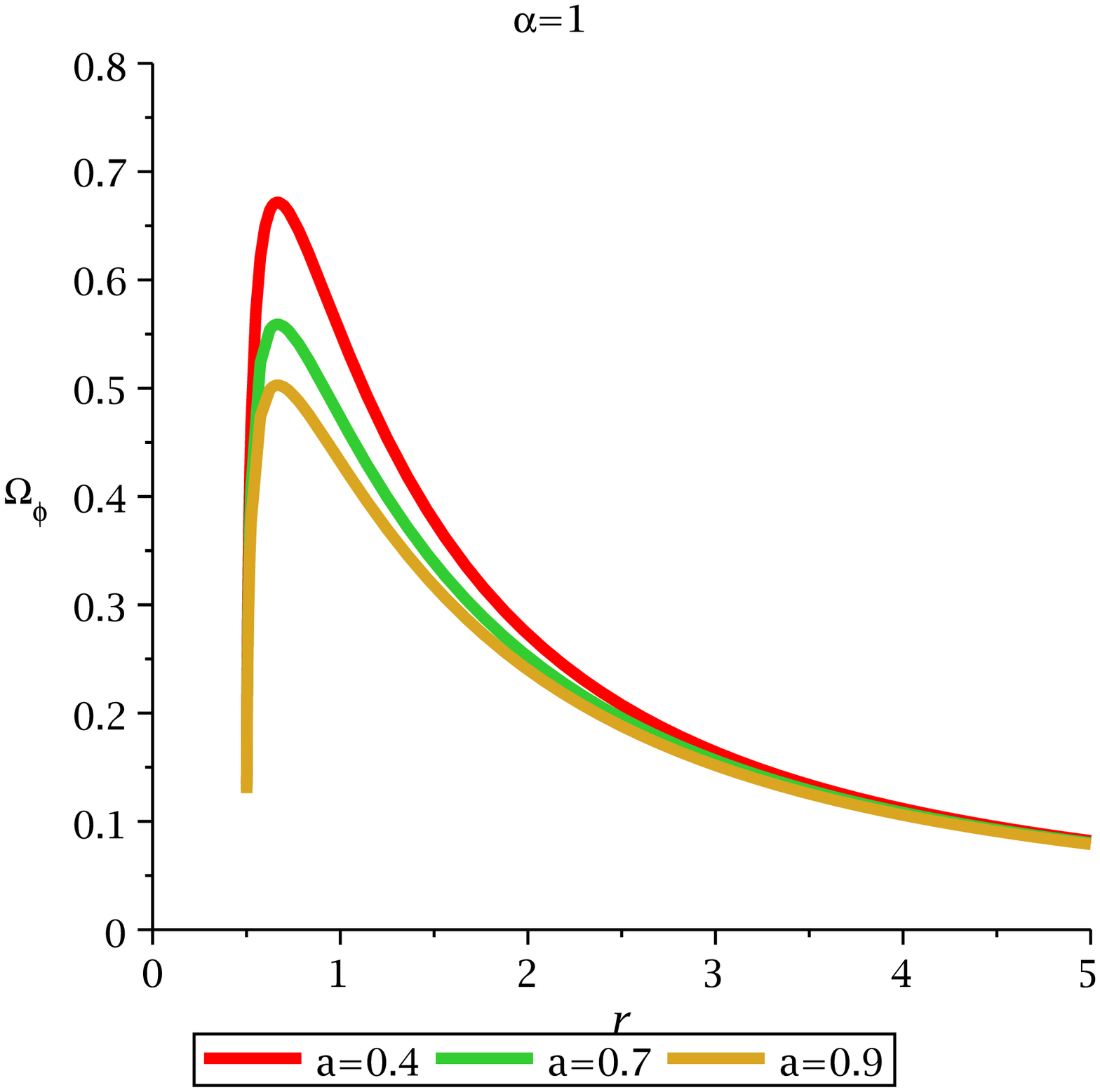}} 
\caption{ The figure depicts the variation  of $\Omega_{\phi}$  with $r$ for different MOG parameter and 
spin parameter. Each figure demonstrates the difference between non-extremal BH, extremal BH and NS. Without MOG 
there is no peak while with MOG there exists a peak value in Keplerian frequency.}
\label{xq6}
\end{center}
\end{figure}
%%%%%%%%%%%%%%%%%%%%%%%%%%%%%%%%%%%%%%%%%%%%%%%%%%%%%%Radial Frequency%%%%%%%%%%%%%%%%%%
\begin{figure}
\begin{center}
\subfigure[]{
\includegraphics[width=2in,angle=0]{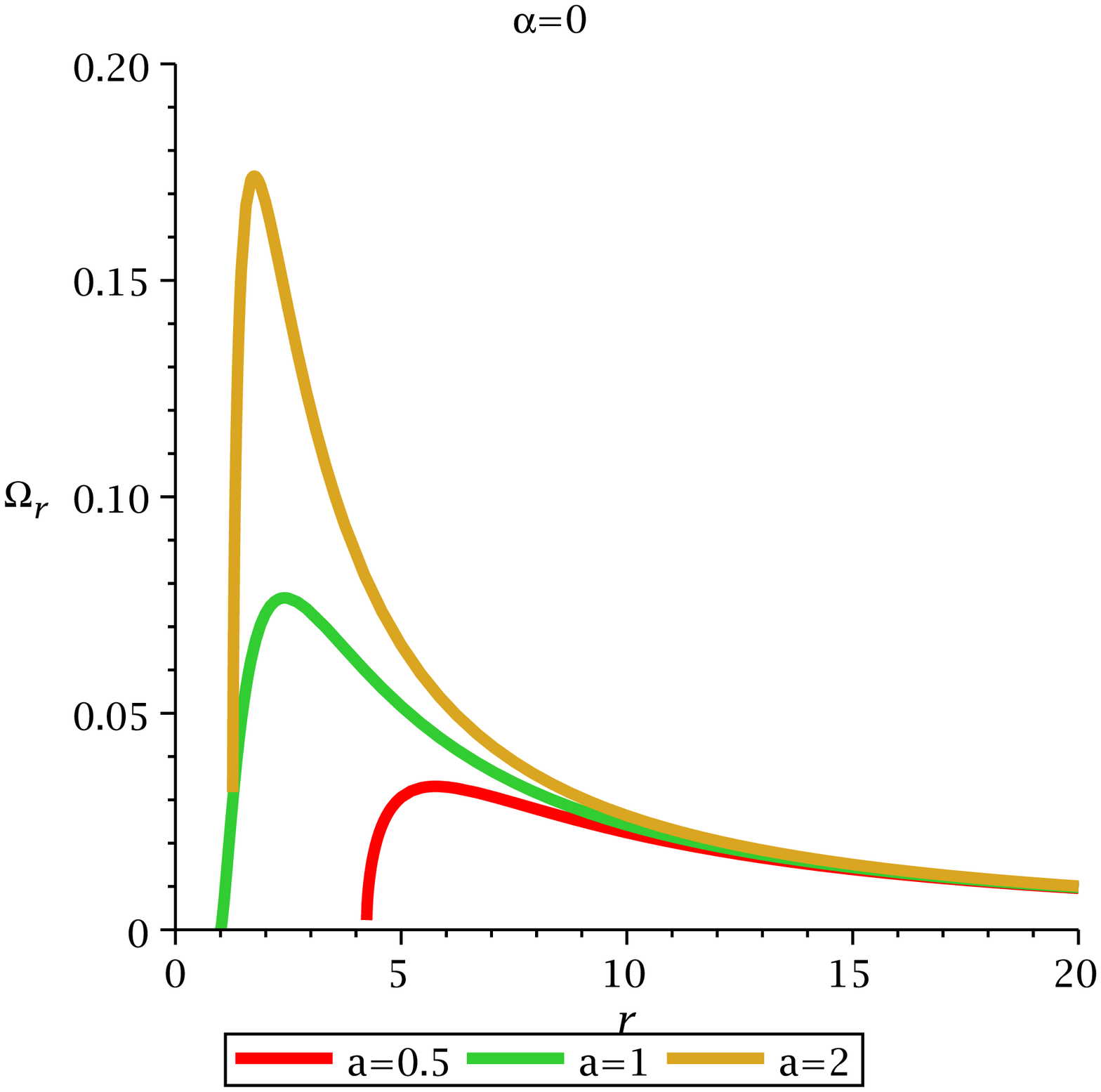}} 
\subfigure[]{
\includegraphics[width=2in,angle=0]{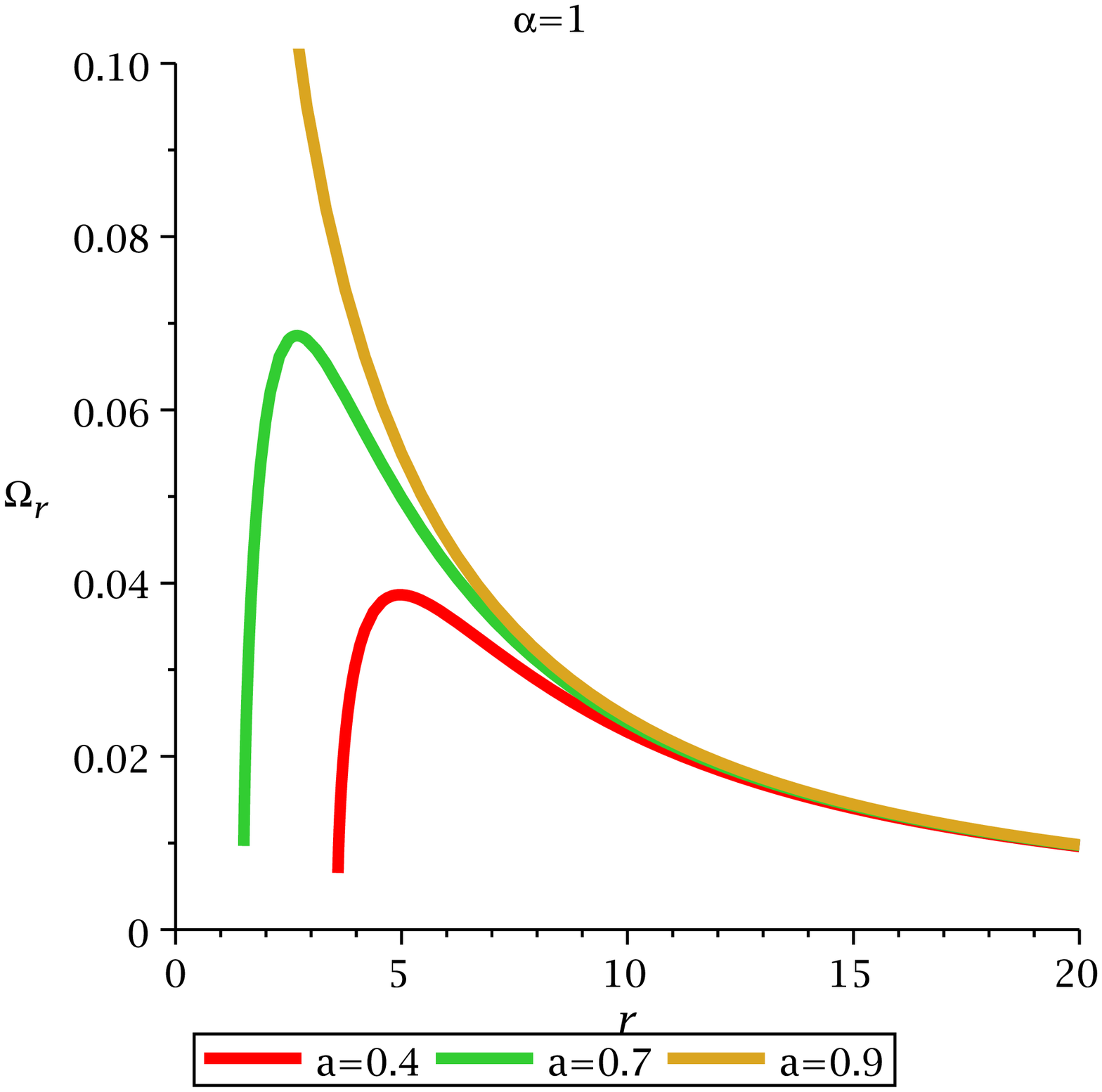}} 
\subfigure[]{
\includegraphics[width=2in,angle=0]{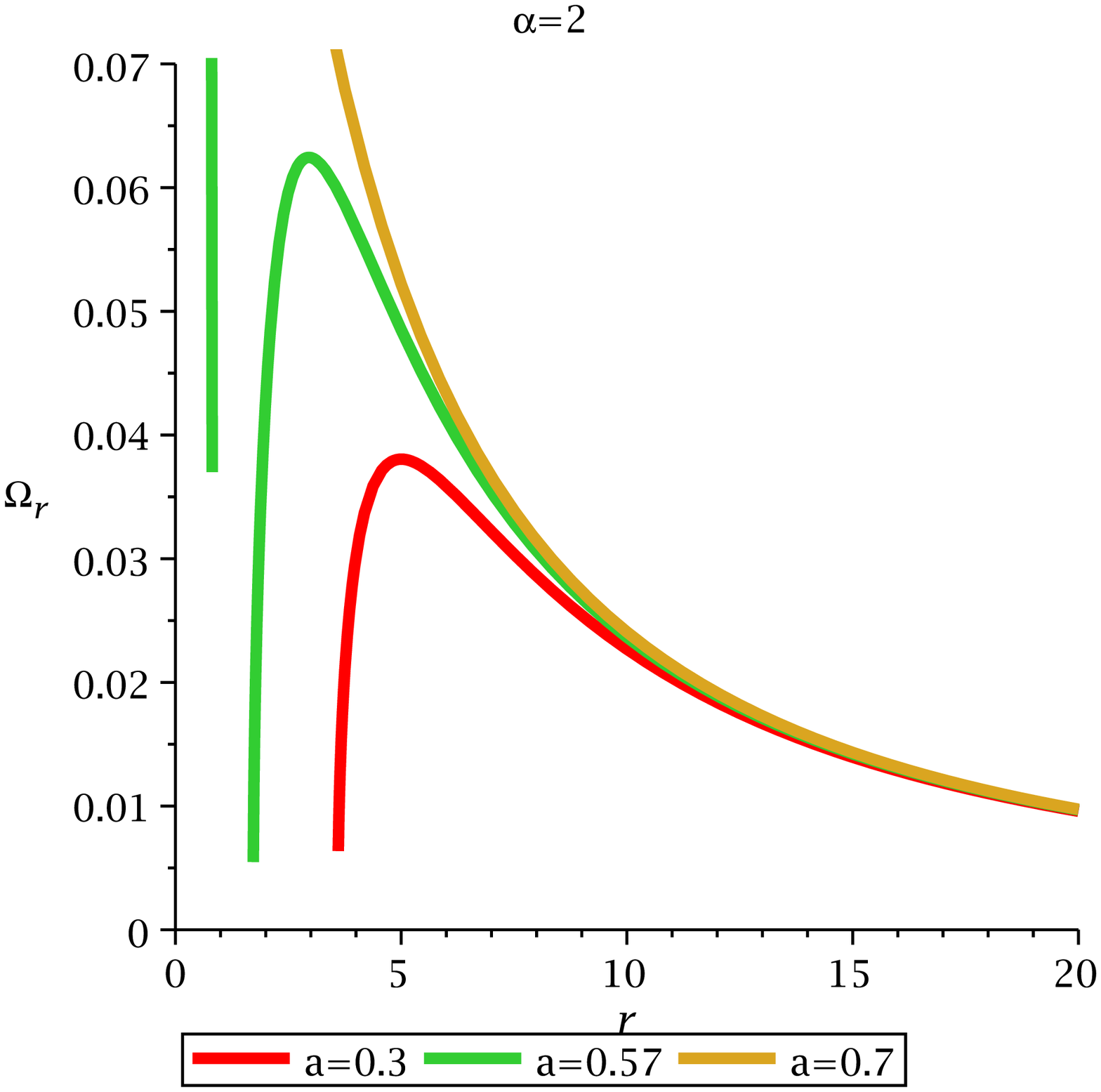}} 
\caption{The figure depicts the variation  of $\Omega_{r}$  with $r$ for different MOG parameter and 
spin parameter. Each figure shows the difference between three compact objects namely,  non-extremal BH, 
extremal BH and NS. }
\label{xq7}
\end{center}
\end{figure}
%%%%%%%%%%%%%%%%%%%%%%%%%%%%%%%%%%%%%%%%%%%Vertical Frequency%%%%%%%%%%%%%%%%%%%%%%%%%%%
\begin{figure}
\begin{center}
\subfigure[]{
\includegraphics[width=2in,angle=0]{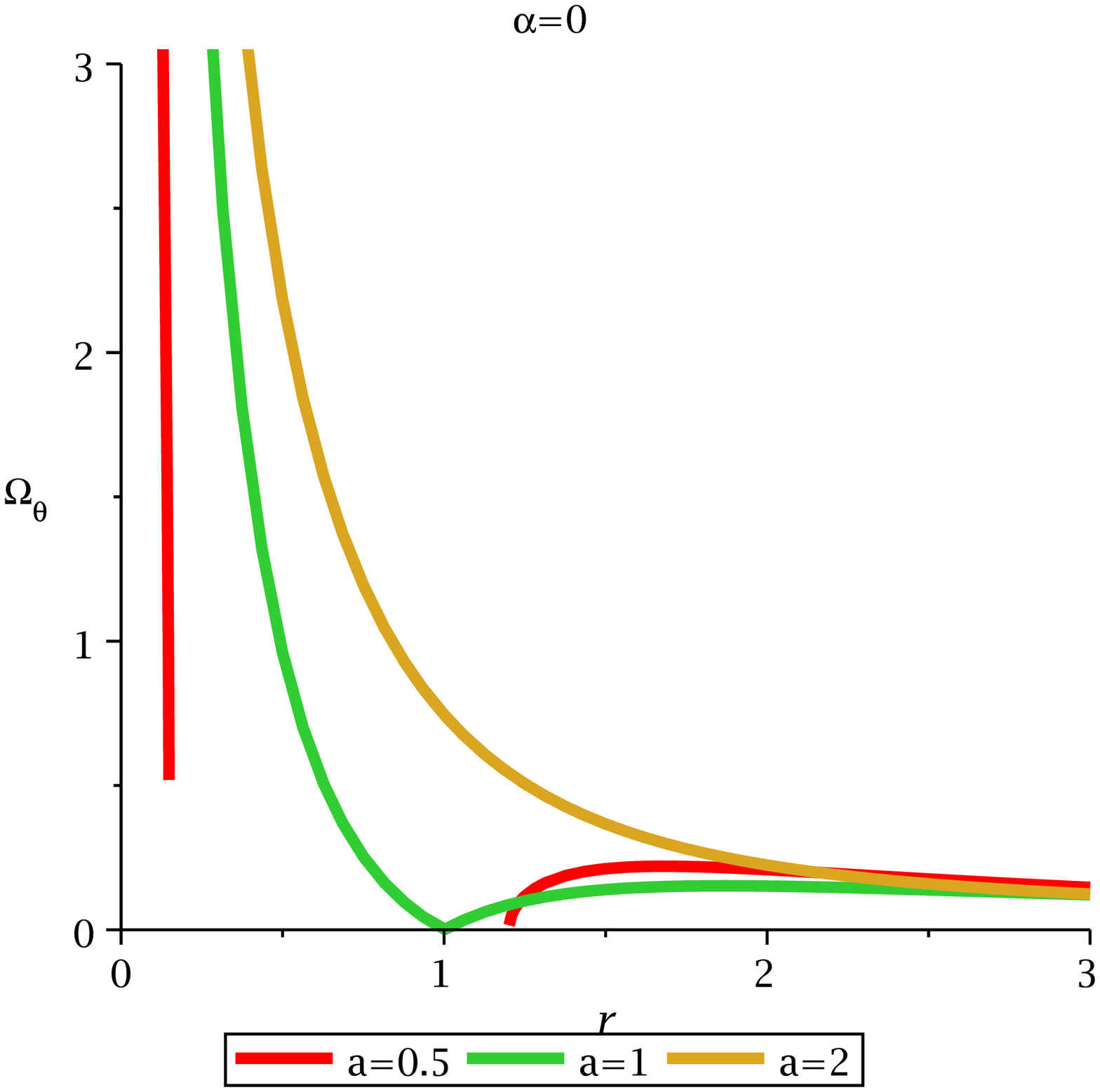}} 
\subfigure[]{
\includegraphics[width=2in,angle=0]{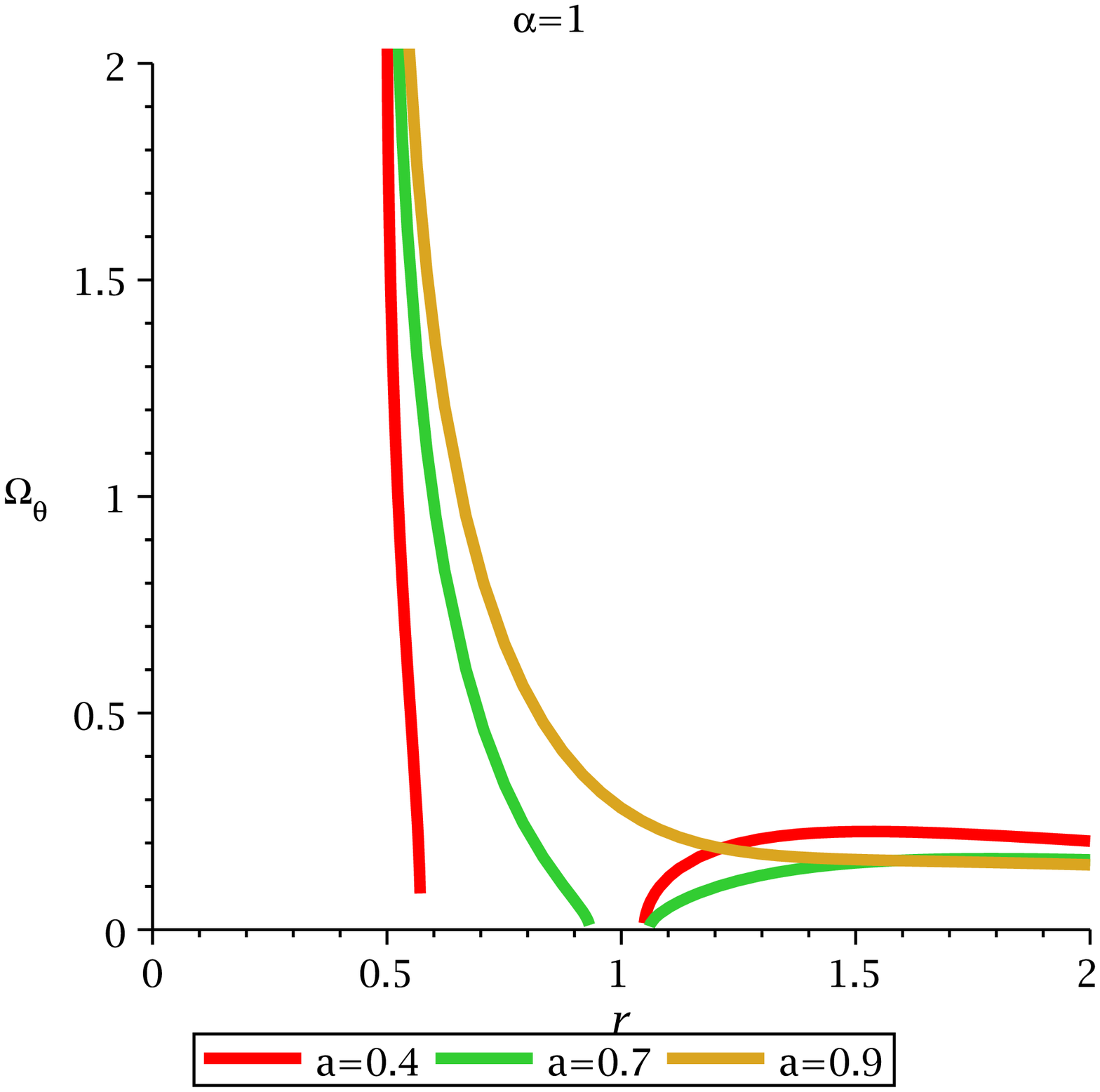}} 
\subfigure[]{
\includegraphics[width=2in,angle=0]{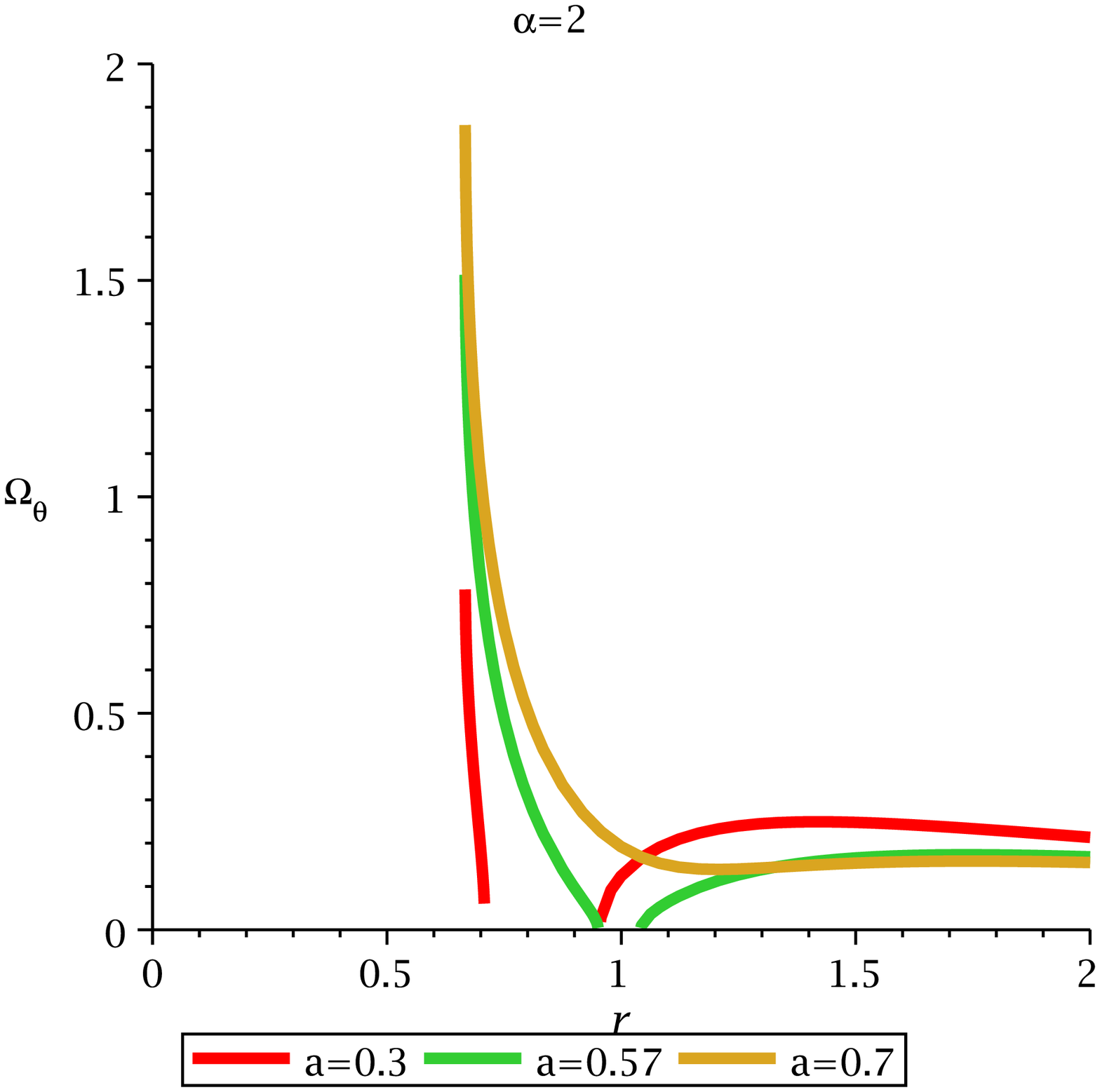}} 
\caption{ The figure depicts the variation  of $\Omega_{\theta}$  with $r$ for different MOG parameter and 
spin parameter. Each figure characterizes the difference between non-extremal BH, extremal BH and NS.}
\label{xq8}
\end{center}
\end{figure}
The other two important frequency namely, the periastron 
precession frequency~($\Omega_{per}$) and nodal precession frequency~($\Omega_{nod}$) could 
be defined as 
\begin{eqnarray}
\Omega_{per} &=& \Omega_{\phi}-\Omega_{r}\\
\Omega_{nod} &=& \Omega_{\phi}-\Omega_{\theta}~\label{ekm3} 
\end{eqnarray}
All these frequencies above that we have defined are related to the precession of the orbit and orbital plane.  
Periastron frequency occurs due to the precession of the orbit while nodal precession frequency occurs due to 
the precession of orbital plane. Sometimes the nodal precession frequency is called as orbital planer precession 
frequency or Lense-Thirring precession frequency~\cite{lt}. The radial variation of these two frequencies could 
be observed from Fig. ~(\ref{xq9}) and Fig. ~(\ref{xq10}).
\begin{figure}
\begin{center}
\subfigure[]{
\includegraphics[width=2in,angle=0]{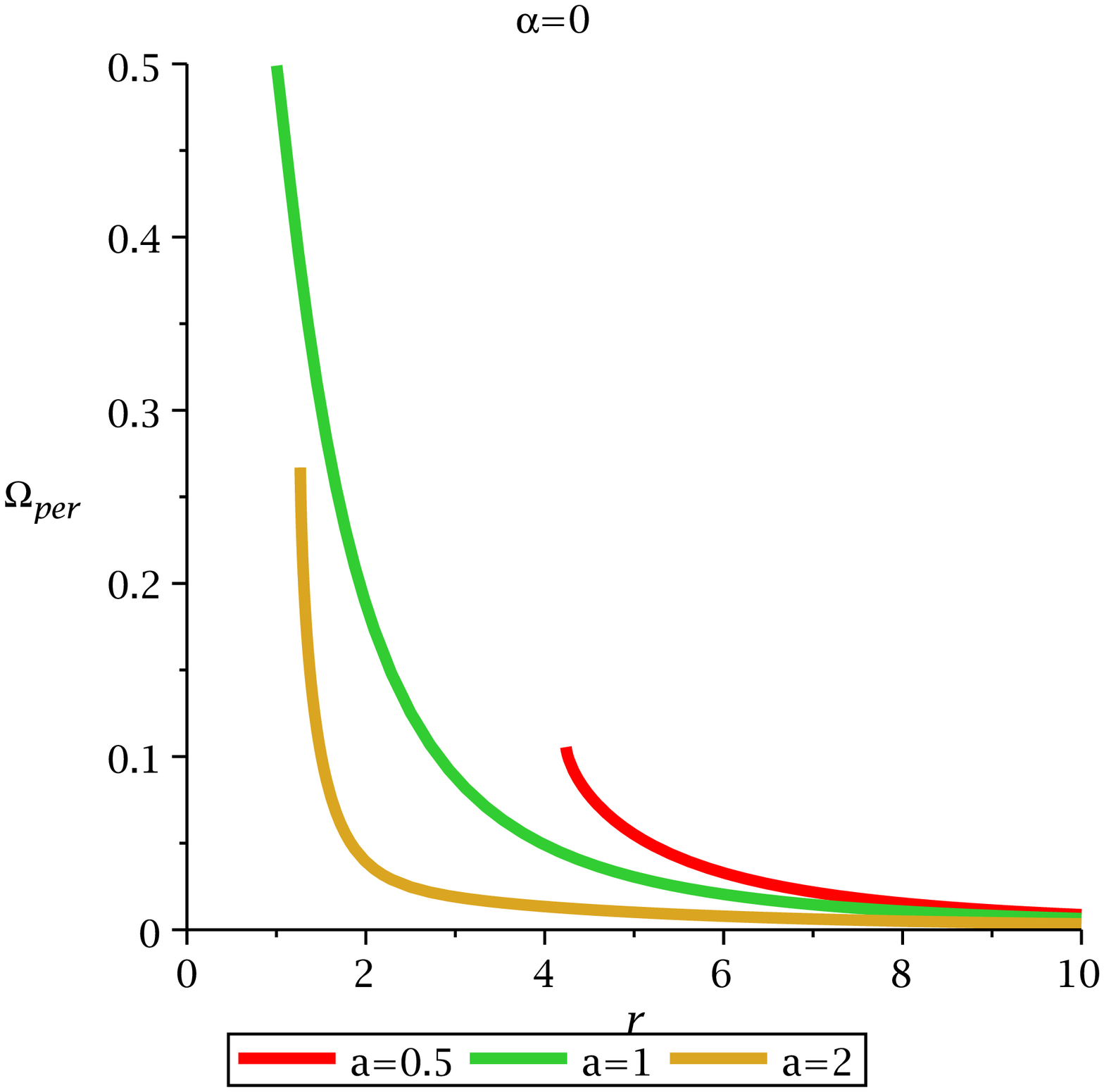}} 
\subfigure[]{
\includegraphics[width=2in,angle=0]{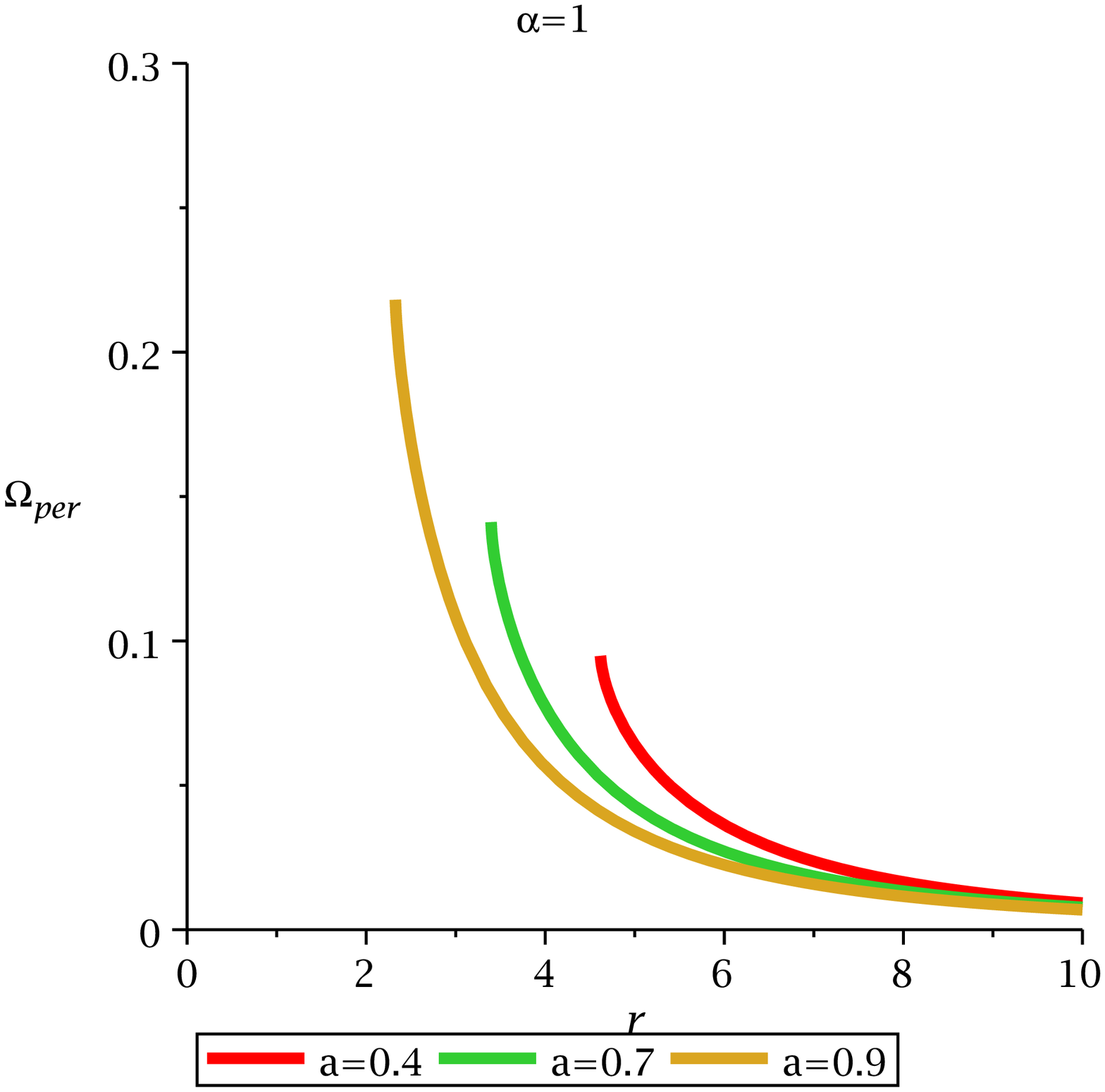}} 
\subfigure[]{
\includegraphics[width=2in,angle=0]{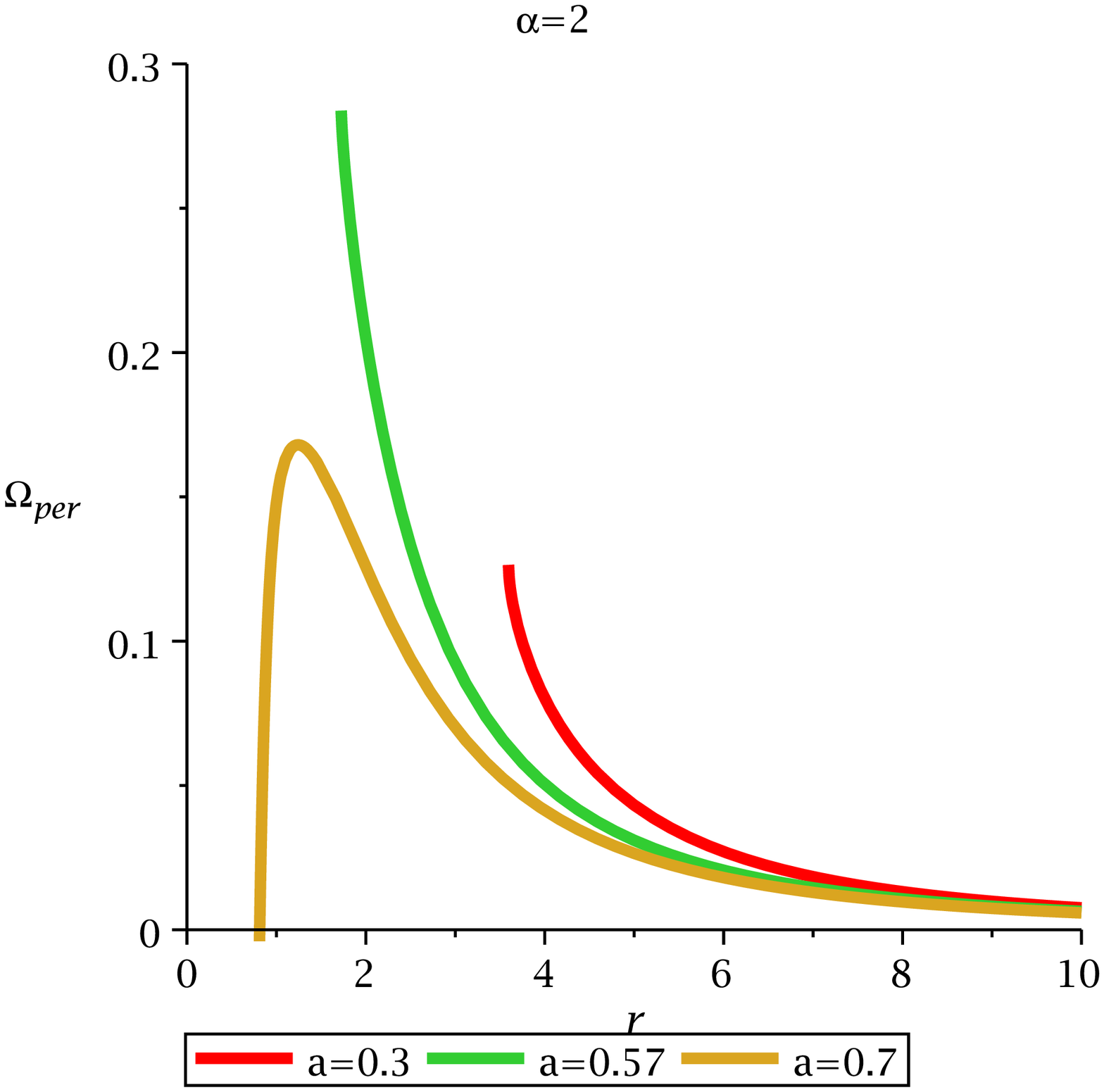}} 
\caption{ The figure depicts the variation  of $\Omega_{per}$  with $r$ for different MOG parameter and 
spin parameter. Each figure shows the difference between non-extremal BH, extremal BH and NS.}
\label{xq9}
\end{center}
\end{figure}
%%%%%%%%%%%%%%%%%%%%%%%%%%%%%%%%%%%%%%%%%%%%%%%%%%%%%%%%%%
\begin{figure}
\begin{center}
\subfigure[]{
\includegraphics[width=2in,angle=0]{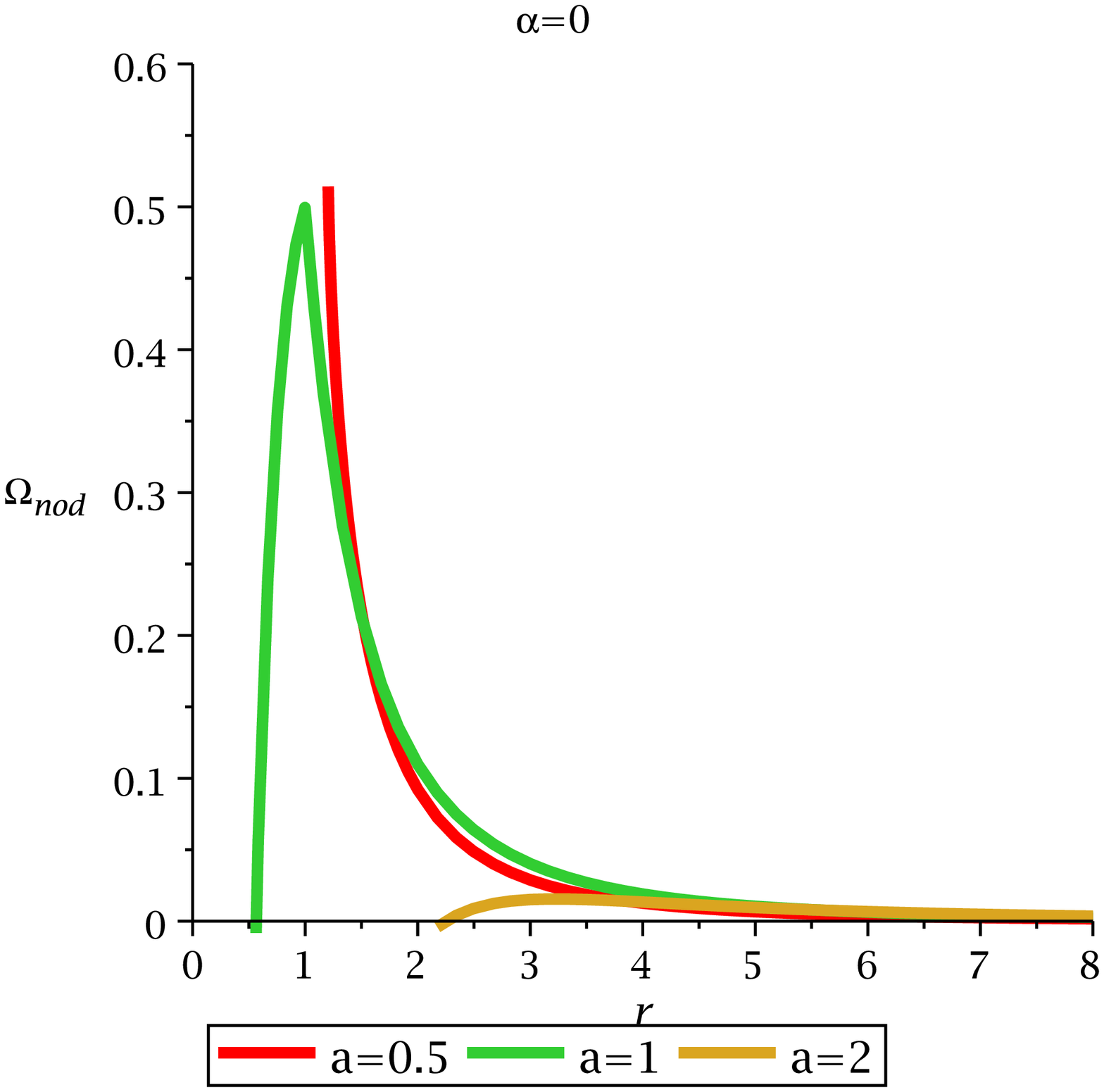}} 
\subfigure[]{
\includegraphics[width=2in,angle=0]{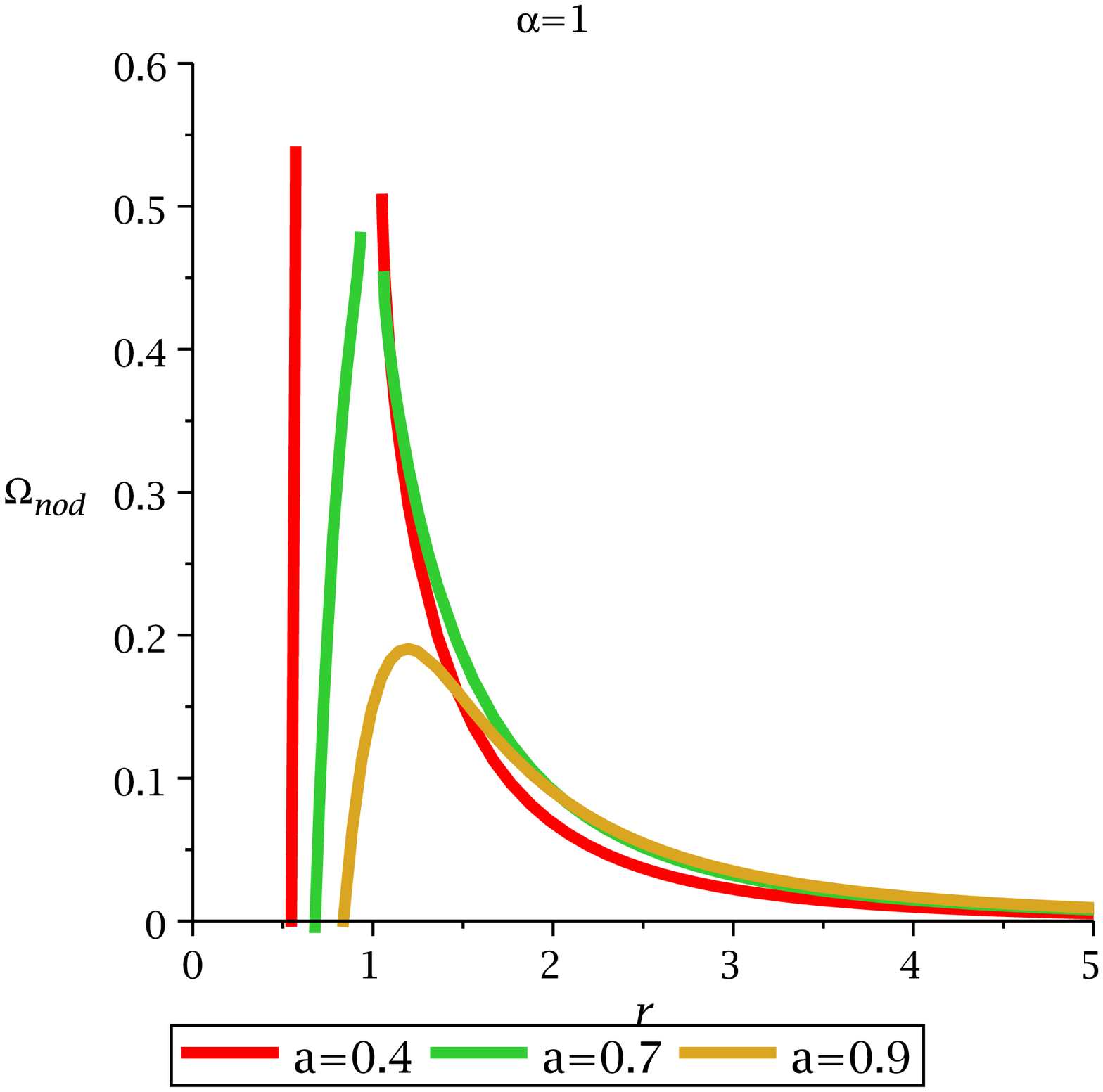}} 
\subfigure[]{
\includegraphics[width=2in,angle=0]{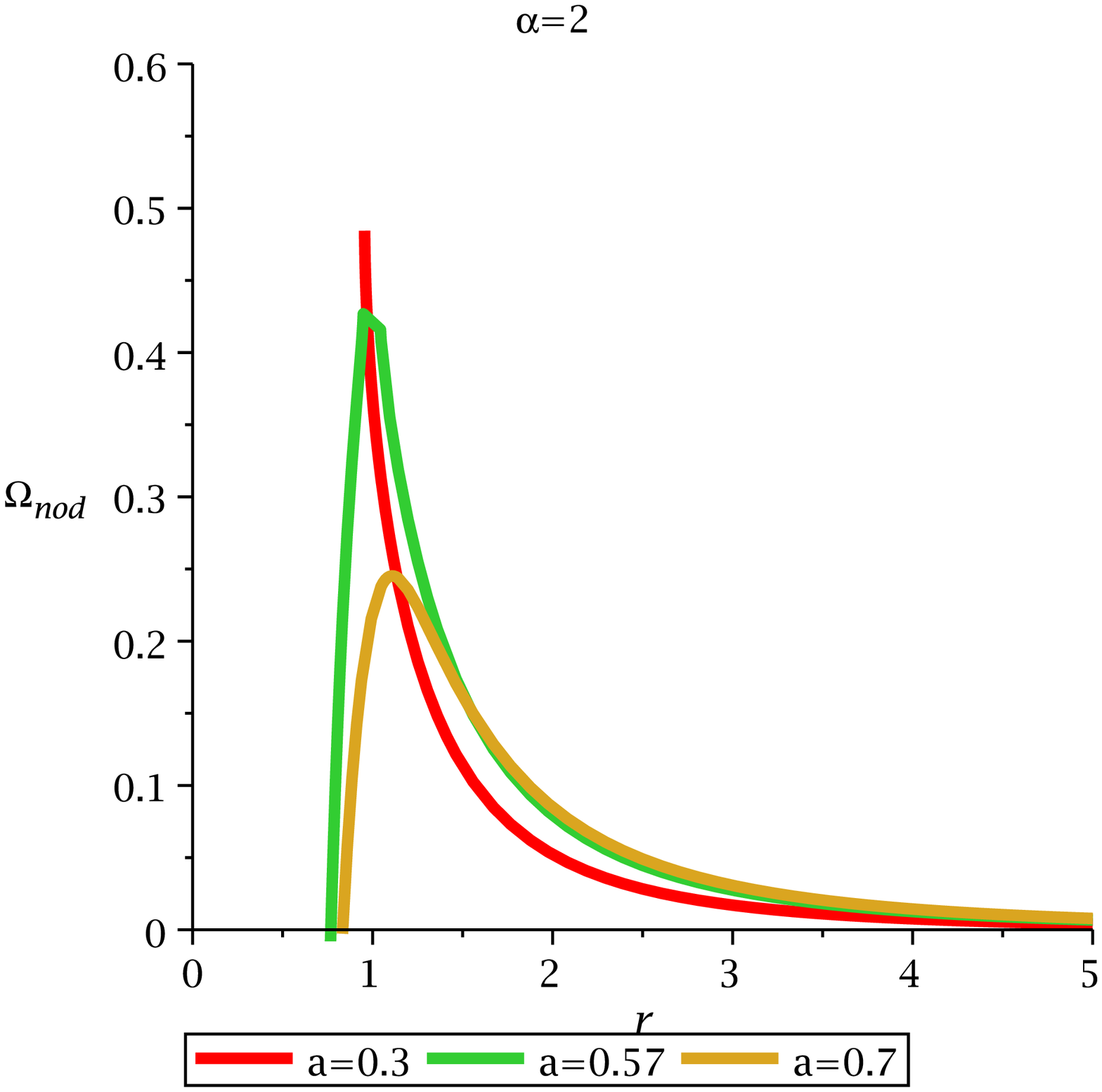}} 
\caption{ The figure depicts the variation  of $\Omega_{nod}$  with $r$ for different MOG parameter and 
spin parameter. Each figure classifies the difference between non-extremal BH, extremal BH and NS.}
\label{xq10}
\end{center}
\end{figure}
%%%%%%%%%%%%%%%%%%%%%%%%%%%%%%5
%%%%%%%%%%%%%%%%%%%%%%%%%%%%%%%%%%%%%%%%%%%%%%%%%%%%%%%%%%%%%%%%%%%%%%%%%%%%%%%%%
\begin{figure}
\begin{center}
\subfigure[]{
\includegraphics[width=2in,angle=0]{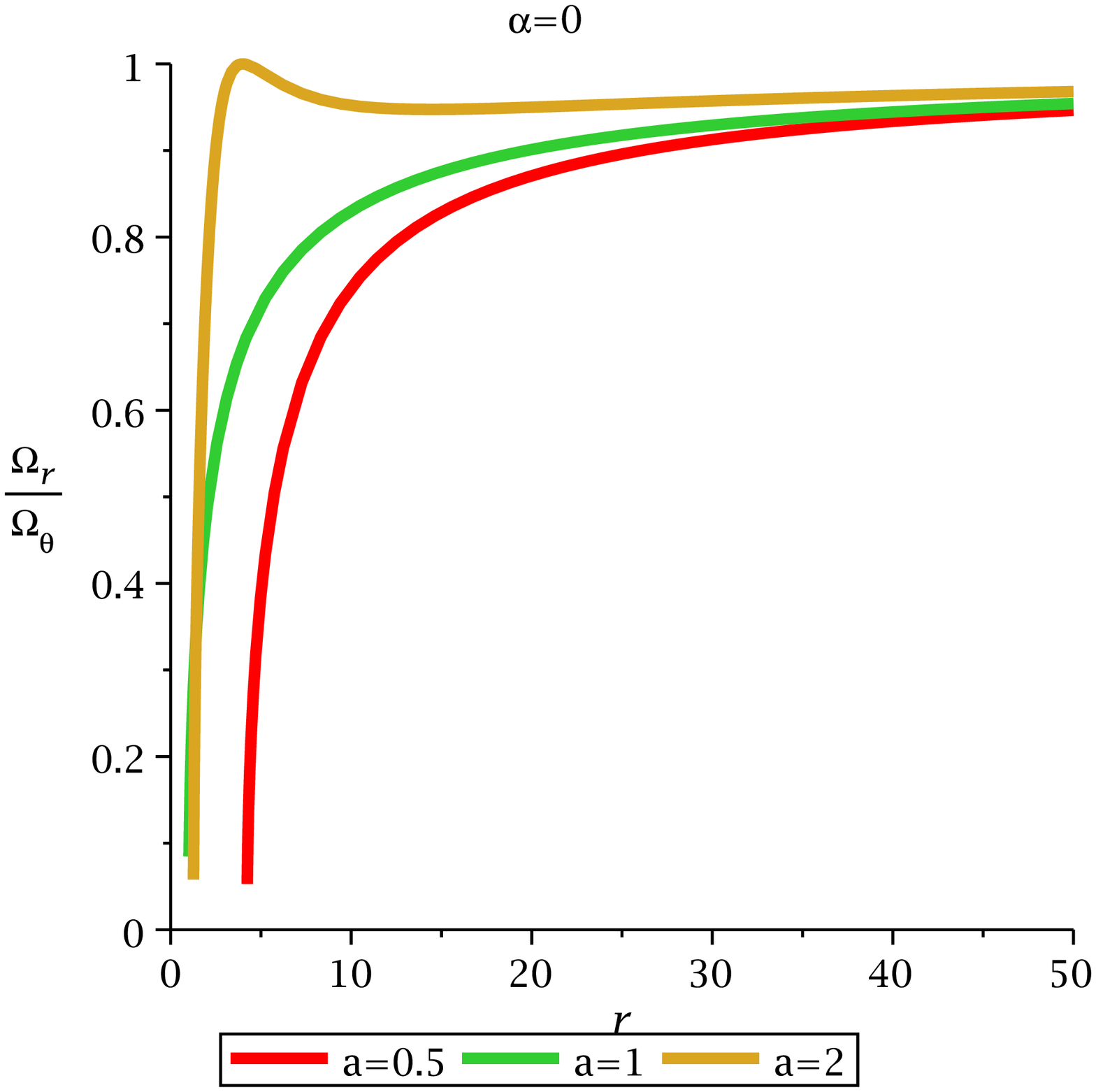}} 
\subfigure[]{
\includegraphics[width=2in,angle=0]{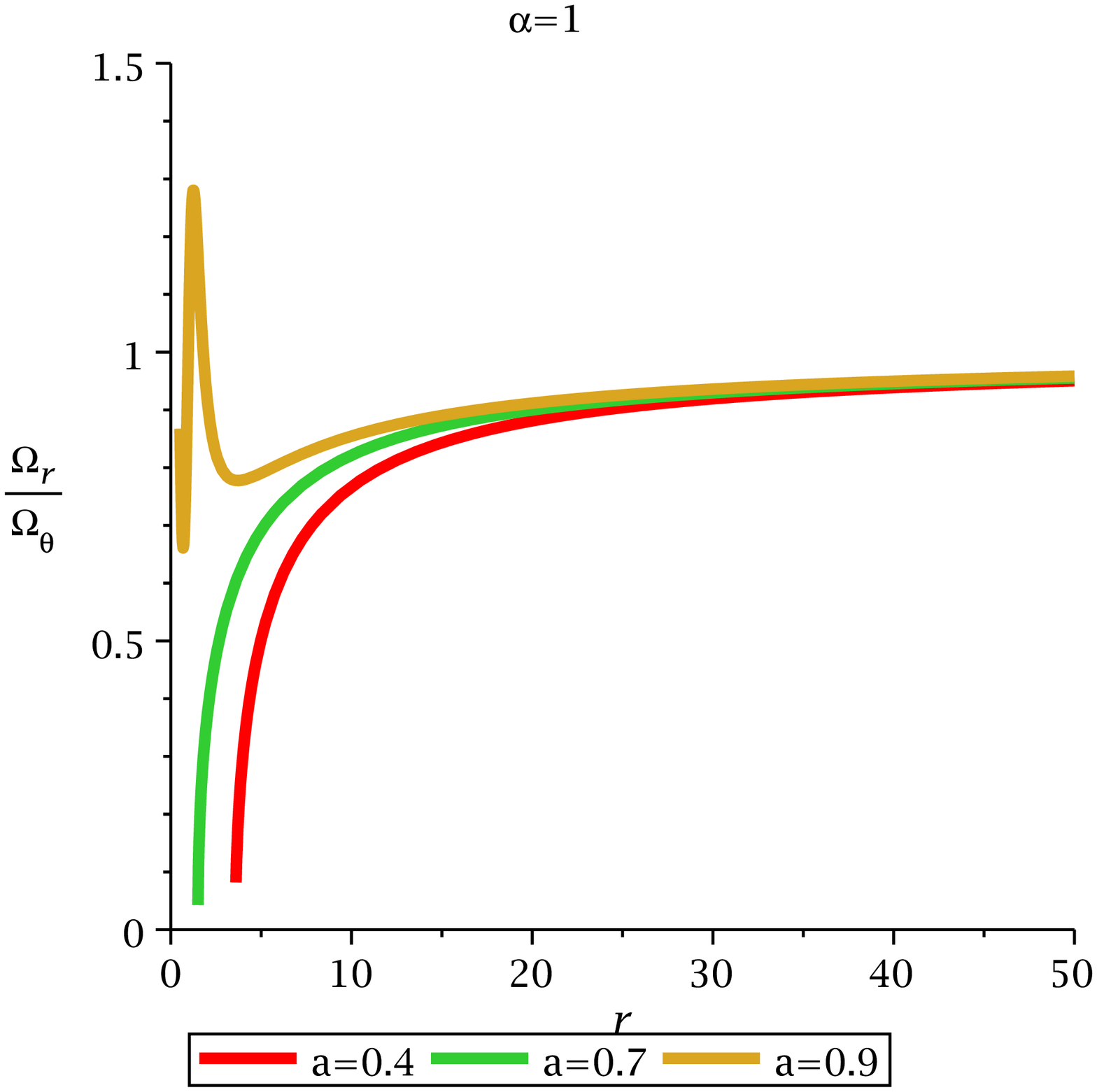}} 
\subfigure[]{
\includegraphics[width=2in,angle=0]{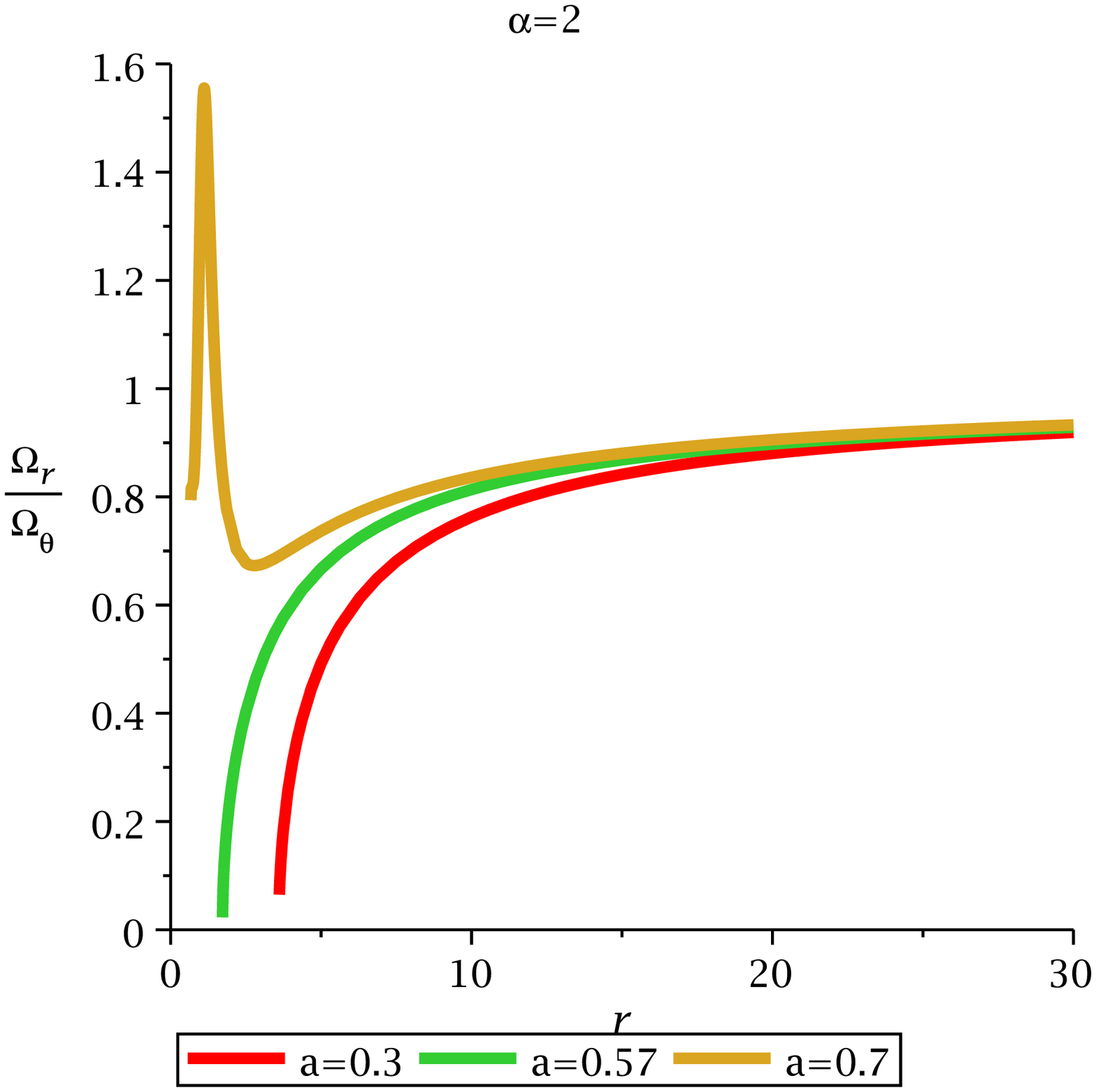}}
\subfigure[]{
\includegraphics[width=2in,angle=0]{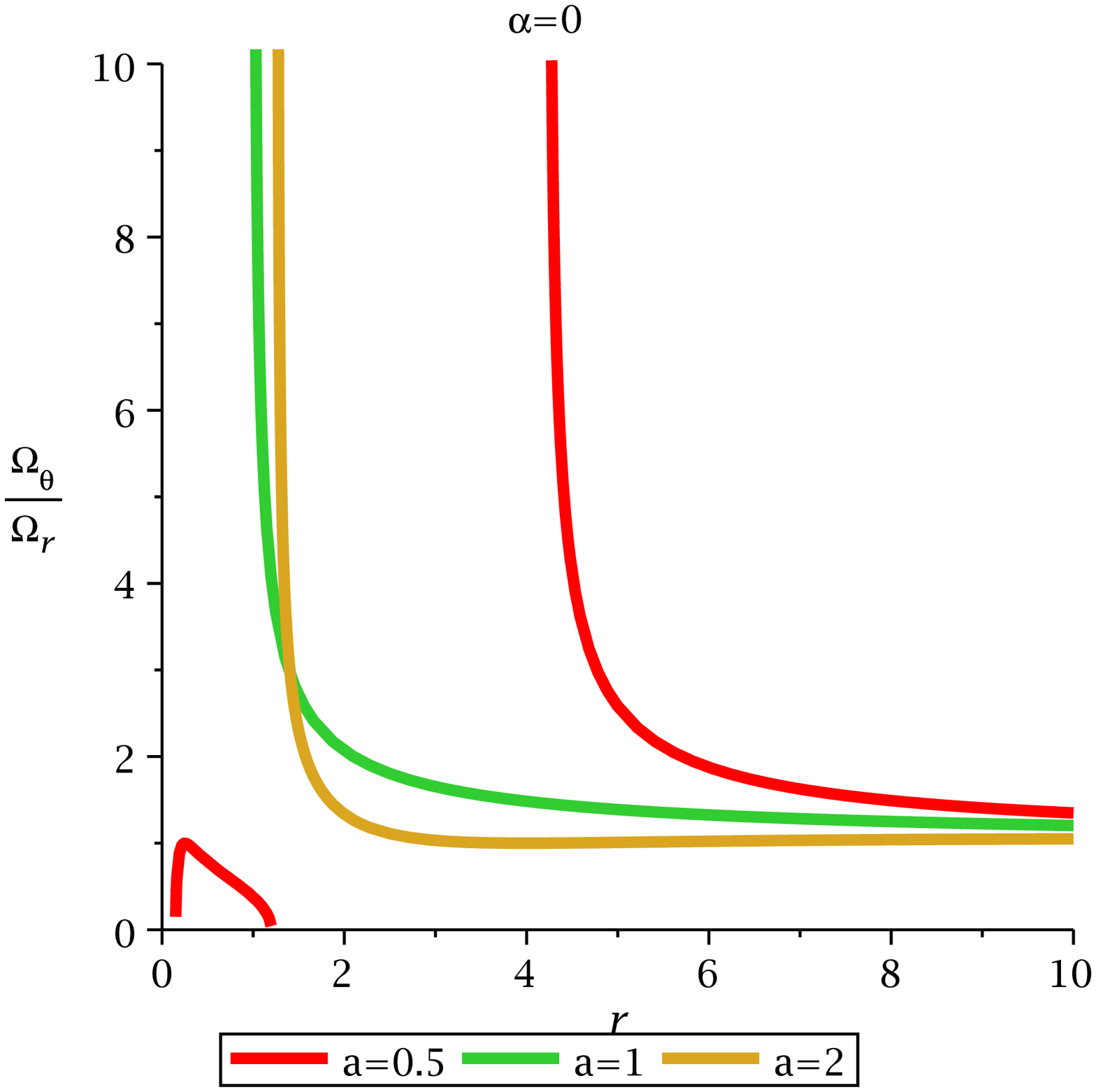}} 
\subfigure[]{
\includegraphics[width=2in,angle=0]{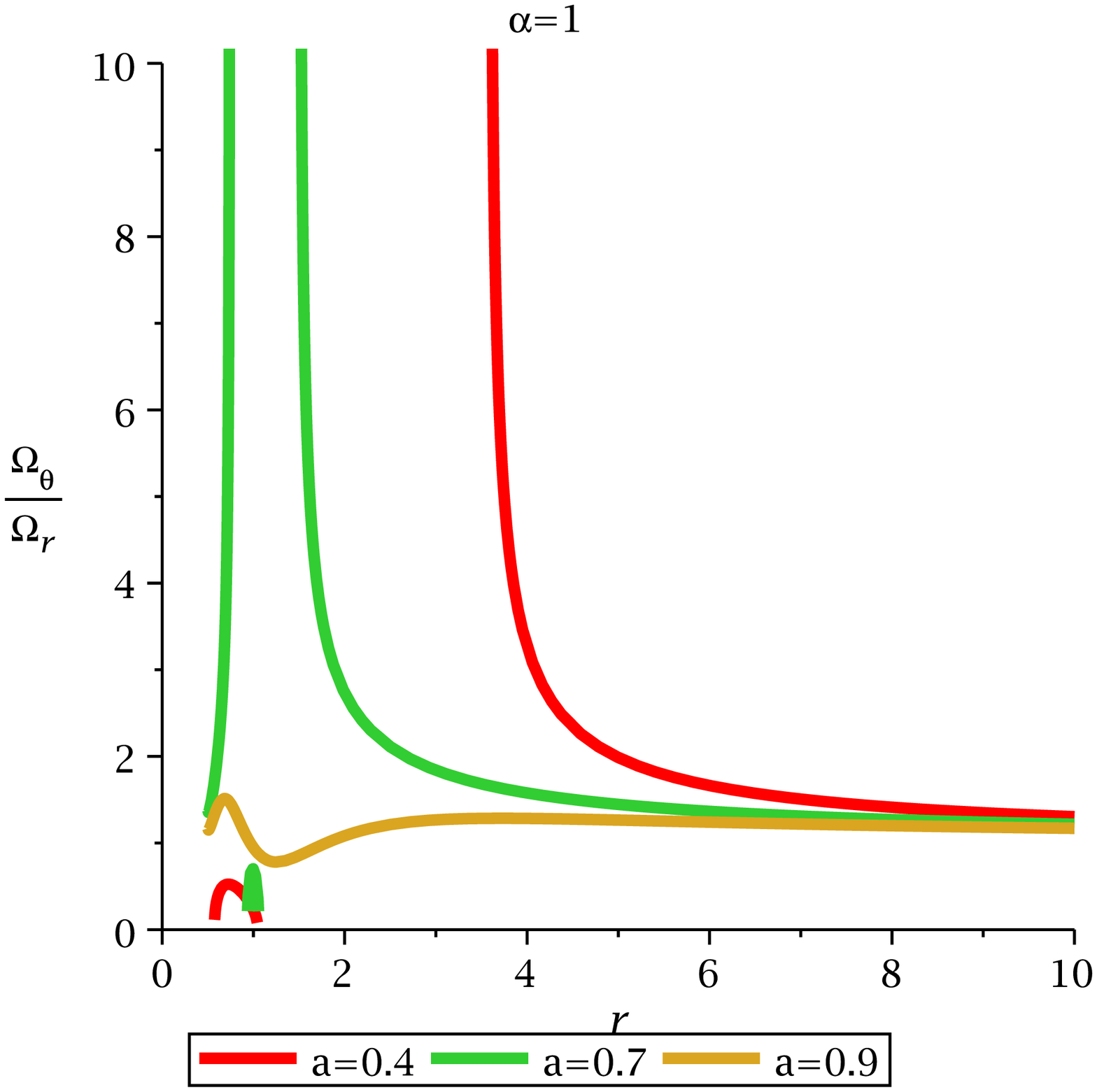}} 
\subfigure[]{
\includegraphics[width=2in,angle=0]{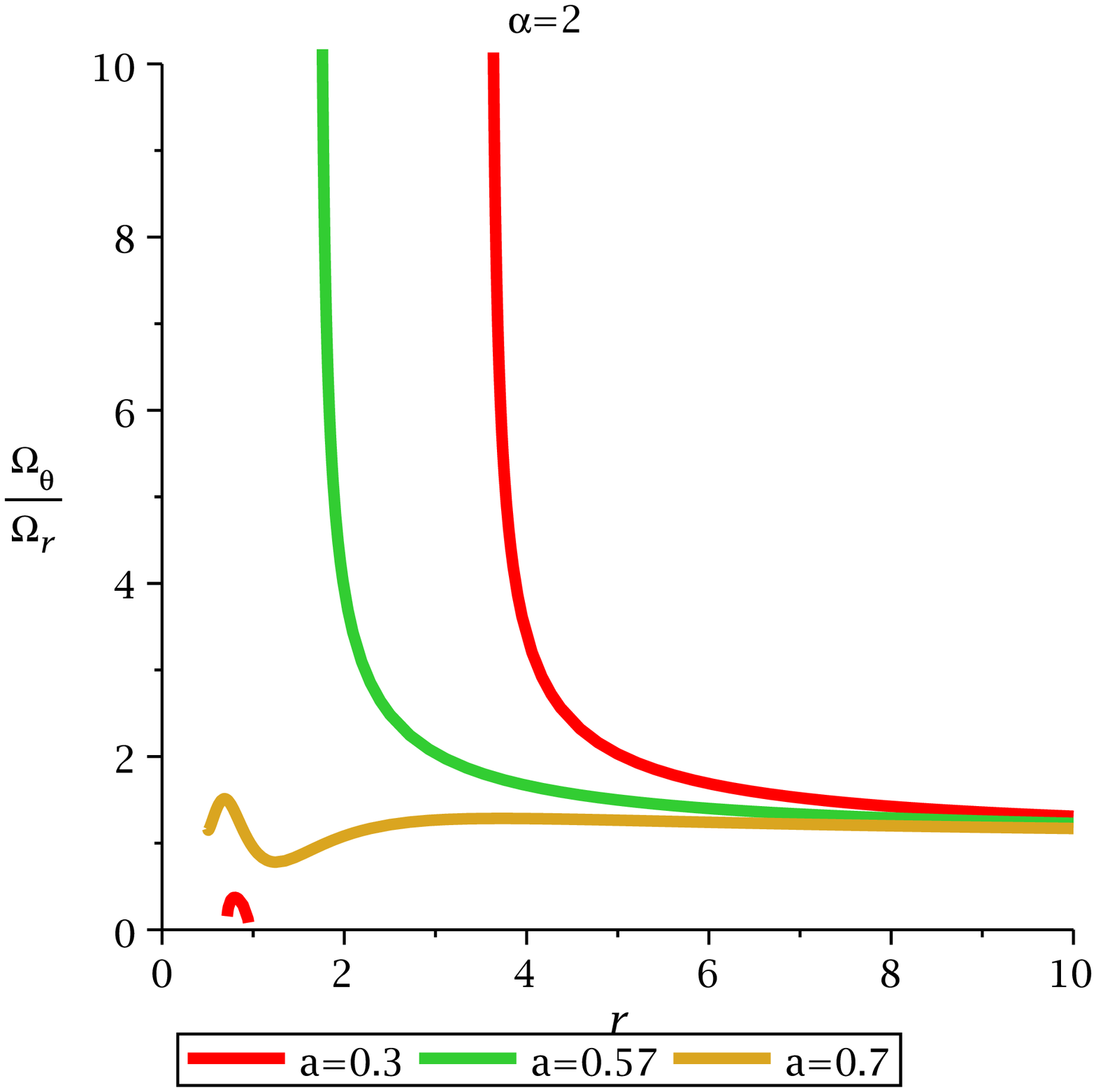}}
\caption{ The figure depicts the variation  of $\frac{\Omega_{r}}{\Omega_{\theta}}$ and 
$\frac{\Omega_{\theta}}{\Omega_{r}}$  with $r$ for different MOG parameter and 
spin parameter. Each figure depicts the difference between non-extremal BH,
extremal BH and NS.}
\label{xq11}
\end{center}
\end{figure}
%%%%%%%%%%%%%%%%%%%%%%%%%%%%%%%%%%%%%%%%%%%%%%%%%%%%%%%%%%%%%%%%%%%%%%%%%%%%%%%%%%%%%%
\begin{figure}
\begin{center}
\subfigure[]{
\includegraphics[width=2in,angle=0]{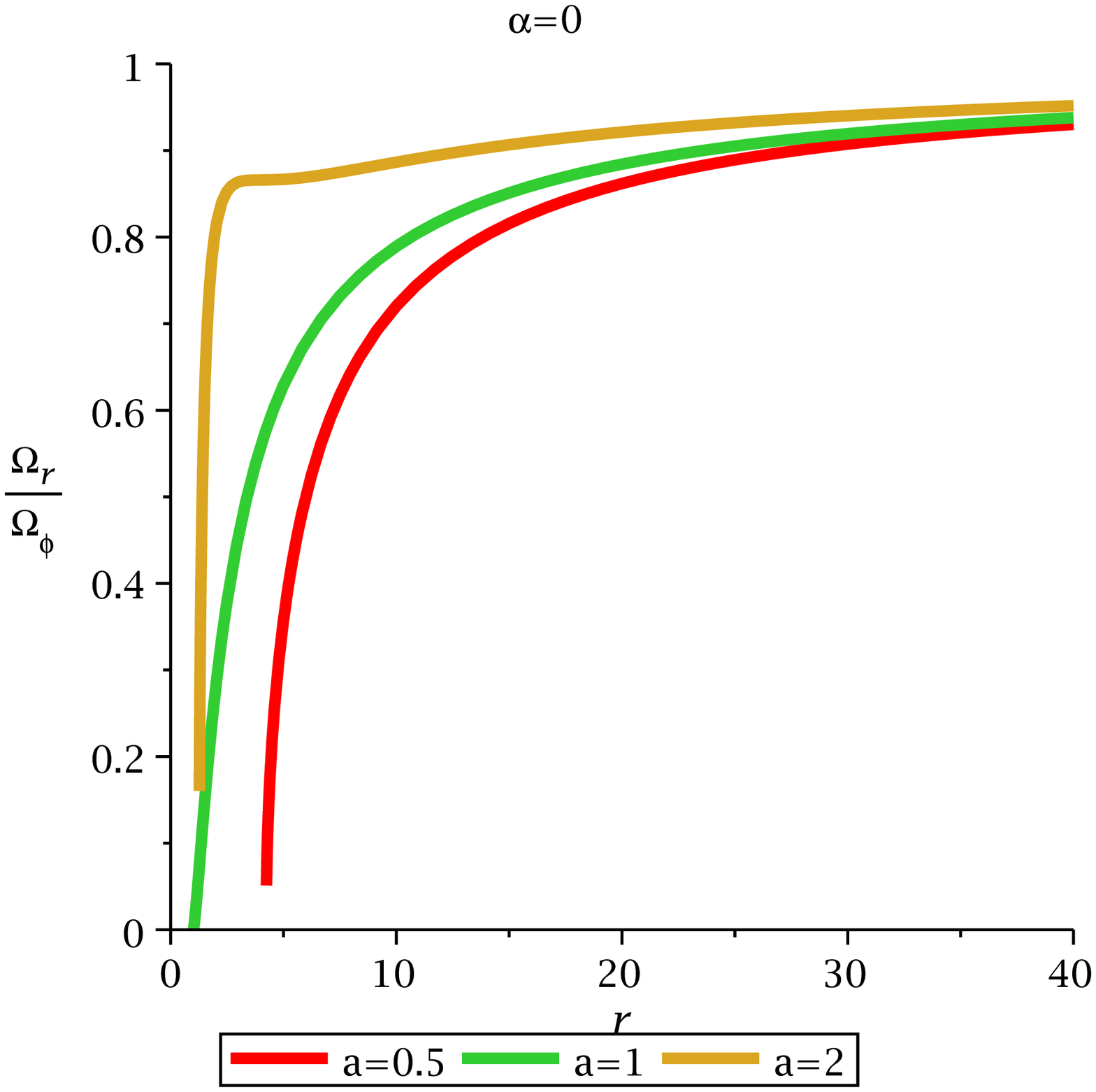}} 
\subfigure[]{
\includegraphics[width=2in,angle=0]{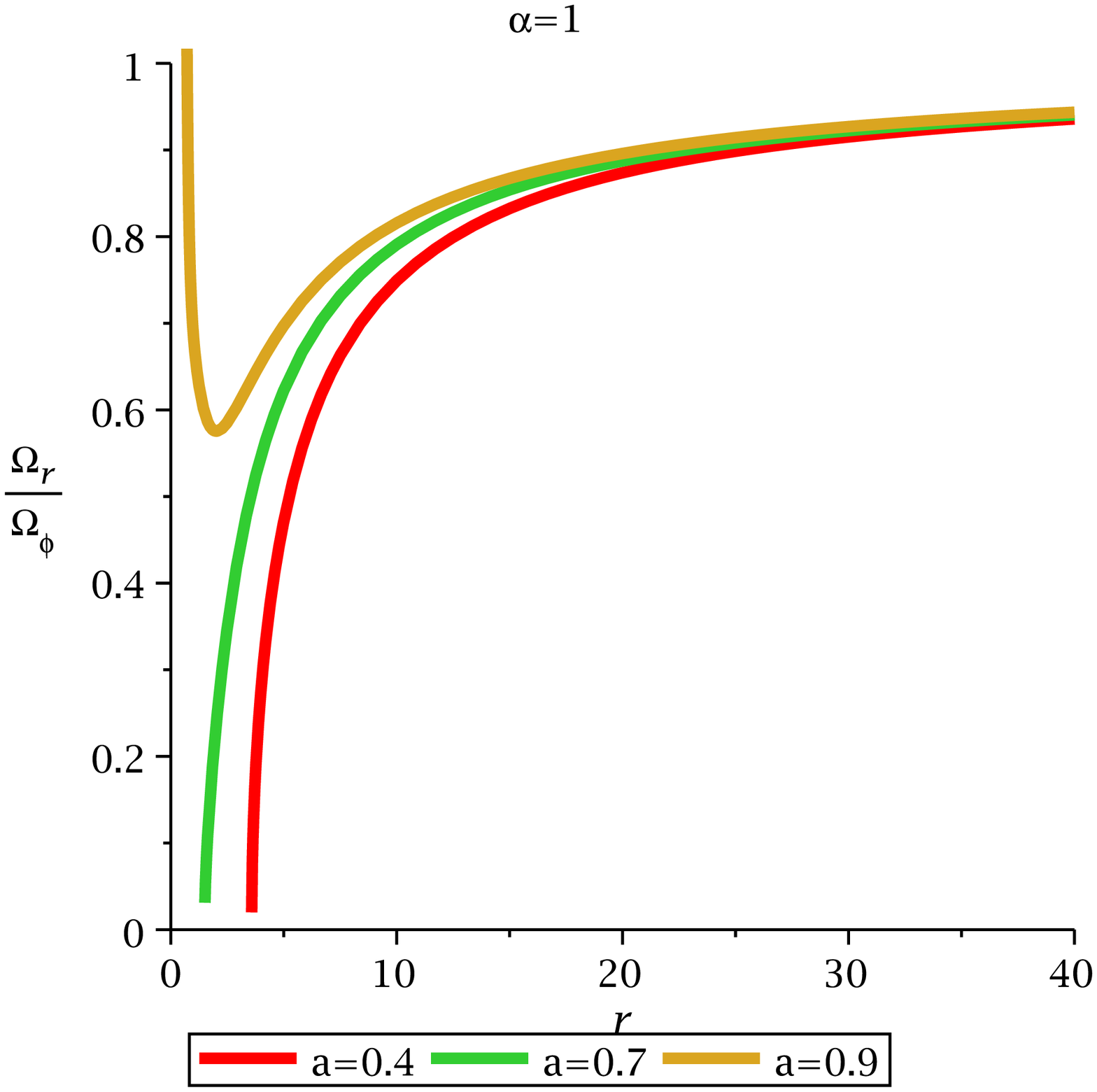}} 
\subfigure[]{
\includegraphics[width=2in,angle=0]{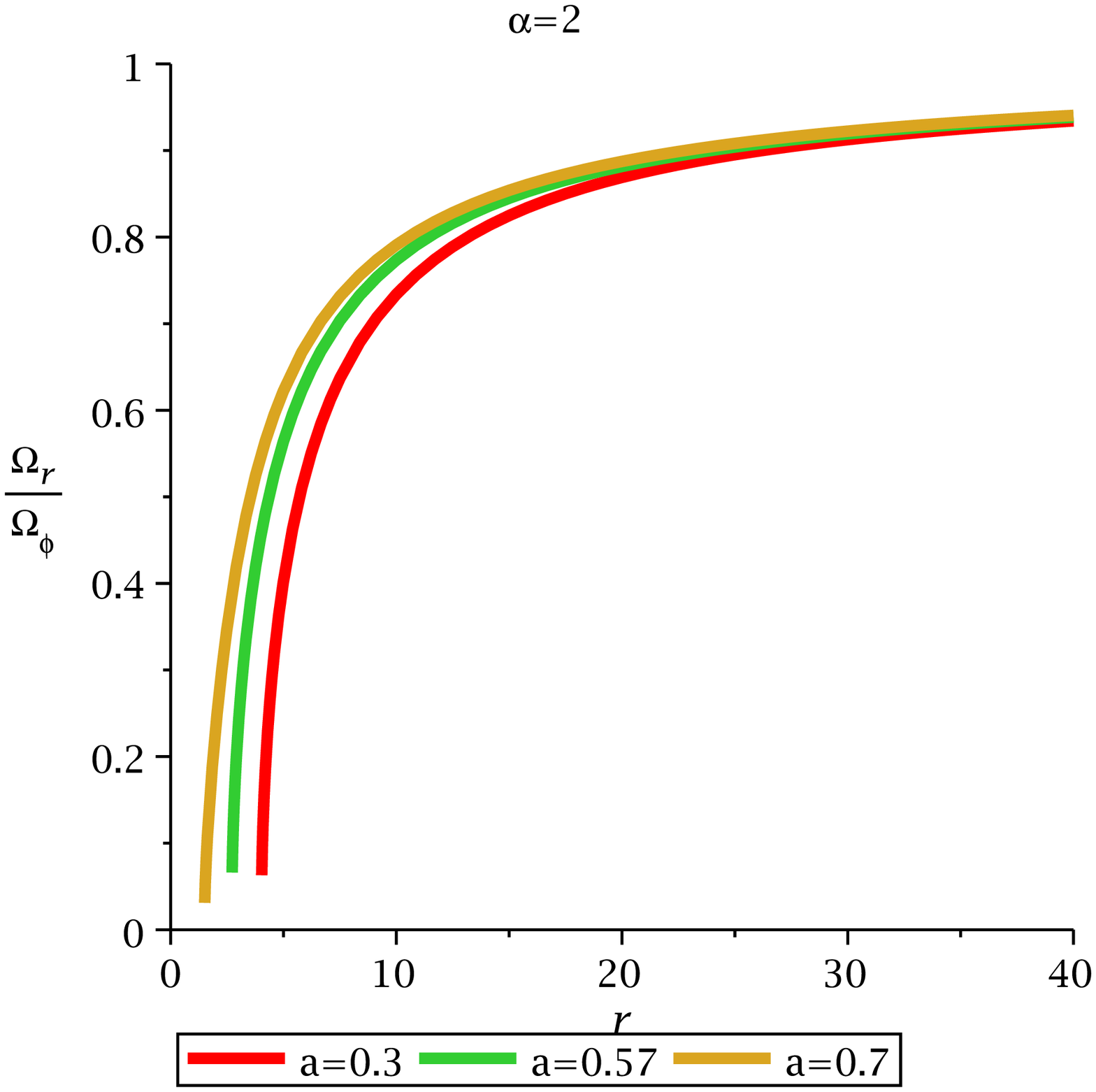}}
\caption{The figure depicts the variation  of 
$\frac{\Omega_{r}}{\Omega_{\phi}}$ with $r$ for 
different MOG parameter and spin parameter. 
Each figure depicts the difference between 
non-extremal BH, extremal BH and NS.}
\label{xq12}
\end{center}
\end{figure}
%%%%%%%%%%%%%%%%%%%%%%%%%%%%%%%%%%%%%%%%%%%%%%%%%%%%%%%%%%%%%%%%%%%%%%%%%%%%%%%%%%%%%%%
\begin{figure}
\begin{center}
\subfigure[]{
\includegraphics[width=2in,angle=0]{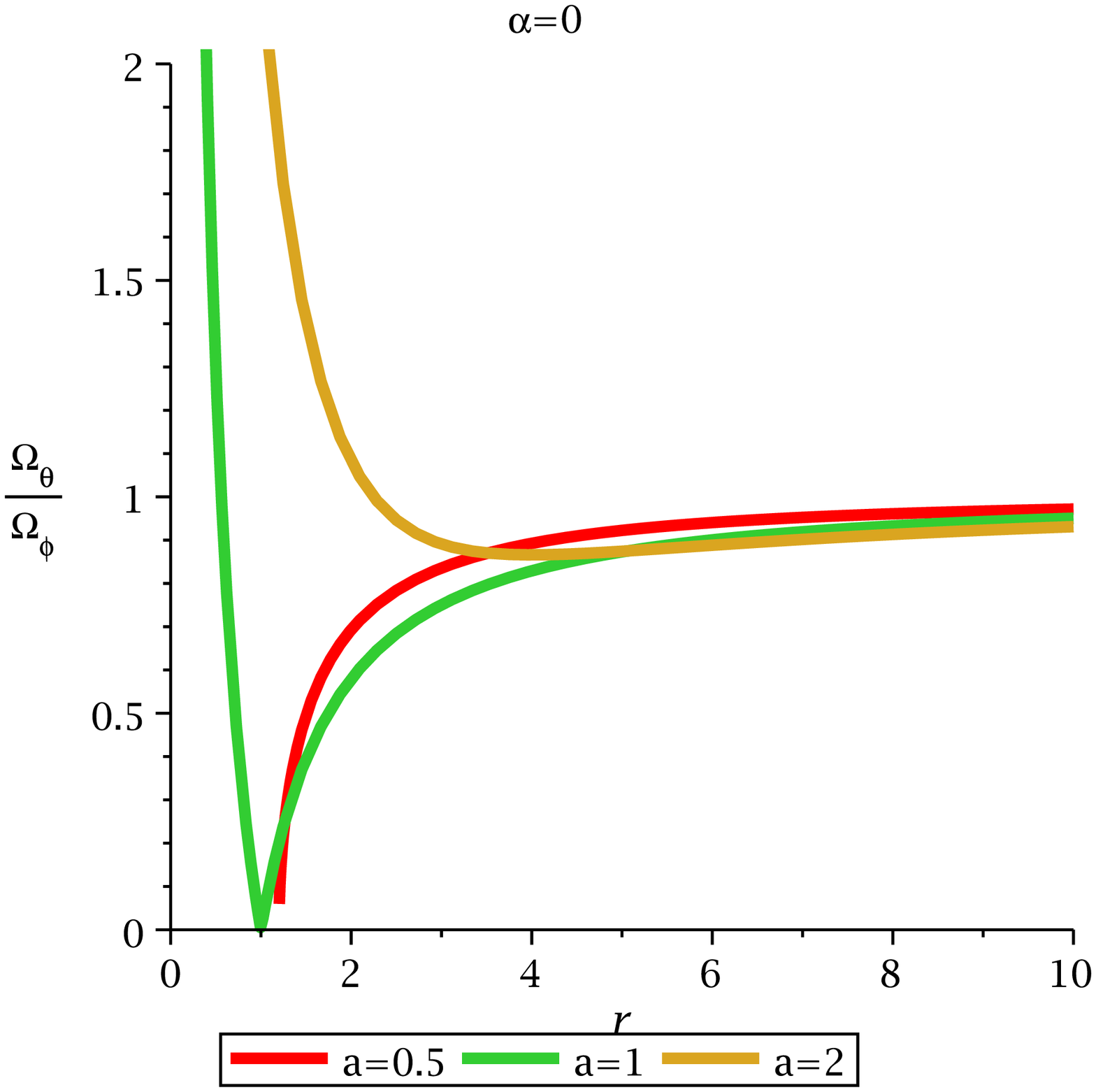}} 
\subfigure[]{
\includegraphics[width=2in,angle=0]{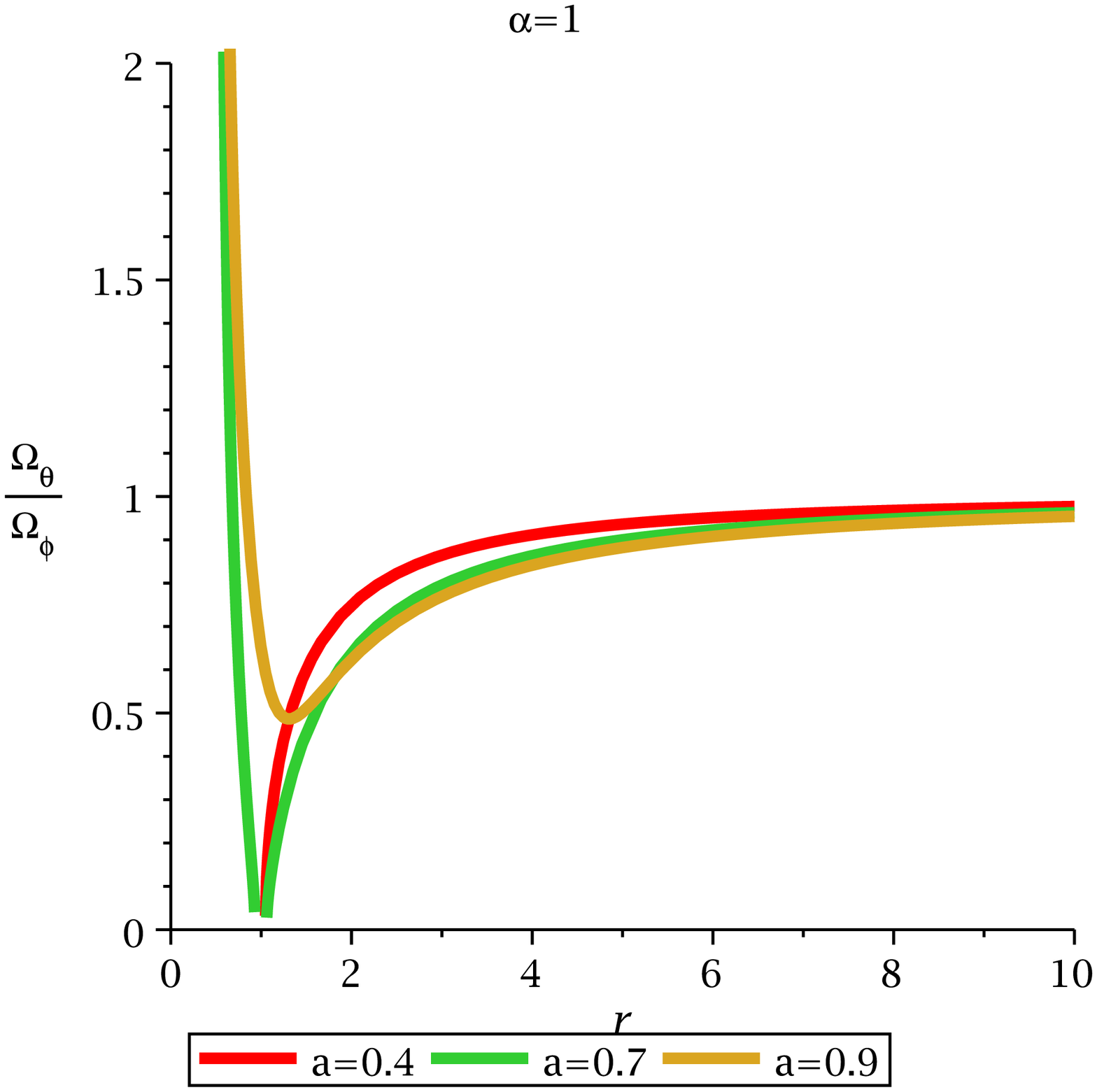}} 
\subfigure[]{
\includegraphics[width=2in,angle=0]{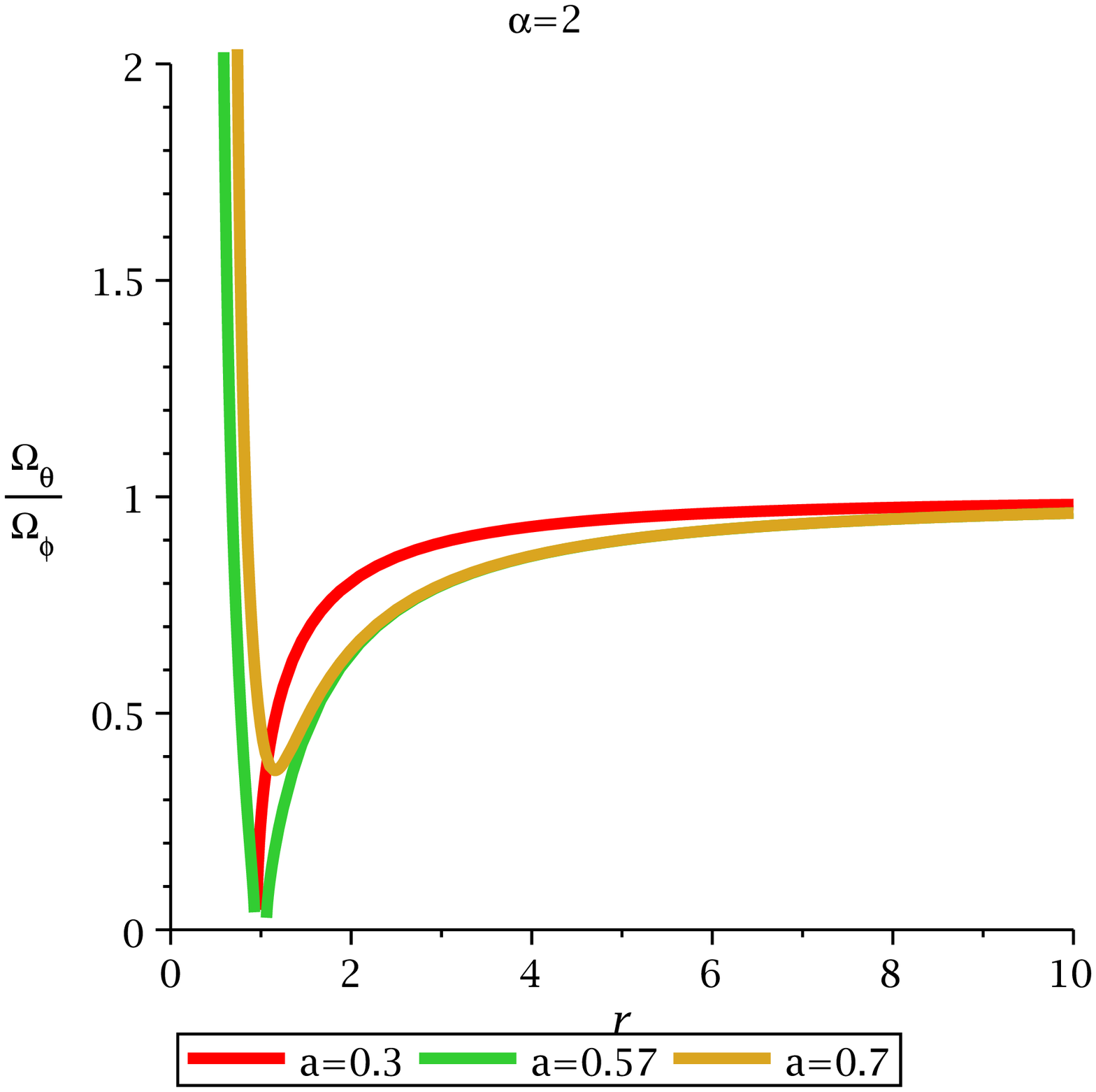}}
\caption{The figure depicts the variation  of 
$\frac{\Omega_{r}}{\Omega_{\phi}}$ with $r$ for 
different MOG parameter and spin parameter. 
Each figure depicts the difference between 
non-extremal BH, extremal BH and NS.}
\label{xq13}
\end{center}
\end{figure}
%%%%%%%%%%%%%%%%%%%%%%%%%%%%%%%%%%%%%%%%%%%%%%%%%%%%%%%%%%%%%%%%%%%%%%%%%%%%%%%%%%%%%%%
 
The fact that $\Omega_{r}^2\geq 0$ and $\Omega_{\theta}^2\geq 0$  determined the stability of 
circular orbits. From the first condition one can compute radii of ISCO. It is well known that 
ISCO of Kerr BH is located at~\cite{bpt}
\begin{eqnarray}
\frac{r_{ISCO}}{\cal M} &=& 3+z_{2}\mp \sqrt{(3-z_{1})(3+z_{1}+2z_{2})}
\end{eqnarray}
where
\begin{eqnarray}
z_{1} &=& 1+\left(1-\frac{a^2}{{\cal M}^2}\right)^{1/3} \left[(1-\frac{a}{{\cal M}})^{1/3}+(1+\frac{a}{{\cal M}})^{1/3}\right]\\
z_{2} &=& \sqrt{3\frac{a^2}{{\cal M}^2}+z_{1}^2}
\end{eqnarray}
Here, upper sign indicates  direct orbit and lower sign indicates retrograde orbit.
For extremal Kerr BH, the ISCO is situated at $r_{ISCO}={\cal M}$ for direct orbit while 
$r_{ISCO}=9{\cal M}$ for retrograde orbit~\cite{bpt}. The non-negativeness of $\Omega_{\theta}$ 
implies that the geodesic motion is stable under small oscillations in the vertical 
direction. While in KMOG BH, the ISCO radii can be calculated via the equation $\Omega_{r}^2=0$:
$$
G_{N}{\cal M}r \Delta-4\left(\Pi_{\alpha}-G_{N}{\cal M}r\right) 
\left(\sqrt{\Pi_{\alpha}-G_{N}{\cal M}r} \mp a \right)^2 = 0
$$
In spherically symmetric spacetime where the value of spin parameter $a=0$ means that 
$\Omega_{\phi}=\Omega_{\theta}$, which implies that the Lense-Thirring precession is 
absent  while in axisymmetric spacetime $\Omega_{\phi} \neq \Omega_{\theta}$. This 
will be vanish for a particular value of $r=r_{0}$ i.e. 
$$
\Omega_{nod}|_{r=r_{0}} =0
$$
which implies that
$$
4\Pi_{\alpha}^2 \frac{(\Pi_{\alpha}-G_{N}{\cal M}r)}{(2\Pi_{\alpha}-G_{N}{\cal M}r)^2}|_{r=r_{0}}=a^2
$$
when MOG parameter vanishes it reduces to 
$$
\frac{16}{9} \frac{r_{0}}{M} =a^2
$$
We can differentiate three compact objects via ratio of two epicyclic frequencies~
(plot for direct orbit only) which is defined to be as follows 
$$
\frac{\Omega_{r}}{\Omega_{\theta}} =\frac{\sqrt{G_{N}{\cal M}r \Delta-4\left(\Pi_{\alpha}-G_{N}{\cal M}r\right) 
\left(\sqrt{\Pi_{\alpha}-G_{N}{\cal M}r} \mp a \right)^2}}
{\sqrt{r^2\left(\Pi_{\alpha}-G_{N}{\cal M}r\right) \mp 2 a \Pi_{\alpha}\sqrt{\Pi_{\alpha}-G_{N}{\cal M}r}
+a^2\left(2\Pi_{\alpha}-G_{N}{\cal M}r \right)}}
$$

$$
\frac{\Omega_{r}}{\Omega_{\phi}} =\pm \frac{\sqrt{G_{N}{\cal M}r \Delta-4\left(\Pi_{\alpha}-G_{N}{\cal M}r\right) 
\left(\sqrt{\Pi_{\alpha}-G_{N}{\cal M}r} \mp a \right)^2}}{r\sqrt{\Pi_{\alpha}-G_{N}{\cal M}r}}
$$

$$
\frac{\Omega_{\theta}}{\Omega_{\phi}} =\pm \frac{\sqrt{r^2\left(\Pi_{\alpha}-G_{N}{\cal M}r\right) 
\mp 2 a \Pi_{\alpha}\sqrt{\Pi_{\alpha}-G_{N}{\cal M}r}
+a^2\left(2\Pi_{\alpha}-G_{N}{\cal M}r \right)}} {r\sqrt{\Pi_{\alpha}-G_{N}{\cal M}r}}
$$
The variation of these ratio may be seen from the following diagram~[(\ref{xq11}),
(\ref{xq12}), (\ref{xq13})].

\section{Discussion}
We {performed a detailed analysis of} the inertial frame-dragging effect {of 
stationary, axisymmetric} KMOG BH. 
Specifically, we computed {the} generalized spin precession of a test gyroscope 
{due to frame-dragging effect when it is placed in the domain of the BH as well as NS.} 
By {computing} 
this frequency, we distinguished the behaviour of three astrophysical compact objects 
namely non-extremal BH, extremal BH and {NS}. We showed {in principle} a clear distinction 
between these three compact objects visually. We also compared this result with {the} Kerr 
BH. We studied different features in MOG by using spin precession versus radial profile.
In each spin precession vs. radial diagram, we {observed distinguished features of}
three compact objects for various spin limits.

{Perhaps the most promising result we obtained is that the presence of the MOG parameter deformed 
the geometric construction of different essential parameters like 
event horizon~($r_{+}$), Cauchy horizon~($r_{-}$), outer ergosphere~($r^{+}_{e}$), 
inner ergosphere~($r^{-}_{e}$), generalized spin frequency~($\Omega_{p}$), 
LT frequency~($\Omega_{LT}$) etc.  in contrast to zero MOG parameter.} 
 
We also investigated the generalized spin precession frequency for various angular 
limits: $\theta=0$, $\theta=\frac{\pi}{6}$, $\theta=\frac{\pi}{4}$, $\theta=\frac{\pi}{3}$ and 
finally $\theta=\frac{\pi}{2}$. Moreover, we  studied the generalized spin frequency for ring 
singularity. Then we studied the frame-dragging effect with vanishing angular velocity. 
This means that we derived the LT frequency only by ommiting the other {frequencies}. We showed 
that the LT frequency is affected by the MOG parameter. It {is} clearly  observed 
from LT frequency vs. radial diagram. For vanishing $\Omega=0$, we also studied the LT 
frequency for angular values: $\theta=0$, $\theta=\frac{\pi}{6}$, $\theta=\frac{\pi}{4}$, 
$\theta=\frac{\pi}{3}$ and $\theta=\frac{\pi}{2}$. Each diagram clearly showed the distinction 
between three compact objects namely, non-extremal BH, extremal BH and NS.

Moreover, we studied the LT frequency particularly for extremal KMOG BH in compared with extremal 
Kerr BH. In this case we also examined the LT frequency for various angular values. From each diagram 
of $\Omega_{LT}$ vs. $r$, we observed a diverging value of LT frequency at $r=G_{N}M$. It {has been} also 
observed that the presence of MOG parameter {significantly} changes the geometry of the BH spacetime 
in contrast to zero MOG parameter.

Finally, we studied the accretion disk properties by computing three {fundamental} epicyclic frequencies, 
namely the Keplerian frequency, radial epicyclic frequency and vertical epicyclic frequency. We also studied 
periastron frequency and nodal frequency. Using these properties {of epicyclic frequencies}, we differentiated 
{three compact objects}. From different frequency diagram, it {should} be clearly observed 
that three geometries are distinct. Furthermore, we calculated the ratio $\frac{\Omega_{r}}{\Omega_{\phi}}$, 
$\frac{\Omega_{r}}{\Omega_{\theta}}$ and $\frac{\Omega_{\theta}}{\Omega_{\phi}}$. Using these features, we 
differentiated three compact objects. 

In summary, to test the strong gravity in MOG and to distinguish three compact objects 
namely the non-extremal BH, extremal BH and NS we have computed different essential 
parameters i. e. $\Omega_{p}$, $\Omega_{LT}$, $\Omega_{r}$, $\Omega_{\theta}$, 
$\Omega_{\phi}$, $\Omega_{per}$, $\Omega_{nod}$, $\frac{\Omega_{r}}{\Omega_{\phi}}$, 
$\frac{\Omega_{r}}{\Omega_{\theta}}$ and $\frac{\Omega_{\theta}}{\Omega_{\phi}}$. 
By using the features of these parameters, one can distinguish 
between BH~(non-extremal \& extremal) and NS.

\section*{Acknowledgements}
I would like to thank Dr. Chandrachur Chakraborty of KIAA, China for helpful suggestions.
\\
\appendix
\begin{appendix}
\section{\label{app} Epicyclic frequencies in a general axisymmetric and stationary spacetime}
We consider a general stationary and axisymmetric spacetime as
\begin{eqnarray}
ds^2=g_{tt}\,dt^2+g_{rr}\,dr^2+g_{\theta\theta}\,d\theta^2+g_{\phi\phi}\,d\phi^2+2g_{t\phi}\,d\phi dt, \label{a1}
\end{eqnarray}
where $g_{\mu\nu}=g_{\mu\nu}(r, \theta)$. For this spacetime the proper angular momentum ~($l$) of 
a test particle can be defined as 
\begin{eqnarray}
 l=-\frac{g_{t\phi}+\Omega_{\phi} g_{\phi\phi}}{g_{tt}+\Omega_{\phi} g_{t\phi}},
\end{eqnarray}
where, $\Omega_{\phi}$ is the orbital frequency of a test particle. Now the $\Omega_{\phi}$
can be defined as
\begin{eqnarray}
\Omega_{\phi} =\frac{\dot{\phi}}{\dot{t}}=\frac{(\frac{d\phi}{d\tau})}{(\frac{dt}{d\tau})} 
=\frac{d\phi}{dt}
=\frac{-\partial_{r}g_{t\phi}\pm \sqrt{(\partial_{r}g_{t\phi})^2-(\partial_{r}g_{tt})(\partial_{r}g_{\phi\phi})}}
{\partial_{r}g_{\phi\phi}} 
 \label{a2}
\end{eqnarray}
The upper sign is for corotating orbit and the lower sign is for counterrotating orbit. 
The general expressions for computing the radial~($\Omega_{r}$) and vertical
($\Omega_{\theta}$) epicyclic frequencies are~\cite{doneva,jcapcpp}
\begin{eqnarray}
\Omega_{r}^2 &=& \frac{(g_{tt}+\Omega_{\phi}g_{t\phi})^2}{2~g_{rr}}~\partial_{r}^2~U\nonumber\\
 &=& \frac{(g_{tt}+\Omega_{\phi}g_{t\phi})^2}{2~g_{rr}}
\left[\partial_{r}^2\left(\frac{g_{\phi\phi}}{Y}\right) +2l~\partial_{r}^2\left(\frac{g_{t\phi}}{Y}\right)
+l^2~\partial_{r}^2\left(\frac{g_{tt}}{Y}\right)\right]|_{r=const.,~\theta=\frac{\pi}{2}}.~ \label{a3}
\end{eqnarray}
and
\begin{eqnarray}
\Omega_{\theta}^2 &=& \frac{(g_{tt}+\Omega_{\phi}g_{t\phi})^2}{2~g_{\theta\theta}}~\partial_{\theta}^2~U \nonumber\\
 &=& \frac{(g_{tt}+\Omega_{\phi}g_{t\phi})^2}{2~g_{\theta\theta}}\left[\partial_{\theta}^2\left(\frac{g_{\phi\phi}}{Y}\right)
 +2l~\partial_{\theta}^2\left(\frac{g_{t\phi}}{Y}\right)+l^2~\partial_{\theta}^2\left(\frac{g_{tt}}{Y}\right)\right]|_{r=const.,
~\theta=\frac{\pi}{2}}.~\label{a4}
\end{eqnarray}
respectively and $Y$ can be defined as
\begin{eqnarray}
 Y=g_{tt}g_{\phi\phi}-g_{t\phi}^2 .
\end{eqnarray}
\end{appendix}


\begin{thebibliography}{99}
\bibitem{de} W. de Sitter, 
``On the bearing of the Principle of Relativity on Gravitational Astronomy'', 
\textit{Mon. Not. Roy. Astron. Soc. }, {\bf 77} {155} (1916).

\bibitem{lt} J. Lense and H. Thirring, 
``On the effect of rotating distant masses in Einstein's theory of gravitation'', 
\textit{Physikalische Zeitschrift}, {\bf 19} {156-163} (1918).

\bibitem{schiff} L. I. Schiff, 
`` Possible New Experimental Test of General Relativity Theory'', 
\textit{Phys. Rev. Lett }, {\bf 4} {215} (1960).

\bibitem{prl11} C. W. F. Everitt et al., 
``Gravity probe B: final results of a space experiment to test general relativity'', 
\textit{Phys. Rev. Lett }, {\bf 106 } {221101}  (2011).

\bibitem{gpb} GP-B website: http://einstein.stanford.edu.

\bibitem{rossi} ``The Rossi X-ray Timing Explorer Mission~(1995-2012)'', 
https://heasarc.gsfc.nasa.gov/docs/xte.

\bibitem{eht} Event Horizon Telescope, https://eventhorizontelescope.org.

\bibitem{saha} R. Kannan and P. Saha, 
``Frame dragging and the kinematics of galactic-center stars'', 
\textit{The Astrophys. J.}, {\bf 690} {1553} (2009). 

\bibitem{defelice93} F. de Felice, S. U. Tommaset, 
``Schwarzschild spacetime: measurements in orbiting space-stations'', 
\textit{Class. Quant. Grav.}, {\bf 10} {353} (1993).

\bibitem{wheeler} I. Ciufolini \&  J. A. Wheeler, 
``Gravitation and Inertia'', 
\textit{Princeton Univ. Press, Princeton, New Jersey, (1995)}.

\bibitem{will} C. M. Will, 
``Theory and Experiment in Gravitational Physics., 2nd edn.'',  
\textit{Cambridge Univ. Press, Cambridge, UK, (1993)}.

\bibitem{ciufo} I. Ciufolini,   
``Dragging of inertial frames'',
\textit{Nature}, {\bf 449}, {41} (2007) and refrences therein.

\bibitem{rio} L. Iorio, 
``An Assessment of the Systematic Uncertainty in Present
and Future Tests of the Lense-Thirring Effect with Satellite Laser Ranging'', 
\textit{Space Sci. Rev.}  \textbf{148}: {363–381}, (2009).

\bibitem{jh} J. B. Hartle,  
\textit{Gravity: An introduction to Einstein's General relativity}, Pearson (2009).

\bibitem{herbert} H. Pfister, 
``On the history of the so-called Lense-Thirring effect'', 
\textit{Gen. Rel. Grav. }  \textbf{39: 1735-1748} (2007).

\bibitem{bini16}  D. Bini, A. Geralico, R. T. Jantzen,
``Gyroscope precession along bound equatorial plane orbits around a Kerr black hole''
\textit{ Phys. Rev. D} {\bf 94}, 064066 (2016).

\bibitem{chp} C. Chakraborty \& P. Majumdar, 
`` Strong gravity Lense-Thirring precession in Kerr and Kerr-Taub-NUT spacetimes'' , 
\textit{Class. Quantum Grav.}, {\bf 31} {075006} (2014).

\bibitem{chppp} C. Chakraborty \& P. Pradhan, 
``Lense-Thirring precession in Pleba\'{n}ski-Demia\'{n}ski Spacetimes'', 
\textit{Eur. Phys. J. C }, {\bf 73} {2536} (2013).

\bibitem{chppp1} C. Chakraborty \& P. Pradhan,  
`` Behavior of a test gyroscope moving towards a rotating traversable wormhole '', 
\textit{JCAP} {\bf 003 }, {035} (2017).

\bibitem{ckd} C. Chakraborty et al., 
`` Dragging of inertial frames inside the rotating neutron stars '',
\textit{The Astrophysical J.}, {\bf 790:2} (2014).

\bibitem{ccb} D. Chatterjee et al., 
`` Gravitomagnetic effect in magnetized neutron stars '', 
\textit{JCAP} {\bf 01}, 062 (2017).

\bibitem{ms99} S. M. Morsink \& L. Stella,  
`` Relativistic precession around rotating neutron stars: Effects due to frame-dragging and stellar oblateness '', 
\textit{The Astrophysical J.}, {\bf 513} {827} (1999).

\bibitem{ch17} C. Chakraborty et al., 
`` Distinguishing Kerr naked singularities and black holes using the spin precession 
of a test gyro in strong gravitational fields'',
\textit{Phys. Rev.} {\bf D 95}, {084024} (2017).

\bibitem{ch18} C. Chakraborty et al.,  
``Inertial Frame Dragging in an Acoustic Analogue Spacetime'', 
\textit{Ann. Phys. (Berlin)}, {1700231}, (2017).

\bibitem{mf} J. W. Moffat, 
``Black Holes in Modified Gravity'', 
\textit{Eur. Phys. J. C} {75} {175} (2015).

\bibitem{mfjcap} J. W. Moffat, 
``Scalar-Tensor-Vector Gravity Theory'', 
\textit{JCAP} {\bf 0603}, {004} (2006).

\bibitem{mf1} J. W. Moffat, 
``Modified Gravity Black Holes and their Observable Shadows'',
\textit{Eur. Phys. J. C} {75} {130} (2015).

\bibitem{mf3} J. R. Mureika et al., 
``Black Hole Thermodynamics in Modified Gravity'',
\textit{ Phys. Lett. B} {\bf 757}, {528} (2016).

\bibitem{mf4} M. F. Wondrak et al., 
``Superradiance in Modified Gravity (MOG)'', 
\textit{JCAP} {\bf 021}, (2018).

\bibitem{mf5} J. W. Moffat \& V. T. Toth, 
``The bending of light and lensing in modified gravity'', 
\textit{MNRAS} {\bf 397}, {1885} (2009).

\bibitem{mf6} L. Manfredi et al., 
``Quasinormal modes of modified gravity (MOG) black holes'', 
\textit{ Phys. Lett. B} {\bf 779}, {492} (2018).

\bibitem{pp18} P. Pradhan, 
``Area (or Entropy) Products in Modified Gravity and Kerr-MOG/CFT Correspondence'', 
\textit{The Euro. Phys. J. Plus}, {\bf 133}, 187 (2018).

\bibitem{liu18} S. W. Wei \& Y. X. Liu, 
``Merger estimates for rotating Kerr black holes in modified gravity'', 
\textit{Phys. Rev. D}{\bf 98}, 024042 (2018).

\bibitem{ps} P. Sheoren et al., 
``Mass and spin of a Kerr black hole in modified gravity and a test of the 
Kerr black hole hypothesis '', 
\textit{Phys. Rev. D}, {\bf 97} {124049} (2018).

\bibitem{epjc19} P. Pradhan, 
``Study of energy extraction and epicyclic frequencies in Kerr-MOG~(Modified Gravity) black hole'', 
\textit{Eur. Phys. J. C }, {\bf 79} {401} (2019).

\bibitem{ch17a} C. Chakraborty et al., 
``Spin precession in a black hole and naked singularity spacetime'',
\textit{Phys. Rev.} {\bf D 95}, {044006} (2017).

\bibitem{pugliese13} D. Pugliese et al., 
``Equatorial circular orbits of neutral test particles in the Kerr-Newman spacetime'',
\textit{Phys. Rev.} {\bf D 88}, {024042} (2013).

\bibitem{defelice78} F. de Felice, 
``Classical instability of a naked singularity'', 
\textit{Nature} {\bf 273}, {8} (1978).

\bibitem{calvini78}   M. Calvini et al., 
``Are Naked Singularities Really Visible?'',
\textit{Lett. Nuovo Cimento}, {\bf 23} {15} (1978).

\bibitem{zhang16} H. Zhang, 
``Naked Singularity, firewall, and Hawking radiation'', 
\textit{Scientific Reports}, {\bf 7} {4000} (2017).

\bibitem{pug19} D. Pugliese, H. Quevedo,
``Disclosing connections between black holes and naked
singularities: horizon remnants, Killing throats and bottlenecks'', 
\textit{Eur. Phys. J. C }, {\bf 79} {209} (2019).


\bibitem{pug20} D. Pugliese, H. Quevedo,
``Kerr metric bundles. Killing horizons confinement, light-surfaces and horizons replicas'', 
arXiv: 2005.04130.


\bibitem{ns} N. Straumann, 
\textit{ General Relativity with applications to Astrophysics}, Springer (2009).

\bibitem{vafa} A. Strominger and C. Vafa, `` Microscopic origin of the Bekenstein-Hawking entropy''
\textit{Phys. Lett. B.} {\bf 379} {99} (1996).


%\bibitem{doneva} D. D. Doneva, S. S. Yazadjiev, N. Stergioulas, K. D. Kokkotas \&
%T. M. Athanasiadis, {\it Phys. Rev.} {\bf D 90}, {044004} (2014).

\bibitem{jantzen} R. T. Jantzen, P. Carini \& D. Bini, 
``The Many Faces of Gravitoelectromagnetism, 
\textit{ Annals Phys.} {\bf 215}, 1 (1992).

\bibitem{bd}  J. M. Bardeen, 
``A Variational Principle for Rotating Stars in General Relativity'', 
\emph{Astrophys. J. } {\bf 162}, 71 (1970).

\bibitem{bpt} J. M. Bardeen, W. H. Press \& S. A. Teukolsky, 
``Rotating Black Holes: Locally Non-rotating Frames, Energy Extraction, and Scalar Synchrotron Radiation'', 
\emph{Astrophys. J.} {\bf 178}, 347 (1972).


\bibitem{maselli17} A. Maselli et al., 
``Geodesic models of quasi-periodic-oscillations as probes of quadratic gravity'', 
\textit{Astrophysical Journal}, {\bf 1}, 843 (2017).


\bibitem{stella99} L. Stella et al., ``Correlations in the QPOs frequencies of low mass 
x-ray binaries and the relativistic precession model'',  
\textit{Astrophysical Journal}, {\bf L63-L66}, 524 (1999).


\bibitem{torok05} G. T\"or\"ok \& Z. Stuchlik,
``Radial and vertical epicyclic frequencies of Keplerian motion in the field of Kerr naked singularities'', 
\textit{Astronomy and Astrophysics}{\bf 437}, 775 (2005).



\bibitem{chiba} K. Sakina \& J. Chiba, 
``Parallel transport of a vector along a circular orbit in Schwarzschild space-time'',
\textit{ Phys. Rev. D} {\bf 19}, 2280 (1979).

\bibitem{zas17} I. V. Tanatarov, O. B. Zaslavskii, 
``Collisional super-Penrose Process and Wald inequalities'',
\textit{Gen. Relat. Gravit.} {\bf 49}, {119} (2017).


\bibitem{mtw} C. W. Misner, K. S. Thorne \& J. A. Wheeler, 
\textit{Gravitation}, \emph{W. H.  Freeman \& Company} (1973).

\bibitem{doneva} D. D. Doneva et al., ``Orbital and epicyclic frequencies around rapidly rotating 
compact stars in scalar-tensor theories of gravity'',
\textit{ Phys. Rev.} {\bf D 90}, {044004} (2014).

\bibitem{jcapcpp} C. Chakraborty \& P. Pradhan, `` Behavior of a test gyroscope moving towards 
a rotating traversable wormhole'', \textit{JCAP} {\bf 03}, {035} (2017).

\end{thebibliography}
\end{document}